\documentclass[
pof,
preprint,
superscriptaddress,
longbibliography,
]{revtex4-1}

\usepackage[utf8]{inputenc} 				% unicode formatted text
\usepackage[main=english, ngerman]{babel}	% correct hyphenation

\usepackage[T1]{fontenc} 				% font encoding for umlauts
\usepackage{textcomp} 					% various symbols (e.g. ° and €)
\usepackage{amsmath}					% extended mathematics environment
\usepackage{amssymb} 					% extended mathematics symbol collection
\usepackage{mathtools}					% extended mathematics symbol collection
\usepackage{cases}                      % case in math environment (curly braces spanning multiple lines)
\usepackage[hidelinks]{hyperref}
\usepackage[noabbrev]{cleveref}			% automatically print the type of reference
\usepackage[final]{microtype}			% improved justification
\usepackage{csquotes}					% quotes using \enquote{}
\usepackage{graphicx}                   % including and adjusting graphics of various formats
\usepackage{overpic}					% overlapping figures
\usepackage{grffile}					% allows dots (prior to file-ending) in graphics files' names
\usepackage{subcaption}					% includes subfigure-environment
\usepackage{multirow}                   % tables with multiple rows
\usepackage{array}						% vertical centered tabular cells
\usepackage{makecell}					% line breaks within tabular cells
\usepackage{booktabs}					% thicker lines in tables
\usepackage{caption}
\usepackage[section]{placeins}			% force floats to be print at \FloatBarrier before continuing
\usepackage[export]{adjustbox}			% vertical alignement of figures via \includegraphics[valign=c]
\usepackage{fp} 						% floating point computations

\makeatletter
\newcommand*{\getlength}[1]{\strip@pt#1}
\makeatother

\usepackage{pgfplots}
\pgfplotsset{compat=newest}
\usepgfplotslibrary{groupplots,dateplot,fillbetween}
\usetikzlibrary{arrows,arrows.meta,patterns,shapes.arrows,calc,angles,quotes,babel}
\usetikzlibrary{calc,angles,quotes}
\newlength\figureheight
\newlength\figurewidth
\newlength\mdist 	% distance of measure line and object
\newlength\radius	% radius of the Taylor bubble in the setup figure

\newcommand{\cpp}{C\nolinebreak\hspace{-.05em}\raisebox{.4ex}{\tiny\textbf{+}}\nolinebreak\hspace{-.02em}\raisebox{.4ex}{\tiny\textbf{+}}} 	% nice way to typeset C++ via \cpp

\newcommand{\walberla}{\textsc{waLBerla}}

\makeatletter
\def\@email#1#2{%
 \endgroup
 \patchcmd{\titleblock@produce}
  {\frontmatter@RRAPformat}
  {\frontmatter@RRAPformat{\produce@RRAP{*#1\href{mailto:#2}{#2}}}\frontmatter@RRAPformat}
  {}{}
}%
\makeatother
\begin{document}

\preprint{AIP/123-QED}

\title{Comparison of refilling schemes in the free-surface lattice Boltzmann method}
\author{Christoph Schwarzmeier}
\affiliation{Chair for System Simulation, Friedrich-Alexander-Universität Erlangen-Nürnberg, Cauerstraße 11, 91058 Erlangen, Germany}
\email{christoph.schwarzmeier@fau.de}

\author{Ulrich Rüde}%
\affiliation{Chair for System Simulation, Friedrich-Alexander-Universität Erlangen-Nürnberg, Cauerstraße 11, 91058 Erlangen, Germany}
\affiliation{CERFACS, 42 Avenue Gaspard Coriolis, 31057 Toulouse Cedex 1, France}

\date{November 17, 2022}

\begin{abstract}
	Simulating mobile liquid--gas interfaces with the free-surface lattice Boltzmann method (FSLBM) requires frequent re-initialization of fluid flow information in computational cells that convert from gas to liquid.
	The corresponding algorithm, here referred to as the refilling scheme, is crucial for the successful application of the FSLBM in terms of accuracy and numerical stability.
	This study compares five refilling schemes that extract information from surrounding liquid and interface cells by either averaging, extrapolating, or assuming one of three different equilibrium states.
	Six numerical experiments were performed covering a broad spectrum of possible scenarios. 
	These include a standing gravity wave, a rectangular and cylindrical dam break, a Taylor bubble, a drop impact into liquid, and a bubbly plane Poiseuille flow.
	In some simulations, the averaging, extrapolation, and one equilibrium-based scheme were numerically unstable.
	Overall, the results have shown that the simplest equilibrium-based scheme should be preferred in terms of numerical stability, computational costs, accuracy, and ease of implementation.
\end{abstract}

\maketitle

%!TEX root = ../main.tex

\section{Introduction}\label{sec:int}
The free-surface lattice Boltzmann method (FSLBM)~\cite{korner2005LatticeBoltzmannModel} is a numerical model for simulating free-surface flows combining the latticed Boltzmann method (LBM) for hydrodynamics simulations with the volume of fluid (VOF) approach~\cite{hirt1981VolumeFluidVOF} for interface tracking.
It successfully simulates applications such as rising bubbles~\cite{donath2011VerificationSurfaceTension}, waves~\cite{zhao2013LatticeBoltzmannMethod}, dam break scenarios~\cite{janssen2011FreeSurfaceFlow}, impacts of droplets~\cite{lehmann2021EjectionMarineMicroplastics}, and electron-beam melting~\cite{ammer2014SimulatingFastElectron}.
Free-surface flows relate to immiscible two-fluid flow problems in which the fluid dynamics of the lighter fluid can be neglected.
Therefore, the problem reduces to a single-fluid flow with a free boundary.
In this article, the lighter fluid will be called gas, and the heavier fluid will be called liquid.
The Eulerian computational grid is represented by lattice cells in the LBM. 
In the FSLBM, each lattice cell is categorized as either gas, liquid, or interface type, with the latter separating the former.
In the LBM, information about the flow field is stored in each cell in terms of particle distribution functions (PDFs).
Agreeing with the free-surface definition, in the FSLBM, valid PDFs are only available in liquid and interface cells but not in gas cells.
Gas cells are converted to interface cells during the simulation because of the free interface's motion.
These cells must be refilled with valid flow field information, that is, their PDFs must be reinitialized.
To the authors' knowledge, no other refilling scheme but the one suggested in the original FSLBM from Körner et al.~\cite{korner2005LatticeBoltzmannModel} has yet been tested for the FSLBM.
However, there have been similar studies about moving solid obstacle cells in the LBM. There, analogously, cells are converted from solid to liquid and must be refilled~\cite{peng2016ImplementationIssuesBenchmarking,lallemand2003LatticeBoltzmannMethod,chikatamarla2006GradApproximationMissing,krithivasan2014DiffusedBouncebackCondition,dorschner2015GradApproximationMoving,tao2016InvestigationMomentumExchange,fang2002LatticeBoltzmannMethod}.
Based on the schemes used for this application, five different schemes for refilling cells in the FSLBM are compared in this article.
\par

The manuscript is structured as follows.
First, the numerical foundations of the LBM and FSLBM are introduced.
Then, the different refilling schemes are presented and discussed in terms of mass conservation and computational costs.
The first scheme under investigation initializes the PDFs with their equilibrium constructed with the average fluid velocity and density of non-newly converted neighboring interface and liquid cells~\cite{korner2005LatticeBoltzmannModel}.
The second and third scheme extend the first one by adding a contribution of the non-equilibrium PDFs~\cite{peng2016ImplementationIssuesBenchmarking}, or by including information about the local pressure tensor using Grad's moment system~\cite{grad1949KineticTheoryRarefied,dorschner2015GradApproximationMoving,krithivasan2014DiffusedBouncebackCondition,chikatamarla2006GradApproximationMissing}, respectively.
The fourth refilling scheme initializes PDFs with a second-order extrapolation from neighboring cells' PDFs~\cite{lallemand2003LatticeBoltzmannMethod}.
In contrast to these, in the final scheme tested here, the PDFs are initialized with the average corresponding PDFs from neighboring, non-newly converted interface or liquid cells~\cite{fang2002LatticeBoltzmannMethod}.
Six numerical benchmarks then compare the refilling schemes in terms of accuracy and numerical stability.
These benchmarks include a standing gravity wave, the collapse of a rectangular and cylindrical liquid column, the rise of a Taylor bubble, the impact of a droplet into a thin film of liquid, and a bubbly plane Poiseuille flow.
Finally, it is concluded that for the FSLBM, the simplest equilibrium-based refilling scheme is preferable in terms of numerical stability, computational costs, accuracy, and ease of implementation.
\par

The source code of the implementation used in this study is freely available as part of the open source \cpp software framework \walberla{}~\cite{bauer2021WaLBerlaBlockstructuredHighperformance} (\url{https://www.walberla.net}).
The version of the source code used in this article is provided in the supplementary material.
\par
%!TEX root = ../main.tex

\section{Numerical methods}\label{sec:nm}
This section introduces the foundations of the lattice Boltzmann method and its extension to free-surface flows, the free-surface lattice Boltzmann method.
The section is based on Section 2 from prior articles~\cite{schwarzmeier2022ComparisonFreeSurface,schwarzmeier2022AnalysisComparisonBoundary} but repeated here for completeness.
\par

\subsection{Lattice Boltzmann method}\label{sec:nm-lbm}
The lattice Boltzmann method is a relatively modern approach for simulating computational fluid dynamics.
This article only introduces its fundamental aspects.
A rigorous introduction to the LBM is available in the literature~\cite{kruger2017LatticeBoltzmannMethod}.
\par

The LBM is a discretization of the Boltzmann equation from kinetic gas theory.
It describes the evolution of particle distribution functions on a uniformly discretized Cartesian lattice with spacing $\Delta x \in \mathbb{R^{+}}$.
The macroscopic fluid velocity is discretized with the D$d$Q$q$ velocity set in each cell of the lattice, with $d \in \mathbb{N}$ referring to the lattice's spatial dimension and $q \in \mathbb{N}$ referring to the number of PDFs per cell.
A PDF $f_{i}(\boldsymbol{x}, t) \in \mathbb{R}$ with $i \in \{0, 1, \dots, q-1\}$ describes the probability that there exists a virtual fluid particle population at position $\boldsymbol{x} \in \mathbb{R}^{d}$ and time $t \in \mathbb{R^{+}}$ traveling with lattice velocity $\boldsymbol{c}_{i} \in \Delta x / \Delta t \, \{-1,0, 1\}^{d}$, where $\Delta t \in \mathbb{R^{+}}$ denotes the length of a discrete time step.
The successive steps of collision, also called relaxation,
\begin{equation}\label{eq:nm-lbm-collision}
f_{i}^{\star}(\boldsymbol{x}, t) = f_{i}(\boldsymbol{x}, t) + \Omega_{i}(\boldsymbol{x}, t) + F_{i}(\boldsymbol{x}, t)
\end{equation}
and streaming, also called propagation,
\begin{equation}\label{eq:nm-lbm-streaming}
f_{i}(\boldsymbol{x} + \boldsymbol{c}_{i}\Delta t, t+\Delta t) = f_{i}^{\star}(\boldsymbol{x}, t)
\end{equation}
form the lattice Boltzmann equation.
In the collision step, the collision operator $\Omega_{i}(\boldsymbol{x}, t) \in \mathbb{R}$ relaxes the PDFs towards an equilibrium state $f_{i}^{\text{eq}}(\boldsymbol{x}, t)$, which is influenced by external forces $F_{i}(\boldsymbol{x}, t) \in \mathbb{R}$.
In the streaming step, the post-collision PDFs $f_{i}^{\star}(\boldsymbol{x}, t)$ propagate to neighboring cells.
For the simulations in this article, the single relaxation time (SRT) collision operator
\begin{equation}\label{eq:nm-lbm-collision-operator}
\Omega_{i}(\boldsymbol{x}, t) = \frac{f_{i}(\boldsymbol{x}, t) - f_{i}^{\text{eq}}(\boldsymbol{x}, t)}{\tau} \Delta t
\end{equation}
was used, where $\tau > \Delta t / 2$ is the relaxation time.
The PDF's equilibrium~\cite{bauer2020TruncationErrorsD3Q19a}
\begin{equation}\label{eq:nm-lbm-equilibrium}
f_{i}^{\text{eq}}(\boldsymbol{x}, t) \equiv f_{i}^{\text{eq}}(\boldsymbol{u}, \rho) = w_{i}\rho\left(1 + \frac{\boldsymbol{u}\cdot\boldsymbol{c}_{i}}{c_{s}^{2}} + \frac{(\boldsymbol{u} \cdot \boldsymbol{c}_{i})^{2}}{2 c_{s}^{4}} - \frac{\boldsymbol{u} \cdot \boldsymbol{u}}{2 c_{s}^{2}} \right)
\end{equation}
can be derived from the Maxwell--Boltzmann distribution and includes the lattice weights $w_{i} \in \mathbb{R}$, the lattice speed of sound $c_{s}^{2} \in \mathbb{R^{+}}$, the macroscopic fluid density $\rho \equiv \rho(\boldsymbol{x}, t) \in \mathbb{R^{+}}$, and the macroscopic fluid velocity $\boldsymbol{u} \equiv \boldsymbol{u}(\boldsymbol{x}, t) \in \mathbb{R}^{d}$.
In this article, the well-established D$2$Q$9$ and D$3$Q$19$ lattice models are used.
The corresponding lattice weights are available in the literature~\cite{kruger2017LatticeBoltzmannMethod}.
The lattice speed of sound for these velocity sets is $c_{s}^{2}=\sqrt{1/3} \, \Delta x / \Delta t$.
It relates the macroscopic fluid density $\rho(\boldsymbol{x}, t)$ and pressure $p(\boldsymbol{x}, t)=c_{s}^{2}\rho(\boldsymbol{x}, t)$.
The PDFs' zeroth- and first-order moments yield the fluid's density
\begin{equation}\label{eq:nm-lbm-density}
\rho(\boldsymbol{x}, t) = \sum_{i} f_{i}(\boldsymbol{x}, t)
\end{equation}
and velocity
\begin{equation}\label{eq:nm-lbm-velocity}
\boldsymbol{u}(\boldsymbol{x}, t) = \frac{\boldsymbol{F}(\boldsymbol{x}, t)\Delta t}{2 \rho(\boldsymbol{x}, t)} + \frac{1}{\rho(\boldsymbol{x}, t)}\sum_{i} \boldsymbol{c}_{i} f_{i}(\boldsymbol{x}, t)
\end{equation}
with external force $\boldsymbol{F}(\boldsymbol{x}, t) \in \mathbb{R}^{d}$.
The fluid's kinematic viscosity
\begin{equation}\label{eq:nm-lbm-viscosity}
\nu = c_{s}^{2} \left(\tau - \frac{\Delta t}{2}\right)
\end{equation}
is computed from the relaxation time $\tau$, that is, relaxation rate $\omega=1/\tau$.
In the simulations used in this article, the gravitational force, as part of $F_{i}(\boldsymbol{x}, t)$ in the LBM collision~\eqref{eq:nm-lbm-collision}, was modeled according to Guo et al.~\cite{guo2002DiscreteLatticeEffects} with
\begin{equation}\label{eq:nm-lbm-force-guo}
F_{i}(\boldsymbol{x}, t) =
\left(1 - \frac{\Delta t}{2 \tau} \right)
w_{i}\left(\frac{\boldsymbol{c}_{i} - \boldsymbol{u}}{c_{s}^{2}} + \frac{(\boldsymbol{c}_{i} \cdot \boldsymbol{u}) \boldsymbol{c}_{i}}{c_{s}^4}\right) \cdot \boldsymbol{F}(\boldsymbol{x}, t),
\end{equation}
where again, $\boldsymbol{u} \equiv \boldsymbol{u}(\boldsymbol{x}, t)$ was used.
\par

The rectangular and cylindrical dam break simulations in \Cref{sec:ne-rdb,sec:ne-cdb} were performed using a Smagorinksy-type large eddy simulation turbulence model~\cite{hou1996LatticeBoltzmannSubgrid,yu2005DNSDecayingIsotropic}.
Based on the user-chosen relaxation time $\tau_{0} > \Delta t / 2$, the collision operator's relaxation time $\tau(\boldsymbol{x}, t) = \tau_{0} + \tau_{t}(\boldsymbol{x}, t)$ is locally adjusted by the model with a contribution $\tau_{t}(\boldsymbol{x}, t) \in \mathbb{R}$ from the turbulence viscosity
\begin{equation}\label{eq:nm-lbm-turbulence-viscosity}
\nu_{t}(\boldsymbol{x}, t) \coloneqq \tau_t(\boldsymbol{x}, t) c_{s}^{2} = \left(C_{S}\Delta x_{\text{LES}}\right)^{2} \bar{S}(\boldsymbol{x}, t).
\end{equation}
The turbulence viscosity is obtained from the filtered strain rate tensor
\begin{equation}\label{eq:nm-lbm-strain-rate}
\bar{S}(\boldsymbol{x}, t) = \frac{\bar{Q}(\boldsymbol{x}, t)}{2\rho c_{s}^2 \tau_{0}},
\end{equation}
where the filtered mean momentum flux
\begin{equation}\label{eq:nm-lbm-mean-momentum-flux}
\bar{Q}(\boldsymbol{x}, t) = \sqrt{2\sum_{\alpha, \beta}\bar{Q}_{\alpha, \beta}(\boldsymbol{x}, t)\bar{Q}_{\alpha, \beta}(\boldsymbol{x}, t)}
\end{equation}
is computed from the momentum fluxes
\begin{equation}\label{eq:nm-lbm-momentum-fluxes}
\bar{Q}_{\alpha, \beta}(\boldsymbol{x}, t) = \sum_{i} c_{i, \alpha} c_{i, \beta}\Bigl( f_{i}(\boldsymbol{x}, t) - f_{i}^{\text{eq}}(\boldsymbol{x}, t)\Bigr).
\end{equation}
The index notation with $\alpha$ and $\beta$ refers to the components of a vector or tensor.
The moment fluxes are given by the second-order moments of the PDFs' non-equilibrium parts.
The turbulence model's contribution to the relaxation time is then~\cite{hou1996LatticeBoltzmannSubgrid}
\begin{equation}\label{eq:nm-lbm-turublence-relaxation-time}
\tau_{t}(\boldsymbol{x}, t) = \frac{1}{2}\sqrt{\tau_{0}^{2} + 2\sqrt{2}(C_{S}\Delta x_{\text{LES}})^{2}(\rho c_{s}^4)^{-1}\bar{Q}(\boldsymbol{x}, t)}-\tau_{0},
\end{equation}
where $\Delta x_{\text{LES}}$ is the filter length and $C_{S}$ is the Smagorinsky constant.
For the simulations in this article, these parameters were chosen
$\Delta x_{\text{LES}}=\Delta x$ and $C_{S}=0.1$, as suggested by Yu et al.~\cite{yu2005DNSDecayingIsotropic}.
\par

At solid obstacles, a no-slip boundary condition were realized using the bounce-back approach, where PDFs streaming into solid obstacle cells are reflected reversely.
The PDF's original direction with index $i$ is reversed, denoted as $\bar{i}$, with lattice velocity $\boldsymbol{c}_{\bar{i}}=-\boldsymbol{c}_{i}$~\cite{kruger2017LatticeBoltzmannMethod}.
Free-slip boundary conditions are modeled similarly with the PDFs being reflected specularly.
In the resulting lattice velocity $\boldsymbol{c}_{j}$, the normal velocity component of the incoming velocity $\boldsymbol{c}_{i}$ is reversed with $c_{j,n} = - c_{i,n}$ at a free-slip boundary~\cite{kruger2017LatticeBoltzmannMethod}.
\par

In the remainder of this article, $\Delta x=1$ and $\Delta t=1$ are assumed as this is common practice in the LBM~\cite{kruger2017LatticeBoltzmannMethod}.
All quantities are denoted in the LBM unit system if not explicitly stated otherwise.
The LBM reference density $\rho_{0}=1$ and pressure $p_{0} = c_{s}^{2} \rho_{0} = 1/3$ were set in all simulations.
The relaxation time $\tau$ or relaxation rate $\omega$ specified for the numerical experiments in \Cref{sec:ne} refer to the constant user-chosen values that the Smagorinsky turbulence model did not yet adjust.
\par

\subsection{Free-surface lattice Boltzmann method}\label{sec:nm-fslbm}
The free-surface lattice Boltzmann method as presented by Körner et al.~\cite{korner2005LatticeBoltzmannModel} is used in this article.
The FSLBM extends the LBM by simulating the interface between two immiscible fluids.
It assumes that the heavier fluid governs the entire flow dynamics of the system with the lighter fluid's influence being negligible.
Consequently, the immiscible two-fluid flow problem reduces to a single-fluid flow with a free boundary.
Therefore, the hydrodynamics of the lighter fluid are not simulated in the FSLBM.
A simplification such as this is valid if the fluids' densities and viscosities differ substantially, for example as in liquid--gas flows.
In what follows, the heavier fluid is called liquid, whereas the lighter fluid is called gas.
\par

The free interface between the liquid and gas is treated as in the VOF approach~\cite{hirt1981VolumeFluidVOF}.
A fill level $\varphi (\boldsymbol{x}, t)$ is assigned to each lattice cell, acting as an indicator that describes the affiliation to one of the phases.
Cells can be of liquid ($\varphi(\boldsymbol{x}, t)=1$), gas ($\varphi(\boldsymbol{x}, t)=0$), or interface type ($\varphi(\boldsymbol{x}, t) \in \left(0, 1\right)$).
A sharp and closed layer of interface cells separates liquid and gas cells.
Interface and liquid cells are treated like regular LBM cells, which contain PDFs and participate in the LBM  collision~\eqref{eq:nm-lbm-collision} and streaming~\eqref{eq:nm-lbm-streaming}.
In contrast, conforming with the free-surface definition, gas cells neither contain PDFs nor participate in the LBM update.
\par

The liquid mass of each cell
\begin{equation}\label{eq:nm-fslbm-mass}
m\left(\boldsymbol{x}, t\right) = \varphi\left(\boldsymbol{x}, t\right) \rho\left(\boldsymbol{x}, t\right) \Delta x^{3}
\end{equation}
is determined by the cell's fill level $\varphi(\boldsymbol{x}, t)$, fluid density $\rho(\boldsymbol{x}, t)$, and volume $\Delta x^{3}$.
Note that in two-dimensional simulations, the cell's volume is also given by $\Delta x^{3}$.
The domain is then assumed to have an extension of a single lattice cell in the third direction.
The mass flux between an interface cell and cells of other types is computed from the LBM streaming step via
\begin{equation}\label{eq:nm-fslbm-mass-flux}
\frac{\Delta m_{i}\left(\boldsymbol{x}, t\right)}{\Delta x^{3}} = 
\begin{cases}
0 & \boldsymbol{x} + \boldsymbol{c}_{i}\Delta t \in \text{gas} \\

f_{\overline{i}}^{\star}\left(\boldsymbol{x} + \boldsymbol{c}_{i}\Delta t, t\right) - f_{i}^{\star}\left(\boldsymbol{x}, t\right) & \boldsymbol{x} + \boldsymbol{c}_{i}\Delta t \in \text{liquid}\\

\frac{1}{2}\Bigl(\varphi\left(\boldsymbol{x}, t\right) + \varphi\left(\boldsymbol{x} + \boldsymbol{c}_{i}\Delta t, t\right) \Bigr)
\Bigl(f_{\overline{i}}^{\star}\left(\boldsymbol{x} + \boldsymbol{c}_{i}\Delta t, t\right) - 
f_{i}^{\star}\left(\boldsymbol{x}, t\right)\Bigr) & \boldsymbol{x} + \boldsymbol{c}_{i}\Delta t \in \text{interface}.
\end{cases}
\end{equation}
The simplicity of this mass flux computation is an advantage of the FSLBM when compared to non-LBM-based VOF approaches.
In these methods, the advection of mass commonly requires solving a partial differential equation that describes the evolution of the mass.
\par

In the implementation used here, interface cells are not immediately converted to gas or liquid when their fill level becomes $\varphi(\boldsymbol{x}, t)=0$ or $\varphi(\boldsymbol{x}, t)=1$, respectively.
Instead, they are converted with respect to the heuristically chosen threshold $\varepsilon_{\varphi}=10^{-2}$ that prevents oscillatory conversions~\cite{pohl2008HighPerformanceSimulation}.
Consequently, an interface cell is converted to gas or liquid if its fill level becomes below zero with $\varphi(\boldsymbol{x}, t)< 0-\varepsilon_{\varphi}$ or above one with $\varphi(\boldsymbol{x}, t)>1+\varepsilon_{\varphi}$, respectively.

When an interface cell converts to gas or liquid, surrounding gas or liquid cells may convert to interface cells to maintain a closed interface layer.
It is important to point out that neither liquid nor gas cells can directly convert into one another.
Instead, both cell types can only convert to interface cells.
The separation of liquid and gas is prioritized in case of conflicting conversions.
When converting an interface cell with fill level $\varphi^{\text{conv}}(\boldsymbol{x}, t)$ to gas or liquid, the fill level is forcefully set to $\varphi(\boldsymbol{x}, t)=0$ or $\varphi(\boldsymbol{x}, t)=1$.
The resulting excess mass
\begin{equation}\label{eq:nm-fslbm-excess-mass}
\frac{m_{\text{ex}}\left(\boldsymbol{x}, t\right)}{\rho\left(\boldsymbol{x}, t\right)\Delta x^{3}} = 
\begin{cases}
\varphi^{\text{conv}}\left(\boldsymbol{x}, t\right) - 1 & \text{if } \boldsymbol{x} \text{ is converted to liquid} \\
\varphi^{\text{conv}}\left(\boldsymbol{x}, t\right) & \text{if } \boldsymbol{x} \text{ is converted to gas}
\end{cases}
\end{equation}
is distributed evenly among all surrounding interface cells to ensure mass conservation.
\par

During a simulation, unnecessary interface cells may appear, which do not have neighboring gas or liquid cells.
In the implementation used in this study, these cells are forced to fill or empty by adjusting the mass flux~\eqref{eq:nm-fslbm-mass-flux}, as suggested by Thürey~\cite{thurey2007PhysicallyBasedAnimation}.
\par

The cells' PDFs are not modified when cells are converted from interface to liquid or vice versa.
If a cell converts from interface to gas type, the cell's PDFs need not be considered further and can therefore be invalidated.
Note that this does not affect mass conservation as any excess mass~\eqref{eq:nm-fslbm-excess-mass} will be distributed accordingly.
However, no valid PDF information is available when cells convert from gas to interface type.
The PDFs of these cells must be initialized with one of the schemes presented in \Cref{sec:rs}.
\par

The LBM collision \eqref{eq:nm-lbm-collision} and streaming \eqref{eq:nm-lbm-streaming} are performed in all interface and liquid cells.
Körner et al.~\cite{korner2005LatticeBoltzmannModel} proposed to weight the gravitational acceleration with an interface cell's fill level in the LBM collision.
Conforming with the work of other authors~\cite{bogner2017DirectNumericalSimulation,pohl2008HighPerformanceSimulation,donath2011WettingModelsParallel}, the implementation used here did not weight the gravitational force with the fill level.
\par

The macroscopic boundary condition at the free surface is given by
~\cite{bogner2017DirectNumericalSimulation,scardovelli1999DirectNumericalSimulation}
\begin{equation}\label{eq:nm-fslbm-boundary-condition-macroscopic}
\begin{aligned}
p\left(\boldsymbol{x}, t\right) - p^{\text{G}}\left(\boldsymbol{x}, t\right) + p^{\text{L}}\left(\boldsymbol{x}, t\right) &=  2\mu\partial_{n}u_{n}\left(\boldsymbol{x}, t\right)\\
0 &= \partial_{t_{1}}u_{n}\left(\boldsymbol{x}, t\right) + \partial_{n}u_{t_{1}}\left(\boldsymbol{x}, t\right)\\
0 &= \partial_{t_{2}}u_{n}\left(\boldsymbol{x}, t\right) + \partial_{n}u_{t_{2}}\left(\boldsymbol{x}, t\right)
\end{aligned}
\end{equation}
where $p^{\text{G}}\left(\boldsymbol{x}, t\right)$ is the gas pressure, $p^{\text{L}}\left(\boldsymbol{x}, t\right)$ is the Laplace pressure, $\boldsymbol{t}_{1}(\boldsymbol{x}, t)$ and $\boldsymbol{t}_{2}(\boldsymbol{x}, t)$ are interface-tangent vectors, and $\boldsymbol{n}(\boldsymbol{x}, t)$ is the interface-normal vector.
As shown by Bogner et al.~\cite{bogner2015BoundaryConditionsFree}, this macroscopic boundary condition is approximated by the LBM anti-bounce-back pressure boundary condition
\begin{equation}\label{eq:nm-fslbm-boundary-condition}
f_{i}^{\star}\left(\boldsymbol{x} - \boldsymbol{c}_{i}\Delta t, t\right)
= f_{i}^\text{eq}\left(\rho^{\text{G}}, \boldsymbol{u}\right)
+ f_{\overline{i}}^\text{eq}\left(\rho^{\text{G}}, \boldsymbol{u}\right)
- f_{\overline{i}}^{\star}\left(\boldsymbol{x}, t\right),
\end{equation} 
which Körner et al.~\cite{korner2005LatticeBoltzmannModel} suggested to use.
In this equation, $\boldsymbol{u} \equiv \boldsymbol{u}\left(\boldsymbol{x}, t\right)$ is the interface cell's velocity and $\rho^{\text{G}} \equiv \rho^{\text{G}}\left(\boldsymbol{x}, t\right)=p^{\text{G}}\left(\boldsymbol{x}, t\right)/c_{s}^{2}$ is the gas density.
Other formulations of the boundary condition have been investigated in the literature~\cite{thies2005LatticeBoltzmannModeling,bogner2015BoundaryConditionsFree}.
The free-surface boundary condition~\eqref{eq:nm-fslbm-boundary-condition} is applied to all PDFs streaming from the gas towards the interface as these PDFs are not available.
However, in the original FSLBM~\cite{korner2005LatticeBoltzmannModel}, this boundary condition is not only used to reconstruct missing PDFs.
It is also used to reconstruct some PDFs that are already available.
It should be pointed out that this approach overwrites existing information about the flow field.
In the implementation used in the study presented here, no information is overwritten, and only missing PDFs are reconstructed at the free boundary.
This scheme was found to be of superior accuracy~\cite{schwarzmeier2022AnalysisComparisonBoundary}.
Note that the free-surface boundary condition~\eqref{eq:nm-fslbm-boundary-condition} must also be applied at free-slip boundaries for specularly reflected PDFs that originate from gas cells.
\par

The gas pressure
\begin{equation} \label{eq:nm-fslbm-pressure-gas}
p^{\text{G}}\left(\boldsymbol{x}, t\right) = p^{\text{V}}\left(t\right) - p^{\text{L}}\left(\boldsymbol{x}, t\right),
\end{equation}
incorporates the volume pressure $p^{\text{V}}(t)$ and Laplace pressure $p^{\text{L}}(\boldsymbol{x}, t)$.
The volume pressure stays constant in case of atmospheric pressure or results from changes in the volume $V(t)$ of an enclosed gas volume, that is, a bubble, according to
\begin{equation} \label{eq:nm-fslbm-pressure-volume}
p^{\text{V}}\left(t\right)=p^{\text{V}}\left(0\right) \frac{V\left(0\right)}{V\left(t\right)}.
\end{equation}
The Laplace pressure
\begin{equation} \label{eq:nm-fslbm-pressure-laplace}
p^{\text{L}}\left(\boldsymbol{x}, t\right)=2 \sigma \kappa\left(\boldsymbol{x}, t\right)
\end{equation}
is defined by the surface tension $\sigma \in \mathbb{R^{+}}$ and the interface curvature $\kappa(\boldsymbol{x}, t) \in \mathbb{R}$.
As suggested by Bogner et al.~\cite{bogner2016CurvatureEstimationVolumeoffluid}, in the simulations shown in this article, a finite difference approximation of
\begin{equation}
\kappa(\boldsymbol{x}, t) =-\nabla \cdot \boldsymbol{\hat{n}}(\boldsymbol{x}, t)
\end{equation}
was used, where $\boldsymbol{\hat{n}}(\boldsymbol{x}, t) \in \mathbb{R}^{d}$ is the normalized interface normal vector.
The interface normal 
\begin{equation}
\boldsymbol{n}(\boldsymbol{x}, t) = \nabla \varphi(\boldsymbol{x}, t)
\end{equation}
was computed with central finite differences according to Parker and Youngs~\cite{parker1992TwoThreeDimensional}.
Near obstacle cells, the computation of the normal is modified as proposed by Donath~\cite{donath2011WettingModelsParallel}.
This modification narrows the access pattern of the finite differences such that obstacle cells are excluded from the computation.
A bubble model algorithm is used to keep track of the bubbles' volume pressure~\cite{pohl2008HighPerformanceSimulation,anderl2014FreeSurfaceLattice}.
This is required because bubbles might coalesce or segment during the simulation.
\par
%!TEX root = ../main.tex

\section{Refilling schemes}\label{sec:rs}
In the FSLBM, gas cells do not contain valid PDF information as described in \Cref{sec:nm-fslbm}.
Therefore, when a gas cell converts to an interface cell, its PDFs must be reinitialized.
This reinitialization is commonly referred to as refilling.
While refilling has not yet been studied in the context of the FSLBM, it has been investigated for moving solid obstacle cells~\cite{peng2016ImplementationIssuesBenchmarking,lallemand2003LatticeBoltzmannMethod,chikatamarla2006GradApproximationMissing,krithivasan2014DiffusedBouncebackCondition,dorschner2015GradApproximationMoving,tao2016InvestigationMomentumExchange}.
Analogously to the gas cells in the FSLBM, solid cells do not carry valid PDFs, such that these PDFs must be refilled after conversion.
In what follows, the schemes developed for refilling moving solid cells are introduced and adapted to the FSLBM.
Then, their influence on the conservation of mass and their computational costs are briefly discussed.
\par

\subsection{Scheme definitions}\label{sec:rs-sd}
In the FSLBM proposed by Körner et al.~\cite{korner2005LatticeBoltzmannModel}, PDFs are refilled with their equilibrium~\eqref{eq:nm-lbm-equilibrium} according to
\begin{equation}
\text{EQ:} \quad
f_{i}\left(\boldsymbol{x}, t\right)
\coloneqq
f_{i}^\text{eq}\left(\bar{\rho}, \bar{\boldsymbol{u}}\right).
\end{equation}
The average local fluid density $\bar{\rho} \equiv \bar{\rho}\left(\boldsymbol{x}, t\right)$ and velocity $\bar{\boldsymbol{u}} \equiv \bar{\boldsymbol{u}}\left(\boldsymbol{x}, t\right)$ are computed from all surrounding, non-newly created interface and liquid cells.
Note that this is in contrast to the same scheme applied for solid obstacle cells, where the velocity of the solid object is used instead~\cite{lallemand2003LatticeBoltzmannMethod,peng2016ImplementationIssuesBenchmarking,tao2016InvestigationMomentumExchange}.
\par

The EQ scheme can be extended by adding the non-equilibrium contribution from neighboring fluid cells with~\cite{peng2016ImplementationIssuesBenchmarking}
\begin{equation}
\text{EQ+NEQ:} \quad
f_{i}\left(\boldsymbol{x}, t\right)
\coloneqq
f_{i}^\text{eq}\left(\bar{\rho}, \bar{\boldsymbol{u}}\right)
+
f_{i}^\text{neq}\left(\boldsymbol{x} + \boldsymbol{c}_{i}^{n}\Delta t, t\right),
\end{equation}
where $f_{i}^\text{neq}\left(\boldsymbol{x}, t\right) = f_{i}\left(\boldsymbol{x}, t\right) - f_{i}^\text{eq}\left(\bar{\rho}, \bar{\boldsymbol{u}}\right)$ is the PDF's non-equilibrium part.
The lattice direction $\boldsymbol{c}_{i}^{n} \coloneqq \boldsymbol{c}_{i} \, | \, \boldsymbol{c}_{i} \cdot \boldsymbol{n} \geq \boldsymbol{c}_{j} \cdot \boldsymbol{n} \,~ \forall j \in \{0, 1, \dots, q - 1\}$
is used to access the cell that corresponds best to the local interface normal $\boldsymbol{n} \equiv \boldsymbol{n}(\boldsymbol{x}, t)$, that is, for which the scalar product $\boldsymbol{c}_{i} \cdot \boldsymbol{n}$ is the largest.
Note again that only non-newly created neighboring interface and liquid cells are valid directions for $\boldsymbol{c}_{i}^{n}$.
\par

Another variant extends the EQ scheme with information from the local pressure tensor~\cite{dorschner2015GradApproximationMoving,krithivasan2014DiffusedBouncebackCondition,chikatamarla2006GradApproximationMissing} using Grad's moment system~\cite{grad1949KineticTheoryRarefied}, leading to
\begin{equation}
\text{GEQ:} \quad
f_{i}\left(\boldsymbol{x}, t\right)
\coloneqq
f_{i}^\text{eq}\left(\bar{\rho}, \bar{\boldsymbol{u}}\right)
+
\frac{w_{i} \bar{\rho}}{2 c_{s}^{2} \nu}
\sum_{\alpha,\beta}
\left(
\frac{\partial u_{\alpha}}{\partial x_{\beta}}
+
\frac{\partial u_{\beta}}{\partial x_{\alpha}}
\right)
\left(
c_{s}^{2} \delta_{\alpha \beta} - c_{i,\alpha}c_{i,\beta}
\right),
\end{equation}
where $\delta_{\alpha\beta}$ is the Kronecker delta.
In the implementation used in this study, the velocity derivatives are approximated by second-order finite differences.
If not enough neighboring fluid cells are available, the derivatives are approximated by first-order backward- or forward-finite differences.
\par

Instead of using the PDF's equilibrium, Lallemand and Luo~\cite{lallemand2003LatticeBoltzmannMethod} suggested refilling based on a second-order extrapolation scheme.
The PDFs are extrapolated from the lattice direction closest to the surface normal $\boldsymbol{c}_{i}^{n}$ by
\begin{equation}
\text{EXT:} \quad
f_{i}\left(\boldsymbol{x}, t\right)
\coloneqq
3 f_{i}\left(\boldsymbol{x} + \boldsymbol{c}_{i}^{n}\Delta t, t\right)
-
3 f_{i}\left(\boldsymbol{x} + 2\boldsymbol{c}_{i}^{n}\Delta t, t\right)
+
f_{i}\left(\boldsymbol{x} + 3\boldsymbol{c}_{i}^{n}\Delta t, t\right).
\end{equation}
In the implementations used for this study, a corresponding lower-order extrapolation is used if the number of neighboring cells in direction $\boldsymbol{c}_{i}^{n}$ is not sufficient.
If no neighboring cell is available in this direction, the EQ scheme is employed as a fallback.
A first-order extrapolation scheme led to the same numerical instabilities as will be discussed in \Cref{sec:ne} for the second-order scheme.
These numerical instabilities were not observed with a zeroth-order extrapolation, that is, copying PDFs from a direct neighboring cell.
However, this zeroth-order extrapolation scheme was more inaccurate than other refilling schemes, and it even led to physically implausible results in some tests.
For brevity reasons, these results are not included in this article.
\par

The AVG scheme uses the average $\bar{f}_{i}\left(\boldsymbol{x}, t\right)$ of the identically oriented PDFs of surrounding, non-newly created interface and liquid cells.
This is denoted by
\begin{equation}
\text{AVG:} \quad
f_{i}\left(\boldsymbol{x}, t\right)
\coloneqq
\bar{f}_{i}\left(\boldsymbol{x}, t\right),
\end{equation}
and has already been used by Fang et al.~\cite{fang2002LatticeBoltzmannMethod} for solid obstacle cells.
However, Fang et al. averaged the PDFs resulting from a higher-order extrapolation scheme, including a larger neighborhood of cells.
In this work, only the PDFs from direct neighboring cells are used.
\par

\subsection{Effect on mass conservation}\label{sec:rs-eomc}
It is important to point out that the choice of the refilling scheme does not affect the system's total mass and its conservation.
When a cell is refilled, it has converted from gas to interface and is initially empty with fill level $\varphi(\boldsymbol{x}, t)=0$.
Therefore, the refilled cell's mass~\eqref{eq:nm-fslbm-mass} is initially $m(\boldsymbol{x}, t)=0$ and is independent of the cell's refilled PDFs.\par

In the following time steps, the interface cell's mass then changes according to
\begin{equation}\label{eq:rs-mass-change}
m(\boldsymbol{x}, t + \Delta t) = m(\boldsymbol{x}, t) + \sum_{i} \Delta m_{i}(\boldsymbol{x}, t),
\end{equation}
where $\Delta m_{i}(\boldsymbol{x}, t)$ is the mass flux \eqref{eq:nm-fslbm-mass-flux}.
If the cell $\boldsymbol{x} + \boldsymbol{c}_{i}\Delta t$ is also an interface cell, the same $\Delta m_{i}(\boldsymbol{x}, t)$ affects this interface cell's change in mass~\eqref{eq:rs-mass-change}.
If the cell $\boldsymbol{x} + \boldsymbol{c}_{i}\Delta t$ is of liquid type, the corresponding $\Delta m_{i}(\boldsymbol{x}, t)$ is implicitly considered in the liquid cell's density $\rho(\boldsymbol{x}, t)$ and, therefore, in its mass, because the mass flux $\Delta m_{i}(\boldsymbol{x}, t)$ is computed directly from the balance of the streaming PDFs.
Consequently, $\Delta m_{i}(\boldsymbol{x}, t)$ is present in the change of the PDFs' values.
The density $\rho(\boldsymbol{x}, t)$ is obtained by taking the PDF's zeroth-order moment~\eqref{eq:nm-lbm-density}.
Since the zeroth-order moment is the sum of the cell's PDFs, $\Delta m_{i}(\boldsymbol{x}, t)$ leads to an according change of $\rho(\boldsymbol{x}, t)$. 

\subsection{Computational costs}\label{sec:rs-cc}
The computational costs of the individual refilling schemes strongly depend on the implementation of the FSLBM and the test case simulated.
For instance, the EQ, EQ+NEQ, and GEQ schemes require the neighboring cells' density $\rho(\boldsymbol{x}, t)$ and velocity $u(\boldsymbol{x}, t)$.
However, these macroscopic quantities are not necessarily available in any LBM implementation but might have to be computed only when required.
The LBM algorithm of collision~\eqref{eq:nm-lbm-collision} and streaming~\eqref{eq:nm-lbm-streaming} works on PDFs but does not involve $\rho(\boldsymbol{x}, t)$ or $u(\boldsymbol{x}, t)$.
Therefore, depending on the implementation, these values might be computed explicitly for the refilling schemes via the moments~\eqref{eq:nm-lbm-density} and~\eqref{eq:nm-lbm-velocity}, increasing the computational costs.\par

As another example, the EQ+NEQ and EXT schemes involve the interface normal $\boldsymbol{n}$.
The interface curvature $\kappa(\boldsymbol{x}, t)$ must be computed if the Laplace pressure~\eqref{eq:nm-fslbm-pressure-laplace} is relevant in a test case.
The curvature computation algorithm employed in this work also uses $\boldsymbol{n}$ so that no additional computations may be required to obtain $\boldsymbol{n}$ when refilling cells.\par

It should also be pointed out that the computational costs are affected by the refilled cell's neighboring cells, as only non-newly created interface and liquid cells are considered by the refilling schemes.\par

These examples show that it is not generally possible to present the specific computational cost of each refilling scheme.
Only the EQ, EQ+NEQ, and GEQ refilling schemes can be put into perspective, as the EQ+NEQ and GEQ build upon $f_{i}^\text{eq}(\bar{\rho}, \bar{\boldsymbol{u}})$ as computed by the EQ scheme but add additional computations.
Therefore, the EQ+NEQ and GEQ schemes are computationally more expensive than the EQ scheme.\par

Note that although cell conversions appear frequently, the computational costs for refilling might be insignificant compared to the costs of other algorithmic parts in the FSLBM.
However, this likewise strongly depends on the test case under investigation.\par
%!TEX root = ../main.tex

\section{Numerical experiments}\label{sec:ne}
The refilling schemes introduced in \Cref{sec:rs} are compared in six numerical experiments in this section.
The chosen test cases are vastly similar to those suggested in prior work~\cite{schwarzmeier2022ComparisonFreeSurface,schwarzmeier2022AnalysisComparisonBoundary}.
Therefore, the corresponding description of the test cases, simulation setups, and figures are based one those from these articles but are repeated here for completeness.
The numerical benchmarks include the simulation of a standing gravity wave, the collapse of a rectangular and cylindrical liquid column, the rise of a Taylor bubble, the impact of a drop into a thin film of liquid, and a bubbly plane Poiseuille flow.
All simulations were performed with double-precision floating-point arithmetic.
\par

\subsection{Gravity wave}\label{sec:ne-gw}
A gravity wave is a standing wave oscillating at the phase boundary between two immiscible fluids.
Surface tension forces are neglected, and gravitational forces entirely govern the wave's flow dynamics.
The analytical model~\cite{dingemans1997WaterWavePropagation,lamb1975Hydrodynamics} is used as reference data for assessing the simulation results.
\par

\subsubsection{Simulation setup}\label{sec:ne-gw-ss}
\begin{figure}[htbp]
	\centering
	\setlength{\figureheight}{0.35\textwidth}
	\setlength{\figurewidth}{0.35\textwidth}
	\setlength\mdist{0.02\textwidth}
	\begin{tikzpicture}
\definecolor{darkorange24213334}{RGB}{242,133,34}
\definecolor{dodgerblue0154222}{RGB}{0,154,222}

\begin{axis}%
[width=\figurewidth,
height=\figureheight,
xmin=0,
xmax=1,
ymin=0,
ymax=1,
ticks=none,
axis lines=none,
clip=false,
scale only axis
]
\addplot[thick, name path=f, domain=0:1,samples=50,smooth,dodgerblue0154222] {0.5+0.1*cos(deg(pi*x*2))};

\path[name path=axis] (axis cs:0,0) -- (axis cs:1,0);

\addplot [thick, color=dodgerblue0154222, fill=dodgerblue0154222!30] fill between[of=f and axis,	soft clip={domain=0:1}];
\end{axis}

% borders left and right
\draw[very thick, loosely dashed, black!50] (0,0)--(0,\figureheight);
\draw[very thick, loosely dashed, black!50] (\figurewidth,0)--(\figurewidth,\figureheight);

% borders top and bottom
\draw[very thick,, black!50] (0,\figureheight)--(\figurewidth,\figureheight);
\draw[very thick,, black!50] (0,0)--(\figurewidth,0);

% domain height
\draw[<->, >=Latex] (\figurewidth+\mdist,0)--(\figurewidth+\mdist,\figureheight) node [pos=0.5,right] {$L$};
\draw[-] (\figurewidth,0)--(\figurewidth+1.5\mdist,0);
\draw[-] (\figurewidth,\figureheight)--(\figurewidth+1.5\mdist,\figureheight);

% domain width
\draw[<->, >=Latex] (0,-\mdist)--(\figurewidth,-\mdist) node [pos=0.5,below] {$L$};
\draw[-] (0,0)--(0,-1.5\mdist) node [below] {$x=0$};
\draw[-] (\figurewidth,0)--(\figurewidth,-1.5\mdist);

% liquid height
\draw[dotted] (0,0.5\figureheight)--(\figurewidth,0.5\figureheight);
\draw[<->, >=Latex] (-\mdist,0)--(-\mdist,0.5\figureheight) node [pos=0.5,left] {$d$};
\draw[-] (-1.5\mdist,0)--(0,0);
\draw[-] (-1.5\mdist,0.5\figureheight)--(0,0.5\figureheight);

% initial amplitude
\draw[<->, >=Latex] (-\mdist,0.5\figureheight)--(-\mdist,0.6\figureheight) node [pos=0.5,left] {$a_{0}$};
\draw[-] (-1.5\mdist,0.6\figureheight)--(0,0.6\figureheight);

% initial profile label
\node[
rectangle,
anchor=south,
dodgerblue0154222] at (0.5\figurewidth,0.54\figureheight) {$y(x) = d + a_{0} \cos\left(kx \right)$};

% vertical line of symmetry
%\draw[loosely dashdotted] (0.5\figurewidth,-0.025\figureheight)--(0.5\figurewidth,1.025\figureheight);

% gravity
\draw[thick, ->, >=Latex, darkorange24213334] (\figurewidth-2\mdist,\figureheight-\mdist)--(\figurewidth-2\mdist,\figureheight-4\mdist) node [pos=0.5,right] {$g$};

% coordinate system
\draw[->, >=Latex] (2\mdist,\figureheight-3\mdist)--(4\mdist,\figureheight-3\mdist) node [below] {$x$};
\draw[->, >=Latex] (2\mdist,\figureheight-3\mdist)--(2\mdist,\figureheight-1\mdist) node [left] {$y$};
\end{tikzpicture}%
	\caption{
	\label{fig:ne-gw-setup}
		Simulation setup of the two-dimensional gravity wave test case with wavelength $L$, liquid depth $d$, initial wave amplitude $a_{0}$, wavenumber $k=2\pi/L$, and gravitational acceleration $g$.
		The domain was periodic in the $x$-direction, whereas no-slip boundary conditions were set at the domain walls in $y$-direction.
		C.~Schwarzmeier, M.~Holzer, T.~Mitchell, M.~Lehmann, F.~Häusl, U.~Rüde, Comparison of free-surface and conservative Allen--Cahn phase-field lattice Boltzmann method, arXiv preprint~\cite{schwarzmeier2022ComparisonFreeSurface}, 2022; licensed under a Creative Commons Attribution (CC BY) license; the colors were changed from the original.
	}
\end{figure}
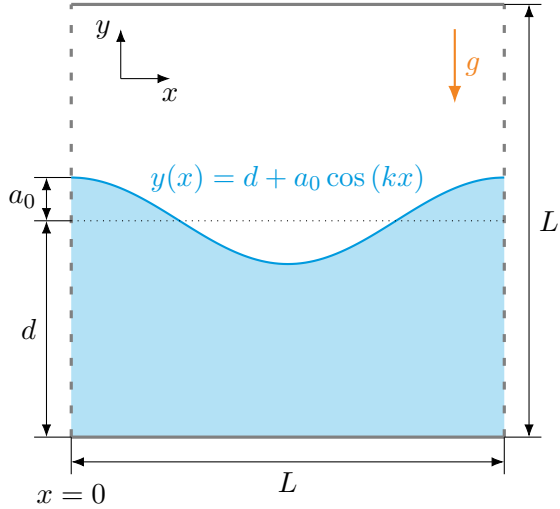

As illustrated in \Cref{fig:ne-gw-setup}, a gravity wave of wavelength $L$ was simulated in a two-dimensional quadratic domain of size $L\times L \times 1$ ($x$-, $y$-, $z$-direction) with $L \in \{200, 400, 800\}$ lattice cells.
The interface at the phase boundary was initialized with the profile $y(x) = d + a_{0} \cos\left(kx \right)$ with liquid depth $d=0.5L$, initial amplitude $a_{0}=0.01L$, and wavenumber $k=2\pi/L$.
Walls confined the domain with periodic and no-slip boundary conditions in the $x$- and $y$-direction, respectively.
The liquid was initialized with hydrostatic pressure according to the gravitational acceleration $g$, so the LBM pressure at $y=d$ was equal to the constant atmospheric volume pressure $p^{\text{V}}(t) = p_{0}$.
The relaxation rate $\omega=1.8$ was chosen for all simulations of all computational domain resolutions to conform with what is referred to as diffusive scaling in the LBM~\cite{kruger2017LatticeBoltzmannMethod}.
The system is characterized by the Reynolds number
\begin{equation}\label{eq:re-wave}
\mathrm{Re} \coloneqq \frac{a_{0}\omega_{0}L}{\nu} = 10
\end{equation}
where
\begin{equation}
\omega_{0} = \sqrt{g k \, \mathrm{tanh} \left(k d \right)},
\end{equation}
is the angular frequency of the wave, and $\nu$ is the kinematic fluid viscosity.
Owing to the gravitational acceleration $g$, the initial profile evolved into a standing wave oscillating around $d$.
It was dampened because of viscous forces.
The dimensionless surface elevation $a^{*}(x, t) \coloneqq a(x, t)/a_{0}$ and non-dimensionalized time $t^{*} \coloneqq t\omega_{0}$ were monitored at $x=0$ every $t^{*} = 0.01$.
The simulations were performed until $t^{*} = 40$, which was found to be sufficient for the wave's motion to be decayed in the simulations.
\par

\subsubsection{Analytical model}\label{sec:ne-gw-am}
The analytical model for the gravity wave is derived by linearization of the continuity and Euler equations with a free-surface boundary condition~\cite{dingemans1997WaterWavePropagation}.
The standing wave's amplitude
\begin{equation}
a(x,t) = a_{D}(t) \cos \left( kx - \omega_{0} t \right) + d,
\end{equation}
is obtained assuming an inviscid fluid with zero damping, such that $a_{D}(t)=a_{0}$.
Viscous damping is considered by~\cite{lamb1975Hydrodynamics}
\begin{equation}
a_{D}(t) = a_{0} \mathrm{e}^{-2 \nu k^{2} t}.
\end{equation}
The analytical model applies if $k |a_{0}| \ll 1$ and $k |a_{0}| \ll kd$~\cite{dingemans1997WaterWavePropagation}, which is true in this study with $k |a_{0}| = 0.02\pi \ll 1 < kd = \pi$.
\par

\subsubsection{Results and discussion}\label{sec:ne-gw-res}
\begin{figure}[htbp]
	\centering
	\setlength{\figureheight}{0.5\textwidth}
	\setlength{\figurewidth}{0.95\textwidth}
	\input{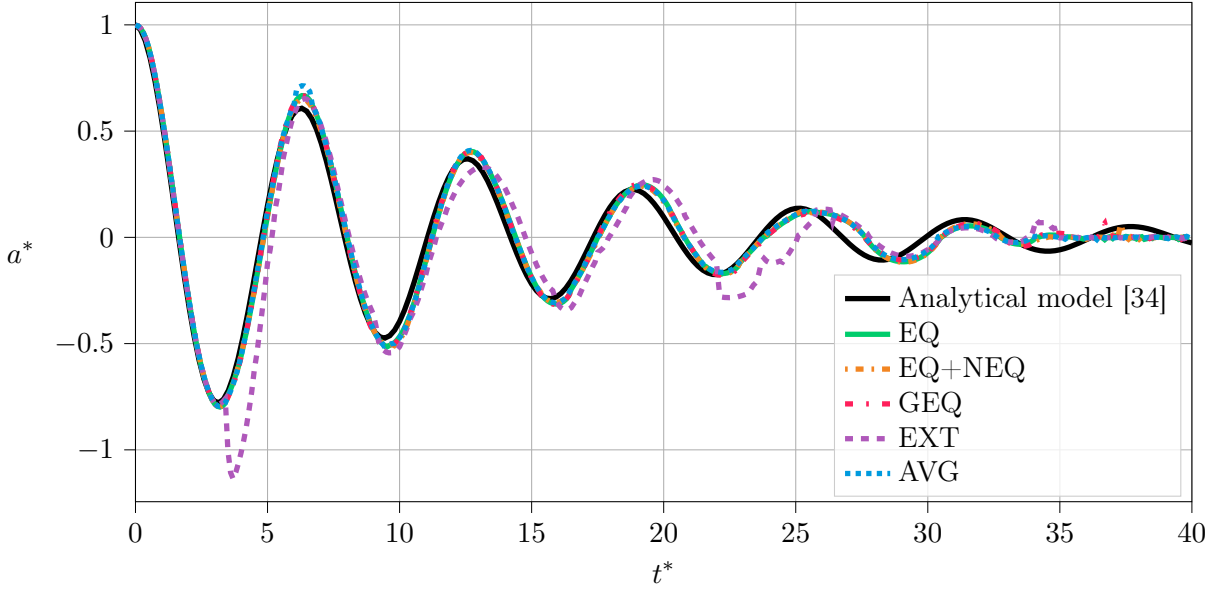}%
	\caption{
		\label{fig:gravity-wave-results}
		Simulated surface elevation of the gravity wave in terms of non-dimensional amplitude $a^{*}(0,t^{*})$ and time $t^{*}$.
		The simulations were performed with a computational domain resolution, that is, wavelength of $L=800$ lattice cells.
		All but the EXT refilling scheme differ only slightly and generally agree well with the analytical model~\cite{dingemans1997WaterWavePropagation}.
	}
\end{figure}

\begin{figure}[htbp]
	\centering
	\setlength{\figureheight}{0.4\textwidth}
	\setlength{\figurewidth}{0.95\textwidth}
	\begin{tikzpicture}

\definecolor{crimson2553191}{RGB}{255,31,91}
\definecolor{darkgray176}{RGB}{176,176,176}
\definecolor{darkorange24213334}{RGB}{242,133,34}
\definecolor{dodgerblue0154222}{RGB}{0,154,222}
\definecolor{lightgray204}{RGB}{204,204,204}
\definecolor{mediumorchid17588186}{RGB}{175,88,186}
\definecolor{springgreen0205108}{RGB}{0,205,108}

\begin{axis}[
height=\figureheight,
legend cell align={left},
legend style={
	fill opacity=0.8,
	draw opacity=1,
	text opacity=1,
	at={(0.01,0.99)},
	anchor=north west,
	draw=lightgray204,
	legend columns=2,
},
tick align=outside,
tick pos=left,
width=\figurewidth,
x grid style={darkgray176},
xlabel={Index of half-period},
xmajorgrids,
xmin=0.5, xmax=9.5,
xtick style={color=black},
y grid style={darkgray176},
ylabel style={rotate=-90.0},
ylabel={\(\displaystyle T^{*}/2\)},
ymajorgrids,
ymin=2.6264207043744, ymax=3.94746139750478,
ytick style={color=black}
]
\addplot [very thick, black, mark=*, mark size=2, mark options={solid}]
table {%
	1 3.14159274101257
	2 3.14159274101257
	3 3.14159274101257
	4 3.14159274101257
	5 3.14159274101257
	6 3.14159274101257
	7 3.14159274101257
	8 3.14159274101257
	9 3.14159274101257
};
\addlegendentry{Analytical model~\cite{dingemans1997WaterWavePropagation}~~~}
\addplot [very thick, springgreen0205108, mark=triangle, mark size=2, mark options={solid,rotate=180,fill opacity=0}]
table {%
	1 3.18385434150696
	2 3.16068649291992
	3 3.21545457839966
	4 3.16537165641785
	5 3.10675048828125
	6 3.28842210769653
	7 2.99571084976196
	8 3.84516572952271
	9 2.87851285934448
};
\addlegendentry{EQ}
\addplot [very thick, darkorange24213334, dash pattern=on 1pt off 3pt on 3pt off 3pt, mark=triangle, mark size=2, mark options={solid,rotate=90,fill opacity=0}]
table {%
	1 3.18696022033691
	2 3.13654327392578
	3 3.24770665168762
	4 3.14031219482422
	5 3.15828704833984
	6 3.22582244873047
	7 3.06003832817078
	8 3.79544591903687
	9 2.85946941375732
};
\addlegendentry{EQ+NEQ~~~}
\addplot [very thick, crimson2553191, dash pattern=on 3pt off 5pt on 1pt off 5pt, mark=square, mark size=2, mark options={solid,fill opacity=0}]
table {%
	1 3.17171812057495
	2 3.15871787071228
	3 3.21299624443054
	4 3.17011666297913
	5 3.11679315567017
	6 3.28922843933105
	7 3.07327389717102
	8 3.74162864685059
	9 2.79292488098145
};
\addlegendentry{GEQ}
\addplot [very thick, mediumorchid17588186, dashed, mark=triangle, mark size=2, mark options={solid,rotate=270,fill opacity=0}]
table {%
	1 3.49511384963989
	2 2.9355788230896
	3 3.29159307479858
	4 3.37442135810852
	5 3.16021871566772
	6 3.35549116134644
	7 3.79096460342407
	8 2.7202250957489
	9 2.68646812438965
};
\addlegendentry{EXT~~~}
\addplot [very thick, dodgerblue0154222, dotted, mark=triangle, mark size=2, mark options={solid,fill opacity=0}]
table {%
	1 3.16091871261597
	2 3.19429349899292
	3 3.20161843299866
	4 3.158616065979
	5 3.11355566978455
	6 3.27169036865234
	7 2.97507786750793
	8 3.88741397857666
	9 2.75283074378967
};
\addlegendentry{AVG}
\end{axis}

\end{tikzpicture}%
	\caption{
		\label{fig:gravity-wave-period}
		Simulated non-dimensionalized duration of the gravity wave's half-period $T^{*}/2$, measured at when the oscillation's amplitude $a^{*}(0,t^{*}) = 0$.
		The simulations were performed with a wavelength of $L=800$ lattice cells.
		Apart from the EXT scheme's larger deviations, no clear trend could be identified.
	}
\end{figure}
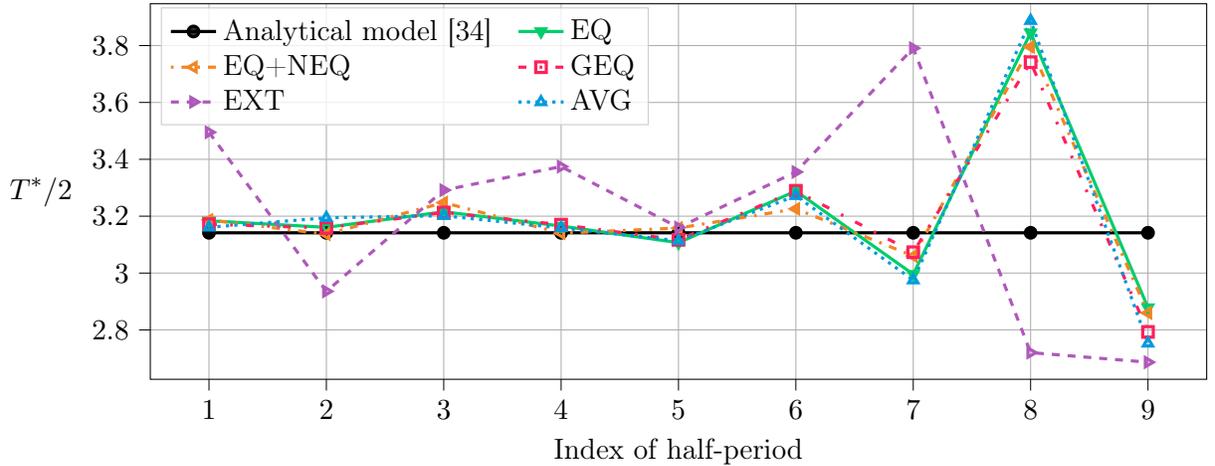

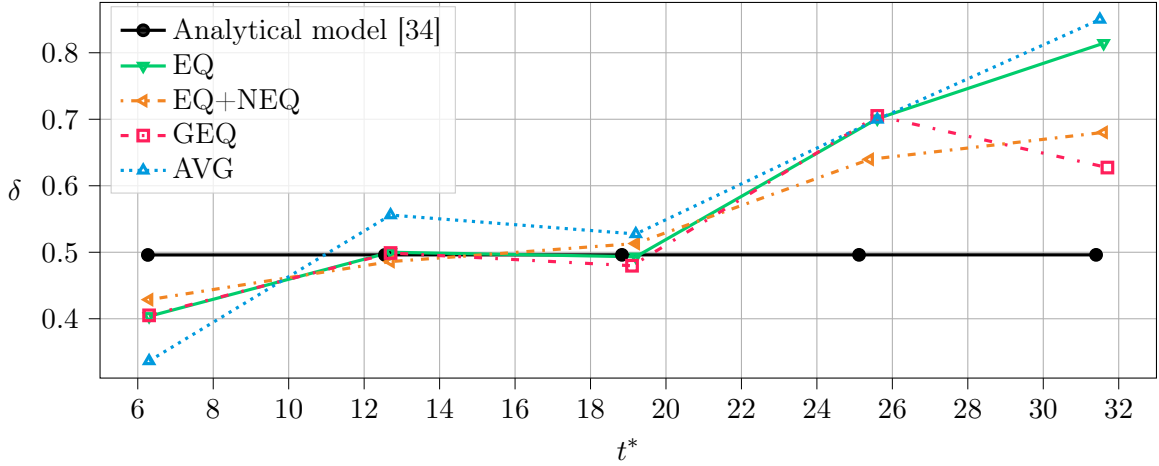
\begin{figure}[htbp]
	\centering
	\setlength{\figureheight}{0.4\textwidth}
	\setlength{\figurewidth}{0.95\textwidth}
	\begin{tikzpicture}

\definecolor{crimson2553191}{RGB}{255,31,91}
\definecolor{darkgray176}{RGB}{176,176,176}
\definecolor{darkorange24213334}{RGB}{242,133,34}
\definecolor{dodgerblue0154222}{RGB}{0,154,222}
\definecolor{lightgray204}{RGB}{204,204,204}
\definecolor{springgreen0205108}{RGB}{0,205,108}

\begin{axis}[
height=\figureheight,
legend cell align={left},
legend style={
	fill opacity=0.8,
	draw opacity=1,
	text opacity=1,
	at={(0.01,0.99)},
	anchor=north west,
	draw=lightgray204
},
tick align=outside,
tick pos=left,
width=\figurewidth,
x grid style={darkgray176},
xlabel={\(\displaystyle t^{*}\)},
xmajorgrids,
xmin=5, xmax=33,
xtick style={color=black},
y grid style={darkgray176},
ylabel style={rotate=-90.0},
ylabel={\(\displaystyle \delta\)},
ymajorgrids,
ymin=0.310808031862565, ymax=0.875680459871436,
ytick style={color=black}
]
\addplot [very thick, black, mark=*, mark size=2, mark options={solid}]
table {%
	6.26740741729736 0.495972394943237
	12.5503702163696 0.496100425720215
	18.8333339691162 0.496100425720215
	25.117036819458 0.496100425720215
	31.3999996185303 0.496100425720215
};
\addlegendentry{Analytical model~\cite{dingemans1997WaterWavePropagation}}
\addplot [very thick, springgreen0205108, mark=triangle, mark size=2, mark options={solid,rotate=180,fill opacity=0}]
table {%
	6.30000019073486 0.403699517250061
	12.6999998092651 0.500024318695068
	19.2000007629395 0.493222117424011
	25.6000003814697 0.700831174850464
	31.6000003814697 0.814314603805542
};
\addlegendentry{EQ}
\addplot [very thick, darkorange24213334, dash pattern=on 1pt off 3pt on 3pt off 3pt, mark=triangle, mark size=2, mark options={solid,rotate=90,fill opacity=0}]
table {%
	6.30000019073486 0.428741455078125
	12.6999998092651 0.486005544662476
	19.2000007629395 0.513132572174072
	25.3999996185303 0.639747381210327
	31.6000003814697 0.6802659034729
};
\addlegendentry{EQ+NEQ}
\addplot [very thick, crimson2553191, dash pattern=on 3pt off 5pt on 1pt off 5pt, mark=square, mark size=2, mark options={solid,fill opacity=0}]
table {%
	6.30000019073486 0.405120253562927
	12.6999998092651 0.498883008956909
	19.1000003814697 0.479717254638672
	25.6000003814697 0.705025672912598
	31.7000007629395 0.627374887466431
};
\addlegendentry{GEQ}
\addplot [very thick, dodgerblue0154222, dotted, mark=triangle, mark size=2, mark options={solid,fill opacity=0}]
table {%
	6.30000019073486 0.33648407459259
	12.6999998092651 0.555904150009155
	19.2000007629395 0.527507305145264
	25.6000003814697 0.699696063995361
	31.5 0.850004434585571
};
\addlegendentry{AVG}
\end{axis}

\end{tikzpicture}%
	\caption{
		\label{fig:gravity-wave-logarithmic-decrement}
		Simulated logarithmic decrement $\delta$, computed from the gravity wave's positive maximum amplitudes.
		The time $t^{*}$ is non-dimensionalized and the simulations were performed with a wavelength of $L=800$ lattice cells.
		The EXT refilling scheme is not included as it was subject to irregularities in the amplitude that deviated the evaluation algorithm.
		The EQ+NEQ scheme was arguably most accurate in this comparison.
	}
\end{figure}

\Cref{fig:gravity-wave-results} shows the gravity wave's non-dimensionalized amplitude $a^{*}(0,t^{*})$ over time $t^{*}$ as simulated with $L = 800$ and the refilling schemes from \Cref{sec:rs}.
All refilling schemes but the EXT generally agreed well with the analytical model.
The other refilling schemes showed only minor differences.
Most notably is a slight overestimation of the wave's first positive amplitude when using the AVG scheme.
The duration of a gravity wave's half-period $T^{*}/2$ is shown in \Cref{fig:gravity-wave-period}.
The EXT refilling scheme had the largest deviations when compared to the analytical model.
All other refilling schemes did not follow a clear trend.
\Cref{fig:gravity-wave-logarithmic-decrement} shows the logarithmic decrement
\begin{equation}
\delta = \mathrm{ln}\frac{a^{*}(0,t^{*})}{a^{*}(0,t^{*} + T^{*})}
\end{equation}
of the gravity wave's oscillations.
Although the differences are relatively small, the EQ+NEQ scheme could arguably be considered most accurate in this comparison.
\par

The simulation results presented here have converged, as illustrated in \Cref{fig:ne-gravity-wave-convergence}.
As pointed out in prior work~\cite{schwarzmeier2022ComparisonFreeSurface}, the FSLBM can only predict the wave's motion sufficiently well, if the amplitude spans over at least one but preferably multiple cells.
This is also shown in \Cref{fig:ne-gravity-wave-convergence}, where less wave periods could be simulated when using lower computational domain resolutions.
\par

\subsection{Rectangular dam break}\label{sec:ne-rdb}
In the rectangular dam break benchmark case, a rectangular liquid column collapses and spreads at the bottom surface.
This test case is regularly used as a numerical benchmark to validate free-surface flow simulations~\cite{janssen2011FreeSurfaceFlow,sato2022ComparativeStudyCumulant,moraga2015VOFFVMPrediction}.
The experiments from Martin and Moyce~\cite{martin1952PartIVExperimental} were used as reference data for the simulations in this section.
\par

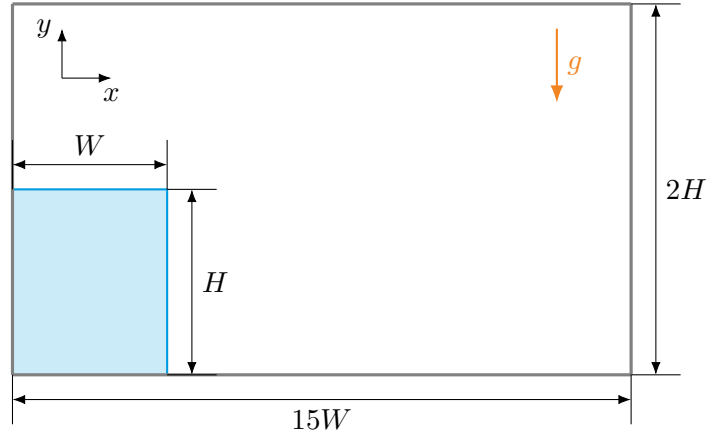
\begin{figure}[htbp]
	\centering
	\setlength{\figureheight}{0.3\textwidth}
	\setlength{\figurewidth}{0.5\textwidth}
	\setlength\mdist{0.02\textwidth}
	\begin{tikzpicture}
\definecolor{darkorange24213334}{RGB}{242,133,34}
\definecolor{dodgerblue0154222}{RGB}{0,154,222}

% liquid column
\draw[thick, fill=dodgerblue0154222!20, draw=none] (0,0) rectangle (0.25\figurewidth,0.5\figureheight);
\draw[thick, dodgerblue0154222] (0,0.5\figureheight)--(0.25\figurewidth,0.5\figureheight);
\draw[thick, dodgerblue0154222] (0.25\figurewidth,0)--(0.25\figurewidth,0.5\figureheight);

% borders left and right
\draw[very thick, black!50] (0,0)--(0,\figureheight);
\draw[very thick, black!50] (\figurewidth,0)--(\figurewidth,\figureheight);

% borders top and bottom
\draw[very thick, black!50] (0,\figureheight)--(\figurewidth,\figureheight);
\draw[very thick, black!50] (0,0)--(\figurewidth,0);

% domain height
\draw[<->, >=Latex] (\figurewidth+\mdist,0)--(\figurewidth+\mdist,\figureheight) node [pos=0.5,right] {$2H$};
\draw[-] (\figurewidth,0)--(\figurewidth+2\mdist,0);
\draw[-] (\figurewidth,\figureheight)--(\figurewidth+2\mdist,\figureheight);

% domain width
\draw[<->, >=Latex] (0,-\mdist)--(\figurewidth,-\mdist) node [pos=0.5,below] {$15 W$};
\draw[-] (0,0)--(0,-2\mdist);
\draw[-] (\figurewidth,0)--(\figurewidth,-2\mdist);

% liquid column height
\draw[<->, >=Latex] (0.25\figurewidth+\mdist,0)--(0.25\figurewidth+\mdist,0.5\figureheight) node [pos=0.5,right] {$H$};
\draw[-] (0.25\figurewidth,0.5\figureheight)--(0.25\figurewidth+2\mdist,0.5\figureheight);
\draw[-] (0.25\figurewidth,0)--(0.25\figurewidth+2\mdist,0);

% liquid column diameter
\draw[<->, >=Latex] (0,0.5\figureheight+\mdist)--(0.25\figurewidth,0.5\figureheight+\mdist) node [pos=0.5,above] {$W$};
\draw[-] (0,0.5\figureheight)--(0,0.5\figureheight+2\mdist);
\draw[-] (0.25\figurewidth,0.5\figureheight)--(0.25\figurewidth,0.5\figureheight+2\mdist);

% gravity
\draw[thick, ->, >=Latex, darkorange24213334] (\figurewidth-3\mdist,\figureheight-\mdist)--(\figurewidth-3\mdist,\figureheight-4\mdist) node [pos=0.5,right] {$g$};

% coordinate system
\draw[->, >=Latex] (2\mdist,\figureheight-3\mdist)--(4\mdist,\figureheight-3\mdist) node [below] {$x$};
\draw[->, >=Latex] (2\mdist,\figureheight-3\mdist)--(2\mdist,\figureheight-1\mdist) node [left] {$y$};
\end{tikzpicture}%
	\caption{\label{fig:dam-break-rectangular-setup}
		Simulation setup of the two-dimensional rectangular dam break test case with the liquid column's initial width $W$ and height $H$.
		The gravitational acceleration $g$ acted in negative $y$-direction and led to the liquid column's collapse.
		Free-slip boundary conditions were set at all domain walls.
		C.~Schwarzmeier, U.~Rüde, Analysis and comparison of boundary condition variants in the free-surface lattice Boltzmann method, arXiv preprint~\cite{schwarzmeier2022AnalysisComparisonBoundary}, 2022; licensed under a Creative Commons Attribution (CC BY) license.
	}
\end{figure}

\subsubsection{Simulation setup}\label{sec:ne-rdb-ss}
The simulation setup resembled that of the reference experiments~\cite{martin1952PartIVExperimental}, and is illustrated in \Cref{fig:dam-break-rectangular-setup}.
A liquid column of width $W \in \{50, 100, 200\}$ lattice cells and height $H=2W$ was positioned at the domain's left wall.
The domain was two-dimensional and of size $15W \times 2H \times 1$ ($x$-, $y$-, $z$-direction). 
Owing to the gravitational acceleration $g$ acting in negative $y$-direction, the liquid was initialized with hydrostatic pressure.
Therefore, the LBM pressure at $y=H$ was initially equal to the constant atmospheric gas pressure $p^{\text{V}}(t) = p_{0}$.
Wetting effects were not considered, and free-slip boundary conditions were set at all domain walls.
Conforming with diffusive scaling, the relaxation rate $\omega=1.9995$ was kept constant for all tested computational domain resolutions.
The simulations were performed using the turbulence model from \Cref{sec:nm-lbm} with Smagorinsky constant $C_{S}=0.1$~\cite{yu2005DNSDecayingIsotropic}.
Two dimensionless numbers describe the fluid mechanics of the system.
The Galilei number
\begin{equation}\label{eq:ne-rdb-ss-ga}
\text{Ga} \coloneqq \frac{g W^{3}}{\nu^{2}} = 1.83\cdot 10^{9}
\end{equation}
with kinematic viscosity $\nu$, relates the gravitational to viscous forces.
The Bond number
\begin{equation}\label{eq:ne-rdb-ss-bo}
\text{Bo} \coloneqq \frac{\Delta \rho \, g  W^{2}}{\sigma} = 445
\end{equation}
defines the relation between gravitational and surface tension forces.
It is defined by the surface tension $\sigma$, and the density difference between the liquid and gas phase $\Delta \rho = \rho - \rho^{\text{G}}$.
Note that in a free-surface system, the gas phase density is assumed to be zero so that $\Delta \rho = \rho$.
The reference experiments~\cite{martin1952PartIVExperimental} were performed with liquid water, but the authors did not provide fluid properties.
With given initial column width $W=0.05715$\,m, Ga and Bo were computed assuming $g=9.81$\,m/s\textsuperscript{2} and liquid water at 25\,\textdegree C with density $\rho \approx 1000$\,kg/m\textsuperscript{3}, kinematic viscosity $\nu \approx 10^{-6}$\,m/{s}\textsuperscript{2}, and surface tension $\sigma \approx 7.2\cdot 10^{-2}$\,kg/s\textsuperscript{2}~\cite{rumble2021CRCHandbookChemistry}.
\par

The liquid column's residual height $h(t)$ and width $w(t)$ were monitored during the simulation.
The former was obtained by finding the uppermost interface cell at the left domain wall, that is, $x=0$.
The width $w(t)$ was obtained by searching for the rightmost interface cell at the bottom domain wall, that is, at $y=0$.
As suggested by Martin and Moyce~\cite{martin1952PartIVExperimental}, the height $h^{*}(t) \coloneqq h(t) / H$, width $w^{*}(t) \coloneqq w(t) / W$, and time $t^{*} \coloneqq t \sqrt{2g/W}$ were non-dimensionalized.
The height and width were monitored every $t^{*}=0.01$.
The simulations were performed until $w^{*}(t^{*}) \geq 14$, conforming with the experimental data.
\par

\begin{figure}[htbp]
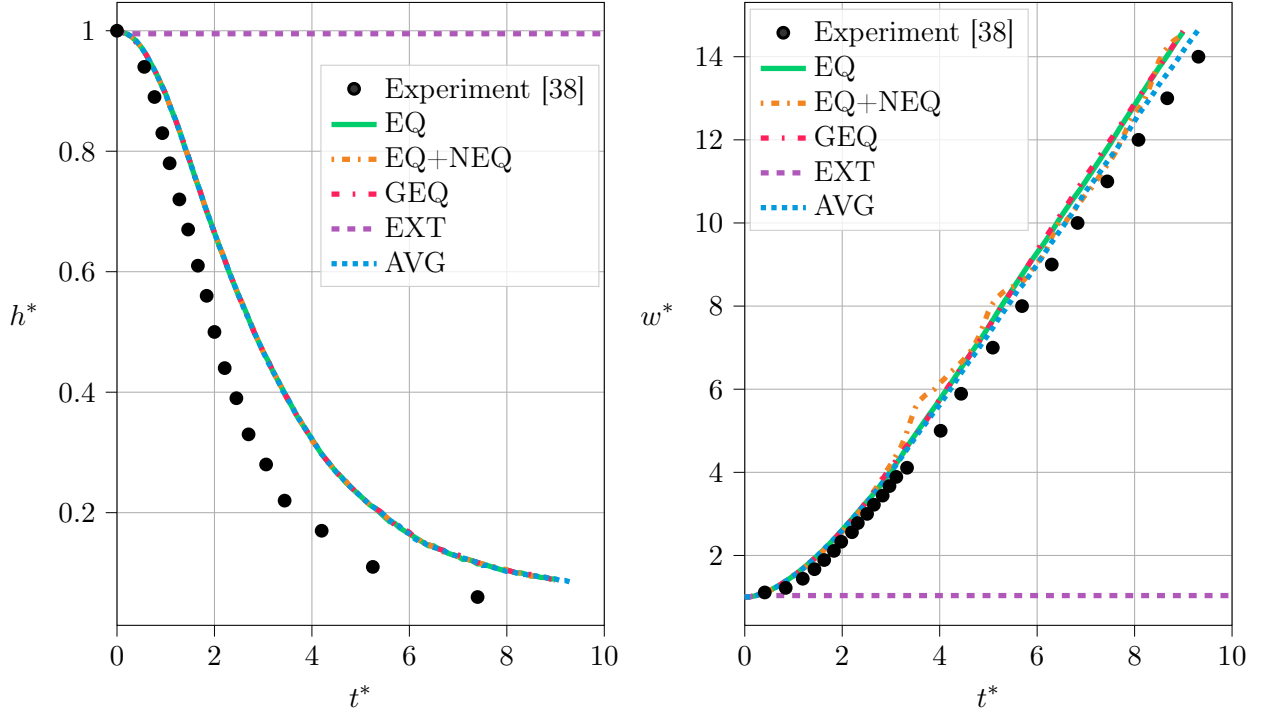

	\centering
	\setlength{\figureheight}{0.6\textwidth}
	\setlength{\figurewidth}{0.49\textwidth}
	\begin{subfigure}[htbp]{0.49\textwidth}
		\centering
		\input{figures/dam-break-rectangular/w-200-h.tex}%
	\end{subfigure}
	\hfill
	\begin{subfigure}[htbp]{0.49\textwidth}
		\centering
		\input{figures/dam-break-rectangular/w-200-w.tex}%
	\end{subfigure}
	\caption{
		\label{fig:dam-break-rectangular-200}
		Simulated rectangular dam break with non-dimensionalized residual dam height $h^{*}(t^{*})$ (a), width $w^{*}(t^{*})$ (b), and time $t^{*}$.
		The simulations were performed with a computational domain resolution, that is, initial dam width of $W=200$ lattice cells.
		All refilling schemes but the EXT generally agreed with the experimental data~\cite{martin1952PartIVExperimental} and are of comparable accuracy.
		The EXT scheme was numerically unstable because of too high lattice velocities.
	}
\end{figure}

\subsubsection{Results and discussion}\label{sec:ne-rdb-res}
\begin{figure}[htbp]
	\centering
	\setlength\mdist{0.02\textwidth}
	\begin{tabular}{
			>{\centering\arraybackslash}m{0.1\textwidth}
			>{\centering\arraybackslash}m{0.6\textwidth}
		}	
		
		EQ & 
		\parbox[m]{0.6\textwidth}{\includegraphics[width=0.5\textwidth]{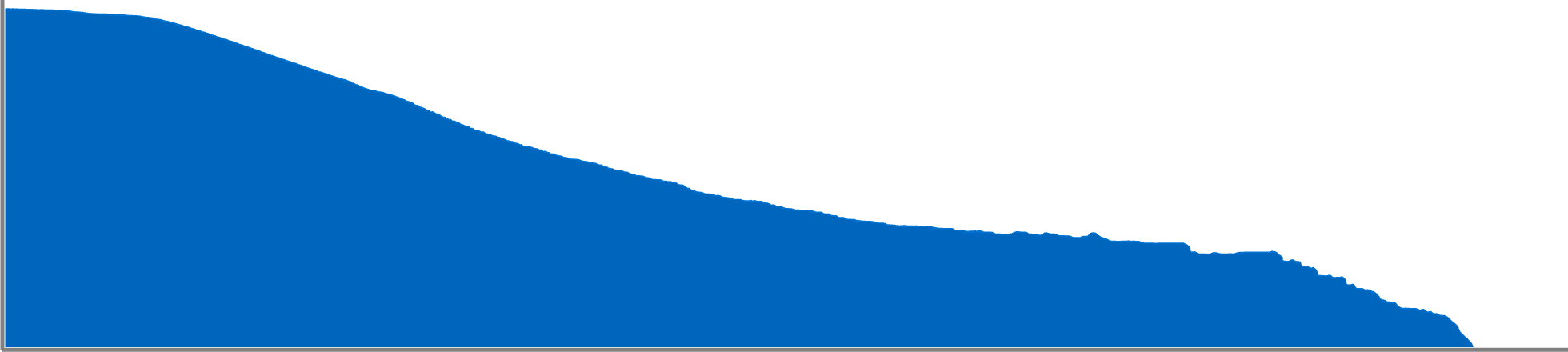}}\\[0.05\textwidth]
		
		EQ+NEQ & 
		\parbox[m]{0.6\textwidth}{\includegraphics[width=0.5\textwidth]{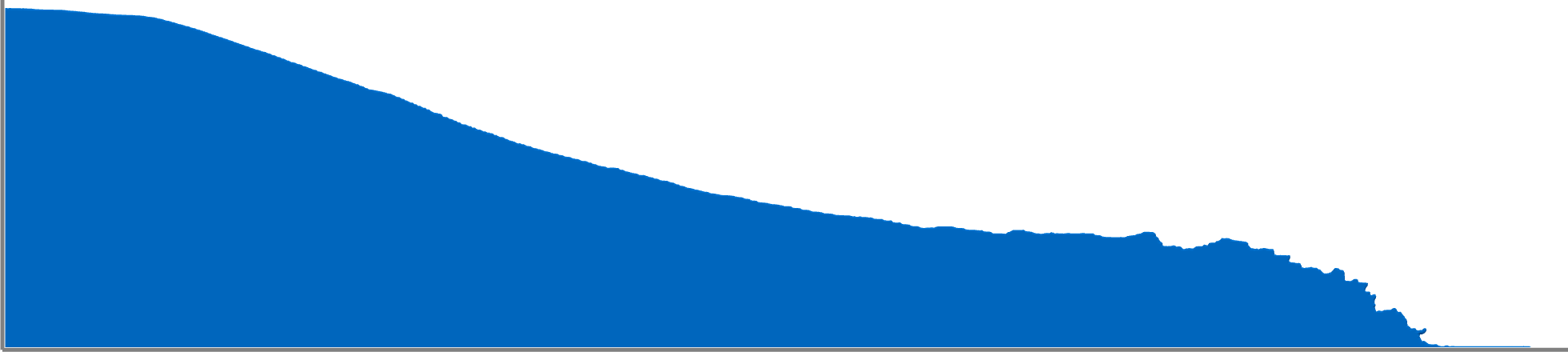}}\\[0.05\textwidth]
		
		GEQ &
		\parbox[m]{0.6\textwidth}{\includegraphics[width=0.5\textwidth]{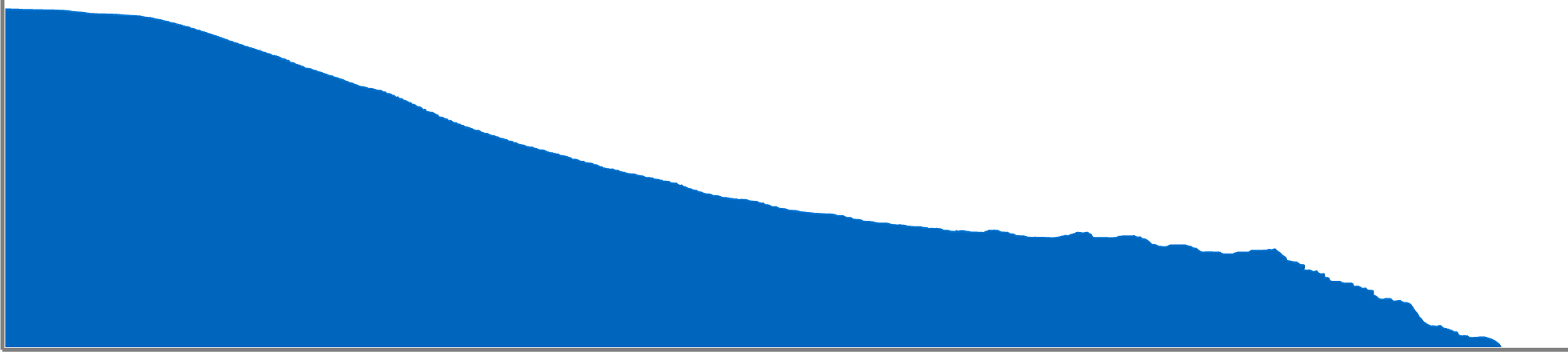}}\\[0.05\textwidth]
		
		AVG &
		\parbox[m]{0.6\textwidth}{\includegraphics[width=0.5\textwidth]{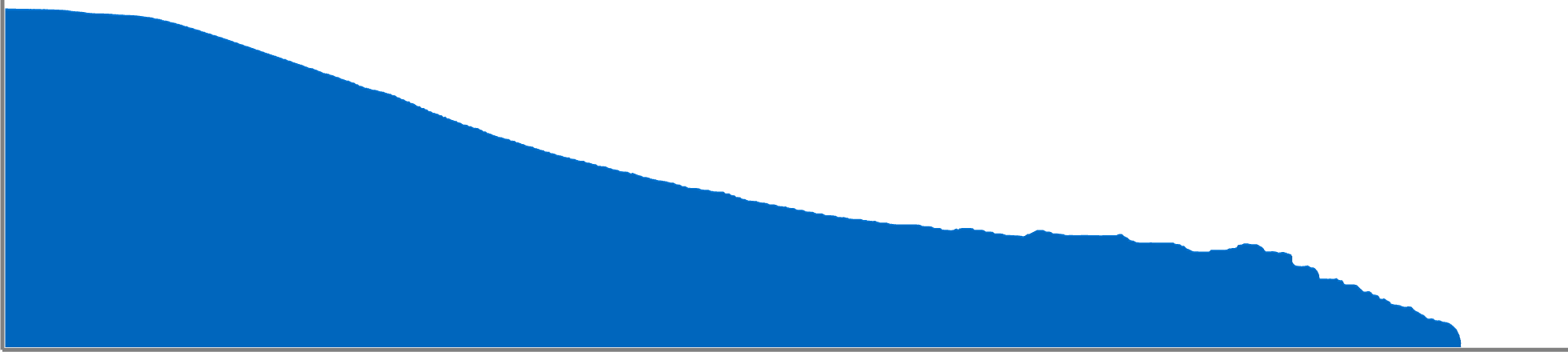}}\\[0.05\textwidth]
	\end{tabular}
	\caption{
		\label{fig:dam-break-rectangular-mesh}
		Contour of the simulated rectangular dam break at time $t^{*}=3$ with an initial dam width of $W=200$ lattice cells with the EQ (a), EQ+NEQ (b), GEQ (c), and AVG (d) refilling schemes.
		The simulation with the EXT refilling scheme is not included here because the simulation was numerically unstable.
		There are noticeable differences between the individual refilling schemes.
		Owing to the lack of reference data, no accuracy assessment could be made for the contours.
	}
\end{figure}

\begin{figure}[htbp]
	\centering
	\includegraphics[width=0.2\textwidth]{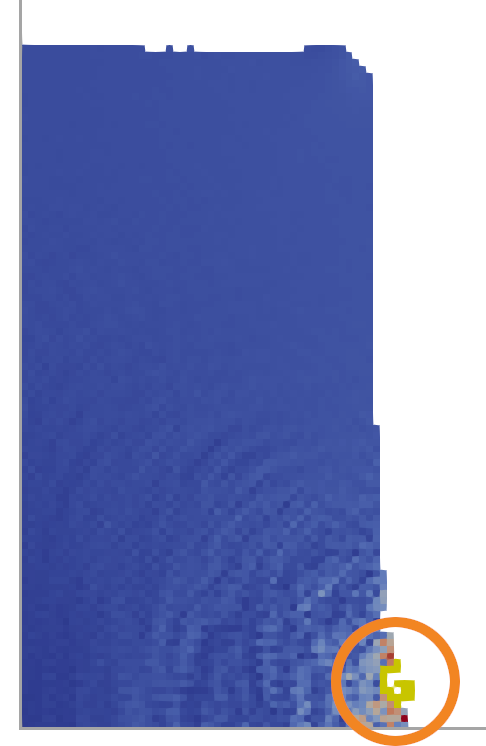}%
	\caption{
		\label{fig:dam-break-rectangular-ext-unstable}
		Contour of the simulated rectangular dam break with an initial dam width of $W=50$ lattice cells and EXT refilling scheme.
		At non-dimensionalized time $t^{*} \approx 0.43$, the simulation became numerically unstable.
		These instabilities were caused by the lattice velocity exceeding the lattice speed of sound $c_{s}$ in the area marked with the orange circle.
	}
\end{figure}

\Cref{fig:dam-break-rectangular-200} compares the simulated dam break with the experimental measurements~\cite{martin1952PartIVExperimental} in terms of the non-dimensionalized width $w^{*}(t^{*})$ and height $h^{*}(t^{*})$.
The simulations were performed with a computational domain resolution, that is, initial dam width of $W=200$.
The simulations with the EXT refilling scheme became numerically unstable because the macroscopic velocity gradually increased after refilling, and eventually locally exceeded the lattice speed of sound $c_{s}$, as illustrated in \Cref{fig:dam-break-rectangular-ext-unstable}.
Exceeding the lattice speed of sound is often an effect of a scheme being numerically unstable in the LBM~\cite{kruger2017LatticeBoltzmannMethod}. These numerical instabilities eventually lead to the collapse of the simulation.
Note that these instabilities were not an immediate consequence of a certain cell being refilled with the EXT scheme.
Instead, high macroscopic velocities appeared in the later course of the simulation.
All other refilling schemes produced results of similar accuracy and agreed with the trend of the experimental observations for $h^{*}(t^{*})$.
The simulated dam width $w^{*}(t^{*})$ generally agreed better with the experimental measurements than the simulated height.
However, the choice of the refilling scheme had more effect on $w^{*}(t^{*})$.
The AVG scheme was the most accurate, while the EQ and GEQ schemes were less accurate.
Although the EQ+NEQ scheme also produced accurate results, it temporarily deviated more significantly from the experimental data at $t^{*}\approx 3.7$.
Except for the unstable simulation with the EXT refilling scheme, the simulated dam contours at $t^{*} = 3$ are visualized in \Cref{fig:dam-break-rectangular-mesh}.
While there are noticeable differences between the schemes, no accuracy assessment could be made due to the lack of suitable experimental reference data.
\par

As shown in \Cref{fig:dam-break-rectangular-convergence}, the simulation results presented in this section were converged in terms of computational domain resolution.
A resolution equivalent to $W=50$ was sufficient to reasonably agree with the experimental data.
\par

\subsection{Cylindrical dam break}\label{sec:ne-cdb}
In this section, the simulation setup and results for a cylindrical dam break are presented.
The numerical simulations resemble the laboratory experiments from Martin and Moyce~\cite{martin1952PartIVExperimental}.
This test case was chosen to evaluate whether the model's isotropy is affected by the choice of the refilling scheme.
\par

\subsubsection{Simulation setup}\label{sec:ne-cdb-ss}
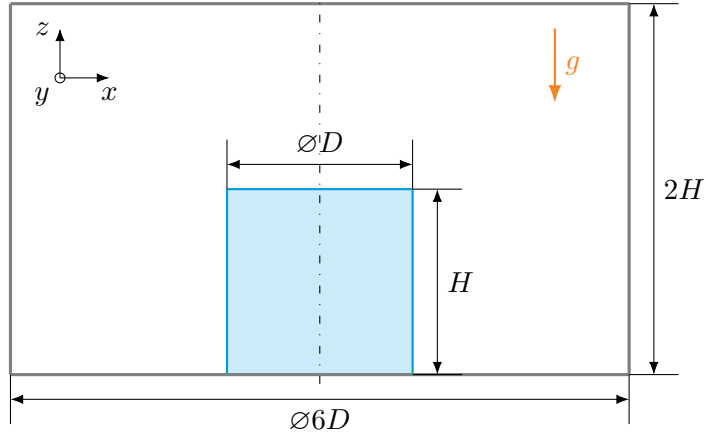
\begin{figure}[htbp]
	\centering
	\setlength{\figureheight}{0.3\textwidth}
	\setlength{\figurewidth}{0.5\textwidth}
	\setlength\mdist{0.02\textwidth}
	\begin{tikzpicture}
\definecolor{darkorange24213334}{RGB}{242,133,34}
\definecolor{dodgerblue0154222}{RGB}{0,154,222}

% liquid column
\draw[thick, fill=dodgerblue0154222!20, draw=none] (0.5\figurewidth-0.25\figureheight,0) rectangle (0.5\figurewidth+0.25\figureheight,0.5\figureheight);
\draw[thick, dodgerblue0154222] (0.5\figurewidth-0.25\figureheight,0.5\figureheight)--(0.5\figurewidth+0.25\figureheight,0.5\figureheight);
\draw[thick, dodgerblue0154222] (0.5\figurewidth-0.25\figureheight,0)--(0.5\figurewidth-0.25\figureheight,0.5\figureheight);
\draw[thick, dodgerblue0154222] (0.5\figurewidth+0.25\figureheight,0)--(0.5\figurewidth+0.25\figureheight,0.5\figureheight);

% borders left and right
\draw[very thick, black!50] (0,0)--(0,\figureheight);
\draw[very thick, black!50] (\figurewidth,0)--(\figurewidth,\figureheight);

% borders top and bottom
\draw[very thick, black!50] (0,\figureheight)--(\figurewidth,\figureheight);
\draw[very thick, black!50] (0,0)--(\figurewidth,0);

% domain height
\draw[<->, >=Latex] (\figurewidth+\mdist,0)--(\figurewidth+\mdist,\figureheight) node [pos=0.5,right] {$2H$};
\draw[-] (\figurewidth,0)--(\figurewidth+2\mdist,0);
\draw[-] (\figurewidth,\figureheight)--(\figurewidth+2\mdist,\figureheight);

% domain width
\draw[<->, >=Latex] (0,-\mdist)--(\figurewidth,-\mdist) node [pos=0.5,below] {$\varnothing 6D$};
\draw[-] (0,0)--(0,-2\mdist);
\draw[-] (\figurewidth,0)--(\figurewidth,-2\mdist);

% liquid column height
\draw[<->, >=Latex] (0.5\figurewidth+0.25\figureheight+\mdist,0)--(0.5\figurewidth+0.25\figureheight+\mdist,0.5\figureheight) node [pos=0.5,right] {$H$};
\draw[-] (0.5\figurewidth+0.25\figureheight,0.5\figureheight)--(0.5\figurewidth+0.25\figureheight+2\mdist,0.5\figureheight);
\draw[-] (0.5\figurewidth+0.25\figureheight,0)--(0.5\figurewidth+0.25\figureheight+2\mdist,0);

% liquid column diameter
\draw[<->, >=Latex] (0.5\figurewidth-0.25\figureheight,0.5\figureheight+\mdist)--(0.5\figurewidth+0.25\figureheight,0.5\figureheight+\mdist) node [pos=0.5,above] {$\varnothing D$};
\draw[-] (0.5\figurewidth-0.25\figureheight,0.5\figureheight)--(0.5\figurewidth-0.25\figureheight,0.5\figureheight+2\mdist);
\draw[-] (0.5\figurewidth+0.25\figureheight,0.5\figureheight)--(0.5\figurewidth+0.25\figureheight,0.5\figureheight+2\mdist);

% vertical line of symmetry
\draw[loosely dashdotted] (0.5\figurewidth,-0.025\figureheight)--(0.5\figurewidth,1.025\figureheight);

% gravity
\draw[thick, ->, >=Latex, darkorange24213334] (\figurewidth-3\mdist,\figureheight-\mdist)--(\figurewidth-3\mdist,\figureheight-4\mdist) node [pos=0.5,right] {$g$};

% coordinate system
\draw[->, >=Latex] (2\mdist,\figureheight-3\mdist)--(4\mdist,\figureheight-3\mdist) node [pos=1,below] {$x$};
\draw[->, >=Latex] (2\mdist,\figureheight-3\mdist)--(2\mdist,\figureheight-1\mdist) node [pos=1,left] {$z$};
\draw[draw=black] (2\mdist,\figureheight-3\mdist) circle [radius=0.2\mdist] node[opacity=1, below left] {$y$};
\end{tikzpicture}%
	\caption{
		\label{fig:dam-break-cylindrical-setup}
		Simulation setup of the three-dimensional cylindrical dam break test case.
		A cylindrical liquid column of diameter $D$ and height $H$ was initialized in the domain's center.
		It collapsed due to the gravitational acceleration $g$ that acted in negative $z$-direction.
		Free-slip boundary conditions were set at all domain borders.
		C.~Schwarzmeier, U.~Rüde, Analysis and comparison of boundary condition variants in the free-surface lattice Boltzmann method, arXiv preprint~\cite{schwarzmeier2022AnalysisComparisonBoundary}, 2022; licensed under a Creative Commons Attribution (CC BY) license.
	}
\end{figure}

As illustrated in \Cref{fig:dam-break-cylindrical-setup}, a cylindrical liquid column of diameter $D \in \{50, 100, 200\}$ lattice cells and height $H=D$ was placed in the center of a three-dimensional domain of size $6D \times 6D \times 2H$ ($x$-, $y$-, $z$-direction).
The setup's remaining configuration was similar to the one of the rectangular dam break in \Cref{sec:ne-rdb-ss}.
However, in the definitions of the Galilei~\eqref{eq:ne-rdb-ss-ga} and Bond number~\eqref{eq:ne-rdb-ss-bo}, the characteristic length $0.5D$ was used.
\par

The liquid column's radius $r(t)$ was monitored during the simulation. 
It was obtained by finding the distance of the liquid front to the column's initial center of symmetry.
The liquid column's collapse can not be assumed to be perfectly symmetric.
Consequently, $r(t)$ was computed for every interface cell detected by a \textit{seed-fill} algorithm~\cite{pavlidis1982AlgorithmsGraphicsImage}.
The starting point of this algorithm was set to an arbitrary domain boundary.
With this configuration, the algorithm only detected the outermost interface cells, that is, only interface cells at the spreading liquid's front.
A statistical sample was used to evaluate $r(t)$ by computing the maximum, minimum, and mean values of $r(t)$ at every $t^{*}=0.01$.
The radius $r^{*}(t) \coloneqq 2r(t)/D$ and time $t^{*} \coloneqq t \sqrt{4g/D}$ were non-dimensionalized as suggested in the reference data~\cite{martin1952PartIVExperimental}.
Conforming with the experimental data~\cite{martin1952PartIVExperimental}, the simulations were stopped at $r_{\text{max}}^{*}(t^{*}) \geq 4.33$ with the non-dimensionalized maximum liquid front radius $r_{\text{max}}^{*}(t^{*})$.
\par

\subsubsection{Results and discussion}\label{sec:ne-cdb-res}
\begin{figure}[htbp]
	\centering
	\setlength{\figureheight}{0.5\textwidth}
	\setlength{\figurewidth}{0.95\textwidth}
	\input{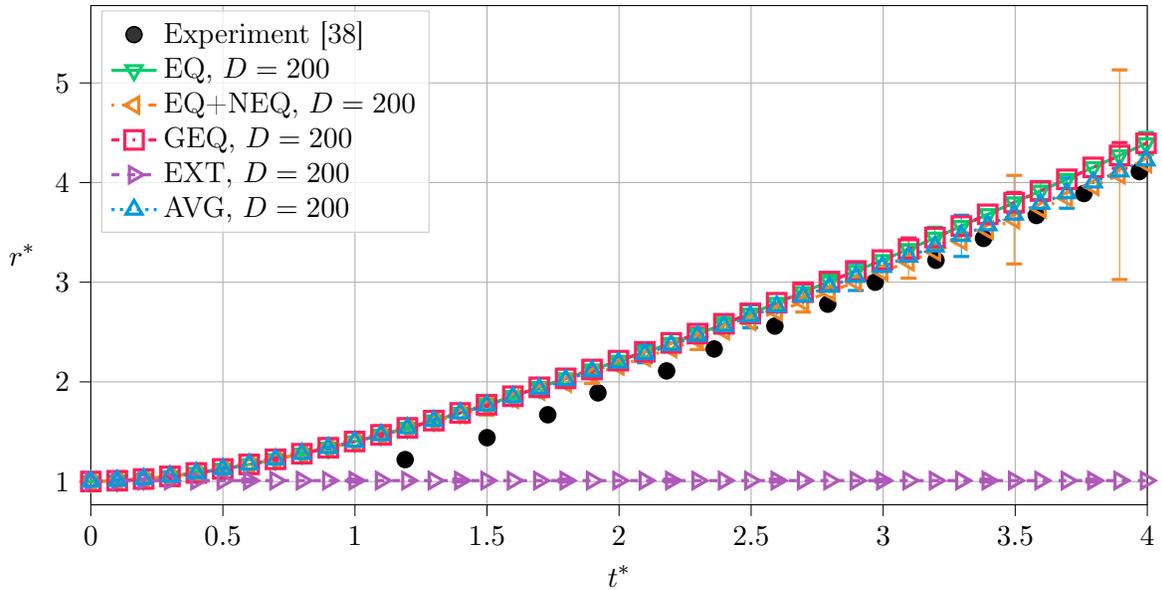}%
	\caption{
		\label{fig:rot-dam-break-200}
		Simulated cylindrical dam break with non-dimensionalized liquid column radius $r^{*}(t^{*})$ and time $t^{*}$.
		The simulations were performed with a computational domain resolution, that is, initial column diameter of $D=200$ lattice cells.
		The markers represent the mean value of $r^{*}(t^{*})$, and the error bars indicate its maximum and minimum.
		The EXT refilling scheme was numerically unstable.
		The large error bars with the EQ+NEQ scheme were caused by tiny droplets detaching from the liquid column.
		These droplets were detected by the evaluation algorithm and considered part of the liquid surge front.
		All other schemes generally agreed well with the experimental data~\cite{martin1952PartIVExperimental} and were of comparable accuracy.
	}
\end{figure}

\begin{figure}[htbp]
	\centering
	\setlength\mdist{0.02\textwidth}
	\begin{subfigure}[htbp]{0.49\textwidth}
		\centering
		\includegraphics[width=\textwidth]{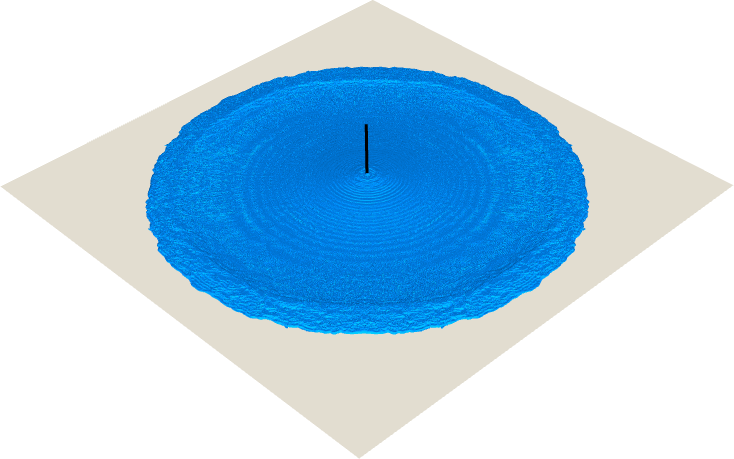}%
		\caption{\label{fig:rot-dam-break-mesh-eq}EQ}
	\end{subfigure}
	\hfill
	\begin{subfigure}[htbp]{0.49\textwidth}
		\centering
		\includegraphics[width=\textwidth]{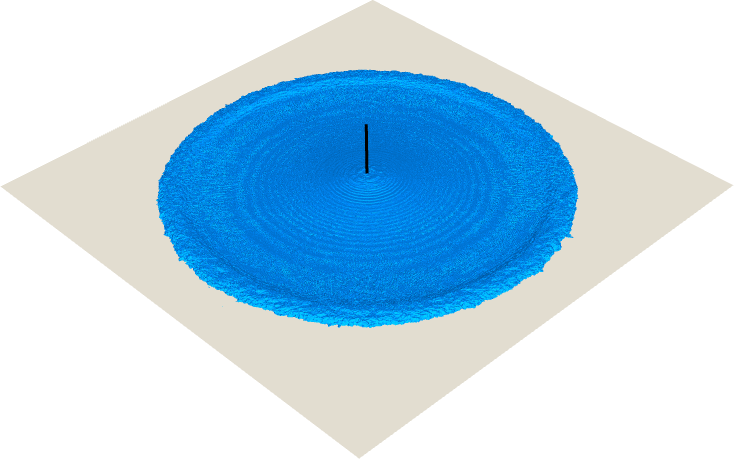}%
		\caption{\label{fig:rot-dam-break-mesh-eq-neq}EQ+NEQ}
	\end{subfigure}%
	\vspace{0.5cm}
	\begin{subfigure}[htbp]{0.49\textwidth}
		\centering
		\includegraphics[width=\textwidth]{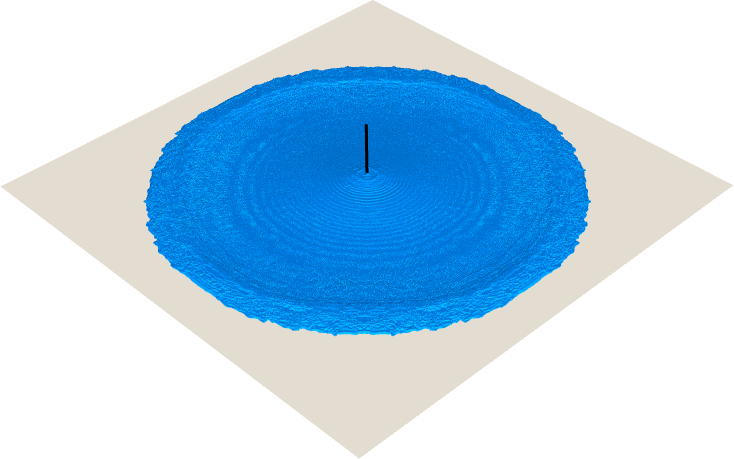}%
		\caption{\label{fig:rot-dam-break-mesh-geq}GEQ}
	\end{subfigure}
	\hfill
	\begin{subfigure}[htbp]{0.49\textwidth}
		\centering
		\includegraphics[width=\textwidth]{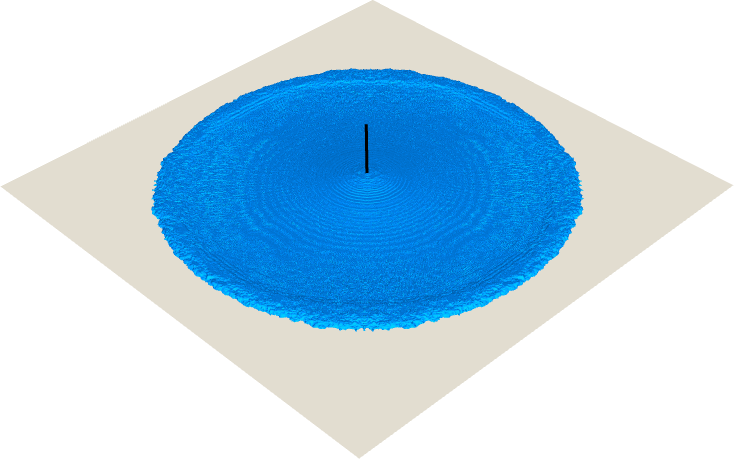}%
		\caption{\label{fig:rot-dam-break-mesh-avg}AVG}
	\end{subfigure}
	\caption{
		\label{fig:rot-dam-break-mesh}
		Shape of the simulated cylindrical dam break at non-dimensionalized time $t^{*}=4$ with the EQ (a), EQ+NEQ (b), GEQ (c), and AVG (d) refilling schemes.
		The simulations were performed with an initial column diameter of $D=200$ lattice cells.
		The black line indicates the column's initial center of symmetry.
		The simulation with the EXT refilling scheme was numerically unstable and is not included here.
		All other schemes kept their initial center of symmetry and remained rotationally symmetric.
		There are slight differences between the individual refilling schemes.
		Owing to the lack of reference data, no accuracy assessment could be made in terms of shape.
	}
\end{figure}

\Cref{fig:rot-dam-break-200} shows the simulation results with an initial liquid column diameter of $D=200$ lattice cells.
The markers show the mean value of the non-dimensionalized radius $r^{*}(t^{*})$.
The error bars represent the maximum and minimum value of $r^{*}(t^{*})$ over time $t^{*}$, respectively.
Large error bars indicate a deviation from rotational symmetry.
As for the breaking dam test case in \Cref{sec:ne-rdb}, the EXT refilling scheme was numerically unstable.
All other refilling schemes generally agreed well with the experimental data~\cite{martin1952PartIVExperimental}.
Although the error bars for the EQ+NEQ scheme indicate asymmetry, the liquid spread's front remained qualitatively symmetric as shown in \Cref{fig:rot-dam-break-mesh}.
However, tiny droplets detached from the main liquid spread.
The evaluation algorithm detected these droplets as part of the liquid surge front.
The EQ and GEQ schemes are approximately of equal accuracy, while being less accurate than the AVG scheme.
The shape of the collapsing liquid column at time $t^{*}=4$ is visualized in \Cref{fig:rot-dam-break-mesh}.
The solid black lines indicate the initial center of symmetry.
In general, all refilling schemes remained rotationally symmetric and did not move from their initial center of symmetry.
\par

As illustrated in \Cref{fig:dam-break-cylindrical-convergence}, the simulation results shown in this section have converged moderately well with increasing computational domain resolution.
For $D=100$, the EQ+NEQ and AVG schemes became numerically unstable.
These instabilities were caused by droplets detaching from the liquid front and exceeding the lattice speed of sound, as discussed in \Cref{sec:ne-rdb-res}.
\par

\FloatBarrier

\subsection{Taylor bubble}\label{sec:ne-tb}
A Taylor bubble is a gas bubble that rises in a cylindrical tube filled with a stagnant liquid due to buoyancy forces.
It has an elongated shape and a round leading edge with a length of multiple times its diameter.
The simulation results were compared to the experimental data from Bugg and Saad~\cite{bugg2002VelocityFieldTaylor}.
\par

\subsubsection{Simulation setup}\label{sec:ne-tb-ss}
The simulation setup is illustrated in \Cref{fig:taylor-bubble-setup} and resembled that of the reference experiments~\cite{bugg2002VelocityFieldTaylor}.
A cylindrical tube of diameter $D = \{32, 64, 128\}$ lattice cells pointed in the $z$-direction in a three-dimensional computational domain of size $1D \times 1D \times 10D$ ($x$-, $y$-, $z$-direction).
The tube's walls and all domain walls were set to no-slip boundary conditions.
The inner part of the tube was filled with a stagnant liquid.
In the liquid, there was a cylindrical gas bubble oriented in the $z$-direction with a diameter of $0.75D$ and a length of $3D$.
It was initially located $D$ above the domain's bottom wall with the volumetric gas pressure $p^{\text{V}}(t) = p_{0}$.
The gravitational acceleration $g$ acted in the negative $z$-direction.
Hydrostatic pressure was initialized according to $g$, so the LBM pressure was equivalent to $p_{0}$ at $5D$ in the $z$-direction.
All simulations were performed with the relaxation rate $\omega = 1.8$, conforming with diffusive scaling.
The Morton and Bond number define the fluid mechanics of the system.
The Morton number
\begin{equation}\label{eq:taylor-mo}
\text{Mo} \coloneqq \frac{g \mu^{4}}{\rho \sigma^{3}} = 0.015
\end{equation}
describes the ratio of viscous to surface tension forces with surface tension $\sigma$, dynamic fluid viscosity $\mu$, and liquid density $\rho$.
The Bond number $\text{Bo}=100$ \eqref{eq:ne-rdb-ss-bo}, is used with characteristic length $D$.
The evaluations were performed in terms of the non-dimensionalized bubble radius $r^{*} \coloneqq r/(0.5D)$, axial location $z^{*} \coloneqq z/D$, and time $t^{*} \coloneqq t\sqrt{g/D}$.
\par

\begin{figure}[htbp]
	\centering
	\setlength{\figureheight}{0.65\textwidth}
	\setlength{\figurewidth}{0.175\textwidth}
	\setlength\mdist{0.02\textwidth}
	\begin{tikzpicture}
\definecolor{dodgerblue0154222}{RGB}{0,154,222}
\definecolor{darkorange24213334}{RGB}{242,133,34}
\setlength\radius{0.375\figurewidth}

% background filling
\draw [fill=dodgerblue0154222!30, draw=none] (0,0) rectangle (\figurewidth,\figureheight);

% borders left and right
\draw[very thick, black!50] (0,0)--(0,\figureheight);
\draw[very thick, black!50] (\figurewidth,0)--(\figurewidth,\figureheight);

% borders top and bottom
\draw[very thick, black!50] (0,\figureheight)--(\figurewidth,\figureheight);
\draw[very thick, black!50] (0,0)--(\figurewidth,0);

% domain height
\draw[<->, >=Latex] (\figurewidth+\mdist,0)--(\figurewidth+\mdist,\figureheight) node [pos=0.5,right] {$10D$};
\draw[-] (\figurewidth,0)--(\figurewidth+2\mdist,0);
\draw[-] (\figurewidth,\figureheight)--(\figurewidth+2\mdist,\figureheight);

% domain width
\draw[<->, >=Latex] (0,-\mdist)--(\figurewidth,-\mdist) node [pos=0.5,below] {$\varnothing D$};
\draw[-] (0,0)--(0,-2\mdist);
\draw[-] (\figurewidth,0)--(\figurewidth,-2\mdist);

% cylindrical bubble
\draw [fill=white, draw=dodgerblue0154222, thick] (0.5\figurewidth-\radius,0.1\figureheight) rectangle (0.5\figurewidth+\radius,0.4\figureheight);

% cylinder diameter
\draw[<->, >=Latex] (0.5\figurewidth-\radius,0.4\figureheight+\mdist)--(0.5\figurewidth+\radius,0.4\figureheight+\mdist) node [pos=0.5,above] {$\varnothing 0.75 D$};
\draw[-] (0.5\figurewidth-\radius,0.4\figureheight)--(0.5\figurewidth-\radius,0.4\figureheight+2\mdist);
\draw[-] (0.5\figurewidth+\radius,0.4\figureheight)--(0.5\figurewidth+\radius,0.4\figureheight+2\mdist);

% cylinder position x
\draw[<->, >=Latex] (0.5\figurewidth,0.1\figureheight-\mdist)--(\figurewidth,0.1\figureheight-\mdist) node [pos=0.5,below] {$0.5 D$};
\draw[-] (0.5\figurewidth,0.1\figureheight)--(0.5\figurewidth,0.1\figureheight-2\mdist);
\draw[-] (\figurewidth,0.1\figureheight)--(\figurewidth,0.1\figureheight-2\mdist);

% cylinder position y
\draw[<->, >=Latex] (-\mdist,0)--(-\mdist,0.1\figureheight) node [pos=0.5,left] {$D$};
\draw[-] (-2\mdist,0.1\figureheight)--(0.5\figurewidth-\radius,0.1\figureheight);
\draw[-] (-2\mdist,0)--(0.5\figurewidth-\radius,0);

% cylinder length
\draw[<->, >=Latex] (-\mdist,0.1\figureheight)--(-\mdist,0.4\figureheight) node [pos=0.5,left] {$3D$};
\draw[-] (-2\mdist,0.4\figureheight)--(0.5\figurewidth-\radius,0.4\figureheight);

% vertical line of symmetry
\draw[loosely dashdotted] (0.5\figurewidth,-0.025\figureheight)--(0.5\figurewidth,1.025\figureheight);

% gravity
\draw[thick, ->, >=Latex, darkorange24213334] (\figurewidth-1.5\mdist,\figureheight-\mdist)--(\figurewidth-1.5\mdist,\figureheight-4\mdist) node [pos=0.5,right] {$g$};

% coordinate system
\draw[->, >=Latex] (2\mdist,\figureheight-3\mdist)--(4\mdist,\figureheight-3\mdist) node [below] {$x$};
\draw[->, >=Latex] (2\mdist,\figureheight-3\mdist)--(2\mdist,\figureheight-1\mdist) node [left] {$z$};
\draw[draw=black] (2\mdist,\figureheight-3\mdist) circle [radius=0.175\mdist] node[opacity=1, below left] {$y$};
\end{tikzpicture}%
	\caption{
		\label{fig:taylor-bubble-setup}
		Simulation setup of the three-dimensional Taylor bubble test case.
		A cylindrical gas bubble  was initialized in a cylindrical tube of diameter $D$.
		The gravitational acceleration $g$ acted in the negative $z$-direction.
		No-slip boundary conditions were applied at the tube and all domain walls.
		C.~Schwarzmeier, M.~Holzer, T.~Mitchell, M.~Lehmann, F.~Häusl, U.~Rüde, Comparison of free-surface and conservative Allen--Cahn phase-field lattice Boltzmann method, arXiv preprint~\cite{schwarzmeier2022ComparisonFreeSurface}, 2022; licensed under a Creative Commons Attribution (CC BY) license; the colors were changed from the original.
	}
\end{figure}

\subsubsection{Results and discussion}\label{sec:ne-tb-res}
\begin{figure}[htbp]
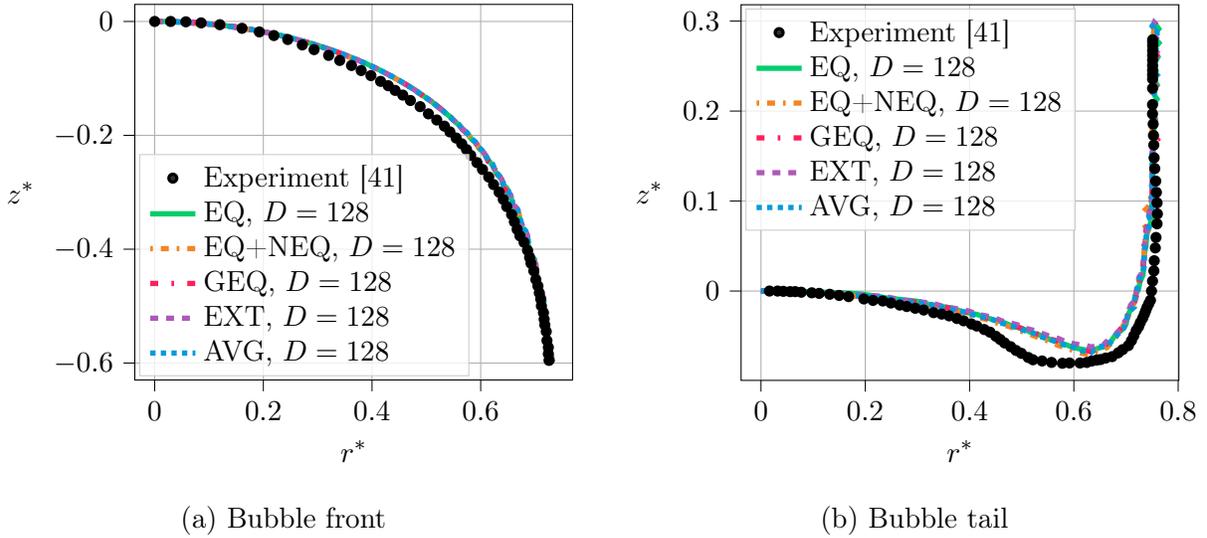

	\centering
	\setlength{\figureheight}{0.4\textwidth}
	\setlength{\figurewidth}{0.45\textwidth}
	\begin{subfigure}[htbp]{0.49\textwidth}
		\centering
		\input{figures/taylor-bubble/d-128-shape-front.tex}%
		\caption{\label{fig:taylor-bubble-shape-front}Bubble front}
	\end{subfigure}
	\hfill
	\begin{subfigure}[htbp]{0.49\textwidth}
		\centering
		\input{figures/taylor-bubble/d-128-shape-tail.tex}%
		\caption{\label{fig:taylor-bubble-shape-tail}Bubble tail}
	\end{subfigure}
	\caption{
		\label{fig:taylor-bubble-shape}
		Simulated shape of the Taylor bubble's front (a) and tail (b).
		The simulations were performed with a computational domain resolution, that is, tube diameter of $D=128$ lattice cells.
		The comparison with experimental data~\cite{bugg2002VelocityFieldTaylor} is drawn in terms of the non-dimensionalized axial location $z^{*}$ and radial location $r^{*}$ at time $t^{*}=15$.
		The refilling schemes had almost no impact on the shape of the simulated Taylor bubble.
	}
\end{figure}

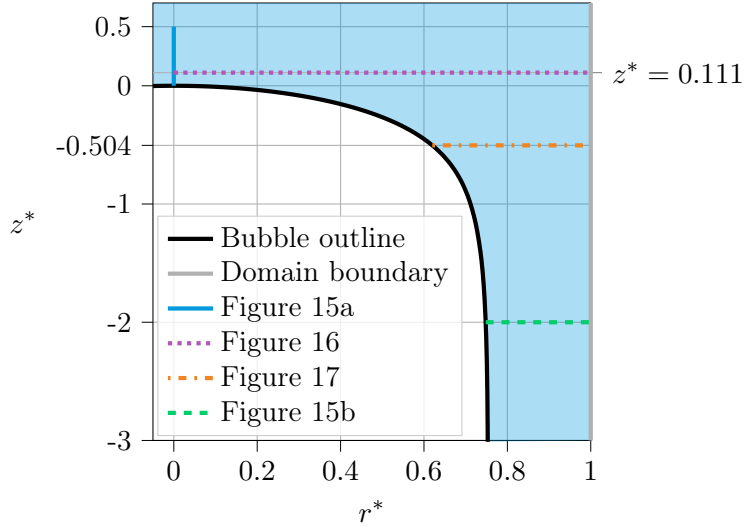
\begin{figure}[htbp]
	\centering
	\setlength{\figureheight}{0.45\textwidth}
	\setlength{\figurewidth}{0.45\textwidth}
	\begin{tikzpicture}
	
	\definecolor{crimson2553191}{RGB}{255,31,91}
	\definecolor{darkgray176}{RGB}{176,176,176}
	\definecolor{darkorange24213334}{RGB}{242,133,34}
	\definecolor{dodgerblue0154222}{RGB}{0,154,222}
	\definecolor{lightgray204}{RGB}{204,204,204}
	\definecolor{mediumorchid17588186}{RGB}{175,88,186}
	\definecolor{springgreen0205108}{RGB}{0,205,108}
	
	\begin{axis}[
	height=\figureheight,
	width=\figurewidth,
	axis y line*=right,
	axis x line=none,
	ylabel style={rotate=-90.0},
	xmin=0, xmax=0,
	ymin=-3, ymax=0.7,
	ytick={0.111},
	yticklabels = {$z^{*}=0.111$},
	tick align=outside,
	tick pos=right,
	y grid style={darkgray176},
	ymajorgrids,
	]
	\end{axis}
	
	\begin{axis}[
	height=\figureheight,
	legend cell align={left},
	legend style={
		fill opacity=0.8,
		draw opacity=1,
		text opacity=1,
		at={(0.01,0.01)},
		anchor=south west,
		draw=lightgray204
	},
	tick align=outside,
	tick pos=left,
	width=\figurewidth,
	x grid style={darkgray176},
	xlabel={\(\displaystyle r^{*}\)},
	xmajorgrids,
	xmin=-0.05, xmax=1,
	xtick style={color=black},
	y grid style={darkgray176},
	ylabel style={rotate=-90.0},
	ylabel={\(\displaystyle z^{*}\)},
	ymajorgrids,
	ymin=-3, ymax=0.7,
	ytick style={color=black},
	ytick={-3, -2, -1, -0.504, 0, 0.5},
	yticklabels = {-3,-2,-1,-0.504, 0, 0.5},
	yticklabel style={
		/pgf/number format/fixed,
		/pgf/number format/precision=3},
	clip=false,
	]
	\addplot [ultra thick, black, name path=bubble]
	table {%
		-0.05 -0.00331425666809082
		-0.0468618869781494 -0.0018768310546875
		-0.0312368869781494 -0.000843405723571777	
		-0.0156118869781494 -0.000216245651245117
		-1.31130218505859e-05 0
		0.0156118869781494 -0.000216245651245117
		0.0312368869781494 -0.000843405723571777
		0.0468618869781494 -0.0018768310546875
		0.0624868869781494 -0.00331425666809082
		0.0781118869781494 -0.00516211986541748
		0.0937368869781494 -0.00744259357452393
		0.11590051651001 -0.0115705728530884
		0.124986886978149 -0.0134730339050293
		0.140611886978149 -0.0170856714248657
		0.156236886978149 -0.0210834741592407
		0.176718354225159 -0.0271955728530884
		0.203111886978149 -0.0361255407333374
		0.2204270362854 -0.0428205728530884
		0.234361886978149 -0.0486605167388916
		0.255831003189087 -0.0584455728530884
		0.265611886978149 -0.0632587671279907
		0.286195278167725 -0.0740705728530884
		0.296861886978149 -0.0800961256027222
		0.312974691390991 -0.0896955728530884
		0.336896896362305 -0.105320572853088
		0.343736886978149 -0.110080242156982
		0.359361886978149 -0.121415257453918
		0.378740310668945 -0.136570572853088
		0.397185802459717 -0.152195572853088
		0.414325714111328 -0.167820572853088
		0.430325031280518 -0.183445572853088
		0.445314407348633 -0.199070572853088
		0.459401726722717 -0.214695572853088
		0.472667455673218 -0.230320572853088
		0.485145807266235 -0.245945572853088
		0.499986886978149 -0.265821099281311
		0.515611886978149 -0.288233757019043
		0.518711566925049 -0.292820572853088
		0.531236886978149 -0.312572002410889
		0.538240194320679 -0.324070572853088
		0.547301888465881 -0.339695572853088
		0.564130306243896 -0.370945572853088
		0.579358339309692 -0.402195572853088
		0.593736886978149 -0.434839367866516
		0.599634408950806 -0.449070572853088
		0.611665010452271 -0.480320572853088
		0.62250554561615 -0.511570572853088
		0.627724647521973 -0.527195572853088
		0.637151956558228 -0.558445572853088
		0.645975351333618 -0.589695572853088
		0.657675385475159 -0.636570572853088
		0.664606094360352 -0.667820572853088
		0.674072861671448 -0.714695572853088
		0.679645299911499 -0.745945572853088
		0.687486886978149 -0.794471502304077
		0.691862821578979 -0.824070572853088
		0.69793701171875 -0.870945572853088
		0.705211281776428 -0.933445572853088
		0.709868550300598 -0.980320572853088
		0.715261220932007 -1.04282057285309
		0.720094203948975 -1.10532057285309
		0.724194288253784 -1.16782057285309
		0.727686405181885 -1.23032057285309
		0.731468200683594 -1.30844557285309
		0.735384106636047 -1.40219557285309
		0.738070726394653 -1.48032057285309
		0.740710735321045 -1.57407057285309
		0.743194937705994 -1.68344557285309
		0.745447278022766 -1.80844557285309
		0.747605800628662 -1.96469557285309
		0.749356508255005 -2.1365704536438
		0.750810861587524 -2.3396954536438
		0.75194787979126 -2.5896954536438
		0.752784967422485 -2.9178204536438
		0.752936363220215 -3.0115704536438
	};
	\addlegendentry{Bubble outline}
	\addplot [ultra thick, darkgray176]
	table {%
		1 -3
		1 0.7
	};
	\addlegendentry{Domain boundary}
	\addplot [ultra thick, dodgerblue0154222]
	table {%
		0 0
		0 0.5
	};
	\addlegendentry{\Cref{fig:taylor-bubble-0.5}}
	\addplot [ultra thick, mediumorchid17588186, dotted]
	table {%
		0 0.111000061035156
		1 0.111000061035156
	};
	\addlegendentry{\Cref{fig:taylor-bubble-0.111}}
	\addplot [ultra thick, darkorange24213334, dash pattern=on 1pt off 3pt on 3pt off 3pt]
	table {%
		0.620000004768372 -0.503999948501587
		1 -0.503999948501587
	};
	\addlegendentry{\Cref{fig:taylor-bubble-0.504}}
	\addplot [ultra thick, springgreen0205108, dashed]
	table {%
		0.748000025749207 -2
		1 -2
	};
	\addlegendentry{\Cref{fig:taylor-bubble-2}}

	\path[name path=xaxis] (-0.05,0.7) -- (1,0.7) -- (1,-3);
	\addplot [fill=dodgerblue0154222, fill opacity=0.33] fill between[of=bubble and xaxis];
	\end{axis}
\end{tikzpicture}%	
	\caption{
		\label{fig:taylor-bubble-schematic}
		Evaluation locations of the velocity profiles at the Taylor bubble's front, as presented in the subsequent figures.
		The axial location $z^{*} = z/D$ and radial location $r^{*} = r/(0.5D)$ are non-dimensionalized.
		The evaluations were performed at a centrally located cross-section with normal in the $x$-direction.
		C.~Schwarzmeier, M.~Holzer, T.~Mitchell, M.~Lehmann, F.~Häusl, U.~Rüde, Comparison of free-surface and conservative Allen--Cahn phase-field lattice Boltzmann method, arXiv preprint~\cite{schwarzmeier2022ComparisonFreeSurface}, 2022; licensed under a Creative Commons Attribution (CC BY) license; the colors were changed from the original.
	}
\end{figure}

\begin{figure}[htbp]
	\centering
	\setlength{\figureheight}{0.5\textwidth}
	\setlength{\figurewidth}{0.45\textwidth}
	\begin{subfigure}[htbp]{0.49\textwidth}
		\centering
		\input{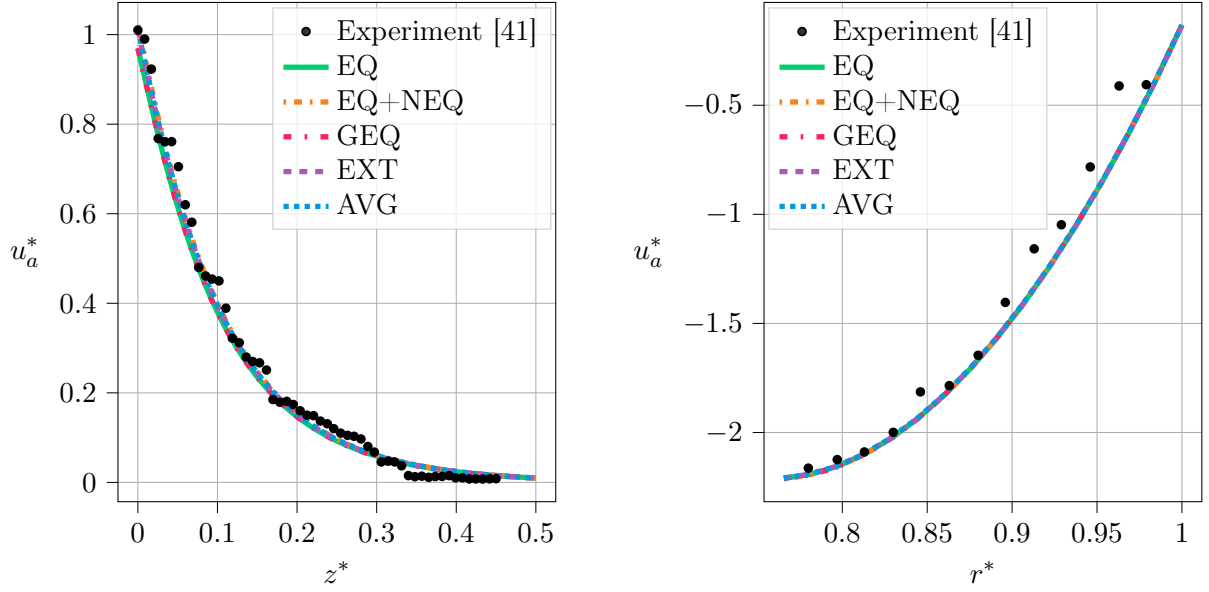}%
		\caption{\label{fig:taylor-bubble-0.5}Axial line with length $0.5D$ from the Taylor bubble's front.}
	\end{subfigure}%
	\hfill
	\begin{subfigure}[htbp]{0.49\textwidth}
		\centering
		% This file was created with tikzplotlib v0.10.1.
\begin{tikzpicture}

\definecolor{crimson2553191}{RGB}{255,31,91}
\definecolor{darkgray176}{RGB}{176,176,176}
\definecolor{darkorange24213334}{RGB}{242,133,34}
\definecolor{dodgerblue0154222}{RGB}{0,154,222}
\definecolor{lightgray204}{RGB}{204,204,204}
\definecolor{mediumorchid17588186}{RGB}{175,88,186}
\definecolor{springgreen0205108}{RGB}{0,205,108}

\begin{axis}[
height=\figureheight,
legend cell align={left},
legend style={
	fill opacity=0.8,
	draw opacity=1,
	text opacity=1,
	at={(0.01,0.99)},
	anchor=north west,
	draw=lightgray204
},
tick align=outside,
tick pos=left,
width=\figurewidth,
x grid style={darkgray176},
xlabel style={align=center},
xlabel={\(\displaystyle r^{*}\)},
xmajorgrids,
xmin=0.75390625, xmax=1.01171875,
xtick style={color=black},
y grid style={darkgray176},
ylabel style={rotate=-90.0},
ylabel={\(\displaystyle u_{a}^{*}\)},
ymajorgrids,
ymin=-2.31709904868623, ymax=-0.0287602760835631,
ytick style={color=black}
]
\addplot [semithick, black, mark=*, mark size=1.5, mark options={solid}, only marks]
table {%
	0.779999971389771 -2.16319990158081
	0.796999931335449 -2.12409996986389
	0.812999963760376 -2.0890998840332
	0.829999923706055 -1.99909996986389
	0.845999956130981 -1.81369996070862
	0.86299991607666 -1.7849999666214
	0.879999995231628 -1.64610004425049
	0.896000027656555 -1.40359997749329
	0.912999987602234 -1.15820002555847
	0.929000020027161 -1.04789996147156
	0.945999979972839 -0.782799959182739
	0.963000059127808 -0.411499977111816
	0.978999972343445 -0.405900001525879
};
\addlegendentry{Experiment~\cite{bugg2002VelocityFieldTaylor}}
\addplot [line width=2pt, springgreen0205108, mark=triangle, mark size=0, mark options={solid,rotate=180,fill opacity=0}]
table {%
	0.765625 -2.20958638191223
	0.78125 -2.19142127037048
	0.796875 -2.15509653091431
	0.8125 -2.10124754905701
	0.828125 -2.02944684028625
	0.84375 -1.94004201889038
	0.859375 -1.83373749256134
	0.875 -1.71063494682312
	0.890625 -1.57067549228668
	0.90625 -1.41432905197144
	0.921875 -1.24187326431274
	0.9375 -1.05325746536255
	0.953125 -0.848390817642212
	0.96875 -0.627115726470947
	0.984375 -0.388885140419006
	1 -0.132942318916321
};
\addlegendentry{EQ}
\addplot [line width=2pt, darkorange24213334, dash pattern=on 1pt off 3pt on 3pt off 3pt, mark=triangle, mark size=0, mark options={solid,rotate=90,fill opacity=0}]
table {%
	0.765625 -2.21308374404907
	0.78125 -2.19469690322876
	0.796875 -2.15834832191467
	0.8125 -2.10378360748291
	0.828125 -2.03177237510681
	0.84375 -1.94230329990387
	0.859375 -1.83566248416901
	0.875 -1.71208465099335
	0.890625 -1.571906208992
	0.90625 -1.415367603302
	0.921875 -1.24262619018555
	0.9375 -1.05382871627808
	0.953125 -0.848855257034302
	0.96875 -0.62745213508606
	0.984375 -0.389074683189392
	1 -0.132996201515198
};
\addlegendentry{EQ+NEQ}
\addplot [line width=2pt, crimson2553191, dash pattern=on 3pt off 5pt on 1pt off 5pt, mark=square, mark size=0, mark options={solid,fill opacity=0}]
table {%
	0.765625 -2.20936727523804
	0.78125 -2.1947877407074
	0.796875 -2.15796113014221
	0.8125 -2.10133290290833
	0.828125 -2.02994084358215
	0.84375 -1.94068014621735
	0.859375 -1.83348119258881
	0.875 -1.70992743968964
	0.890625 -1.56997871398926
	0.90625 -1.4134373664856
	0.921875 -1.24081337451935
	0.9375 -1.05225098133087
	0.953125 -0.84759259223938
	0.96875 -0.626523017883301
	0.984375 -0.388564825057983
	1 -0.13286828994751
};
\addlegendentry{GEQ}
\addplot [line width=2pt, mediumorchid17588186, dashed, mark=triangle, mark size=0, mark options={solid,rotate=270,fill opacity=0}]
table {%
	0.765625 -2.20881199836731
	0.78125 -2.19073247909546
	0.796875 -2.15449571609497
	0.8125 -2.10011005401611
	0.828125 -2.02808260917664
	0.84375 -1.938800573349
	0.859375 -1.83223450183868
	0.875 -1.70888090133667
	0.890625 -1.56896948814392
	0.90625 -1.41268718242645
	0.921875 -1.24029278755188
	0.9375 -1.05182909965515
	0.953125 -0.847269535064697
	0.96875 -0.626333475112915
	0.984375 -0.388425946235657
	1 -0.132793307304382
};
\addlegendentry{EXT}
\addplot [line width=2pt, dodgerblue0154222, dotted, mark=triangle, mark size=0, mark options={solid,fill opacity=0}]
table {%
	0.765625 -2.20914578437805
	0.78125 -2.19003033638
	0.796875 -2.15388965606689
	0.8125 -2.09965085983276
	0.828125 -2.02708411216736
	0.84375 -1.93746840953827
	0.859375 -1.83087027072906
	0.875 -1.70757269859314
	0.890625 -1.5677307844162
	0.90625 -1.4116085767746
	0.921875 -1.2394437789917
	0.9375 -1.05118572711945
	0.953125 -0.846839666366577
	0.96875 -0.6260826587677
	0.984375 -0.388327121734619
	1 -0.132775664329529
};
\addlegendentry{AVG}
\end{axis}
\end{tikzpicture}%
		\caption{\label{fig:taylor-bubble-2}Radial line at $-2D$ from the Taylor bubble's front.}
	\end{subfigure}%
	\caption{
		\label{fig:taylor-bubble-axial-radial}
		Simulated non-dimensionalized axial velocity $u_{a}^{*}$ along the axial (a) and radial (b) monitoring-lines as defined in \Cref{fig:taylor-bubble-schematic}.
		The comparison with experimental data~\cite{bugg2002VelocityFieldTaylor} is drawn in terms of the non-dimensionalized axial location $z^{*}$ and radial location $r^{*}$ at dimensionless time $t^{*}=15$ with tube diameter $D=128$ lattice cells.
	}
\end{figure}

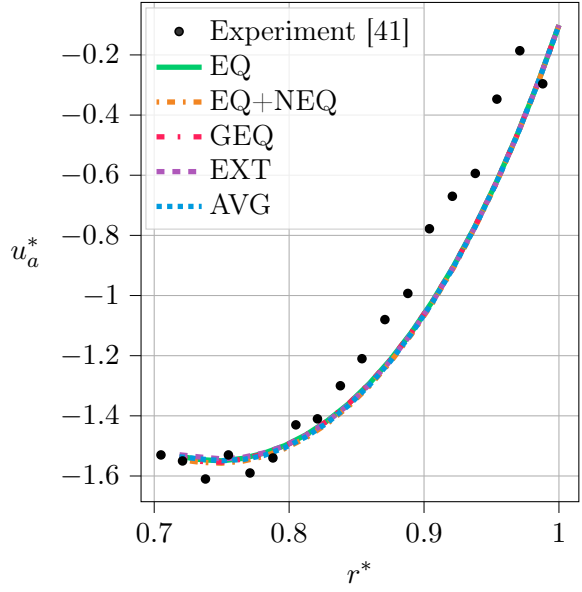
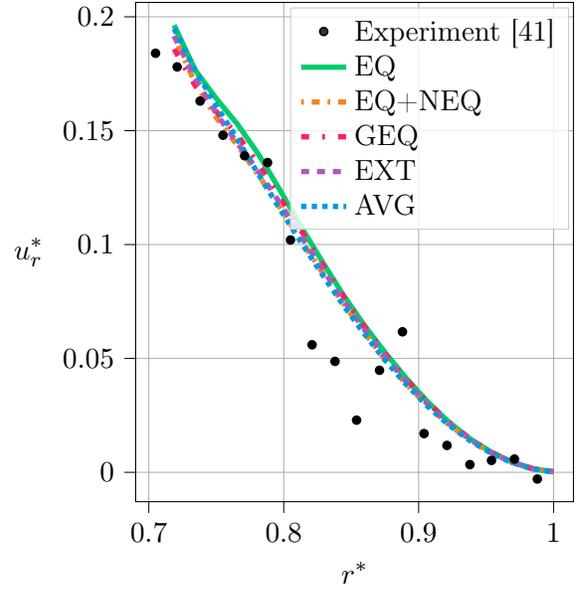
\begin{figure}[htbp]
	\centering
	\setlength{\figureheight}{0.5\textwidth}
	\setlength{\figurewidth}{0.45\textwidth}
	\begin{subfigure}[htbp]{0.49\textwidth}
		\centering
		% This file was created with tikzplotlib v0.10.1.
\begin{tikzpicture}

\definecolor{crimson2553191}{RGB}{255,31,91}
\definecolor{darkgray176}{RGB}{176,176,176}
\definecolor{darkorange24213334}{RGB}{242,133,34}
\definecolor{dodgerblue0154222}{RGB}{0,154,222}
\definecolor{lightgray204}{RGB}{204,204,204}
\definecolor{mediumorchid17588186}{RGB}{175,88,186}
\definecolor{springgreen0205108}{RGB}{0,205,108}

\begin{axis}[
height=\figureheight,
legend cell align={left},
legend style={
	fill opacity=0.8,
	draw opacity=1,
	text opacity=1,
	at={(0.01,0.99)},
	anchor=north west,
	draw=lightgray204
},
tick align=outside,
tick pos=left,
width=\figurewidth,
x grid style={darkgray176},
xlabel style={align=center},
xlabel={\(\displaystyle r^{*}\)},
xmajorgrids,
xmin=0.69025, xmax=1.01475,
xtick style={color=black},
y grid style={darkgray176},
ylabel style={rotate=-90.0},
ylabel={\(\displaystyle u_{a}^{*}\)},
ymajorgrids,
ymin=-1.68545291342337, ymax=-0.0254888181093138,
ytick style={color=black}
]
\addplot [semithick, black, mark=*, mark size=1.5, mark options={solid}, only marks]
table {%
	0.704999923706055 -1.52999997138977
	0.720999956130981 -1.54999995231628
	0.73799991607666 -1.61000001430511
	0.754999995231628 -1.52999997138977
	0.771000027656555 -1.5900000333786
	0.787999987602234 -1.53999996185303
	0.805000066757202 -1.42999994754791
	0.820999979972839 -1.4099999666214
	0.838000059127808 -1.29999995231628
	0.853999972343445 -1.21000003814697
	0.871000051498413 -1.08000004291534
	0.888000011444092 -0.993000030517578
	0.904000043869019 -0.777999997138977
	0.921000003814697 -0.670000076293945
	0.937999963760376 -0.593999981880188
	0.953999996185303 -0.347000002861023
	0.970999956130981 -0.185999989509583
	0.98799991607666 -0.296000003814697
};
\addlegendentry{Experiment~\cite{bugg2002VelocityFieldTaylor}}
\addplot [line width=2pt, springgreen0205108, mark=triangle, mark size=0, mark options={solid,rotate=180,fill opacity=0}]
table {%
	0.71875 -1.53462839126587
	0.734375 -1.54563581943512
	0.75 -1.54960572719574
	0.765625 -1.54152166843414
	0.78125 -1.52435517311096
	0.796875 -1.4988135099411
	0.8125 -1.46304142475128
	0.828125 -1.41723906993866
	0.84375 -1.36088848114014
	0.859375 -1.29299116134644
	0.875 -1.21380603313446
	0.890625 -1.12266314029694
	0.90625 -1.01910316944122
	0.921875 -0.902506828308105
	0.9375 -0.772350668907166
	0.953125 -0.627924680709839
	0.96875 -0.46860682964325
	0.984375 -0.293307781219482
	1 -0.100941777229309
};
\addlegendentry{EQ}
\addplot [line width=2pt, darkorange24213334, dash pattern=on 1pt off 3pt on 3pt off 3pt, mark=triangle, mark size=0, mark options={solid,rotate=90,fill opacity=0}]
table {%
	0.71875 -1.54545426368713
	0.734375 -1.55465698242188
	0.75 -1.55743610858917
	0.765625 -1.55079114437103
	0.78125 -1.53500747680664
	0.796875 -1.50891327857971
	0.8125 -1.47303438186646
	0.828125 -1.42653250694275
	0.84375 -1.36974942684174
	0.859375 -1.30150496959686
	0.875 -1.2215484380722
	0.890625 -1.1295884847641
	0.90625 -1.02510750293732
	0.921875 -0.907571315765381
	0.9375 -0.776502132415771
	0.953125 -0.631200551986694
	0.96875 -0.470937490463257
	0.984375 -0.294647932052612
	1 -0.101392149925232
};
\addlegendentry{EQ+NEQ}
\addplot [line width=2pt, crimson2553191, dash pattern=on 3pt off 5pt on 1pt off 5pt, mark=square, mark size=0, mark options={solid,fill opacity=0}]
table {%
	0.71875 -1.53975427150726
	0.734375 -1.55142378807068
	0.75 -1.554074883461
	0.765625 -1.5437616109848
	0.78125 -1.52668583393097
	0.796875 -1.50041270256042
	0.8125 -1.46385037899017
	0.828125 -1.41769564151764
	0.84375 -1.36113941669464
	0.859375 -1.29335689544678
	0.875 -1.21419584751129
	0.890625 -1.12303423881531
	0.90625 -1.01941287517548
	0.921875 -0.902773857116699
	0.9375 -0.772644281387329
	0.953125 -0.62828254699707
	0.96875 -0.468897819519043
	0.984375 -0.293534755706787
	1 -0.101093053817749
};
\addlegendentry{GEQ}
\addplot [line width=2pt, mediumorchid17588186, dashed, mark=triangle, mark size=0, mark options={solid,rotate=270,fill opacity=0}]
table {%
	0.71875 -1.52941429615021
	0.734375 -1.53704190254211
	0.75 -1.54299008846283
	0.765625 -1.53812313079834
	0.78125 -1.52328455448151
	0.796875 -1.49936616420746
	0.8125 -1.46480083465576
	0.828125 -1.41977262496948
	0.84375 -1.36375308036804
	0.859375 -1.29625630378723
	0.875 -1.21716201305389
	0.890625 -1.12585616111755
	0.90625 -1.0219304561615
	0.921875 -0.904920339584351
	0.9375 -0.77437162399292
	0.953125 -0.629521250724792
	0.96875 -0.469655990600586
	0.984375 -0.293820381164551
	1 -0.101122498512268
};
\addlegendentry{EXT}
\addplot [line width=2pt, dodgerblue0154222, dotted, mark=triangle, mark size=0, mark options={solid,fill opacity=0}]
table {%
	0.71875 -1.53614211082458
	0.734375 -1.54358279705048
	0.75 -1.5491064786911
	0.765625 -1.54539024829865
	0.78125 -1.53179574012756
	0.796875 -1.50557076931
	0.8125 -1.46991872787476
	0.828125 -1.42379069328308
	0.84375 -1.3666604757309
	0.859375 -1.29855763912201
	0.875 -1.21875309944153
	0.890625 -1.12689745426178
	0.90625 -1.02259790897369
	0.921875 -0.905291318893433
	0.9375 -0.774462461471558
	0.953125 -0.629433155059814
	0.96875 -0.469559907913208
	0.984375 -0.293799042701721
	1 -0.101072311401367
};
\addlegendentry{AVG}
\end{axis}
\end{tikzpicture}%
		\caption{\label{fig:taylor-bubble-0.504-axial}Axial velocity}
	\end{subfigure}%
	\hfill
	\begin{subfigure}[htbp]{0.49\textwidth}
		\centering
		% This file was created with tikzplotlib v0.10.1.
\begin{tikzpicture}

\definecolor{crimson2553191}{RGB}{255,31,91}
\definecolor{darkgray176}{RGB}{176,176,176}
\definecolor{darkorange24213334}{RGB}{242,133,34}
\definecolor{dodgerblue0154222}{RGB}{0,154,222}
\definecolor{lightgray204}{RGB}{204,204,204}
\definecolor{mediumorchid17588186}{RGB}{175,88,186}
\definecolor{springgreen0205108}{RGB}{0,205,108}

\begin{axis}[
height=\figureheight,
legend cell align={left},
legend style={fill opacity=0.8, draw opacity=1, text opacity=1,
	at={(0.99,0.99)},
	anchor=north east,
	draw=lightgray204},
tick align=outside,
tick pos=left,
width=\figurewidth,
x grid style={darkgray176},
xlabel style={align=center},
xlabel={\(\displaystyle r^{*}\)},
xmajorgrids,
xmin=0.69025, xmax=1.01475,
xtick style={color=black},
y grid style={darkgray176},
ylabel style={rotate=-90.0},
ylabel={\(\displaystyle u_{r}^{*}\)},
ymajorgrids,
ymin=-0.0129044125520578, ymax=0.206312663593214,
ytick style={color=black},
yticklabel style={
	/pgf/number format/fixed,
	/pgf/number format/precision=5
},
scaled y ticks=false
]
\addplot [semithick, black, mark=*, mark size=1.5, mark options={solid}, only marks]
table {%
	0.704999923706055 0.184000015258789
	0.720999956130981 0.177999973297119
	0.73799991607666 0.162999987602234
	0.754999995231628 0.148000001907349
	0.771000027656555 0.138999938964844
	0.787999987602234 0.136000037193298
	0.805000066757202 0.101999998092651
	0.820999979972839 0.0559999942779541
	0.838000059127808 0.0486999750137329
	0.853999972343445 0.0228999853134155
	0.871000051498413 0.0448000431060791
	0.888000011444092 0.0616999864578247
	0.904000043869019 0.0169999599456787
	0.921000003814697 0.0118000507354736
	0.937999963760376 0.00337004661560059
	0.953999996185303 0.00521004199981689
	0.970999956130981 0.00581002235412598
	0.98799991607666 -0.00294005870819092
};
\addlegendentry{Experiment~\cite{bugg2002VelocityFieldTaylor}}
\addplot [line width=2pt, springgreen0205108, mark=triangle, mark size=0, mark options={solid,rotate=180,fill opacity=0}]
table {%
	0.71875 0.196348190307617
	0.734375 0.176600813865662
	0.75 0.16409707069397
	0.765625 0.152887225151062
	0.78125 0.13964581489563
	0.796875 0.124281883239746
	0.8125 0.108263254165649
	0.828125 0.0929234027862549
	0.84375 0.0782536268234253
	0.859375 0.0648690462112427
	0.875 0.0526858568191528
	0.890625 0.0414456129074097
	0.90625 0.0312329530715942
	0.921875 0.0223314762115479
	0.9375 0.0149909257888794
	0.953125 0.00912237167358398
	0.96875 0.00459420680999756
	0.984375 0.00159800052642822
	1 0.000191688537597656
};
\addlegendentry{EQ}
\addplot [line width=2pt, darkorange24213334, dash pattern=on 1pt off 3pt on 3pt off 3pt, mark=triangle, mark size=0, mark options={solid,rotate=90,fill opacity=0}]
table {%
	0.71875 0.191515445709229
	0.734375 0.170771956443787
	0.75 0.155773282051086
	0.765625 0.143576979637146
	0.78125 0.130455374717712
	0.796875 0.116530179977417
	0.8125 0.102289915084839
	0.828125 0.0882750749588013
	0.84375 0.0744326114654541
	0.859375 0.0615464448928833
	0.875 0.0496220588684082
	0.890625 0.0388147830963135
	0.90625 0.0294121503829956
	0.921875 0.0211281776428223
	0.9375 0.0142805576324463
	0.953125 0.00875258445739746
	0.96875 0.00460290908813477
	0.984375 0.00177884101867676
	1 0.00030362606048584
};
\addlegendentry{EQ+NEQ}
\addplot [line width=2pt, crimson2553191, dash pattern=on 3pt off 5pt on 1pt off 5pt, mark=square, mark size=0, mark options={solid,fill opacity=0}]
table {%
	0.71875 0.185842275619507
	0.734375 0.169761419296265
	0.75 0.158357858657837
	0.765625 0.148099541664124
	0.78125 0.135385632514954
	0.796875 0.121005773544312
	0.8125 0.10649585723877
	0.828125 0.0918846130371094
	0.84375 0.0776721239089966
	0.859375 0.0643243789672852
	0.875 0.0520550012588501
	0.890625 0.0409297943115234
	0.90625 0.0310028791427612
	0.921875 0.0223084688186646
	0.9375 0.0149456262588501
	0.953125 0.00895464420318604
	0.96875 0.00446712970733643
	0.984375 0.00152349472045898
	1 0.000145435333251953
};
\addlegendentry{GEQ}
\addplot [line width=2pt, mediumorchid17588186, dashed, mark=triangle, mark size=0, mark options={solid,rotate=270,fill opacity=0}]
table {%
	0.71875 0.191591501235962
	0.734375 0.172975063323975
	0.75 0.157496571540833
	0.765625 0.144492745399475
	0.78125 0.132111430168152
	0.796875 0.118517279624939
	0.8125 0.104008436203003
	0.828125 0.0897476673126221
	0.84375 0.0759997367858887
	0.859375 0.0630084276199341
	0.875 0.0508958101272583
	0.890625 0.0399147272109985
	0.90625 0.0301762819290161
	0.921875 0.0216913223266602
	0.9375 0.0145628452301025
	0.953125 0.00882852077484131
	0.96875 0.00450766086578369
	0.984375 0.00164389610290527
	1 0.000222444534301758
};
\addlegendentry{EXT}
\addplot [line width=2pt, dodgerblue0154222, dotted, mark=triangle, mark size=0, mark options={solid,fill opacity=0}]
table {%
	0.71875 0.194784283638
	0.734375 0.176572799682617
	0.75 0.160914659500122
	0.765625 0.146249651908875
	0.78125 0.130883574485779
	0.796875 0.11583685874939
	0.8125 0.101284503936768
	0.828125 0.0872255563735962
	0.84375 0.0737099647521973
	0.859375 0.0610958337783813
	0.875 0.0493061542510986
	0.890625 0.0387070178985596
	0.90625 0.0291898250579834
	0.921875 0.0209733247756958
	0.9375 0.0140900611877441
	0.953125 0.00858867168426514
	0.96875 0.00431346893310547
	0.984375 0.00148105621337891
	1 0.000141024589538574
};
\addlegendentry{AVG}
\end{axis}
\end{tikzpicture}%
		\caption{\label{fig:taylor-bubble-0.504-radial}Radial velocity}
	\end{subfigure}%
	\caption{
		\label{fig:taylor-bubble-0.504}
		Simulated non-dimensionalized axial velocity $u_{a}^{*}$ (a) and radial velocity $u_{r}^{*}$ (b) along the radial monitoring-line at $-0.504D$ from the Taylor bubble's front (see \Cref{fig:taylor-bubble-schematic}).
		The comparison with experimental data~\cite{bugg2002VelocityFieldTaylor} is drawn in terms of the non-dimensionalized radial location $r^{*}$ at dimensionless time $t^{*}=15$ with tube diameter $D=128$ lattice cells.
	}
\end{figure}

\Cref{tab:taylor-bubble-re} lists the Reynolds number
\begin{equation}\label{eq:taylor-re}
\text{Re} \coloneqq \frac{D u}{\nu}
\end{equation}
where $\nu$ is the kinematic viscosity, for different computational domain resolutions.
The bubble's rise velocity $u$ was computed from the bubble's center of mass in the $z$-direction at time $t^{*}=10$ and $t^{*}=15$.
The simulations generally agreed well with Re from the reference data~\cite{bugg2002VelocityFieldTaylor}, and there were only minor differences between the refilling schemes.
Similarly, as illustrated in \Cref{fig:taylor-bubble-shape} for $t^{*}=15$, the choice of the refilling scheme had almost no effect on the shape of the simulated Taylor bubble.
\Cref{fig:taylor-bubble-axial-radial,fig:taylor-bubble-0.111,fig:taylor-bubble-0.504} compare the non-dimensionalized axial $u_{a}^{*} = u_{a}/u$ and radial $u_{r}^{*} = u_{r}/u$ velocities at the locations defined in \Cref{fig:taylor-bubble-schematic}.
As for the Taylor bubble's shape, the refilling schemes led to only small differences in the velocity profiles.
Only the radial velocity $u_{r}^{*}$ at a radial line at $0.111D$ from the Taylor bubble's front, was arguably predicted more accurately by the EQ and GEQ schemes, as shown in \Cref{fig:taylor-bubble-0.111}.
\par

\begin{table}[htbp]
	\centering
	\begin{tabular}{
			>{\raggedright}m{0.3\textwidth}
			>{\centering\arraybackslash}m{0.2\textwidth}
			>{\centering\arraybackslash}m{0.2\textwidth}
			>{\centering\arraybackslash}m{0.2\textwidth}
		}
		
		\toprule
		$D$ & $32$ & $64$ & $128$ \\
		\midrule
		
		Re\textsubscript{EQ} & $24.12$ & $25.35$ & $25.89$ \\
		
		Re\textsubscript{EQ+NEQ} & $24.14$ & $25.32$ & $25.88$ \\
		
		Re\textsubscript{GEQ} & $24.16$ & $25.33$ & $25.89$ \\
		
		Re\textsubscript{EXT} & $24.12$ & $25.34$ & $25.86$ \\
		
		Re\textsubscript{AVG} & $24.09$ & $25.33$ & $25.86$ \\
		
		\midrule
		Re\textsubscript{Experiment}~\cite{bugg2002VelocityFieldTaylor} & \multicolumn{3}{c}{$27$} \\
		\bottomrule
	\end{tabular}
	\caption{\label{tab:taylor-bubble-re}
		Reynolds number Re of the simulated Taylor bubble for different computational domain resolutions as specified by the tube diameter $D$.
		The bubble's rise velocity, as used to compute Re, was obtained from the Taylor bubble's location in axial direction at time $t^{*}=10$ and $t^{*}=15$.
	}
\end{table}

As depicted in \Cref{tab:taylor-bubble-re} and \Cref{fig:taylor-bubble-shape-convergence}, the simulation results shown here have sufficiently converged in terms of computational grid resolution.
\par

\begin{figure}[htbp]
	\centering
	\setlength{\figureheight}{0.5\textwidth}
	\setlength{\figurewidth}{0.45\textwidth}
	\begin{subfigure}[htbp]{0.49\textwidth}
		\centering
		\input{figures/taylor-bubble/axial-0.111.tex}%
		\caption{\label{fig:taylor-bubble-0.111-axial}Axial velocity}
	\end{subfigure}%
	\hfill
	\begin{subfigure}[htbp]{0.49\textwidth}
		\centering
		\input{figures/taylor-bubble/radial-0.111.tex}%
		\caption{\label{fig:taylor-bubble-0.111-radial}Radial velocity}
	\end{subfigure}%
	\caption{
		\label{fig:taylor-bubble-0.111}
		Simulated non-dimensionalized axial velocity $u_{a}^{*}$ (a) and radial velocity $u_{r}^{*}$ (b) along the radial monitoring-line at $0.111D$ from the Taylor bubble's front (see \Cref{fig:taylor-bubble-schematic}).
		The comparison with experimental data~\cite{bugg2002VelocityFieldTaylor} is drawn in terms of the non-dimensionalized radial location $r^{*}$ at dimensionless time $t^{*}=15$ with tube diameter $D=128$ lattice cells.
	}
\end{figure}

\FloatBarrier

\subsection{Drop impact}\label{sec:ne-di}
In the fifth test case, the vertical impact of a drop into a pool of liquid was simulated.
As no quantitative experimental measurements are given for the reference experiments from Wang and Chen~\cite{wang2000SplashingImpactSingle}, only a qualitative comparison with photographs could be made.
\par

\subsubsection{Simulation setup}\label{sec:ne-di-ss}
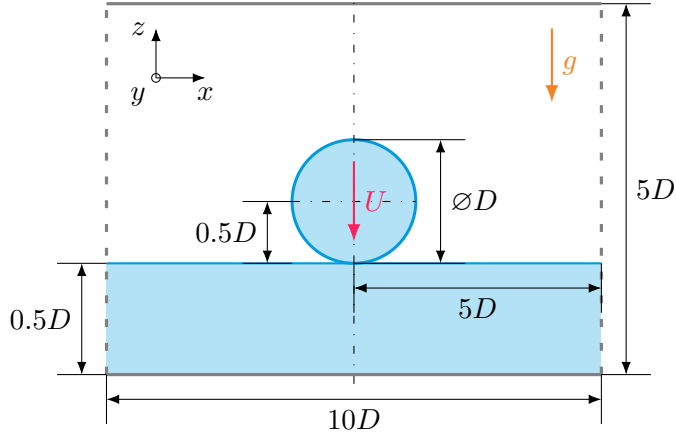
\begin{figure}[htbp]
	\centering
	\setlength{\figureheight}{0.3\textwidth}
	\setlength{\figurewidth}{0.4\textwidth}
	\setlength\mdist{0.02\textwidth}
	\begin{tikzpicture}
\definecolor{dodgerblue0154222}{RGB}{0,154,222}
\definecolor{darkorange24213334}{RGB}{242,133,34}
\definecolor{mediumorchid17588186}{RGB}{175,88,186}
\definecolor{crimson2553191}{RGB}{255,31,91}
\setlength\radius{0.125\figurewidth}

% coordinates used for impact velocity angle
\coordinate (intersectPoint) at (0.5\figurewidth,0.3\figureheight+4\radius);
\coordinate (obliquePoint) at (0.6\figurewidth,0.3\figureheight+2.5\radius);
\coordinate (verticalPoint) at (0.5\figurewidth,0.3\figureheight);

% liquid pool
\draw [thick, fill=dodgerblue0154222!30, draw=none] (0,0) rectangle (\figurewidth,0.3\figureheight);
\draw[thick, dodgerblue0154222] (0,0.3\figureheight)--(\figurewidth,0.3\figureheight);

% borders left and right
\draw[very thick, loosely dashed, black!50] (0,0)--(0,\figureheight);
\draw[very thick, loosely dashed, black!50] (\figurewidth,0)--(\figurewidth,\figureheight);

% borders top and bottom
\draw[very thick, black!50] (0,\figureheight)--(\figurewidth,\figureheight);
\draw[very thick, black!50] (0,0)--(\figurewidth,0);

% domain height
\draw[<->, >=Latex] (\figurewidth+\mdist,0)--(\figurewidth+\mdist,\figureheight) node [pos=0.5,right] {$5D$};
\draw[-] (\figurewidth,0)--(\figurewidth+2\mdist,0);
\draw[-] (\figurewidth,\figureheight)--(\figurewidth+2\mdist,\figureheight);

% domain width
\draw[<->, >=Latex] (0,-\mdist)--(\figurewidth,-\mdist) node [pos=0.5,below] {$10D$};
\draw[-] (0,0)--(0,-2\mdist);
\draw[-] (\figurewidth,0)--(\figurewidth,-2\mdist);

% liquid pool height
\draw[<->, >=Latex] (-\mdist,0)--(-\mdist,0.3\figureheight) node [pos=0.5,left] {$0.5D$};
\draw[-] (-2\mdist,0.3\figureheight)--(0,0.3\figureheight);
\draw[-] (-2\mdist,0)--(0,0);

% drop
\draw[thick, draw=dodgerblue0154222, fill=dodgerblue0154222!30, fill opacity=1, line width=0.4mm] (0.5\figurewidth,0.3\figureheight+\radius) circle [radius=\radius] node {};
\draw[loosely dashdotted] (0.5\figurewidth-\radius,0.3\figureheight+\radius)--(0.5\figurewidth+\radius,0.3\figureheight+\radius);
%\draw[loosely dashdotted] (intersectPoint) -- (verticalPoint);

% drop position y
\draw[<->, >=Latex] (0.5\figurewidth-\radius-\mdist,0.3\figureheight)--(0.5\figurewidth-\radius-\mdist,0.3\figureheight+\radius) node [pos=0.5,left] {$0.5D$};
\draw[-] (0.5\figurewidth-\radius-2\mdist,0.3\figureheight+\radius)--(0.5\figurewidth-\radius,0.3\figureheight+\radius);
\draw[-] (0.5\figurewidth-\radius-2\mdist,0.3\figureheight)--(0.5\figurewidth-\radius,0.3\figureheight);

% drop position x
\draw[<->, >=Latex] (0.5\figurewidth,0.3\figureheight-\mdist)--(\figurewidth,0.3\figureheight-\mdist) node [pos=0.5,below] {$5D$};
\draw[-] (0.5\figurewidth,0.3\figureheight)--(0.5\figurewidth,0.3\figureheight-2\mdist);
\draw[-] (\figurewidth,0.3\figureheight)--(\figurewidth,0.3\figureheight-2\mdist);

% drop diameter
\draw[<->, >=Latex] (0.5\figurewidth+\radius+\mdist,0.3\figureheight)--(0.5\figurewidth+\radius+\mdist,0.3\figureheight+2\radius) node [pos=0.5,right] {$\varnothing D$};
\draw[-] (0.5\figurewidth,0.3\figureheight+2\radius)--(0.5\figurewidth+\radius+2\mdist,0.3\figureheight+2\radius);
\draw[-] (0.5\figurewidth,0.3\figureheight)--(0.5\figurewidth+\radius+2\mdist,0.3\figureheight);

% vertical line of symmetry
\draw[loosely dashdotted] (0.5\figurewidth,-0.025\figureheight)--(0.5\figurewidth,1.025\figureheight);

% impact velocity
\draw[thick, ->, >=Latex, crimson2553191] (0.5\figurewidth,0.3\figureheight+1.65\radius)--(0.5\figurewidth,0.3\figureheight+0.35\radius) node [pos=0.5, right] {$U$};

% gravity
\draw[thick, ->, >=Latex, darkorange24213334] (\figurewidth-2\mdist,\figureheight-\mdist)--(\figurewidth-2\mdist,\figureheight-4\mdist) node [pos=0.5,right] {$g$};

% coordinate system
\draw[->, >=Latex] (2\mdist,\figureheight-3\mdist)--(4\mdist,\figureheight-3\mdist) node [below] {$x$};
\draw[->, >=Latex] (2\mdist,\figureheight-3\mdist)--(2\mdist,\figureheight-1\mdist) node [left] {$z$};
\draw[draw=black] (2\mdist,\figureheight-3\mdist) circle [radius=0.175\mdist] node[opacity=1, below left] {$y$};
\end{tikzpicture}%
	\caption{
		\label{fig:drop-impact-setup}
		Simulation setup of the drop impact test case. 
		A spherical drop of liquid with diameter $D$ was initialized right above the surface of a liquid pool of height $0.5D$ in a domain of size $10D \times 10D \times 5D$.
		The gravitational acceleration $g$ acted in the negative $z$-direction, and the droplet was initialized with impact velocity $U$ in the same direction.
		The domain's side walls in $x$- and $y$-direction were periodic, whereas the domain's top and bottom walls in the $z$-direction were set to no-slip boundary conditions.
		C.~Schwarzmeier, M.~Holzer, T.~Mitchell, M.~Lehmann, F.~Häusl, U.~Rüde, Comparison of free-surface and conservative Allen--Cahn phase-field lattice Boltzmann method, arXiv preprint~\cite{schwarzmeier2022ComparisonFreeSurface}, 2022; licensed under a Creative Commons Attribution (CC BY) license; the colors were changed from the original.
	}
\end{figure}

As illustrated in \Cref{fig:drop-impact-setup}, a spherical droplet with a diameter of $D=80$ lattice cells was initialized in a three-dimensional computational domain of size $10D \times 10D \times 5D$ ($x$-, $y$-, $z$-direction).
The droplet was initially located at the surface of a thin liquid film of height $0.5D$ with impact velocity $U$ in the negative $z$-direction.
The gravitational acceleration $g$ also acted in the negative $z$-direction.
Accordingly, hydrostatic pressure was initialized such that the pressure at the pool's surface was equal to the constant atmospheric volumetric gas pressure $p^{\text{V}}(t) = p_{0}$.
The walls in the $x$- and $y$-direction were periodic, and no-slip boundary conditions were set at the top and bottom domain walls in $z$-direction.
The relaxation rate was chosen $\omega=1.989$.
The droplet's impact is described by the Weber number
\begin{equation}
\text{We} \coloneqq \frac{\rho U^{2}D}{\sigma} = 2010,
\end{equation}
which relates inertial and surface tension forces, and by the Ohnesorge number
\begin{equation}
\text{Oh} \coloneqq \frac{\mu}{\sqrt{\sigma \rho D}} = 0.0384,
\end{equation}
which relates viscous to inertial and surface tension forces. 
These dimensionless numbers include the surface tension $\sigma$, dynamic viscosity $\mu$, and liquid density $\rho$.
Assuming $g=9.81$\,m/s\textsuperscript{2}, and using $\rho = 1200$\, kg/m\textsuperscript{3} and $\mu = 0.022$\,kg/(m$\cdot$s)~\cite{wang2000SplashingImpactSingle}, the Bond number $\text{Bo}=3.18$~\eqref{eq:ne-rdb-ss-bo} with characteristic length $D$, closes the definition of the system.
The non-dimensionalized time $t^{*} \coloneqq t U/D$ is offset by $t^{*}=0.16$~\cite{lehmann2021EjectionMarineMicroplastics} to allow a comparison with the numerical simulations performed in this study.
\par

\subsubsection{Results and discussion}\label{sec:ne-di-res}
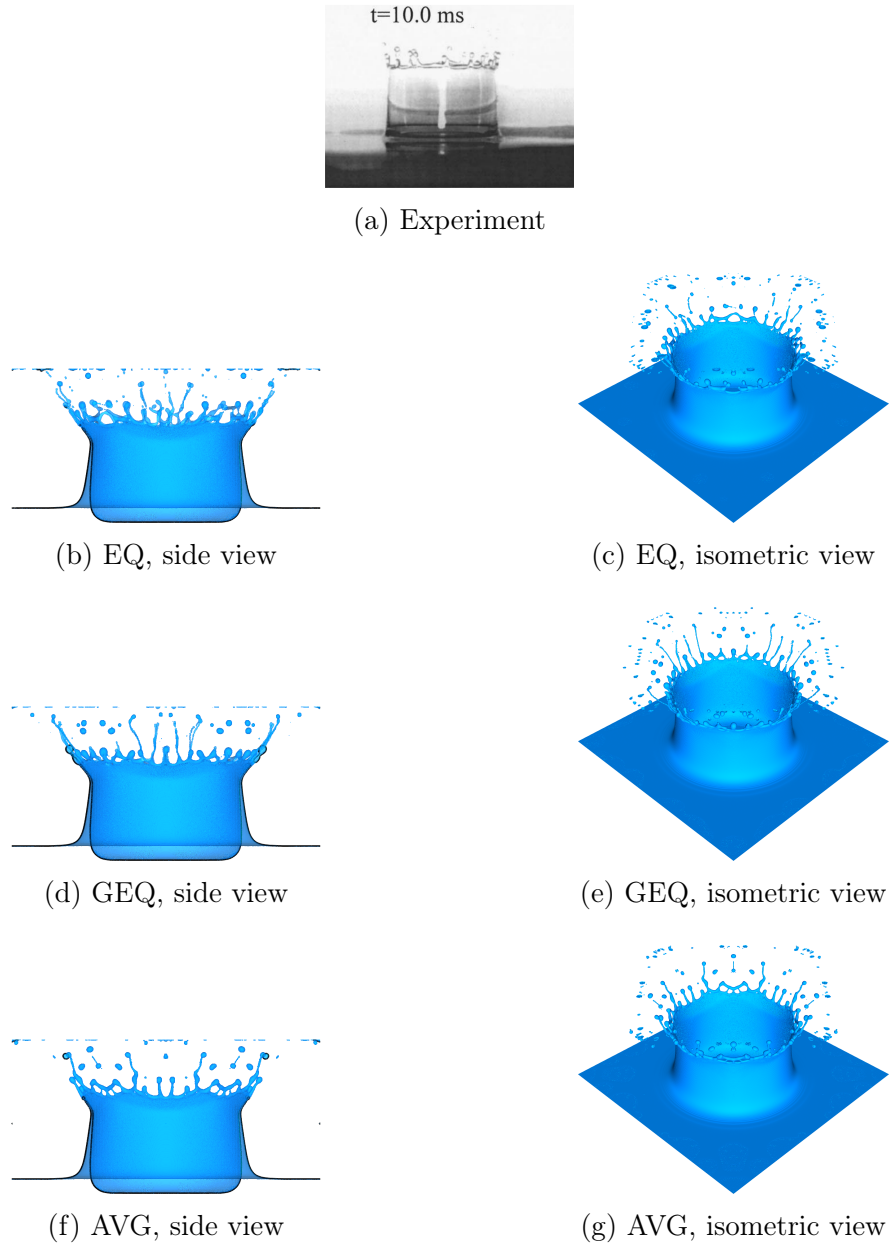
\begin{figure}[htbp]
	\vspace{-1.5\baselineskip}
\centering
\begin{tabular}{
		>{\centering\arraybackslash}m{0.45\textwidth}
		>{\centering\arraybackslash}m{0.45\textwidth}
	}	
	
	\multicolumn{2}{c}{
		\begin{tikzpicture}
			\node[anchor=south] at (0,0) {\includegraphics[width=0.2\textwidth]{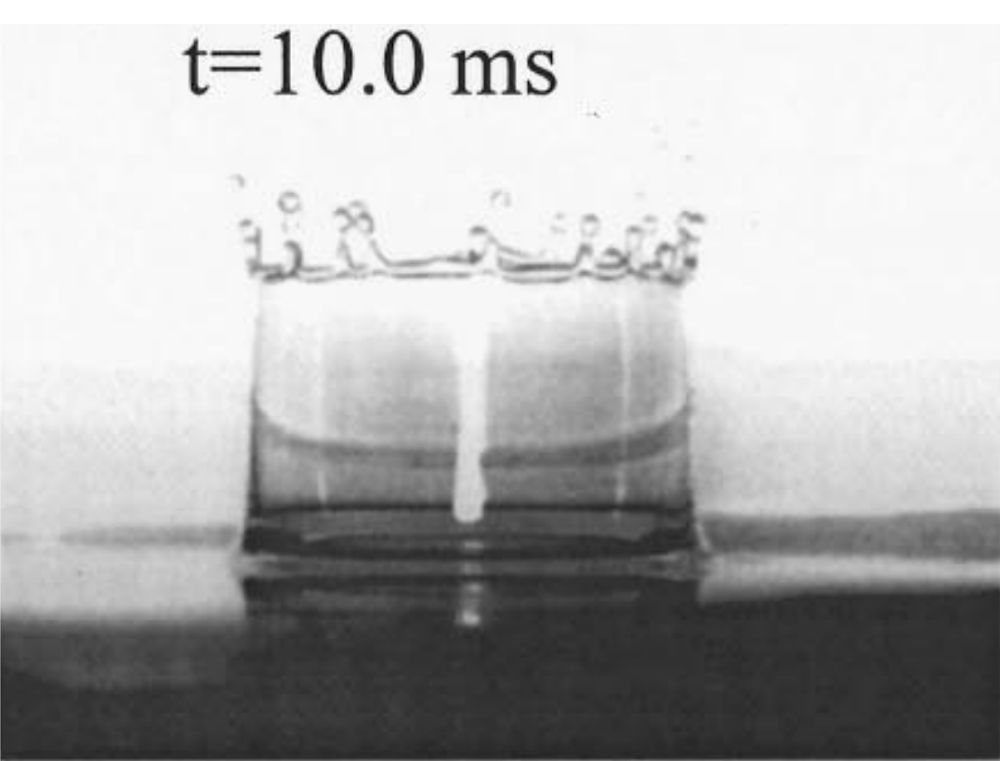}};
			\node[anchor=north,align=center,execute at begin node=\setlength{\baselineskip}{3ex}] at (0,0){(a) Experiment};
		\end{tikzpicture}
	}\\
	
	\begin{tikzpicture}
		\node[anchor=south] at (0,0) {\includegraphics[width=0.25\textwidth]{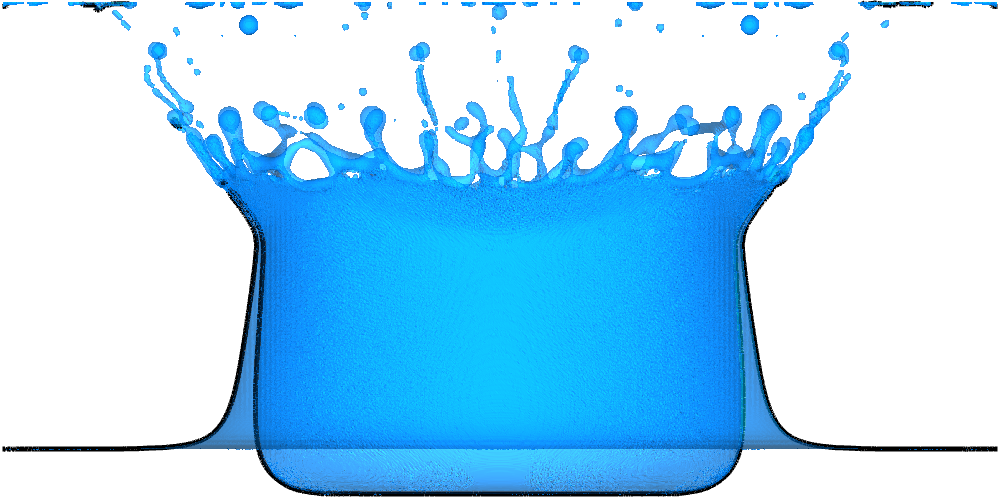}};
		\node[anchor=north] at (0,0){(b) EQ, side view};
	\end{tikzpicture} &
	\begin{tikzpicture}
		\node[anchor=south] at (0,0) {\includegraphics[width=0.25\textwidth]{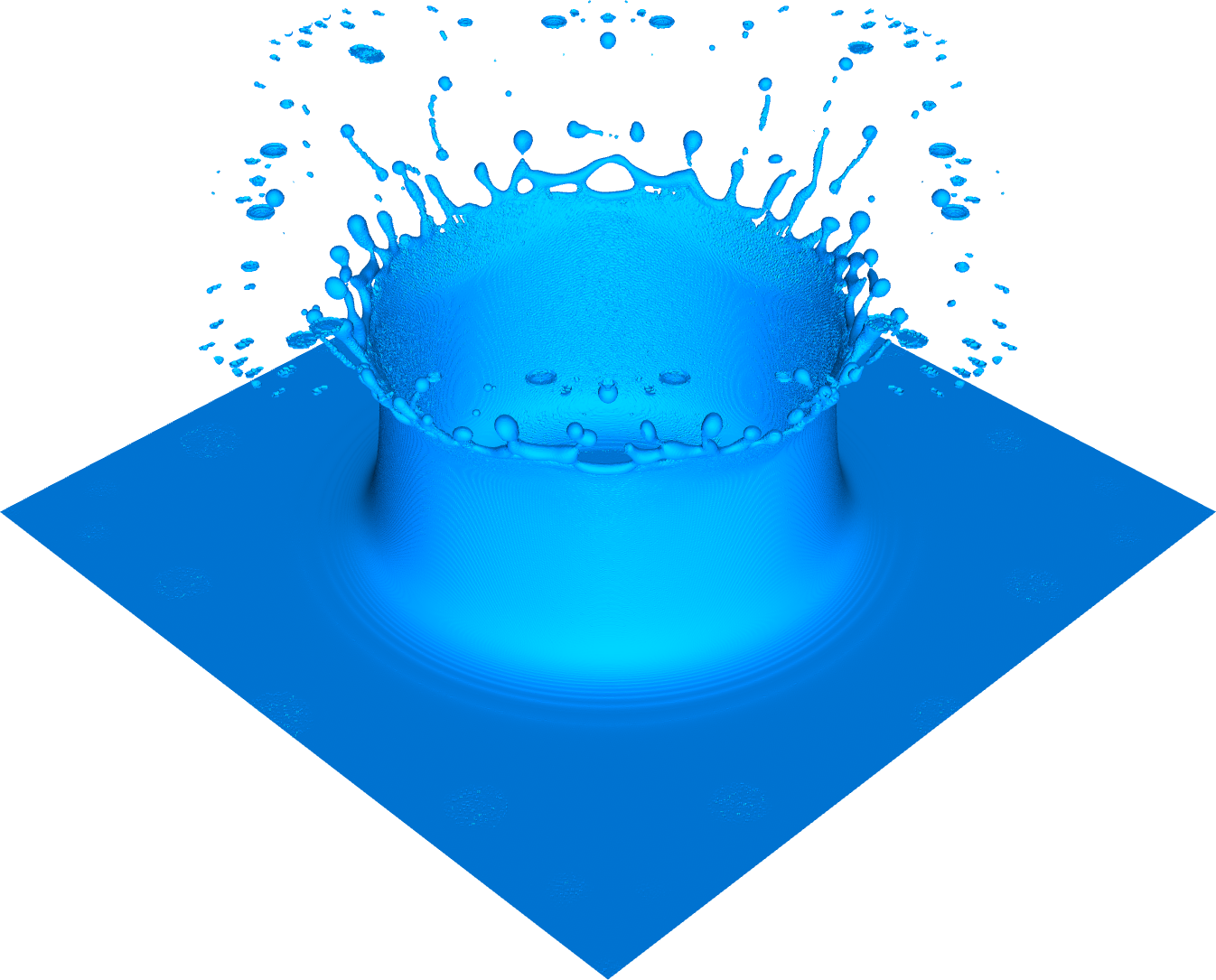}};
		\node[anchor=north] at (0,0){(c) EQ, isometric view};
	\end{tikzpicture} \\

	\begin{tikzpicture}
		\node[anchor=south] at (0,0) {\includegraphics[width=0.25\textwidth]{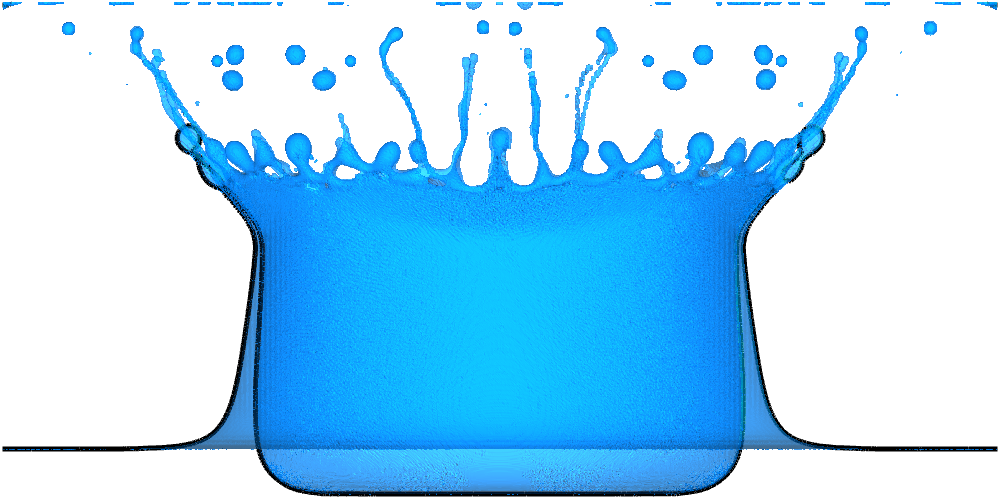}};
		\node[anchor=north] at (0,0){(d) GEQ, side view};
	\end{tikzpicture} &
	\begin{tikzpicture}
		\node[anchor=south] at (0,0) {\includegraphics[width=0.25\textwidth]{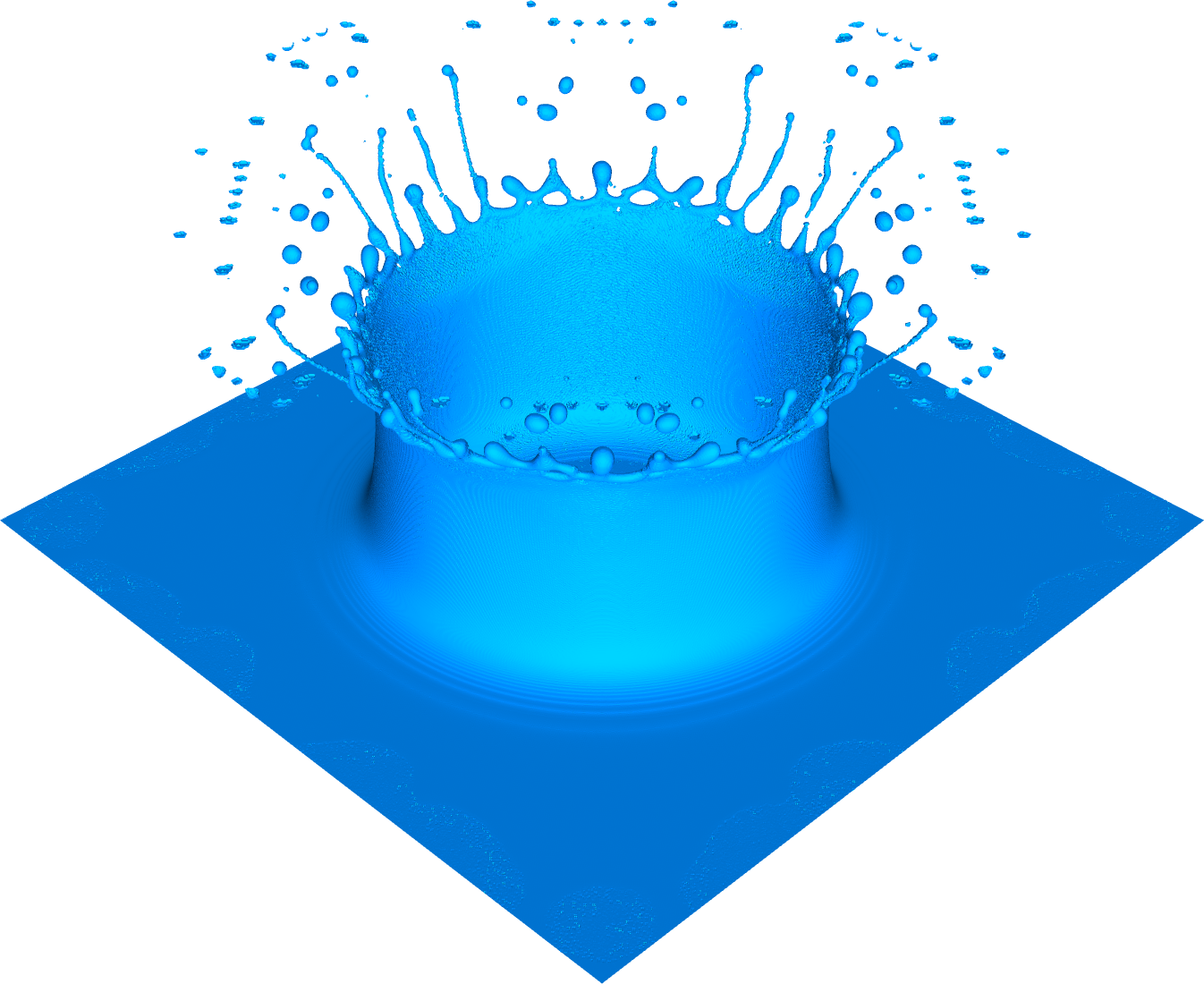}};
		\node[anchor=north] at (0,0){(e) GEQ, isometric view};
	\end{tikzpicture} \\
	
	\begin{tikzpicture}
		\node[anchor=south] at (0,0) {\includegraphics[width=0.25\textwidth]{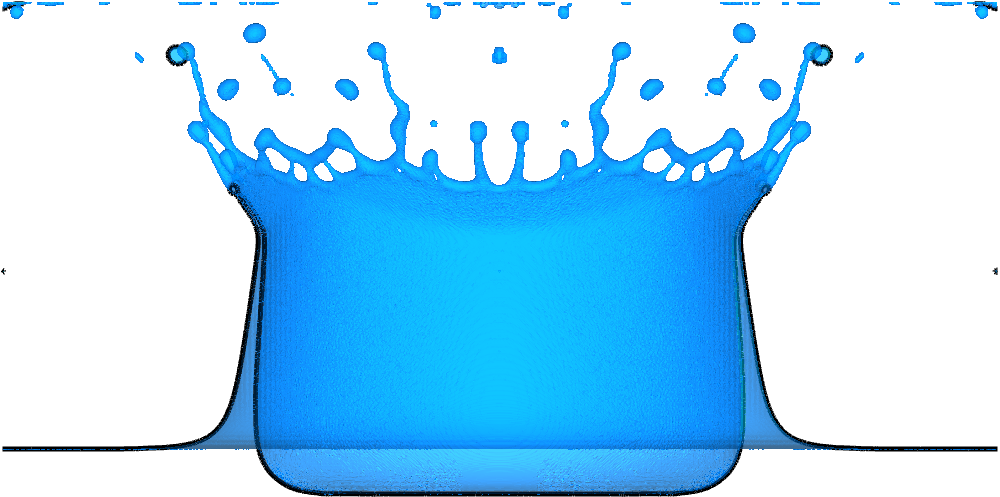}};
		\node[anchor=north] at (0,0){(f) AVG, side view};
	\end{tikzpicture} &
	\begin{tikzpicture}
		\node[anchor=south] at (0,0) {\includegraphics[width=0.25\textwidth]{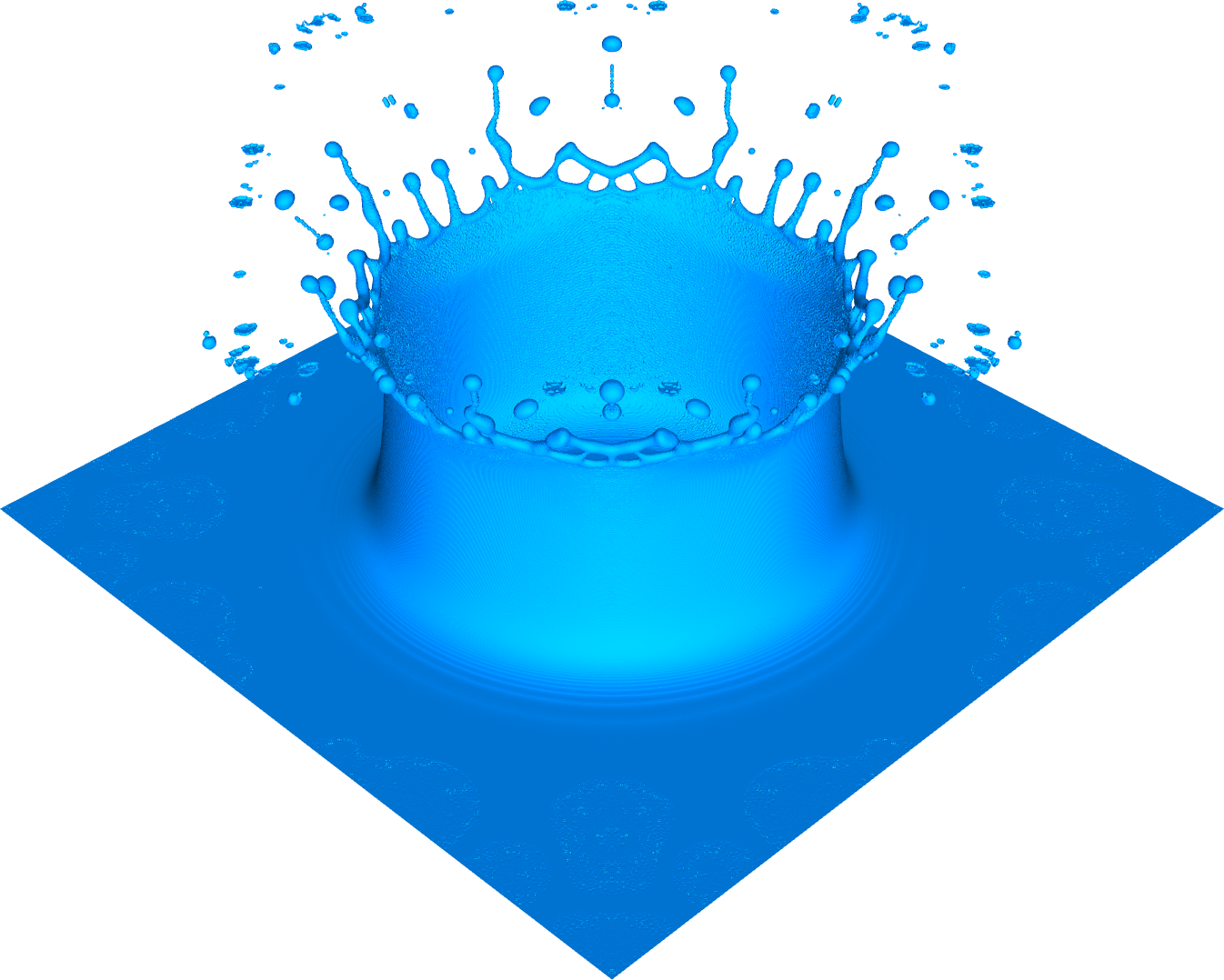}};
		\node[anchor=north] at (0,0){(g) AVG, isometric view};
	\end{tikzpicture}		
\end{tabular}
	\caption{\label{fig:drop-mesh}
		Simulated splash crown of the drop impact at non-dimensionalized time $t^{*}=12$ compared to the laboratory experiment (a)\cite{wang2000SplashingImpactSingle}.
		The simulations were performed with a computational domain resolution, that is, initial drop diameter of $D=80$ lattice cells with the EQ (b and c), GEQ (d and e), and AVG (f and g) refilling schemes.
		The solid black lines illustrate the crown's contour in a centrally located cross-section with normal in the $x$-direction.
		The simulations with the EQ+NEQ and EXT schemes were numerically unstable and are not included here.
		With all other refilling schemes, qualitatively plausible results could be obtained.
		The photograph of the laboratory experiment (a) was reproduced from A.-B.~Wang, C.-C.~Chen, Splashing impact of a single drop onto very thin liquid films~\cite{wang2000SplashingImpactSingle}, Physics of Fluids, 12, 2000, with the permission of AIP Publishing.
	}
\end{figure}

The simulated drop impact, that is, the splash crown formation at $t^{*}=12$, is qualitatively compared with experimental results in \Cref{fig:drop-mesh}.
The solid black line indicates the crown's contour in a central cross-section, oriented with the normal in the $x$-direction.
There was no scale bar provided for the photograph of the laboratory experiment~\cite{wang2000SplashingImpactSingle}.
Therefore, the simulations performed here could only be validated qualitatively.
As in the dam break simulations in \Cref{sec:ne-rdb,sec:ne-cdb}, the EXT refilling scheme became numerically unstable which led to too high macroscopic velocities.
The EQ+NEQ scheme was subject to numerical instabilities for the same reason.
Qualitatively plausible results could be obtained with all other refilling schemes.
However, with the GEQ scheme, the droplets detaching from the crown's top formed thin and long threads of liquid.
In contrast, in the photograph of the experiment, the detaching droplets rather form thicker and shorter liquid threads that then detach as spherical droplets.
This kind of crown formation is arguably resembled best by the EQ scheme.
\par

\begin{figure}[htbp]
	\centering
	\setlength{\figureheight}{0.5\textwidth}
	\setlength{\figurewidth}{\textwidth}
	\setlength\mdist{0.02\textwidth}
	\input{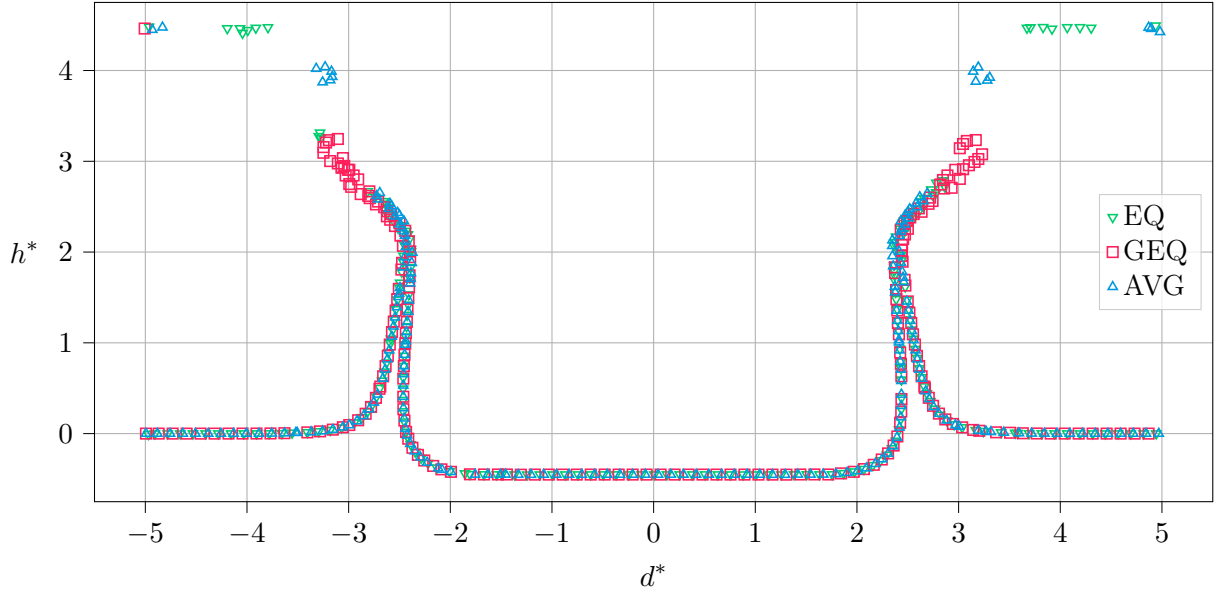}%	
	\caption{
		\label{fig:drop-plot-t-12}
		Simulated non-dimensionalized height $h^{*}=h/D$ and diameter $d^{*} = d/D$ of the drop impact's splash crown at dimensionless time $t^{*}=12$, measured in a centrally located cross-section with normal in the $x$-direction.
		The simulations were performed with a computational domain resolution, that is, an initial drop diameter of $D=80$ lattice cells.
		The simulations with the EQ+NEQ and EXT schemes were numerically unstable and are not included here.
	}
\end{figure}

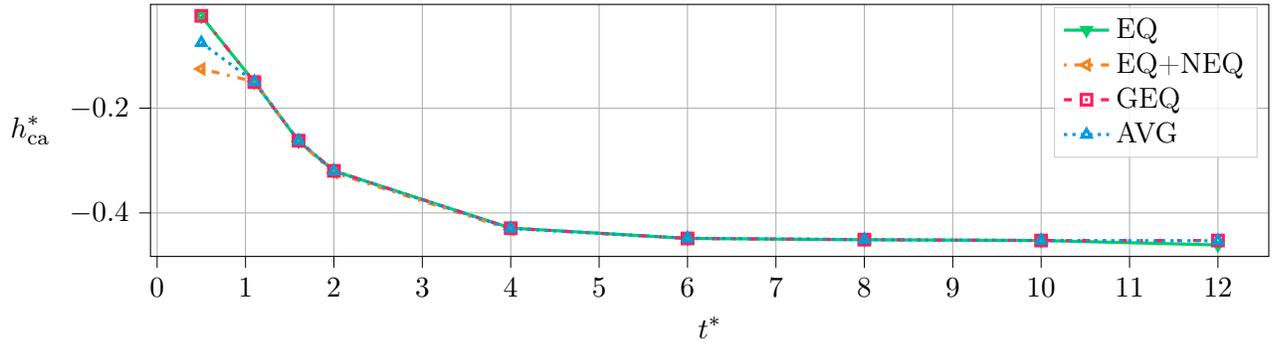
\begin{figure}[htbp]
	\centering
	\setlength{\figureheight}{0.3\textwidth}
	\setlength{\figurewidth}{\textwidth}
	\setlength\mdist{0.02\textwidth}
	% This file was created with tikzplotlib v0.10.1.
\begin{tikzpicture}

\definecolor{crimson2553191}{RGB}{255,31,91}
\definecolor{darkgray176}{RGB}{176,176,176}
\definecolor{darkorange24213334}{RGB}{242,133,34}
\definecolor{dodgerblue0154222}{RGB}{0,154,222}
\definecolor{lightgray204}{RGB}{204,204,204}
\definecolor{springgreen0205108}{RGB}{0,205,108}

\begin{axis}[
height=\figureheight,
legend cell align={left},
legend style={fill opacity=0.8, draw opacity=1, text opacity=1, draw=lightgray204, at={(0.99,0.99)},
	anchor=north east,},
tick align=outside,
tick pos=left,
width=\figurewidth,
x grid style={darkgray176},
xlabel={\(\displaystyle t^{*}\)},
xmajorgrids,
xmin=-0.0750000000000001, xmax=12.575,
xtick style={color=black},
y grid style={darkgray176},
ylabel style={rotate=-90.0},
ylabel={\(\displaystyle h_{\text{ca}}^{*}\)},
ymajorgrids,
ymin=-0.483089452981949, ymax=-0.00179683566093445,
ytick style={color=black}
]
\addplot [very thick, springgreen0205108, mark=triangle, mark size=2, mark options={solid,rotate=180,fill opacity=0}]
table {%
	0.5 -0.0241087675094604
	1.10000002384186 -0.150012493133545
	1.60000002384186 -0.263010025024414
	2 -0.319836258888245
	4 -0.428898811340332
	6 -0.448651671409607
	8 -0.45124626159668
	10 -0.452655673027039
	12 -0.461212515830994
};
\addlegendentry{EQ}
\addplot [very thick, darkorange24213334, dash pattern=on 1pt off 3pt on 3pt off 3pt, mark=triangle, mark size=2, mark options={solid,rotate=90,fill opacity=0}]
table {%
	0.5 -0.12502121925354
	1.10000002384186 -0.149213790893555
	1.60000002384186 -0.263283729553223
	2 -0.322991251945496
	4 -0.431493759155273
};
\addlegendentry{EQ+NEQ}
\addplot [very thick, crimson2553191, dash pattern=on 3pt off 5pt on 1pt off 5pt, mark=square, mark size=2, mark options={solid,fill opacity=0}]
table {%
	0.5 -0.0236737728118896
	1.10000002384186 -0.150285005569458
	1.60000002384186 -0.262145042419434
	2 -0.319623708724976
	4 -0.429231643676758
	6 -0.44874906539917
	8 -0.451058387756348
	10 -0.452688097953796
	12 -0.45270311832428
};
\addlegendentry{GEQ}
\addplot [very thick, dodgerblue0154222, dotted, mark=triangle, mark size=2, mark options={solid,fill opacity=0}]
table {%
	0.5 -0.0750274658203125
	1.10000002384186 -0.149372577667236
	1.60000002384186 -0.26219630241394
	2 -0.319932460784912
	4 -0.429646372795105
	6 -0.448229312896729
	8 -0.450978398323059
	10 -0.4526287317276
	12 -0.453258514404297
};
\addlegendentry{AVG}
\end{axis}

\end{tikzpicture}%
	\caption{
		\label{fig:drop-plot-cavity-depth}
		Simulated non-dimensionalized drop impact splash cavity depth $h_{\text{ca}}^{*} = h_{\text{ca}} / D$ over dimensionless time $t^{*}$.
		The cavity depth $h_{\text{ca}}$ is the maximum distance of the cavity bottom to the initial position of the liquid surface, measured in a centrally located cross-section with normal in the $x$-direction.
		The AVG and EQ+NEQ schemes deviated at $t^{*}=0.5$ because of bubbles located below the cavity bottom that disturbed the evaluation algorithm.
		The EXT refilling scheme is not included as it quickly became numerically unstable, whereas the EQ+NEQ became unstable only at $t^{*} > 4$.
	}
\end{figure}

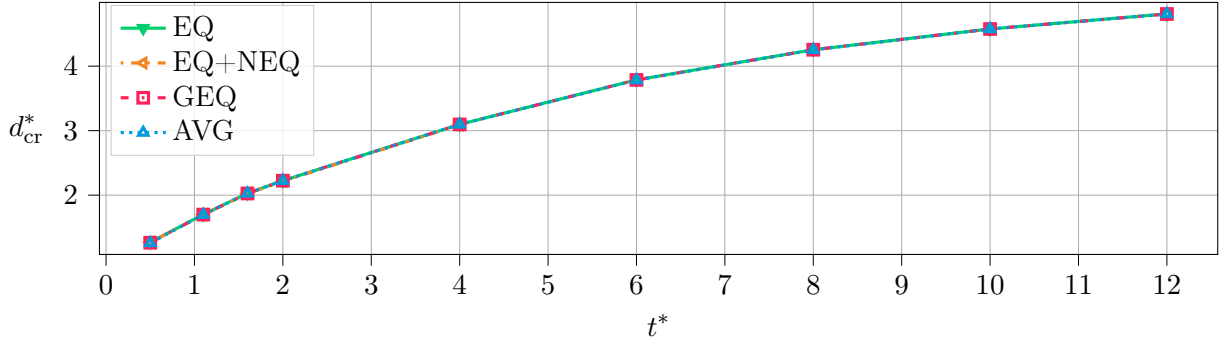
\begin{figure}[htbp]
	\centering
	\setlength{\figureheight}{0.3\textwidth}
	\setlength{\figurewidth}{\textwidth}
	\setlength\mdist{0.02\textwidth}
	% This file was created with tikzplotlib v0.10.1.
\begin{tikzpicture}

\definecolor{crimson2553191}{RGB}{255,31,91}
\definecolor{darkgray176}{RGB}{176,176,176}
\definecolor{darkorange24213334}{RGB}{242,133,34}
\definecolor{dodgerblue0154222}{RGB}{0,154,222}
\definecolor{lightgray204}{RGB}{204,204,204}
\definecolor{springgreen0205108}{RGB}{0,205,108}

\begin{axis}[
height=\figureheight,
legend cell align={left},
legend style={
	fill opacity=0.8,
	draw opacity=1,
	text opacity=1,
	at={(0.01,0.99)},
	anchor=north west,
	draw=lightgray204
},
tick align=outside,
tick pos=left,
width=\figurewidth,
x grid style={darkgray176},
xlabel={\(\displaystyle t^{*}\)},
xmajorgrids,
xmin=-0.0750000000000001, xmax=12.575,
xtick style={color=black},
y grid style={darkgray176},
ylabel style={rotate=-90.0},
ylabel={\(\displaystyle d_{\text{cr}}^{*}\)},
ymajorgrids,
ymin=1.0806557836951, ymax=4.98797417459475,
ytick style={color=black}
]
\addplot [very thick, springgreen0205108, mark=triangle, mark size=2, mark options={solid,rotate=180,fill opacity=0}]
table {%
	0.5 1.26986265182495
	1.10000002384186 1.70046234130859
	1.60000002384186 2.02902507781982
	2 2.22451210021973
	4 3.09714937210083
	6 3.78737497329712
	8 4.25457525253296
	10 4.57643747329712
	12 4.81036901473999
};
\addlegendentry{EQ}
\addplot [very thick, darkorange24213334, dash pattern=on 1pt off 3pt on 3pt off 3pt, mark=triangle, mark size=2, mark options={solid,rotate=90,fill opacity=0}]
table {%
	0.5 1.27038764953613
	1.10000002384186 1.69740009307861
	1.60000002384186 2.0234375
	2 2.21546268463135
	4 3.09012508392334
};
\addlegendentry{EQ+NEQ}
\addplot [very thick, crimson2553191, dash pattern=on 3pt off 5pt on 1pt off 5pt, mark=square, mark size=2, mark options={solid,fill opacity=0}]
table {%
	0.5 1.26233696937561
	1.10000002384186 1.69853734970093
	1.60000002384186 2.02626276016235
	2 2.22395038604736
	4 3.09658765792847
	6 3.78653717041016
	8 4.25460052490234
	10 4.57599925994873
	12 4.80854845046997
};
\addlegendentry{GEQ}
\addplot [very thick, dodgerblue0154222, dotted, mark=triangle, mark size=2, mark options={solid,fill opacity=0}]
table {%
	0.5 1.25826120376587
	1.10000002384186 1.69768762588501
	1.60000002384186 2.02637481689453
	2 2.22356224060059
	4 3.09656286239624
	6 3.78346252441406
	8 4.25309944152832
	10 4.57571268081665
	12 4.80978441238403
};
\addlegendentry{AVG}
\end{axis}

\end{tikzpicture}%
	\caption{
		\label{fig:drop-plot-crown-diameter}
		Simulated non-dimensionalized splash crown diameter $d_{\text{cr}}^{*} = d_{\text{cr}} / D$ over dimensionless time $t^{*}$ of the drop impact.
		The splash crown diameter $d_{\text{cr}}$ is its inner diameter, measured at the initial position of the liquid surface in a centrally located cross-section with normal in the $x$-direction.
		The EXT refilling scheme is not included as it quickly became numerically unstable, whereas the EQ+NEQ became unstable only at $t^{*} > 4$.
	}
\end{figure}

\Cref{fig:drop-plot-t-12,fig:drop-plot-cavity-depth,fig:drop-plot-crown-diameter} compare the simulated splash crowns' shapes quantitatively in a centrally located cross-section with normal in the $x$-direction.
The contours in \Cref{fig:drop-plot-t-12} at $t^{*}=12$ differ at the crown's top, where the droplets detach.
Apart from that, the refilling schemes almost led to the same temporal evolution of the dimensionless cavity depth $h_{\text{ca}}^{*} = h_{\text{ca}} / D$ and crown diameter $d_{\text{cr}}^{*} = d_{\text{cr}} / D$.
Both quantities were evaluated in a centrally located cross-section with normal in the $x$-direction.
The cavity depth $h_{\text{ca}}$ was measured from the cavity's bottom to the initial position of the liquid surface at $t^{*} = 0$.
The crown diameter $d_{\text{cr}}$ is the splash crown's inner diameter, also measured at the initial position of the liquid surface.\par

\FloatBarrier

\subsection{Bubbly plane Poiseuille flow}\label{sec:bppf}
This final benchmark case is inspired by Peng et al.~\cite{peng2016ImplementationIssuesBenchmarking}, where a particle-laden turbulent channel flow was simulated.
The choice of the refilling scheme for solid obstacles affected the particle dynamics, that is, the particles' position during the simulation.
In the study presented here, a similar test case is used with randomly initialized spherical bubbles rather than solid particles.
The flow is force-driven between two parallel plates, also called plane Poiseuille flow.
\par

\begin{figure}[htbp]
	\centering
	\setlength{\figureheight}{0.25\textwidth}
	\setlength{\figurewidth}{0.6\textwidth}
	\setlength\mdist{0.02\textwidth}
	\begin{tikzpicture}
\definecolor{darkorange24213334}{RGB}{242,133,34}
\definecolor{dodgerblue0154222}{RGB}{0,154,222}
\definecolor{crimson2553191}{RGB}{255,31,91}

% liquid
\draw [fill=dodgerblue0154222!30,draw=none] (0,0) rectangle (\figurewidth,\figureheight);

% borders left and right
\draw[very thick, loosely dashed, black!50] (0,0)--(0,\figureheight);
\draw[very thick, loosely dashed, black!50] (\figurewidth,0)--(\figurewidth,\figureheight);

% borders top and bottom
\draw[very thick, black!50] (0,\figureheight)--(\figurewidth,\figureheight);
\draw[very thick, black!50] (0,0)--(\figurewidth,0);

% add bubbles
\draw [fill=white,draw=dodgerblue0154222] (0.1\figurewidth,0.1\figureheight) circle[radius=0.1\figureheight];
\draw [fill=white,draw=dodgerblue0154222] (0.52\figurewidth,0.3\figureheight) circle[radius=0.1\figureheight];
\draw [fill=white,draw=dodgerblue0154222] (0.8\figurewidth,0.15\figureheight) circle[radius=0.1\figureheight];
\draw [fill=white,draw=dodgerblue0154222] (0.89\figurewidth,0.71\figureheight) circle[radius=0.1\figureheight];
\draw [fill=white,draw=dodgerblue0154222] (0.3\figurewidth,0.79\figureheight) circle[radius=0.1\figureheight];
\draw [fill=white,draw=dodgerblue0154222] (0.25\figurewidth,0.35\figureheight) circle[radius=0.1\figureheight];
\draw [fill=white,draw=dodgerblue0154222] (0.6\figurewidth,0.88\figureheight) circle[radius=0.1\figureheight];
\draw [fill=white,draw=dodgerblue0154222] (0.7\figurewidth,0.54\figureheight) circle[radius=0.1\figureheight];

% bubble diameter
\draw[<->, >=Latex] (0.52\figurewidth+0.1\figureheight+\mdist,0.3\figureheight-0.1\figureheight)--(0.52\figurewidth+0.1\figureheight+\mdist,0.3\figureheight+0.1\figureheight) node [pos=0.5,right] {$\varnothing 0.1L$};
\draw[-] (0.52\figurewidth,0.3\figureheight+0.1\figureheight)--(0.52\figurewidth+0.1\figureheight+1.5\mdist,0.3\figureheight+0.1\figureheight);
\draw[-] (0.52\figurewidth,0.3\figureheight-0.1\figureheight)--(0.52\figurewidth+0.1\figureheight+1.5\mdist,0.3\figureheight-0.1\figureheight);

% domain height
\draw[<->, >=Latex] (\figurewidth+\mdist,0)--(\figurewidth+\mdist,\figureheight) node [pos=0.5,right] {$L$};
\draw[-] (\figurewidth,0)--(\figurewidth+1.5\mdist,0) node [right] {$z=0$};
\draw[-] (\figurewidth,\figureheight)--(\figurewidth+1.5\mdist,\figureheight);

% domain width
\draw[<->, >=Latex] (0,-\mdist)--(\figurewidth,-\mdist) node [pos=0.5,below] {$2L$};
\draw[-] (0,0)--(0,-1.5\mdist);
\draw[-] (\figurewidth,0)--(\figurewidth,-1.5\mdist);

% initial profile label
\node[
rectangle,
anchor=east,
crimson2553191] at (0.45\figurewidth,0.57\figureheight) {$u(z) = \frac{F}{2\mu}z(L-z)$};

% initial velocity profile
\begin{axis}%
[width=\figureheight, % width and height exchanged due to rotation
height=0.5\figurewidth,
rotate=90,
xmin=-1,
xmax=1,
ymin=0,
ymax=1,
ticks=none,
axis lines=none,
clip=false,
scale only axis,
]
\addplot[thick, name path=f, domain=-1:1,samples=50,smooth,crimson2553191] {x*x};
\end{axis}

% horizontal line of symmetry
%\draw[loosely dashdotted] (-0.025\figurewidth,0.5\figureheight)--(1.025\figurewidth,0.5\figureheight);

% gravity
\draw[thick, ->, >=Latex, darkorange24213334] (\figurewidth-6\mdist,0.5\figureheight)--(\figurewidth-2\mdist,0.5\figureheight) node [pos=0.5,above] {$F$};

% coordinate system
\draw[->, >=Latex] (2\mdist,\figureheight-3\mdist)--(4\mdist,\figureheight-3\mdist) node [below] {$x$};
\draw[->, >=Latex] (2\mdist,\figureheight-3\mdist)--(2\mdist,\figureheight-1\mdist) node [left] {$z$};
\draw[draw=black] (2\mdist,\figureheight-3\mdist) circle [radius=0.175\mdist] node[opacity=1, below left] {$y$};
\end{tikzpicture}%
	\caption{\label{fig:poiseuille-setup}
		Simulation setup of the three-dimensional bubbly plane Poiseuille flow test case.
		In a domain of size $2L \times L \times L$ filled with liquid, the specified parabolic velocity profile is initialized according to the liquid's dynamics viscosity $\mu$ and the force $F$ that acts in the $x$-direction.
		In the domain, $381$ gas bubbles with diameter $0.1L$ are randomly distributed, leading to a gas volume fraction of approximately 0.1.
		The domain's side walls in $x$- and $y$-direction were periodic, whereas the domain's walls in $z$-direction were set to no-slip.
	}
\end{figure}
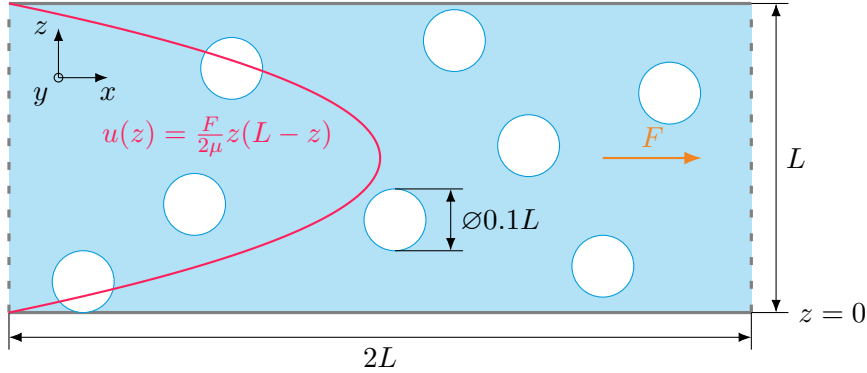

\subsubsection{Simulation setup}\label{sec:bppf-ss}
\begin{figure}[htbp]
	\centering
	\begin{tabular}{
			>{\centering\arraybackslash}m{0.55\textwidth}
			>{\centering\arraybackslash}m{0.4\textwidth}
		}
		
		\parbox[m]{0.55\textwidth}{\includegraphics[width=0.5\textwidth]{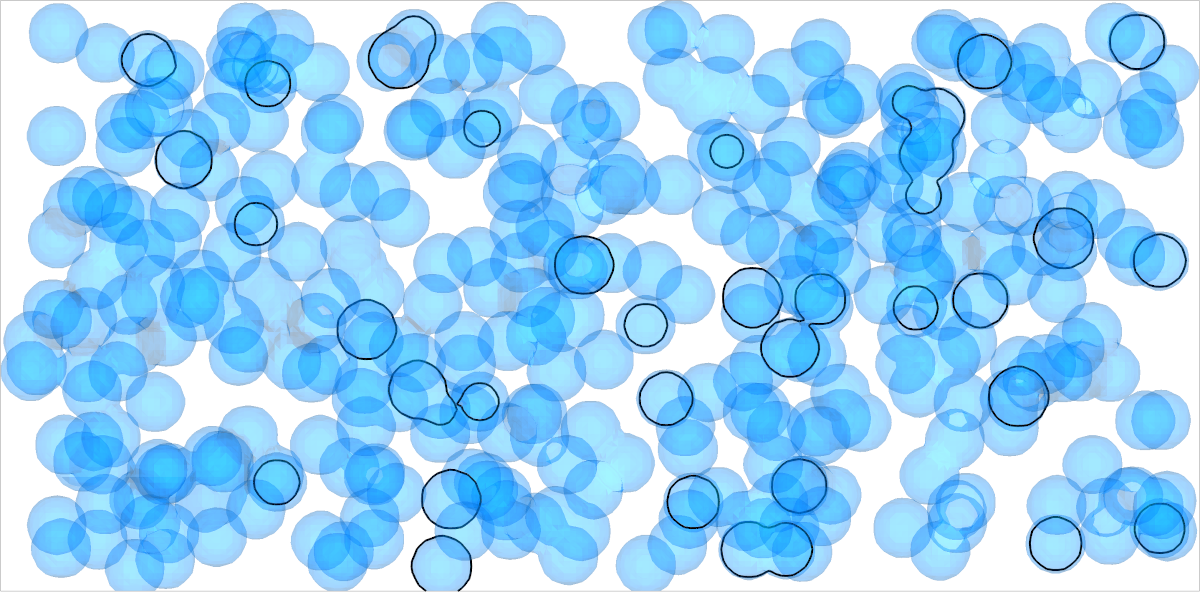}} & \parbox[m]{0.4\textwidth}{\includegraphics[width=0.4\textwidth]{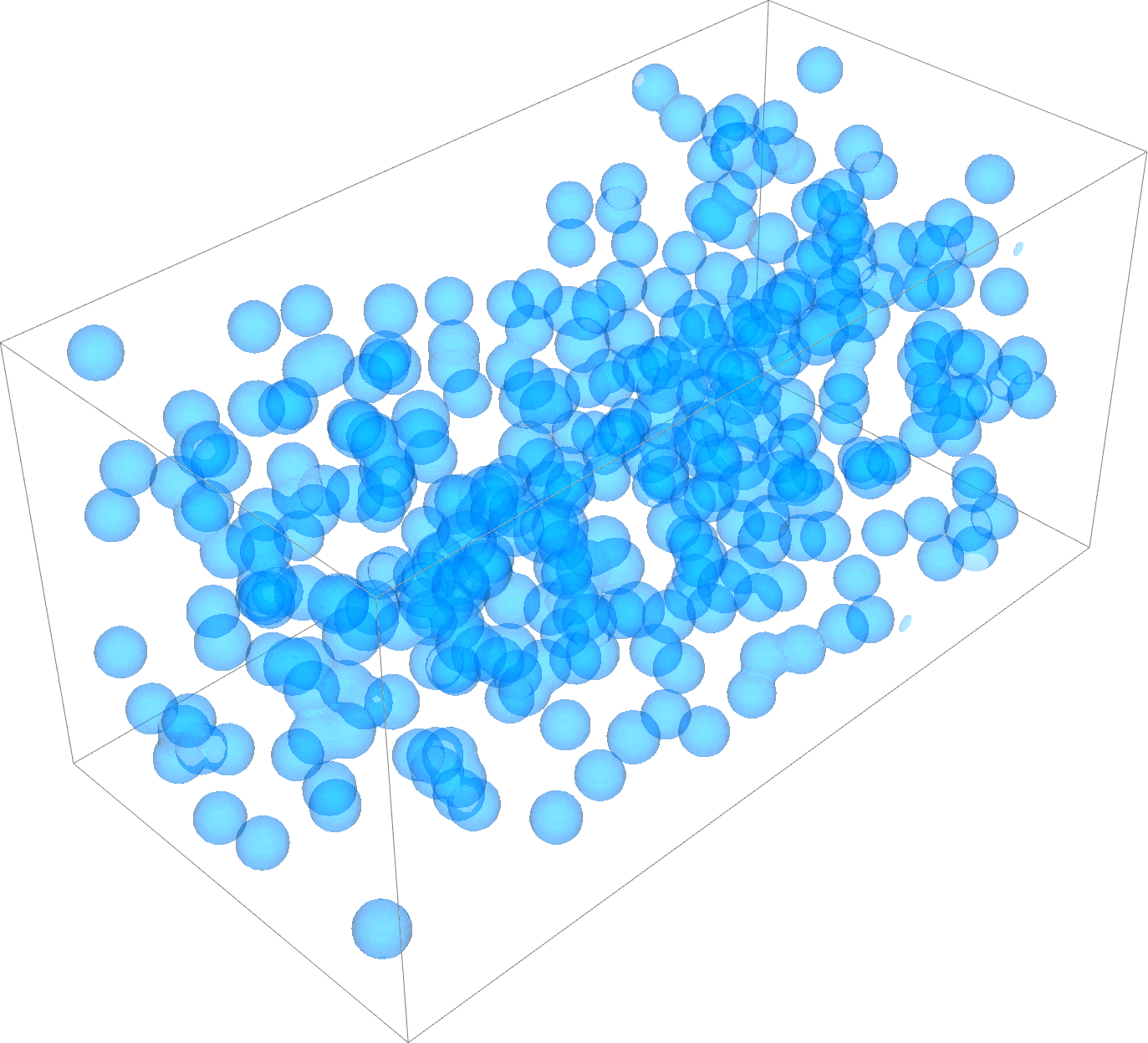}}
	\end{tabular}
	\caption{\label{fig:poiseuille-mesh-t-0}
		Initialized simulation domain of the bubbly plane Poiseuille flow at non-dimensionalized time $t^{*} = 0$ in a side (a) and isometric (b) view.
		The random distribution of the bubbles was chosen once and kept the same for all simulations.
		The solid black lines in the side view show the contour in a centrally located cross-section with normal in the $y$-direction.
	}
\end{figure}

A three-dimensional domain of size $2L \times L \times L$ ($x$-, $y$-, $z$-direction) with a channel width of $L=100$ lattice cells was filled with liquid, as illustrated in \Cref{fig:poiseuille-setup}.
As shown in \Cref{fig:poiseuille-mesh-t-0}, there were $381$ randomly distributed spherical bubbles with a diameter of $0.1L$ in the channel, leading to a gas volume fraction of approximately $0.1$.
The bubbles were arranged so that their center was not closer than $0.05L$ to the domain wall.
The random distribution was chosen once and kept the same for all simulations.
Therefore, the simulation with any refilling scheme started from an identical initial situation.
The domain's walls were periodic in the $x$- and $y$-direction and set to no-slip in the $z$-direction.
With the force $F$ acting in the $x$-direction, the fluid velocity profile took a parabolic shape in the $z$-direction with zero-velocity at the no-slip domain walls.
This velocity profile is commonly referred to as plane Poiseuille flow and analytically given by~\cite{batchelor2000IntroductionFluidDynamics}
\begin{equation}
u(z) = \frac{F}{2\mu}z(L-z),
\end{equation}
where $\mu$ is the dynamic viscosity of the liquid.
The setup is defined by the Morton number $\text{Mo}=10^{-5}$~\eqref{eq:taylor-mo}
and by the Reynolds number $\text{Re}=10^{4}$~\eqref{eq:taylor-re} with characteristic length $L$ and analytical maximum velocity $u_{\text{max}} = u(0.5 L)$.
The relaxation rate was chosen $\omega = 1.989$, and the time $t$ was non-dimensionalized with $t^{*} = t \, u_{\text{max}} / L$.
\par

\subsubsection{Results and discussion}\label{sec:bppf-res}
\begin{figure}[htbp]
	\vspace{-3\baselineskip}
	\centering
	\begin{tabular}{
			>{\centering\arraybackslash}m{0.55\textwidth}
			>{\centering\arraybackslash}m{0.4\textwidth}
		}
		
		\begin{tikzpicture}
			\node[anchor=south] at (0,0) {\includegraphics[width=0.5\textwidth]{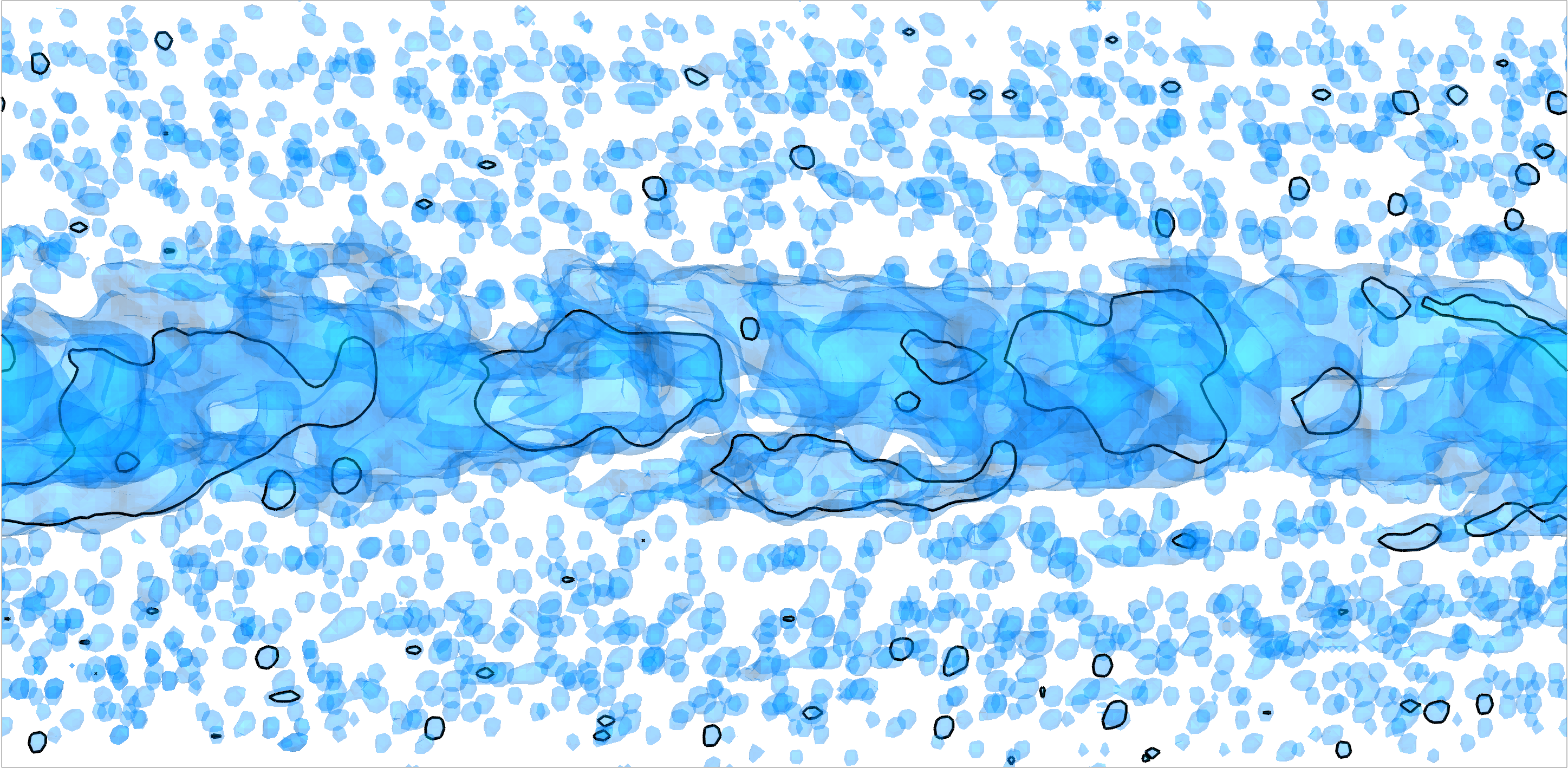}};
			\node[anchor=north] at (0,0){(a) EQ, side view};
		\end{tikzpicture} &
		\begin{tikzpicture}
			\node[anchor=south] at (0,0) {\includegraphics[width=0.4\textwidth]{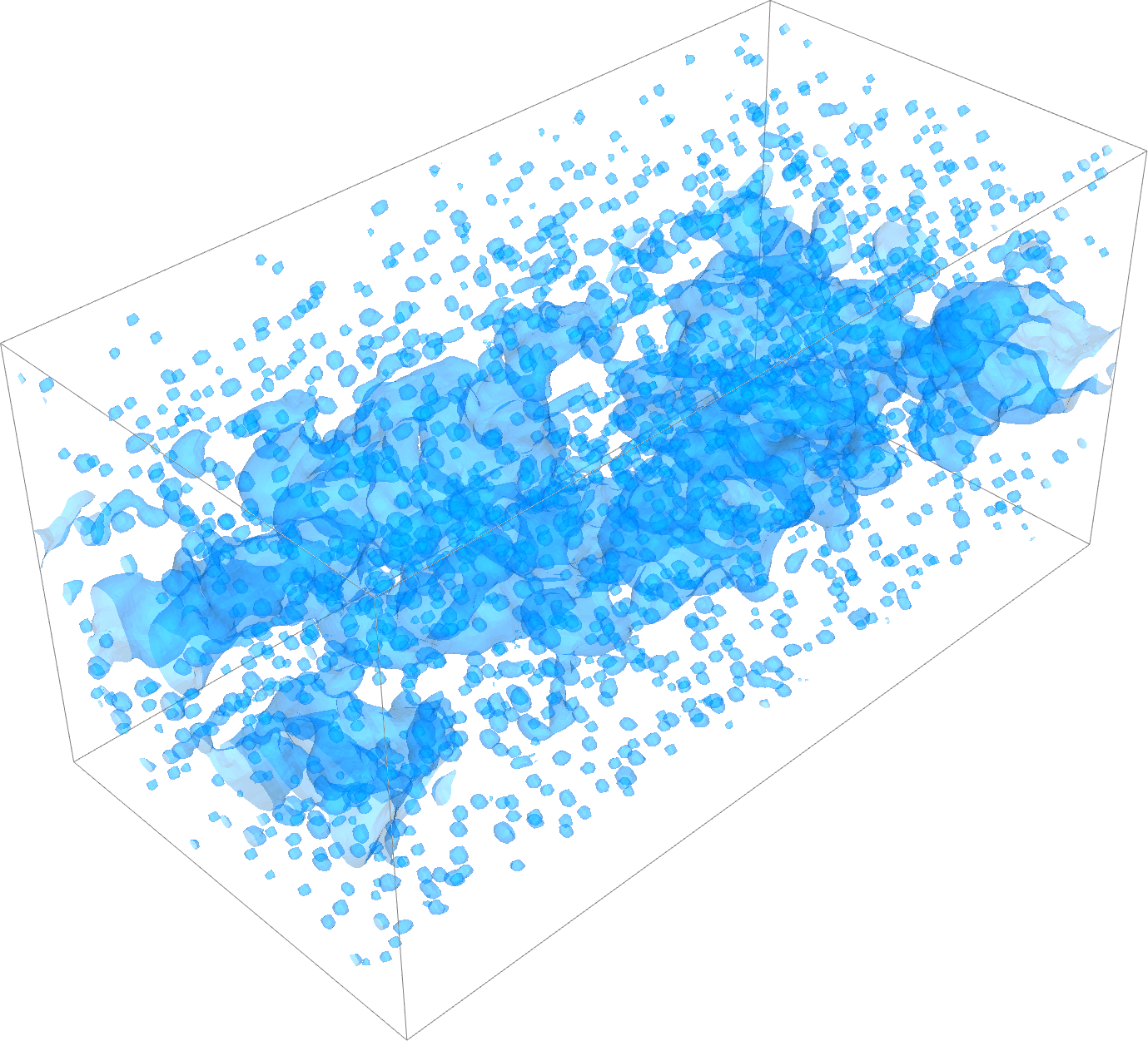}};
			\node[anchor=north] at (0,0){(b) EQ, isometric view};
		\end{tikzpicture} \\

		\begin{tikzpicture}
			\node[anchor=south] at (0,0) {\includegraphics[width=0.5\textwidth]{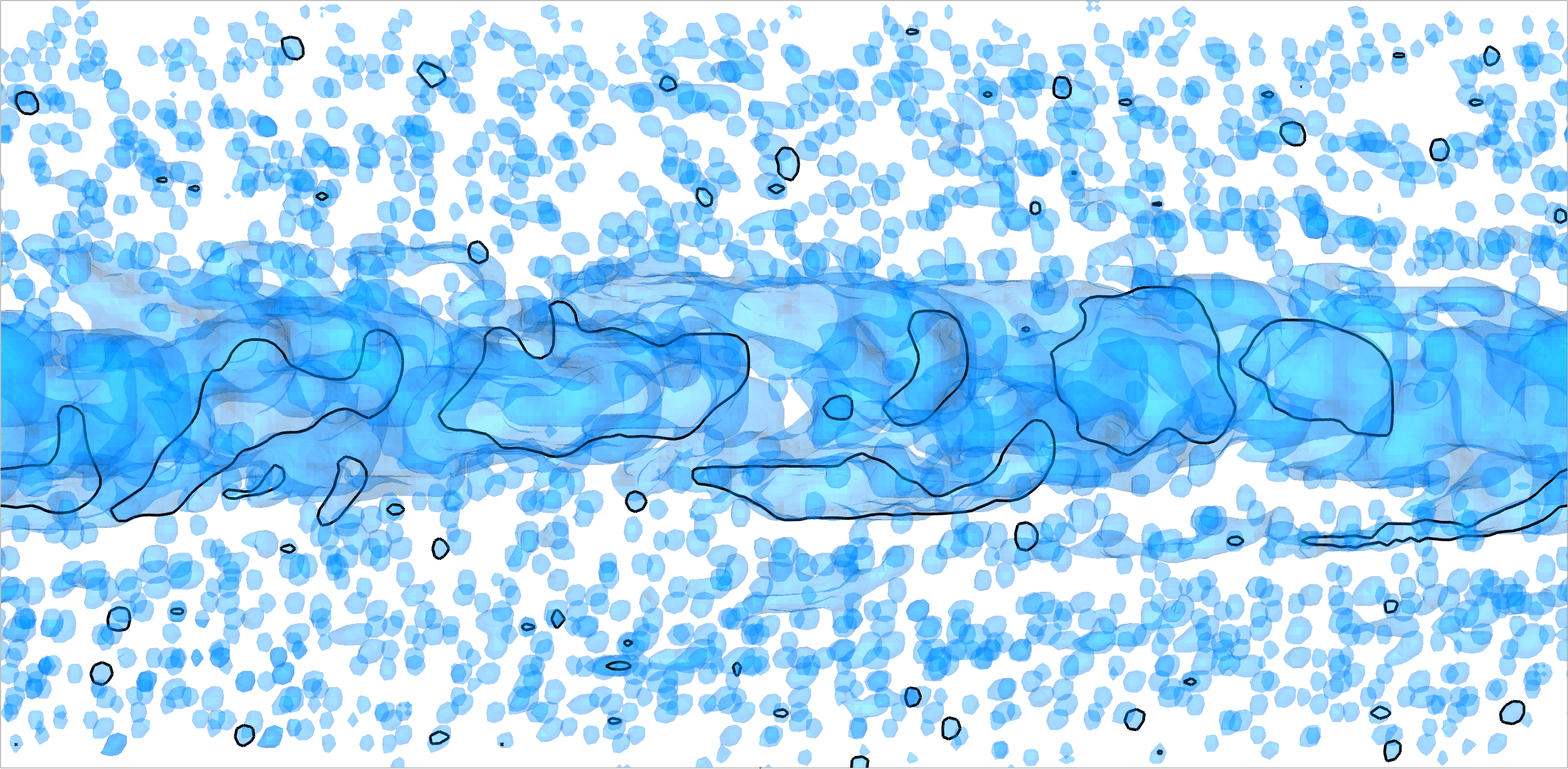}};
			\node[anchor=north] at (0,0){(c) GEQ, side view};
		\end{tikzpicture} &
		\begin{tikzpicture}
			\node[anchor=south] at (0,0) {\includegraphics[width=0.4\textwidth]{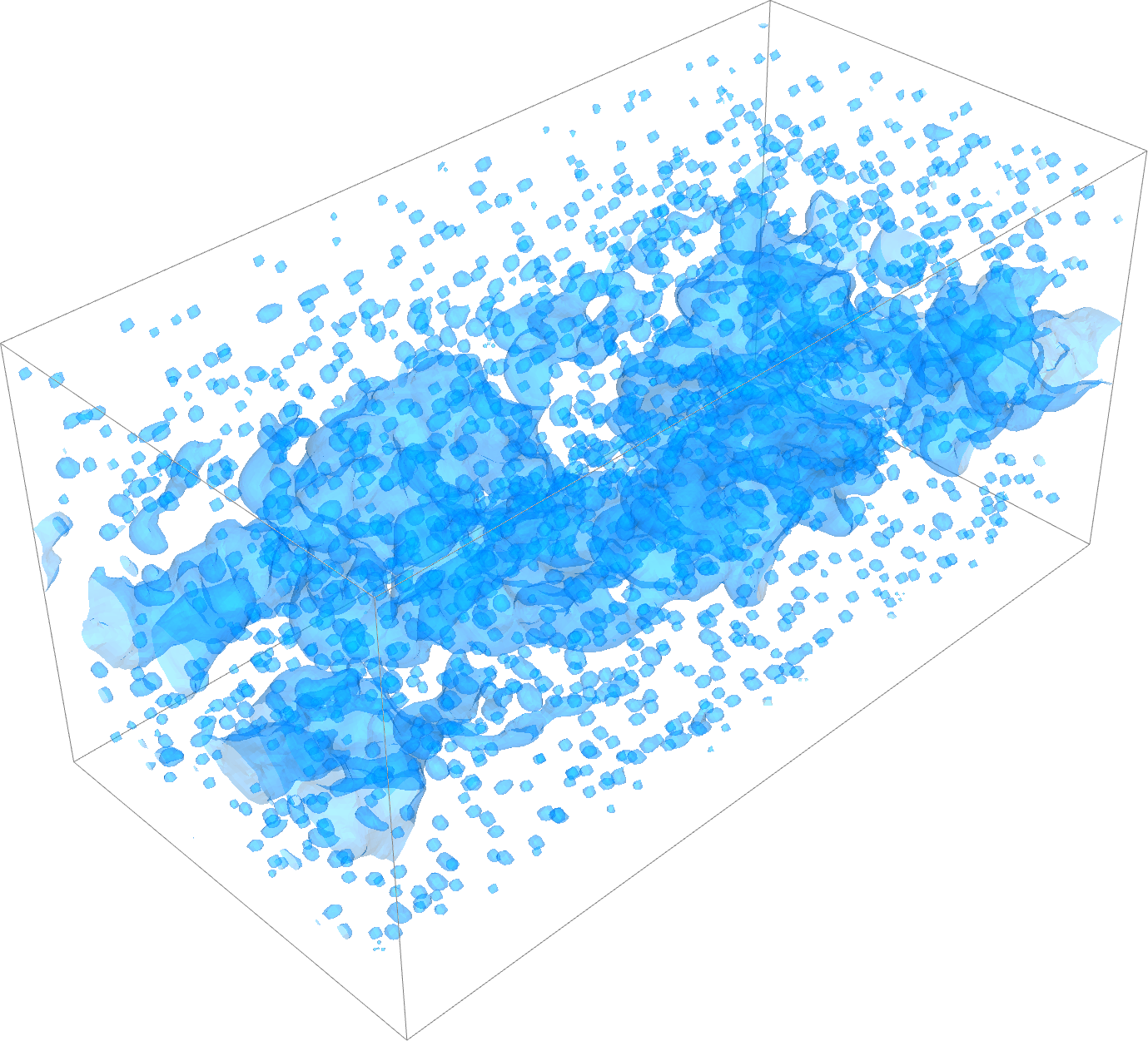}};
			\node[anchor=north] at (0,0){(d) GEQ, isometric view};
		\end{tikzpicture} \\

		\begin{tikzpicture}
			\node[anchor=south] at (0,0) {\includegraphics[width=0.5\textwidth]{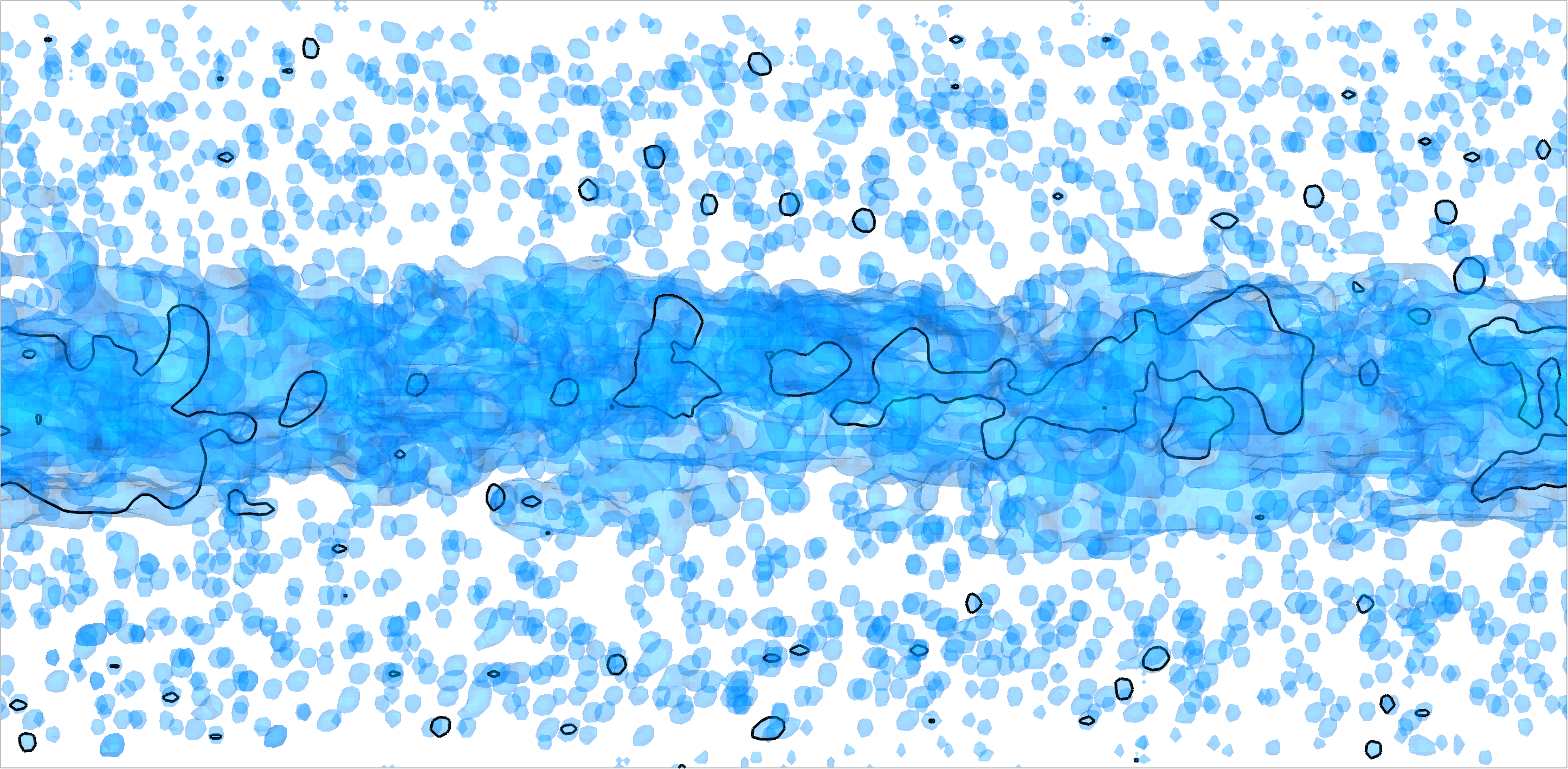}};
			\node[anchor=north] at (0,0){(e) AVG, side view};
		\end{tikzpicture} &
		\begin{tikzpicture}
			\node[anchor=south] at (0,0) {\includegraphics[width=0.4\textwidth]{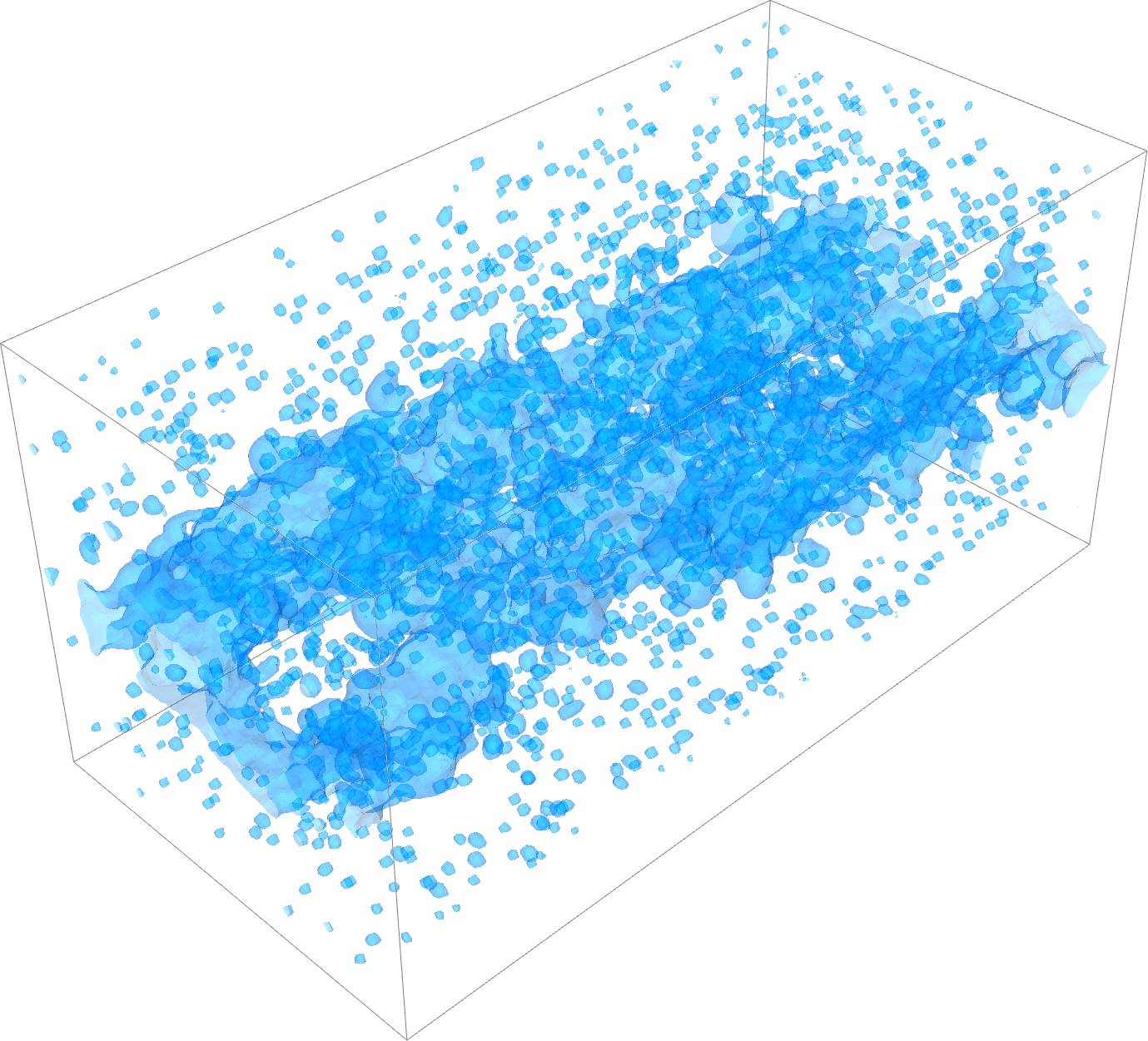}};
			\node[anchor=north] at (0,0){(f) AVG, isometric view};
		\end{tikzpicture}
	\end{tabular}
	\caption{\label{fig:poiseuille-mesh-t-4}
		Simulated bubbly plane Poiseuille flow at non-dimensionalized time $t^{*} = 4$.
		The simulations were performed with a computational domain resolution, that is, initial channel width of $L=100$ lattice cells with the EQ (a and b), GEQ (c and d), and AVG (e and f) refilling schemes.
		The solid black lines show the contour in a centrally located cross-section with normal in the $y$-direction.
		The simulations with the EQ+NEQ and EXT schemes were numerically unstable and are not included here.
		The refilling schemes noticeably affected the bubbles' position and shape.
	}
\end{figure}

As for the drop impact test case in \Cref{sec:ne-di}, the simulations with the EXT and EQ+NEQ refilling schemes were numerically unstable.
The simulation results for the AVG, EQ, and GEQ schemes at $t^{*} = 4$ are shown in \Cref{fig:poiseuille-mesh-t-4}.
The bubbles gathered and coalesced in the center of the domain, where the velocity was the highest.
Although all simulations started from the same initial situation, the refilling schemes led to noticeable differences in the bubble dynamics, that is, the bubbles' positions.
This observation agrees with those made by Peng et al.~\cite{peng2016ImplementationIssuesBenchmarking}, where a turbulent particle-laden channel flow was simulated with different refilling schemes for solid obstacles.
However, due to the lack of missing reference data from experiments, the refilling schemes' accuracy could not be assessed in this test case.
\par

%!TEX root = ../main.tex

\section{Conclusions}\label{sec:conclusion}
This study has compared different refilling schemes for the free-surface lattice Boltzmann method~\cite{korner2005LatticeBoltzmannModel} (FSLBM).
In the FSLBM, it is distinguished between cells belonging to the heavier fluid, lighter fluid, and the interface located between them.
These cells are here referred to as liquid, gas, and interface cells, respectively.
The gas phase is neglected and the interface is treated as a free surface.
Consequently, gas cells neither participate in the LBM flow simulation, nor carry valid information about the flow field.
In the LBM, such information is stored in terms of particle distribution functions (PDFs) in each lattice cell.
Because of the free interface's motion, gas cells regularly convert to interface cells.
As the hydrodynamic LBM simulations are performed in interface cells, those cells' PDFs must be initialized with valid information during the conversion.
This initialization of PDFs is commonly referred to as refilling.
The first refilling scheme under investigation was the one suggested in the original FSLBM as introduced by Körner et al.~\cite{korner2005LatticeBoltzmannModel}.
In this model, PDFs are initialized according to their equilibrium (EQ), which is constructed using the average density and velocity from the neighboring, non-newly converted interface and liquid cells.
This scheme was extended by adding the contribution of the neighboring cells' non-equilibrium PDFs (EQ+NEQ)~\cite{peng2016ImplementationIssuesBenchmarking}, or by including information about the local pressure tensor using Grad's moment system (GEQ)~\cite{dorschner2015GradApproximationMoving,krithivasan2014DiffusedBouncebackCondition,chikatamarla2006GradApproximationMissing,grad1949KineticTheoryRarefied}.
Additionally, the PDFs could also be extrapolated (EXT) from neighboring cells' PDFs~\cite{lallemand2003LatticeBoltzmannMethod} or were taken as the average (AVG) of neighboring, non-newly converted interface and liquid cells' PDFs~\cite{fang2002LatticeBoltzmannMethod}.
\par

These schemes' accuracy and stability properties were investigated in six numerical experiments, with reference data for five of them as either analytical models or laboratory measurements from the literature.
In the experiments conducted here, the EXT and EQ+NEQ schemes often led to numerical instabilities.
These instabilities were caused by the lattice velocity exceeding the lattice speed of sound.
The AVG refilling scheme was also unstable in the cylindrical dam break test case in one of the computational domain resolutions used in the convergence study.
In contrast, the EQ, and GEQ schemes were numerically stable in all simulations performed here.
Although, the AVG scheme was more accurate than the EQ and GEQ schemes in the dam break test cases, it slightly overestimated the gravity wave's amplitude in the first period.
The EQ, and GEQ schemes' simulation results hardly differed when compared in terms of the quantitative reference data available in the literature.
Nevertheless, qualitative differences between these schemes could be observed in the dam break, drop impact, and bubbly Poiseuille flow test cases.
Because of lacking appropriate reference data, a final accuracy comparison could only be made vaguely based on visual comparison.
In the drop impact benchmark, the EQ scheme arguably seemed to be favorable over the GEQ scheme when compared qualitatively.
Additionally the GEQ scheme is computationally more expensive than the EQ scheme.
In summary, for the numerical simulations performed here, the EQ scheme should be preferred in the FSLBM with respect to ease of implementation, computational costs, numerical stability, and accuracy.

\section*{Supplementary material}
The following supplementary material is available as part of the online article:\\
An archive of the C++ source code used in this study.
It is part of the software framework \walberla{}\cite{bauer2021WaLBerlaBlockstructuredHighperformance} (version used here: \url{https://i10git.cs.fau.de/walberla/walberla/-/tree/01a28162ae1aacf7b96152c9f886ce54cc7f53ff}).
The ready-to-run simulation setups for all numerical experiments performed in this article are included in the directory \texttt{apps/showcases/FreeSurface}.

\begin{acknowledgments}
	The authors thank the Deutsche Forschungsgemeinschaft (DFG, German Research Foundation) for funding project 408062554.\\
	This work was supported by the SCALABLE project.
	This project has received funding from the European High-Performance Computing Joint Undertaking (JU) under grant agreement No 956000.
	The JU receives support from the European Union’s Horizon 2020 research and innovation programme and France, Germany, the Czech Republic.\\
	The authors gratefully acknowledge the Gauss Centre for Supercomputing e.V. (www.gauss-centre.eu) for funding this project by providing computing time on the GCS Supercomputer SuperMUC at Leibniz Supercomputing Centre (www.lrz.de).\\
	The authors gratefully acknowledge the scientific support and HPC resources provided by the Erlangen National High Performance Computing Center (NHR@FAU) of the Friedrich-Alexander-Universität Erlangen-Nürnberg (FAU).
	The hardware is funded by the German Research Foundation (DFG).\\
	We acknowledge financial support by Deutsche Forschungsgemeinschaft and Friedrich-Alexander-Universität Erlangen-Nürnberg within the funding programme \enquote{Open Access Publication Funding}.\\
	The authors appreciate the valuable discussions with Christoph Rettinger, and thank Sara Faghih-Naini and Jonas Plewinski for proofreading the manuscript.
\end{acknowledgments}

\section*{AIP publishing data sharing policy}
The data that support the findings of this study can be reproduced with the open-source C++ software framework \walberla{}~\cite{bauer2021WaLBerlaBlockstructuredHighperformance} (\url{https://i10git.cs.fau.de/walberla/walberla/-/tree/01a28162ae1aacf7b96152c9f886ce54cc7f53ff}).

\newpage

\appendix
%!TEX root = ../main.tex

\section{Gravity wave}

\begin{figure}[htbp]
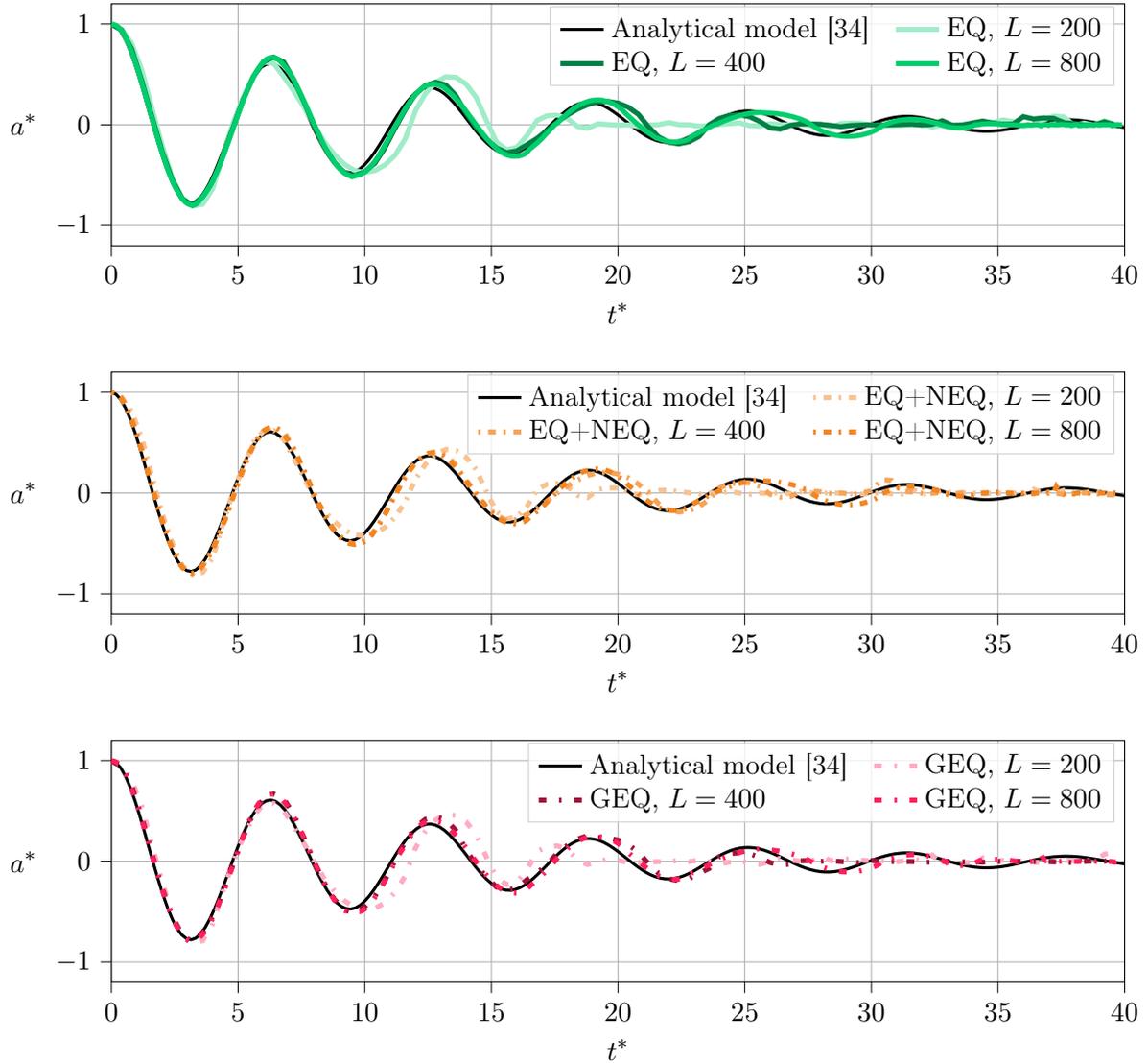

	\centering
	\setlength{\figureheight}{0.3\textwidth}
	\setlength{\figurewidth}{0.95\textwidth}
	\begin{subfigure}{\textwidth}
		\input{figures/gravity-wave/convergence-eq.tex}%
	\end{subfigure}%
	\vspace{0.5cm}
	\begin{subfigure}{\textwidth}
		\input{figures/gravity-wave/convergence-eq-neq.tex}%
	\end{subfigure}%
	\vspace{0.5cm}
	\begin{subfigure}{\textwidth}
		\input{figures/gravity-wave/convergence-geq.tex}%
	\end{subfigure}
	\caption{
		Simulated surface elevation of the gravity wave in terms of non-dimensional amplitude $a^{*}(0,t^{*})$ and time $t^{*}$.
		The simulations were performed with computational domain resolutions, that is, wavelengths of $L \in \{200, 400, 800\}$ lattice cells.
		The simulations converged well, and a higher resolution captures more of the standing wave's oscillations.
	}
	\label{fig:ne-gravity-wave-convergence}
\end{figure}

\begin{figure}[htbp]
	\centering
	\setlength{\figureheight}{0.3\textwidth}
	\setlength{\figurewidth}{0.95\textwidth}
	\begin{subfigure}{\textwidth}
		\input{figures/gravity-wave/convergence-ext.tex}%
	\end{subfigure}%
	\vspace{0.5cm}
	\begin{subfigure}{\textwidth}
		\input{figures/gravity-wave/convergence-avg.tex}%
	\end{subfigure}
	\addtocounter{figure}{-1}
	\caption{(\textit{Continued})}
\end{figure}

\clearpage

\section{Rectangular dam break}

\begin{figure}[htbp]
	\centering
	\setlength{\figureheight}{0.35\textwidth}
	\setlength{\figurewidth}{0.45\textwidth}
	\begin{subfigure}[htbp]{0.49\textwidth}
		\centering
		% This file was created with tikzplotlib v0.10.1.
\begin{tikzpicture}

\definecolor{darkgray176}{RGB}{176,176,176}
\definecolor{lightgray204}{RGB}{204,204,204}
\definecolor{mediumaquamarine128230181}{RGB}{159,236,200}
\definecolor{mediumaquamarine64218145}{RGB}{0,128,68}
\definecolor{springgreen0205108}{RGB}{0,205,108}

\begin{axis}[
height=\figureheight,
legend cell align={left},
legend style={
	fill opacity=0.8,
	draw opacity=1,
	text opacity=1,
	at={(0.99,0.99)},
	anchor=north east,
	draw=lightgray204
},
tick align=outside,
tick pos=left,
width=\figurewidth,
x grid style={darkgray176},
xlabel={\(\displaystyle t^{*}\)},
xmajorgrids,
xmin=0, xmax=10,
xtick style={color=black},
y grid style={darkgray176},
ylabel style={rotate=-90.0},
ylabel={\(\displaystyle h^{*}\)},
ymajorgrids,
ymin=0.013, ymax=1.047,
ytick style={color=black}
]
\addplot [very thick, black, mark=*, mark size=2, mark options={solid}, only marks]
table {%
0 1
0.559999942779541 0.940000057220459
0.769999980926514 0.889999985694885
0.930000066757202 0.829999923706055
1.08000004291534 0.779999971389771
1.27999997138977 0.720000028610229
1.46000003814697 0.670000076293945
1.6599999666214 0.610000014305115
1.8400000333786 0.559999942779541
2 0.5
2.21000003814697 0.440000057220459
2.45000004768372 0.389999985694885
2.70000004768372 0.330000042915344
3.05999994277954 0.279999971389771
3.44000005722046 0.220000028610229
4.19999980926514 0.169999957084656
5.25 0.110000014305115
7.40000009536743 0.059999942779541
};
\addlegendentry{Experiment~\cite{martin1952PartIVExperimental}}
\addplot [line width=2pt, mediumaquamarine128230181]
table {%
0 1
0.0998772382736206 0.990000009536743
0.29963207244873 0.990000009536743
0.599262952804565 0.960000038146973
0.699140071868896 0.940000057220459
0.799018025398254 0.930000066757202
1.29840004444122 0.829999923706055
1.3982800245285 0.799999952316284
1.49816000461578 0.779999971389771
1.59803998470306 0.75
1.79779005050659 0.710000038146973
1.89767003059387 0.680000066757202
2.89644002914429 0.480000019073486
2.99632000923157 0.470000028610229
3.19606995582581 0.430000066757202
3.29594993591309 0.419999957084656
3.39582991600037 0.399999976158142
3.49569988250732 0.389999985694885
3.59558010101318 0.370000004768372
3.79533004760742 0.350000023841858
3.8952100276947 0.330000042915344
4.19483995437622 0.299999952316284
4.29472017288208 0.279999971389771
4.79410982131958 0.230000019073486
4.89398002624512 0.240000009536743
4.99385976791382 0.230000019073486
5.09373998641968 0.230000019073486
5.39337015151978 0.200000047683716
5.59312009811401 0.200000047683716
5.79288005828857 0.180000066757202
5.89275979995728 0.180000066757202
6.09251022338867 0.159999966621399
6.19238996505737 0.159999966621399
6.29226016998291 0.149999976158142
6.39213991165161 0.149999976158142
6.49202013015747 0.139999985694885
6.59189987182617 0.139999985694885
6.69177007675171 0.129999995231628
6.79164981842041 0.129999995231628
6.89153003692627 0.120000004768372
6.99139976501465 0.129999995231628
7.09127998352051 0.129999995231628
7.19116020202637 0.120000004768372
7.29103994369507 0.129999995231628
7.39091014862061 0.120000004768372
7.49078989028931 0.129999995231628
7.69053983688354 0.129999995231628
7.7904200553894 0.120000004768372
7.89029979705811 0.120000004768372
7.99018001556396 0.110000014305115
8.38969039916992 0.110000014305115
8.4895601272583 0.120000004768372
8.58944034576416 0.120000004768372
8.6893196105957 0.110000014305115
8.7891902923584 0.110000014305115
8.9889497756958 0.0900000333786011
9.18869972229004 0.0900000333786011
9.2885799407959 0.100000023841858
};
\addlegendentry{EQ, $W=50$}
\addplot [line width=2pt, mediumaquamarine64218145]
table {%
0 1
0.0998772382736206 0.995000004768372
0.199753999710083 0.995000004768372
0.29963207244873 0.990000009536743
0.599262952804565 0.960000038146973
0.799018025398254 0.930000066757202
0.898895025253296 0.910000085830688
0.998772025108337 0.894999980926514
1.0986499786377 0.875
1.19852995872498 0.850000023841858
1.29840004444122 0.829999923706055
1.49816000461578 0.779999971389771
1.59803998470306 0.759999990463257
1.79779005050659 0.710000038146973
1.89767003059387 0.690000057220459
1.99753999710083 0.664999961853027
2.09741997718811 0.644999980926514
2.19729995727539 0.620000004768372
2.69668006896973 0.519999980926514
2.79656004905701 0.504999995231628
2.89644002914429 0.485000014305115
2.99632000923157 0.470000028610229
3.09618997573853 0.450000047683716
3.19606995582581 0.434999942779541
3.29594993591309 0.414999961853027
3.39582991600037 0.404999971389771
3.8952100276947 0.330000042915344
4.09497022628784 0.309999942779541
4.19483995437622 0.294999957084656
4.39459991455078 0.274999976158142
4.49447011947632 0.269999980926514
4.79410982131958 0.240000009536743
4.89398002624512 0.235000014305115
4.99385976791382 0.225000023841858
5.19361019134521 0.215000033378601
5.29348993301392 0.205000042915344
5.39337015151978 0.200000047683716
5.59312009811401 0.180000066757202
6.09251022338867 0.154999971389771
6.29226016998291 0.154999971389771
6.59189987182617 0.139999985694885
6.69177007675171 0.139999985694885
6.79164981842041 0.129999995231628
6.89153003692627 0.129999995231628
6.99139976501465 0.125
7.19116020202637 0.125
7.39091014862061 0.115000009536743
7.49078989028931 0.115000009536743
7.59067010879517 0.120000004768372
7.7904200553894 0.120000004768372
7.99018001556396 0.110000014305115
8.09004974365234 0.110000014305115
8.28981018066406 0.100000023841858
8.4895601272583 0.100000023841858
8.7891902923584 0.0850000381469727
8.88906955718994 0.0850000381469727
8.9889497756958 0.0800000429153442
};
\addlegendentry{EQ, $W=100$}
\addplot [line width=2pt, springgreen0205108]
table {%
0 1
0.199880957603455 0.995000004768372
0.299821019172668 0.990000009536743
0.399760961532593 0.982500076293945
0.499701023101807 0.972500085830688
0.599642038345337 0.960000038146973
0.799521923065186 0.930000066757202
0.99940299987793 0.894999980926514
1.09933996200562 0.872499942779541
1.19928002357483 0.852499961853027
1.29921996593475 0.829999923706055
1.39916002750397 0.805000066757202
1.49909996986389 0.782500028610229
1.59904003143311 0.757499933242798
1.69897997379303 0.735000014305115
1.79892003536224 0.710000038146973
2.09875011444092 0.642499923706055
2.19868993759155 0.622499942779541
2.29862999916077 0.600000023841858
2.49850988388062 0.559999942779541
2.59844994544983 0.542500019073486
2.79833006858826 0.502500057220459
2.99820995330811 0.467499971389771
3.09815001487732 0.452499985694885
3.19809007644653 0.434999942779541
3.69778990745544 0.360000014305115
4.19749021530151 0.297500014305115
4.49731016159058 0.267500042915344
4.59724998474121 0.259999990463257
4.69718980789185 0.25
5.19688987731934 0.212499976158142
5.29683017730713 0.207499980926514
5.59665012359619 0.184999942779541
5.69658994674683 0.182500004768372
5.79653978347778 0.174999952316284
6.39618015289307 0.144999980926514
6.4961199760437 0.144999980926514
6.59605979919434 0.139999985694885
6.69600009918213 0.137500047683716
6.79593992233276 0.132499933242798
6.99582004547119 0.127500057220459
7.09575986862183 0.122499942779541
7.19570016860962 0.122499942779541
7.29563999176025 0.120000004768372
7.39557981491089 0.115000009536743
7.49552011489868 0.115000009536743
7.89527988433838 0.105000019073486
7.99522018432617 0.105000019073486
8.19509983062744 0.100000023841858
8.29504013061523 0.100000023841858
8.4949197769165 0.0950000286102295
8.5948600769043 0.0950000286102295
8.69480037689209 0.0924999713897705
8.79473972320557 0.0924999713897705
8.89468002319336 0.0900000333786011
8.99462032318115 0.0900000333786011
};
\addlegendentry{EQ, $W=200$}
\end{axis}

\end{tikzpicture}%
	\end{subfigure}
	\hfill
	\begin{subfigure}[htbp]{0.49\textwidth}
		\centering
		% This file was created with tikzplotlib v0.10.1.
\begin{tikzpicture}

\definecolor{darkgray176}{RGB}{176,176,176}
\definecolor{lightgray204}{RGB}{204,204,204}
\definecolor{mediumaquamarine128230181}{RGB}{159,236,200}
\definecolor{mediumaquamarine64218145}{RGB}{0,128,68}
\definecolor{springgreen0205108}{RGB}{0,205,108}

\begin{axis}[
height=\figureheight,
legend cell align={left},
legend style={
  fill opacity=0.8,
  draw opacity=1,
  text opacity=1,
  at={(0.01,0.99)},
  anchor=north west,
  draw=lightgray204, 
},
tick align=outside,
tick pos=left,
width=\figurewidth,
x grid style={darkgray176},
xlabel={\(\displaystyle t^{*}\)},
xmajorgrids,
xmin=0, xmax=10,
xtick style={color=black},
y grid style={darkgray176},
ylabel style={rotate=-90.0},
ylabel={\(\displaystyle w^{*}\)},
ymajorgrids,
ymin=0, ymax=28,
ytick style={color=black}
]
\addplot [very thick, black, mark=*, mark size=2, mark options={solid}, only marks]
table {%
0.409999966621399 1.11000001430511
0.839999914169312 1.22000002861023
1.19000005722046 1.44000005722046
1.42999994754791 1.66999995708466
1.62999999523163 1.88999998569489
1.83000004291534 2.10999989509583
1.98000001907349 2.32999992370605
2.20000004768372 2.55999994277954
2.3199999332428 2.77999997138977
2.50999999046326 3
2.65000009536743 3.22000002861023
2.82999992370605 3.44000005722046
2.97000002861023 3.67000007629395
3.10999989509583 3.89000010490417
3.32999992370605 4.1100001335144
4.01999998092651 5
4.44000005722046 5.8899998664856
5.09000015258789 7
5.69000005722046 8
6.30000019073486 9
6.82999992370605 10
7.44000005722046 11
8.07999992370605 12
8.67000007629395 13
9.3100004196167 14
};
\addlegendentry{Experiment~\cite{martin1952PartIVExperimental}}
\addplot [line width=2pt, mediumaquamarine128230181]
table {%
0 1
0.0998772382736206 1
0.199753999710083 1.01999998092651
0.499386072158813 1.13999998569489
0.699140071868896 1.25999999046326
1.0986499786377 1.58000004291534
1.49816000461578 1.98000001907349
1.59803998470306 2.09999990463257
1.69790995121002 2.20000004768372
1.99753999710083 2.55999994277954
2.09741997718811 2.70000004768372
2.19729995727539 2.8199999332428
2.59681010246277 3.38000011444092
2.69668006896973 3.53999996185303
2.79656004905701 3.6800000667572
2.89644002914429 3.83999991416931
2.99632000923157 3.98000001907349
3.19606995582581 4.30000019073486
3.29594993591309 4.48000001907349
3.39582991600037 4.6399998664856
4.19483995437622 6.07999992370605
4.29472017288208 6.23999977111816
4.39459991455078 6.42000007629395
4.59434986114502 6.73999977111816
4.69423007965088 6.92000007629395
4.79410982131958 7.05999994277954
4.89398002624512 7.21999979019165
4.99385976791382 7.40000009536743
5.29348993301392 7.88000011444092
5.99263000488281 9.14000034332275
6.09251022338867 9.30000019073486
6.29226016998291 9.65999984741211
6.39213991165161 9.81999969482422
6.49202013015747 10
6.69177007675171 10.3199996948242
6.79164981842041 10.5
6.89153003692627 10.6400003433228
7.09127998352051 10.960000038147
7.19116020202637 11.1000003814697
7.49078989028931 11.5799999237061
7.59067010879517 11.7200002670288
8.28981018066406 12.8400001525879
8.38969039916992 13.0200004577637
8.58944034576416 13.3400001525879
8.6893196105957 13.5200004577637
8.7891902923584 13.6800003051758
8.88906955718994 13.8599996566772
8.9889497756958 14.0200004577637
9.08882999420166 14.2200002670288
9.2885799407959 14.5799999237061
};
\addlegendentry{EQ, $W=50$}
\addplot [line width=2pt, mediumaquamarine64218145]
table {%
0 1
0.0998772382736206 1
0.199753999710083 1.01999998092651
0.29963207244873 1.04999995231628
0.499386072158813 1.14999997615814
0.699140071868896 1.26999998092651
0.799018025398254 1.3400000333786
0.898895025253296 1.41999995708466
1.19852995872498 1.69000005722046
1.3982800245285 1.88999998569489
1.59803998470306 2.10999989509583
1.99753999710083 2.58999991416931
2.19729995727539 2.84999990463257
2.29717993736267 2.99000000953674
2.39704990386963 3.11999988555908
2.49692988395691 3.26999998092651
2.59681010246277 3.41000008583069
2.89644002914429 3.85999989509583
3.09618997573853 4.17999982833862
3.19606995582581 4.34999990463257
3.39582991600037 4.71000003814697
3.49569988250732 4.88000011444092
3.79533004760742 5.42000007629395
3.8952100276947 5.59000015258789
3.99509000778198 5.76999998092651
4.19483995437622 6.1100001335144
4.39459991455078 6.46999979019165
4.49447011947632 6.6399998664856
4.79410982131958 7.17999982833862
4.89398002624512 7.34999990463257
4.99385976791382 7.53000020980835
5.09373998641968 7.71999979019165
5.19361019134521 7.90000009536743
5.29348993301392 8.10000038146973
5.39337015151978 8.27999973297119
5.49324989318848 8.47000026702881
5.79288005828857 9.01000022888184
5.99263000488281 9.35000038146973
6.09251022338867 9.5
6.39213991165161 9.97999954223633
6.49202013015747 10.1599998474121
6.59189987182617 10.3299999237061
6.69177007675171 10.5100002288818
6.79164981842041 10.6800003051758
7.09127998352051 11.2200002670288
7.19116020202637 11.4099998474121
7.29103994369507 11.5900001525879
7.49078989028931 11.9700002670288
7.59067010879517 12.1499996185303
7.69053983688354 12.3400001525879
7.89029979705811 12.6800003051758
7.99018001556396 12.8400001525879
8.1899299621582 13.1800003051758
8.58944034576416 13.8999996185303
8.6893196105957 14.0900001525879
8.7891902923584 14.2700004577637
8.88906955718994 14.4700002670288
8.9889497756958 14.6599998474121
};
\addlegendentry{EQ, $W=100$}
\addplot [line width=2pt, springgreen0205108]
table {%
0 1
0.099940299987793 1.00499999523163
0.199880957603455 1.02499997615814
0.299821019172668 1.05999994277954
0.399760961532593 1.10000002384186
0.499701023101807 1.14999997615814
0.599642038345337 1.21000003814697
0.699581980705261 1.27499997615814
0.799521923065186 1.35000002384186
0.99940299987793 1.50999999046326
1.09933996200562 1.60000002384186
1.19928002357483 1.69500005245209
1.39916002750397 1.89499998092651
1.59904003143311 2.11500000953674
1.69897997379303 2.23000001907349
1.89885997772217 2.48000001907349
1.9988100528717 2.60999989509583
2.09875011444092 2.74499988555908
2.19868993759155 2.875
2.29862999916077 3.00999999046326
2.39857006072998 3.15000009536743
2.59844994544983 3.42000007629395
2.69839000701904 3.55999994277954
2.79833006858826 3.71000003814697
2.89826989173889 3.86999988555908
2.99820995330811 4.03999996185303
3.09815001487732 4.19500017166138
3.19809007644653 4.3600001335144
3.39796996116638 4.7350001335144
3.79772996902466 5.40999984741211
3.89767003059387 5.57000017166138
4.09754991531372 5.91499996185303
4.19749021530151 6.07999992370605
4.39736986160278 6.43499994277954
4.49731016159058 6.61499977111816
4.79713010787964 7.13000011444092
4.89706993103027 7.30499982833862
5.0969500541687 7.67500019073486
5.19688987731934 7.86999988555908
5.29683017730713 8.05500030517578
5.4967098236084 8.40999984741211
5.79653978347778 8.94499969482422
5.99641990661621 9.27999973297119
6.096360206604 9.4350004196167
6.19630002975464 9.60499954223633
6.29623985290527 9.78499984741211
6.39618015289307 9.97999954223633
6.4961199760437 10.164999961853
6.69600009918213 10.5100002288818
6.89588022232056 10.8299999237061
7.29563999176025 11.5349998474121
7.49552011489868 11.8950004577637
7.69540023803711 12.2799997329712
7.99522018432617 12.8249998092651
8.09515953063965 13.0150003433228
8.4949197769165 13.7299995422363
8.5948600769043 13.8950004577637
8.69480037689209 14.0699996948242
8.79473972320557 14.2550001144409
8.89468002319336 14.4300003051758
8.99462032318115 14.5950002670288
};
\addlegendentry{EQ, $W=200$}
\end{axis}

\end{tikzpicture}%
	\end{subfigure}%
	\vspace{0.5cm}
	\begin{subfigure}[htbp]{0.49\textwidth}
		\centering
		% This file was created with tikzplotlib v0.10.1.
\begin{tikzpicture}

\definecolor{burlywood248194145}{RGB}{250,209,172}
\definecolor{darkgray176}{RGB}{176,176,176}
\definecolor{darkorange24213334}{RGB}{242,133,34}
\definecolor{lightgray204}{RGB}{204,204,204}
\definecolor{sandybrown24516489}{RGB}{181,100,26}

\begin{axis}[
height=\figureheight,
legend cell align={left},
legend style={
	fill opacity=0.8,
	draw opacity=1,
	text opacity=1,
	at={(0.99,0.99)},
	anchor=north east,
	draw=lightgray204
},
tick align=outside,
tick pos=left,
width=\figurewidth,
x grid style={darkgray176},
xlabel={\(\displaystyle t^{*}\)},
xmajorgrids,
xmin=0, xmax=10,
xtick style={color=black},
y grid style={darkgray176},
ylabel style={rotate=-90.0},
ylabel={\(\displaystyle h^{*}\)},
ymajorgrids,
ymin=0.013, ymax=1.047,
ytick style={color=black}
]
\addplot [very thick, black, mark=*, mark size=2, mark options={solid}, only marks]
table {%
0 1
0.559999942779541 0.940000057220459
0.769999980926514 0.889999985694885
0.930000066757202 0.829999923706055
1.08000004291534 0.779999971389771
1.27999997138977 0.720000028610229
1.46000003814697 0.670000076293945
1.6599999666214 0.610000014305115
1.8400000333786 0.559999942779541
2 0.5
2.21000003814697 0.440000057220459
2.45000004768372 0.389999985694885
2.70000004768372 0.330000042915344
3.05999994277954 0.279999971389771
3.44000005722046 0.220000028610229
4.19999980926514 0.169999957084656
5.25 0.110000014305115
7.40000009536743 0.059999942779541
};
\addlegendentry{Experiment~\cite{martin1952PartIVExperimental}}
\addplot [line width=2pt, burlywood248194145, dash pattern=on 1pt off 3pt on 3pt off 3pt]
table {%
0 1
0.0998772382736206 0.990000009536743
0.29963207244873 0.990000009536743
0.599262952804565 0.960000038146973
0.699140071868896 0.940000057220459
0.799018025398254 0.930000066757202
1.29840004444122 0.829999923706055
1.3982800245285 0.799999952316284
1.59803998470306 0.759999990463257
1.69790995121002 0.730000019073486
1.89767003059387 0.690000057220459
1.99753999710083 0.660000085830688
2.69668006896973 0.519999980926514
2.79656004905701 0.509999990463257
2.99632000923157 0.470000028610229
3.09618997573853 0.460000038146973
3.29594993591309 0.419999957084656
3.39582991600037 0.409999966621399
3.49569988250732 0.389999985694885
3.59558010101318 0.379999995231628
3.69546008110046 0.360000014305115
3.79533004760742 0.350000023841858
3.8952100276947 0.330000042915344
4.19483995437622 0.299999952316284
4.29472017288208 0.279999971389771
4.39459991455078 0.279999971389771
4.49447011947632 0.259999990463257
4.59434986114502 0.259999990463257
4.79410982131958 0.240000009536743
4.89398002624512 0.240000009536743
4.99385976791382 0.230000019073486
5.09373998641968 0.230000019073486
5.29348993301392 0.210000038146973
5.39337015151978 0.210000038146973
5.69299983978271 0.180000066757202
5.79288005828857 0.180000066757202
5.89275979995728 0.169999957084656
5.99263000488281 0.169999957084656
6.19238996505737 0.149999976158142
6.59189987182617 0.149999976158142
6.69177007675171 0.139999985694885
6.79164981842041 0.139999985694885
6.89153003692627 0.129999995231628
7.19116020202637 0.129999995231628
7.29103994369507 0.139999985694885
7.99018001556396 0.139999985694885
8.09004974365234 0.129999995231628
8.1899299621582 0.139999985694885
8.28981018066406 0.129999995231628
8.4895601272583 0.129999995231628
8.58944034576416 0.120000004768372
8.88906955718994 0.120000004768372
8.9889497756958 0.110000014305115
9.48832988739014 0.110000014305115
9.588210105896 0.100000023841858
9.98771953582764 0.100000023841858
};
\addlegendentry{EQ+NEQ, $W=50$}
\addplot [line width=2pt, sandybrown24516489, dash pattern=on 1pt off 3pt on 3pt off 3pt]
table {%
0 1
0.0998772382736206 0.995000004768372
0.199753999710083 0.995000004768372
0.29963207244873 0.990000009536743
0.599262952804565 0.960000038146973
0.799018025398254 0.930000066757202
0.898895025253296 0.910000085830688
0.998772025108337 0.894999980926514
1.0986499786377 0.875
1.19852995872498 0.850000023841858
1.29840004444122 0.829999923706055
1.49816000461578 0.779999971389771
1.59803998470306 0.759999990463257
1.79779005050659 0.710000038146973
1.89767003059387 0.690000057220459
1.99753999710083 0.664999961853027
2.09741997718811 0.644999980926514
2.19729995727539 0.620000004768372
2.59681010246277 0.539999961853027
2.69668006896973 0.524999976158142
2.89644002914429 0.485000014305115
2.99632000923157 0.470000028610229
3.19606995582581 0.430000066757202
3.59558010101318 0.370000004768372
3.69546008110046 0.360000014305115
3.8952100276947 0.330000042915344
4.09497022628784 0.309999942779541
4.19483995437622 0.294999957084656
4.49447011947632 0.264999985694885
4.59434986114502 0.259999990463257
4.79410982131958 0.240000009536743
4.89398002624512 0.235000014305115
4.99385976791382 0.225000023841858
5.29348993301392 0.210000038146973
5.39337015151978 0.200000047683716
5.49324989318848 0.194999933242798
5.59312009811401 0.184999942779541
6.49202013015747 0.139999985694885
6.69177007675171 0.139999985694885
6.79164981842041 0.134999990463257
6.89153003692627 0.134999990463257
6.99139976501465 0.125
7.09127998352051 0.125
7.29103994369507 0.115000009536743
7.49078989028931 0.115000009536743
7.59067010879517 0.110000014305115
7.7904200553894 0.110000014305115
7.89029979705811 0.105000019073486
7.99018001556396 0.105000019073486
8.28981018066406 0.0900000333786011
8.4895601272583 0.0900000333786011
8.58944034576416 0.0950000286102295
8.7891902923584 0.0950000286102295
8.88906955718994 0.0900000333786011
8.9889497756958 0.0900000333786011
9.08882999420166 0.0850000381469727
9.18869972229004 0.0850000381469727
9.2885799407959 0.0800000429153442
};
\addlegendentry{EQ+NEQ, $W=100$}
\addplot [line width=2pt, darkorange24213334, dash pattern=on 1pt off 3pt on 3pt off 3pt]
table {%
0 1
0.199880957603455 0.995000004768372
0.299821019172668 0.990000009536743
0.399760961532593 0.982500076293945
0.499701023101807 0.972500085830688
0.599642038345337 0.960000038146973
0.799521923065186 0.930000066757202
0.99940299987793 0.894999980926514
1.09933996200562 0.872499942779541
1.19928002357483 0.852499961853027
1.29921996593475 0.829999923706055
1.39916002750397 0.805000066757202
1.49909996986389 0.782500028610229
1.59904003143311 0.757499933242798
1.69897997379303 0.735000014305115
1.79892003536224 0.710000038146973
1.9988100528717 0.664999961853027
2.09875011444092 0.644999980926514
2.29862999916077 0.600000023841858
2.49850988388062 0.559999942779541
2.59844994544983 0.542500019073486
2.69839000701904 0.522500038146973
2.79833006858826 0.504999995231628
2.89826989173889 0.485000014305115
2.99820995330811 0.467499971389771
3.09815001487732 0.452499985694885
3.19809007644653 0.434999942779541
3.39796996116638 0.404999971389771
3.4979100227356 0.387500047683716
3.69778990745544 0.362499952316284
3.79772996902466 0.347499966621399
3.89767003059387 0.335000038146973
3.99761009216309 0.319999933242798
4.09754991531372 0.309999942779541
4.19749021530151 0.297500014305115
4.59724998474121 0.257499933242798
4.69718980789185 0.252500057220459
4.79713010787964 0.242500066757202
5.19688987731934 0.212499976158142
5.29683017730713 0.207499980926514
5.39677000045776 0.200000047683716
5.4967098236084 0.194999933242798
5.59665012359619 0.1875
5.79653978347778 0.177500009536743
5.89648008346558 0.169999957084656
6.29623985290527 0.149999976158142
6.39618015289307 0.147500038146973
6.4961199760437 0.142500042915344
6.59605979919434 0.139999985694885
6.69600009918213 0.134999990463257
7.69540023803711 0.110000014305115
7.79534006118774 0.110000014305115
8.19509983062744 0.100000023841858
8.29504013061523 0.100000023841858
8.4949197769165 0.0950000286102295
8.5948600769043 0.0950000286102295
8.69480037689209 0.0924999713897705
8.79473972320557 0.0924999713897705
8.99462032318115 0.0874999761581421
};
\addlegendentry{EQ+NEQ, $W=200$}
\end{axis}

\end{tikzpicture}%
	\end{subfigure}
	\hfill
	\begin{subfigure}[htbp]{0.49\textwidth}
		\centering
		% This file was created with tikzplotlib v0.10.1.
\begin{tikzpicture}

\definecolor{burlywood248194145}{RGB}{250,209,172}
\definecolor{darkgray176}{RGB}{176,176,176}
\definecolor{darkorange24213334}{RGB}{242,133,34}
\definecolor{lightgray204}{RGB}{204,204,204}
\definecolor{sandybrown24516489}{RGB}{181,100,26}

\begin{axis}[
height=\figureheight,
legend cell align={left},
legend style={
  fill opacity=0.8,
  draw opacity=1,
  text opacity=1,
  at={(0.01,0.99)},
  anchor=north west,
  draw=lightgray204
},
tick align=outside,
tick pos=left,
width=\figurewidth,
x grid style={darkgray176},
xlabel={\(\displaystyle t^{*}\)},
xmajorgrids,
xmin=0, xmax=10,
xtick style={color=black},
y grid style={darkgray176},
ylabel style={rotate=-90.0},
ylabel={\(\displaystyle w^{*}\)},
ymajorgrids,
ymin=0, ymax=28,
ytick style={color=black}
]
\addplot [very thick, black, mark=*, mark size=2, mark options={solid}, only marks]
table {%
0.409999966621399 1.11000001430511
0.839999914169312 1.22000002861023
1.19000005722046 1.44000005722046
1.42999994754791 1.66999995708466
1.62999999523163 1.88999998569489
1.83000004291534 2.10999989509583
1.98000001907349 2.32999992370605
2.20000004768372 2.55999994277954
2.3199999332428 2.77999997138977
2.50999999046326 3
2.65000009536743 3.22000002861023
2.82999992370605 3.44000005722046
2.97000002861023 3.67000007629395
3.10999989509583 3.89000010490417
3.32999992370605 4.1100001335144
4.01999998092651 5
4.44000005722046 5.8899998664856
5.09000015258789 7
5.69000005722046 8
6.30000019073486 9
6.82999992370605 10
7.44000005722046 11
8.07999992370605 12
8.67000007629395 13
9.3100004196167 14
};
\addlegendentry{Experiment~\cite{martin1952PartIVExperimental}}
\addplot [line width=2pt, burlywood248194145, dash pattern=on 1pt off 3pt on 3pt off 3pt]
table {%
0 1
0.0998772382736206 1
0.199753999710083 1.01999998092651
0.499386072158813 1.13999998569489
0.699140071868896 1.25999999046326
0.799018025398254 1.3400000333786
0.898895025253296 1.39999997615814
1.0986499786377 1.55999994277954
1.29840004444122 1.75999999046326
1.3982800245285 1.8400000333786
1.59803998470306 2.03999996185303
1.69790995121002 2.09999990463257
1.79779005050659 2.1800000667572
1.89767003059387 2.27999997138977
2.19729995727539 2.70000004768372
2.39704990386963 2.94000005722046
2.49692988395691 3.07999992370605
2.59681010246277 3.20000004768372
2.69668006896973 3.29999995231628
2.79656004905701 3.42000007629395
2.89644002914429 3.55999994277954
3.09618997573853 3.92000007629395
3.39582991600037 4.34000015258789
3.49569988250732 4.46000003814697
3.59558010101318 4.59999990463257
3.8952100276947 5.07999992370605
3.99509000778198 5.26000022888184
4.09497022628784 5.42000007629395
4.19483995437622 5.55999994277954
4.59434986114502 6.19999980926514
4.89398002624512 6.61999988555908
4.99385976791382 6.78000020980835
5.19361019134521 7.05999994277954
5.29348993301392 7.21999979019165
5.39337015151978 7.3600001335144
5.49324989318848 7.48000001907349
5.59312009811401 7.57999992370605
5.69299983978271 7.71999979019165
5.79288005828857 7.82000017166138
5.89275979995728 7.96000003814697
6.19238996505737 8.4399995803833
6.29226016998291 8.57999992370605
6.39213991165161 8.73999977111816
6.59189987182617 9.02000045776367
6.69177007675171 9.14000034332275
6.79164981842041 9.27999973297119
6.89153003692627 9.4399995803833
6.99139976501465 9.61999988555908
7.09127998352051 9.73999977111816
7.39091014862061 10.1599998474121
7.59067010879517 10.4799995422363
7.89029979705811 10.8999996185303
7.99018001556396 11.0600004196167
8.09004974365234 11.1999998092651
8.1899299621582 11.3199996948242
8.28981018066406 11.460000038147
8.4895601272583 11.6999998092651
8.6893196105957 11.8999996185303
8.7891902923584 12.0200004577637
8.88906955718994 12.1599998474121
8.9889497756958 12.3199996948242
9.08882999420166 12.460000038147
9.18869972229004 12.5799999237061
9.2885799407959 12.7600002288818
9.38846015930176 12.9200000762939
9.48832988739014 13.1000003814697
9.588210105896 13.2600002288818
9.7879695892334 13.5
9.98771953582764 13.8199996948242
};
\addlegendentry{EQ+NEQ, $W=50$}
\addplot [line width=2pt, sandybrown24516489, dash pattern=on 1pt off 3pt on 3pt off 3pt]
table {%
0 1
0.0998772382736206 1
0.199753999710083 1.01999998092651
0.29963207244873 1.04999995231628
0.499386072158813 1.14999997615814
0.699140071868896 1.26999998092651
0.799018025398254 1.35000002384186
0.898895025253296 1.44000005722046
0.998772025108337 1.54999995231628
1.0986499786377 1.64999997615814
1.19852995872498 1.76999998092651
1.29840004444122 1.89999997615814
1.3982800245285 2.04999995231628
1.59803998470306 2.32999992370605
1.69790995121002 2.48000001907349
1.79779005050659 2.61999988555908
1.89767003059387 2.74000000953674
1.99753999710083 2.83999991416931
2.09741997718811 2.92000007629395
2.19729995727539 3.00999999046326
2.39704990386963 3.23000001907349
2.49692988395691 3.35999989509583
2.59681010246277 3.47000002861023
2.69668006896973 3.58999991416931
2.79656004905701 3.74000000953674
2.89644002914429 3.9300000667572
3.29594993591309 4.73000001907349
3.39582991600037 4.86999988555908
3.49569988250732 4.98999977111816
3.59558010101318 5.11999988555908
3.69546008110046 5.28000020980835
3.79533004760742 5.42000007629395
3.8952100276947 5.57000017166138
4.09497022628784 5.92999982833862
4.19483995437622 6.07999992370605
4.29472017288208 6.25
4.49447011947632 6.53000020980835
4.59434986114502 6.65000009536743
4.69423007965088 6.82000017166138
4.79410982131958 7.01000022888184
4.89398002624512 7.17000007629395
4.99385976791382 7.3600001335144
5.09373998641968 7.57999992370605
5.19361019134521 7.78999996185303
5.29348993301392 7.98000001907349
5.39337015151978 8.14999961853027
5.49324989318848 8.30000019073486
5.59312009811401 8.43000030517578
5.69299983978271 8.53999996185303
5.79288005828857 8.75
5.89275979995728 8.90999984741211
5.99263000488281 9.09000015258789
6.09251022338867 9.25
6.19238996505737 9.44999980926514
6.39213991165161 9.78999996185303
6.49202013015747 9.93000030517578
6.59189987182617 10.0799999237061
6.79164981842041 10.3999996185303
6.89153003692627 10.6199998855591
6.99139976501465 10.8000001907349
7.09127998352051 10.9499998092651
7.19116020202637 11.0900001525879
7.29103994369507 11.210000038147
7.39091014862061 11.3500003814697
7.49078989028931 11.539999961853
7.59067010879517 11.75
7.69053983688354 11.9499998092651
7.7904200553894 12.1800003051758
7.99018001556396 12.5600004196167
8.09004974365234 12.710000038147
8.28981018066406 12.9499998092651
8.38969039916992 13.0799999237061
8.58944034576416 13.3599996566772
8.6893196105957 13.5500001907349
8.7891902923584 13.6800003051758
8.88906955718994 13.8299999237061
8.9889497756958 14.0100002288818
9.08882999420166 14.1999998092651
9.18869972229004 14.3800001144409
9.2885799407959 14.5699996948242
};
\addlegendentry{EQ+NEQ, $W=100$}
\addplot [line width=2pt, darkorange24213334, dash pattern=on 1pt off 3pt on 3pt off 3pt]
table {%
0 1
0.099940299987793 1.00499999523163
0.199880957603455 1.02499997615814
0.299821019172668 1.05999994277954
0.399760961532593 1.10000002384186
0.499701023101807 1.14999997615814
0.699581980705261 1.26999998092651
0.899461984634399 1.42999994754791
0.99940299987793 1.52499997615814
1.09933996200562 1.60000002384186
1.19928002357483 1.67999994754791
1.29921996593475 1.7849999666214
1.39916002750397 1.88499999046326
1.49909996986389 1.98000001907349
1.59904003143311 2.09500002861023
1.69897997379303 2.20499992370605
1.79892003536224 2.3199999332428
1.89885997772217 2.44000005722046
2.09875011444092 2.69000005722046
2.19868993759155 2.80999994277954
2.29862999916077 2.96000003814697
2.49850988388062 3.28500008583069
2.69839000701904 3.63000011444092
2.79833006858826 3.80999994277954
2.89826989173889 4.01000022888184
2.99820995330811 4.19500017166138
3.09815001487732 4.3899998664856
3.19809007644653 4.59499979019165
3.29802989959717 4.86499977111816
3.39796996116638 5.22499990463257
3.4979100227356 5.55000019073486
3.59785008430481 5.7350001335144
3.69778990745544 5.84999990463257
3.79772996902466 5.92999982833862
3.89767003059387 6.00500011444092
3.99761009216309 6.14499998092651
4.09754991531372 6.25
4.19749021530151 6.3899998664856
4.29743003845215 6.4850001335144
4.39736986160278 6.59499979019165
4.49731016159058 6.71999979019165
4.59724998474121 6.84999990463257
4.69718980789185 7.02500009536743
4.79713010787964 7.25500011444092
4.89706993103027 7.54500007629395
4.99701023101807 7.8600001335144
5.0969500541687 8.07499980926514
5.19688987731934 8.22500038146973
5.29683017730713 8.32999992370605
5.39677000045776 8.40999984741211
5.4967098236084 8.46500015258789
5.59665012359619 8.52499961853027
5.69658994674683 8.59000015258789
5.79653978347778 8.71000003814697
5.89648008346558 8.875
5.99641990661621 9.02999973297119
6.096360206604 9.17500019073486
6.19630002975464 9.39000034332275
6.39618015289307 9.90499973297119
6.4961199760437 10.1049995422363
6.59605979919434 10.0500001907349
6.79593992233276 10.3649997711182
6.99582004547119 10.6850004196167
7.09575986862183 10.8149995803833
7.19570016860962 10.9750003814697
7.29563999176025 11.1599998474121
7.49552011489868 11.4899997711182
7.59545993804932 11.6750001907349
7.79534006118774 12.1549997329712
7.89527988433838 12.4099998474121
7.99522018432617 12.6149997711182
8.19509983062744 12.9399995803833
8.29504013061523 13.2250003814697
8.39498043060303 13.5500001907349
8.4949197769165 13.8900003433228
8.5948600769043 14.1049995422363
8.69480037689209 14.25
8.79473972320557 14.3800001144409
8.89468002319336 14.460000038147
8.99462032318115 14.5349998474121
};
\addlegendentry{EQ+NEQ, $W=200$}
\end{axis}

\end{tikzpicture}%
	\end{subfigure}%
	\caption{\label{fig:dam-break-rectangular-convergence}
		Simulated rectangular dam break with non-dimensionalized residual dam height $h^{*}(t^{*})$, width $w^{*}(t^{*})$, and time $t^{*}$.
		The simulations were performed with computational domain resolutions, that is, initial dam widths of $W \in \{50, 100, 200\}$ lattice cells.
		The EXT scheme was numerically unstable in all tested resolutions.
		All other simulations have converged well.
	}
\end{figure}

\begin{figure}[htbp]
	\centering
	\setlength{\figureheight}{0.35\textwidth}
	\setlength{\figurewidth}{0.45\textwidth}
	\begin{subfigure}[htbp]{0.49\textwidth}
		\centering
		% This file was created with tikzplotlib v0.10.1.
\begin{tikzpicture}

\definecolor{crimson2553191}{RGB}{255,31,91}
\definecolor{darkgray176}{RGB}{176,176,176}
\definecolor{hotpink255143173}{RGB}{255,171,193}
\definecolor{lightcoral25587132}{RGB}{159,19,57}
\definecolor{lightgray204}{RGB}{204,204,204}

\begin{axis}[
height=\figureheight,
legend cell align={left},
legend style={
	fill opacity=0.8,
	draw opacity=1,
	text opacity=1,
	at={(0.99,0.99)},
	anchor=north east,
	draw=lightgray204
},
tick align=outside,
tick pos=left,
width=\figurewidth,
x grid style={darkgray176},
xlabel={\(\displaystyle t^{*}\)},
xmajorgrids,
xmin=0, xmax=10,
xtick style={color=black},
y grid style={darkgray176},
ylabel style={rotate=-90.0},
ylabel={\(\displaystyle h^{*}\)},
ymajorgrids,
ymin=0.013, ymax=1.047,
ytick style={color=black}
]
\addplot [very thick, black, mark=*, mark size=2, mark options={solid}, only marks]
table {%
0 1
0.559999942779541 0.940000057220459
0.769999980926514 0.889999985694885
0.930000066757202 0.829999923706055
1.08000004291534 0.779999971389771
1.27999997138977 0.720000028610229
1.46000003814697 0.670000076293945
1.6599999666214 0.610000014305115
1.8400000333786 0.559999942779541
2 0.5
2.21000003814697 0.440000057220459
2.45000004768372 0.389999985694885
2.70000004768372 0.330000042915344
3.05999994277954 0.279999971389771
3.44000005722046 0.220000028610229
4.19999980926514 0.169999957084656
5.25 0.110000014305115
7.40000009536743 0.059999942779541
};
\addlegendentry{Experiment~\cite{martin1952PartIVExperimental}}
\addplot [line width=2pt, hotpink255143173, dash pattern=on 3pt off 5pt on 1pt off 5pt]
table {%
0 1
0.0998772382736206 0.990000009536743
0.29963207244873 0.990000009536743
0.599262952804565 0.960000038146973
0.699140071868896 0.940000057220459
0.799018025398254 0.930000066757202
1.29840004444122 0.829999923706055
1.3982800245285 0.799999952316284
1.59803998470306 0.759999990463257
1.69790995121002 0.730000019073486
1.89767003059387 0.690000057220459
1.99753999710083 0.660000085830688
2.89644002914429 0.480000019073486
2.99632000923157 0.470000028610229
3.19606995582581 0.430000066757202
3.29594993591309 0.419999957084656
3.39582991600037 0.399999976158142
3.49569988250732 0.389999985694885
3.59558010101318 0.370000004768372
3.79533004760742 0.350000023841858
3.8952100276947 0.330000042915344
4.19483995437622 0.299999952316284
4.29472017288208 0.279999971389771
4.39459991455078 0.279999971389771
4.49447011947632 0.259999990463257
4.59434986114502 0.259999990463257
4.79410982131958 0.240000009536743
4.89398002624512 0.240000009536743
4.99385976791382 0.230000019073486
5.09373998641968 0.230000019073486
5.49324989318848 0.190000057220459
5.59312009811401 0.190000057220459
5.69299983978271 0.180000066757202
5.89275979995728 0.180000066757202
5.99263000488281 0.169999957084656
6.19238996505737 0.169999957084656
6.29226016998291 0.159999966621399
6.49202013015747 0.159999966621399
6.59189987182617 0.149999976158142
6.79164981842041 0.149999976158142
7.09127998352051 0.120000004768372
7.19116020202637 0.120000004768372
7.29103994369507 0.110000014305115
7.39091014862061 0.120000004768372
7.49078989028931 0.120000004768372
7.59067010879517 0.110000014305115
7.89029979705811 0.110000014305115
7.99018001556396 0.100000023841858
8.09004974365234 0.110000014305115
8.28981018066406 0.110000014305115
8.38969039916992 0.100000023841858
8.4895601272583 0.110000014305115
8.58944034576416 0.110000014305115
8.6893196105957 0.120000004768372
9.2885799407959 0.120000004768372
};
\addlegendentry{GEQ, $W=50$}
\addplot [line width=2pt, lightcoral25587132, dash pattern=on 3pt off 5pt on 1pt off 5pt]
table {%
0 1
0.0998772382736206 0.995000004768372
0.199753999710083 0.995000004768372
0.29963207244873 0.990000009536743
0.599262952804565 0.960000038146973
0.799018025398254 0.930000066757202
0.898895025253296 0.910000085830688
0.998772025108337 0.894999980926514
1.0986499786377 0.875
1.19852995872498 0.850000023841858
1.29840004444122 0.829999923706055
1.49816000461578 0.779999971389771
1.59803998470306 0.759999990463257
1.79779005050659 0.710000038146973
1.89767003059387 0.690000057220459
1.99753999710083 0.664999961853027
2.09741997718811 0.644999980926514
2.19729995727539 0.620000004768372
2.69668006896973 0.519999980926514
2.79656004905701 0.504999995231628
2.89644002914429 0.485000014305115
2.99632000923157 0.470000028610229
3.09618997573853 0.450000047683716
3.29594993591309 0.419999957084656
3.39582991600037 0.399999976158142
3.49569988250732 0.389999985694885
3.8952100276947 0.330000042915344
4.09497022628784 0.309999942779541
4.19483995437622 0.294999957084656
4.39459991455078 0.274999976158142
4.49447011947632 0.269999980926514
4.79410982131958 0.240000009536743
4.99385976791382 0.230000019073486
5.09373998641968 0.220000028610229
5.29348993301392 0.210000038146973
5.39337015151978 0.200000047683716
5.49324989318848 0.194999933242798
5.59312009811401 0.184999942779541
6.39213991165161 0.144999980926514
6.49202013015747 0.144999980926514
6.79164981842041 0.129999995231628
6.89153003692627 0.129999995231628
6.99139976501465 0.125
7.09127998352051 0.125
7.29103994369507 0.115000009536743
7.39091014862061 0.115000009536743
7.49078989028931 0.120000004768372
7.69053983688354 0.120000004768372
7.7904200553894 0.115000009536743
7.89029979705811 0.120000004768372
7.99018001556396 0.115000009536743
8.09004974365234 0.105000019073486
8.4895601272583 0.0850000381469727
8.7891902923584 0.0850000381469727
8.88906955718994 0.0900000333786011
8.9889497756958 0.0900000333786011
};
\addlegendentry{GEQ, $W=100$}
\addplot [line width=2pt, crimson2553191, dash pattern=on 3pt off 5pt on 1pt off 5pt]
table {%
0 1
0.199880957603455 0.995000004768372
0.299821019172668 0.990000009536743
0.399760961532593 0.980000019073486
0.499701023101807 0.972500085830688
0.599642038345337 0.960000038146973
0.799521923065186 0.930000066757202
0.99940299987793 0.894999980926514
1.09933996200562 0.872499942779541
1.19928002357483 0.852499961853027
1.39916002750397 0.807500004768372
1.59904003143311 0.757499933242798
1.69897997379303 0.735000014305115
1.79892003536224 0.710000038146973
2.09875011444092 0.642499923706055
2.19868993759155 0.622499942779541
2.29862999916077 0.600000023841858
2.59844994544983 0.539999961853027
2.69839000701904 0.522500038146973
2.79833006858826 0.502500057220459
2.99820995330811 0.467499971389771
3.09815001487732 0.452499985694885
3.19809007644653 0.434999942779541
3.29802989959717 0.419999957084656
3.39796996116638 0.402500033378601
3.4979100227356 0.389999985694885
3.69778990745544 0.360000014305115
4.19749021530151 0.297500014305115
4.49731016159058 0.267500042915344
4.59724998474121 0.259999990463257
4.69718980789185 0.25
5.19688987731934 0.212499976158142
5.39677000045776 0.202499985694885
5.69658994674683 0.180000066757202
5.79653978347778 0.177500009536743
5.99641990661621 0.167500019073486
6.096360206604 0.159999966621399
6.29623985290527 0.149999976158142
6.39618015289307 0.147500038146973
6.4961199760437 0.142500042915344
6.69600009918213 0.137500047683716
6.79593992233276 0.132499933242798
6.89588022232056 0.132499933242798
7.09575986862183 0.127500057220459
7.19570016860962 0.122499942779541
7.89527988433838 0.105000019073486
7.99522018432617 0.105000019073486
8.19509983062744 0.100000023841858
8.29504013061523 0.100000023841858
8.39498043060303 0.0974999666213989
8.4949197769165 0.0974999666213989
8.69480037689209 0.0924999713897705
8.79473972320557 0.0924999713897705
8.89468002319336 0.0900000333786011
8.99462032318115 0.0900000333786011
};
\addlegendentry{GEQ, $W=200$}
\end{axis}

\end{tikzpicture}%
	\end{subfigure}
	\hfill
	\begin{subfigure}[htbp]{0.49\textwidth}
		\centering
		% This file was created with tikzplotlib v0.10.1.
\begin{tikzpicture}

\definecolor{crimson2553191}{RGB}{255,31,91}
\definecolor{darkgray176}{RGB}{176,176,176}
\definecolor{hotpink255143173}{RGB}{255,171,193}
\definecolor{lightcoral25587132}{RGB}{159,19,57}
\definecolor{lightgray204}{RGB}{204,204,204}

\begin{axis}[
height=\figureheight,
legend cell align={left},
legend style={
  fill opacity=0.8,
  draw opacity=1,
  text opacity=1,
  at={(0.01,0.99)},
  anchor=north west,
  draw=lightgray204
},
tick align=outside,
tick pos=left,
width=\figurewidth,
x grid style={darkgray176},
xlabel={\(\displaystyle t^{*}\)},
xmajorgrids,
xmin=0, xmax=10,
xtick style={color=black},
y grid style={darkgray176},
ylabel style={rotate=-90.0},
ylabel={\(\displaystyle w^{*}\)},
ymajorgrids,
ymin=0, ymax=28,
ytick style={color=black}
]
\addplot [very thick, black, mark=*, mark size=2, mark options={solid}, only marks]
table {%
0.409999966621399 1.11000001430511
0.839999914169312 1.22000002861023
1.19000005722046 1.44000005722046
1.42999994754791 1.66999995708466
1.62999999523163 1.88999998569489
1.83000004291534 2.10999989509583
1.98000001907349 2.32999992370605
2.20000004768372 2.55999994277954
2.3199999332428 2.77999997138977
2.50999999046326 3
2.65000009536743 3.22000002861023
2.82999992370605 3.44000005722046
2.97000002861023 3.67000007629395
3.10999989509583 3.89000010490417
3.32999992370605 4.1100001335144
4.01999998092651 5
4.44000005722046 5.8899998664856
5.09000015258789 7
5.69000005722046 8
6.30000019073486 9
6.82999992370605 10
7.44000005722046 11
8.07999992370605 12
8.67000007629395 13
9.3100004196167 14
};
\addlegendentry{Experiment~\cite{martin1952PartIVExperimental}}
\addplot [line width=2pt, hotpink255143173, dash pattern=on 3pt off 5pt on 1pt off 5pt]
table {%
0 1
0.0998772382736206 1
0.199753999710083 1.01999998092651
0.499386072158813 1.13999998569489
0.699140071868896 1.25999999046326
1.0986499786377 1.58000004291534
1.49816000461578 1.98000001907349
1.59803998470306 2.09999990463257
1.69790995121002 2.20000004768372
1.89767003059387 2.44000005722046
1.99753999710083 2.57999992370605
2.19729995727539 2.8199999332428
2.79656004905701 3.66000008583069
2.99632000923157 3.98000001907349
3.09618997573853 4.11999988555908
3.49569988250732 4.76000022888184
3.59558010101318 4.94000005722046
3.8952100276947 5.42000007629395
4.09497022628784 5.78000020980835
4.19483995437622 5.94000005722046
4.79410982131958 7.01999998092651
4.89398002624512 7.17999982833862
5.39337015151978 8.07999992370605
5.49324989318848 8.23999977111816
5.79288005828857 8.77999973297119
5.89275979995728 8.9399995803833
5.99263000488281 9.11999988555908
6.49202013015747 9.92000007629395
6.59189987182617 10.0600004196167
6.99139976501465 10.6999998092651
7.09127998352051 10.8800001144409
7.49078989028931 11.5200004577637
7.59067010879517 11.6999998092651
7.99018001556396 12.3400001525879
8.09004974365234 12.5200004577637
8.28981018066406 12.8400001525879
8.4895601272583 13.1999998092651
8.6893196105957 13.5200004577637
8.7891902923584 13.6999998092651
8.9889497756958 14.0200004577637
9.08882999420166 14.1999998092651
9.2885799407959 14.5200004577637
};
\addlegendentry{GEQ, $W=50$}
\addplot [line width=2pt, lightcoral25587132, dash pattern=on 3pt off 5pt on 1pt off 5pt]
table {%
0 1
0.0998772382736206 1
0.199753999710083 1.01999998092651
0.29963207244873 1.04999995231628
0.499386072158813 1.14999997615814
0.699140071868896 1.26999998092651
0.799018025398254 1.3400000333786
0.898895025253296 1.41999995708466
1.19852995872498 1.69000005722046
1.3982800245285 1.88999998569489
1.59803998470306 2.10999989509583
1.99753999710083 2.58999991416931
2.19729995727539 2.84999990463257
2.49692988395691 3.26999998092651
2.79656004905701 3.72000002861023
2.89644002914429 3.88000011444092
2.99632000923157 4.03000020980835
3.09618997573853 4.19000005722046
3.29594993591309 4.53000020980835
3.69546008110046 5.25
3.79533004760742 5.44000005722046
4.29472017288208 6.34000015258789
4.49447011947632 6.67999982833862
4.59434986114502 6.84000015258789
4.89398002624512 7.38000011444092
4.99385976791382 7.57000017166138
5.09373998641968 7.73999977111816
5.19361019134521 7.92000007629395
5.29348993301392 8.07999992370605
5.59312009811401 8.59000015258789
5.69299983978271 8.75
6.29226016998291 9.77000045776367
6.69177007675171 10.4899997711182
6.99139976501465 11
7.29103994369507 11.4799995422363
7.39091014862061 11.6599998474121
7.59067010879517 12.039999961853
7.7904200553894 12.3999996185303
7.89029979705811 12.5699996948242
7.99018001556396 12.75
8.09004974365234 12.9200000762939
8.58944034576416 13.8199996948242
8.7891902923584 14.1999998092651
8.9889497756958 14.5600004196167
};
\addlegendentry{GEQ, $W=100$}
\addplot [line width=2pt, crimson2553191, dash pattern=on 3pt off 5pt on 1pt off 5pt]
table {%
0 1
0.099940299987793 1.00499999523163
0.199880957603455 1.02499997615814
0.299821019172668 1.05999994277954
0.399760961532593 1.10000002384186
0.499701023101807 1.14999997615814
0.599642038345337 1.21000003814697
0.699581980705261 1.27499997615814
0.799521923065186 1.35000002384186
0.99940299987793 1.50999999046326
1.09933996200562 1.60000002384186
1.19928002357483 1.69500005245209
1.39916002750397 1.89499998092651
1.59904003143311 2.11500000953674
1.69897997379303 2.23000001907349
1.79892003536224 2.34999990463257
1.9988100528717 2.59999990463257
2.09875011444092 2.73000001907349
2.29862999916077 3
2.49850988388062 3.29500007629395
2.59844994544983 3.44000005722046
2.89826989173889 3.96000003814697
3.19809007644653 4.42500019073486
3.29802989959717 4.59499979019165
3.4979100227356 4.8899998664856
3.59785008430481 5.03499984741211
3.69778990745544 5.19999980926514
3.99761009216309 5.75
4.09754991531372 5.92000007629395
4.19749021530151 6.09999990463257
4.39736986160278 6.43499994277954
4.49731016159058 6.59499979019165
4.59724998474121 6.78000020980835
4.99701023101807 7.47499990463257
5.29683017730713 8.04500007629395
5.4967098236084 8.39000034332275
5.59665012359619 8.5649995803833
5.69658994674683 8.75
5.99641990661621 9.32999992370605
6.096360206604 9.51500034332275
6.29623985290527 9.86999988555908
6.59605979919434 10.4049997329712
6.79593992233276 10.7600002288818
6.99582004547119 11.125
7.19570016860962 11.4549999237061
7.39557981491089 11.8149995803833
7.79534006118774 12.5
7.99522018432617 12.8649997711182
8.09515953063965 13.0349998474121
8.29504013061523 13.3999996185303
8.4949197769165 13.7349996566772
8.69480037689209 14.1000003814697
8.99462032318115 14.6149997711182
};
\addlegendentry{GEQ, $W=200$}
\end{axis}

\end{tikzpicture}%
	\end{subfigure}%
	\vspace{0.5cm}
	\begin{subfigure}[htbp]{0.49\textwidth}
		\centering
		% This file was created with tikzplotlib v0.10.1.
\begin{tikzpicture}

\definecolor{darkgray176}{RGB}{176,176,176}
\definecolor{lightgray204}{RGB}{204,204,204}
\definecolor{mediumorchid17588186}{RGB}{175,88,186}
\definecolor{orchid195130203}{RGB}{109,55,116}
\definecolor{plum215172220}{RGB}{225,192,229}

\begin{axis}[
height=\figureheight,
legend cell align={left},
legend style={
  fill opacity=0.8,
  draw opacity=1,
  text opacity=1,
  at={(0.99,0.8)},
  anchor=north east,
  draw=lightgray204
},
tick align=outside,
tick pos=left,
width=\figurewidth,
x grid style={darkgray176},
xlabel={\(\displaystyle t^{*}\)},
xmajorgrids,
xmin=0, xmax=10,
xtick style={color=black},
y grid style={darkgray176},
ylabel style={rotate=-90.0},
ylabel={\(\displaystyle h^{*}\)},
ymajorgrids,
ymin=0.013, ymax=1.047,
ytick style={color=black}
]
\addplot [very thick, black, mark=*, mark size=2, mark options={solid}, only marks]
table {%
0 1
0.559999942779541 0.940000057220459
0.769999980926514 0.889999985694885
0.930000066757202 0.829999923706055
1.08000004291534 0.779999971389771
1.27999997138977 0.720000028610229
1.46000003814697 0.670000076293945
1.6599999666214 0.610000014305115
1.8400000333786 0.559999942779541
2 0.5
2.21000003814697 0.440000057220459
2.45000004768372 0.389999985694885
2.70000004768372 0.330000042915344
3.05999994277954 0.279999971389771
3.44000005722046 0.220000028610229
4.19999980926514 0.169999957084656
5.25 0.110000014305115
7.40000009536743 0.059999942779541
};
\addlegendentry{Experiment~\cite{martin1952PartIVExperimental}}
\addplot [line width=2pt, plum215172220, dashed]
table {%
0 1
0.0998772382736206 0.990000009536743
0.29963207244873 0.990000009536743
0.499386072158813 0.970000028610229
9.98771953582764 0.970000028610229
};
\addlegendentry{EXT, $W=50$}
\addplot [line width=2pt, orchid195130203, dashed]
table {%
0 1
0.0998772382736206 0.995000004768372
0.199753999710083 0.995000004768372
0.29963207244873 0.990000009536743
9.98771953582764 0.990000009536743
};
\addlegendentry{EXT, $W=100$}
\addplot [line width=2pt, mediumorchid17588186, dashed]
table {%
0 1
0.199880957603455 0.995000004768372
9.9940299987793 0.995000004768372
};
\addlegendentry{EXT, $W=200$}
\end{axis}

\end{tikzpicture}%
	\end{subfigure}
	\hfill
	\begin{subfigure}[htbp]{0.49\textwidth}
		\centering
		% This file was created with tikzplotlib v0.10.1.
\begin{tikzpicture}

\definecolor{darkgray176}{RGB}{176,176,176}
\definecolor{lightgray204}{RGB}{204,204,204}
\definecolor{mediumorchid17588186}{RGB}{175,88,186}
\definecolor{orchid195130203}{RGB}{109,55,116}
\definecolor{plum215172220}{RGB}{225,192,229}

\begin{axis}[
height=\figureheight,
legend cell align={left},
legend style={
  fill opacity=0.8,
  draw opacity=1,
  text opacity=1,
  at={(0.01,0.99)},
  anchor=north west,
  draw=lightgray204
},
tick align=outside,
tick pos=left,
width=\figurewidth,
x grid style={darkgray176},
xlabel={\(\displaystyle t^{*}\)},
xmajorgrids,
xmin=0, xmax=10,
xtick style={color=black},
y grid style={darkgray176},
ylabel style={rotate=-90.0},
ylabel={\(\displaystyle w^{*}\)},
ymajorgrids,
ymin=0, ymax=28,
ytick style={color=black}
]
\addplot [very thick, black, mark=*, mark size=2, mark options={solid}, only marks]
table {%
0.409999966621399 1.11000001430511
0.839999914169312 1.22000002861023
1.19000005722046 1.44000005722046
1.42999994754791 1.66999995708466
1.62999999523163 1.88999998569489
1.83000004291534 2.10999989509583
1.98000001907349 2.32999992370605
2.20000004768372 2.55999994277954
2.3199999332428 2.77999997138977
2.50999999046326 3
2.65000009536743 3.22000002861023
2.82999992370605 3.44000005722046
2.97000002861023 3.67000007629395
3.10999989509583 3.89000010490417
3.32999992370605 4.1100001335144
4.01999998092651 5
4.44000005722046 5.8899998664856
5.09000015258789 7
5.69000005722046 8
6.30000019073486 9
6.82999992370605 10
7.44000005722046 11
8.07999992370605 12
8.67000007629395 13
9.3100004196167 14
};
\addlegendentry{Experiment~\cite{martin1952PartIVExperimental}}
\addplot [line width=2pt, plum215172220, dashed]
table {%
0 1
0.0998772382736206 1
0.199753999710083 1.01999998092651
0.29963207244873 1.05999994277954
0.499386072158813 1.10000002384186
9.98771953582764 1.10000002384186
};
\addlegendentry{EXT, $W=50$}
\addplot [line width=2pt, orchid195130203, dashed]
table {%
0 1
0.0998772382736206 1
0.199753999710083 1.01999998092651
0.29963207244873 1.02999997138977
9.98771953582764 1.02999997138977
};
\addlegendentry{EXT, $W=100$}
\addplot [line width=2pt, mediumorchid17588186, dashed]
table {%
0 1
0.099940299987793 1.00499999523163
0.199880957603455 1.0349999666214
9.9940299987793 1.0349999666214
};
\addlegendentry{EXT, $W=200$}
\end{axis}

\end{tikzpicture}%
	\end{subfigure}%
	\vspace{0.5cm}
	\begin{subfigure}[htbp]{0.49\textwidth}
		\centering
		% This file was created with tikzplotlib v0.10.1.
\begin{tikzpicture}

\definecolor{darkgray176}{RGB}{176,176,176}
\definecolor{dodgerblue0154222}{RGB}{0,154,222}
\definecolor{lightgray204}{RGB}{204,204,204}
\definecolor{mediumturquoise64179230}{RGB}{0,116,166}
\definecolor{skyblue128205238}{RGB}{159,217,243}

\begin{axis}[
height=\figureheight,
legend cell align={left},
legend style={
	fill opacity=0.8,
	draw opacity=1,
	text opacity=1,
	at={(0.99,0.99)},
	anchor=north east,
	draw=lightgray204
},
tick align=outside,
tick pos=left,
width=\figurewidth,
x grid style={darkgray176},
xlabel={\(\displaystyle t^{*}\)},
xmajorgrids,
xmin=0, xmax=10,
xtick style={color=black},
y grid style={darkgray176},
ylabel style={rotate=-90.0},
ylabel={\(\displaystyle h^{*}\)},
ymajorgrids,
ymin=0.013, ymax=1.047,
ytick style={color=black}
]
\addplot [very thick, black, mark=*, mark size=2, mark options={solid}, only marks]
table {%
0 1
0.559999942779541 0.940000057220459
0.769999980926514 0.889999985694885
0.930000066757202 0.829999923706055
1.08000004291534 0.779999971389771
1.27999997138977 0.720000028610229
1.46000003814697 0.670000076293945
1.6599999666214 0.610000014305115
1.8400000333786 0.559999942779541
2 0.5
2.21000003814697 0.440000057220459
2.45000004768372 0.389999985694885
2.70000004768372 0.330000042915344
3.05999994277954 0.279999971389771
3.44000005722046 0.220000028610229
4.19999980926514 0.169999957084656
5.25 0.110000014305115
7.40000009536743 0.059999942779541
};
\addlegendentry{Experiment~\cite{martin1952PartIVExperimental}}
\addplot [line width=2pt, skyblue128205238, dotted]
table {%
0 1
0.0998772382736206 0.990000009536743
0.29963207244873 0.990000009536743
0.599262952804565 0.960000038146973
0.699140071868896 0.940000057220459
0.799018025398254 0.930000066757202
1.29840004444122 0.829999923706055
1.3982800245285 0.799999952316284
1.59803998470306 0.759999990463257
1.69790995121002 0.730000019073486
1.89767003059387 0.690000057220459
1.99753999710083 0.660000085830688
2.89644002914429 0.480000019073486
2.99632000923157 0.470000028610229
3.19606995582581 0.430000066757202
3.29594993591309 0.419999957084656
3.39582991600037 0.399999976158142
3.49569988250732 0.389999985694885
3.59558010101318 0.370000004768372
3.79533004760742 0.350000023841858
3.8952100276947 0.330000042915344
4.79410982131958 0.240000009536743
4.89398002624512 0.240000009536743
5.09373998641968 0.220000028610229
5.19361019134521 0.220000028610229
5.29348993301392 0.210000038146973
5.39337015151978 0.210000038146973
5.49324989318848 0.200000047683716
5.59312009811401 0.200000047683716
5.79288005828857 0.180000066757202
5.89275979995728 0.180000066757202
5.99263000488281 0.169999957084656
6.09251022338867 0.169999957084656
6.29226016998291 0.149999976158142
6.49202013015747 0.149999976158142
6.69177007675171 0.129999995231628
6.89153003692627 0.129999995231628
6.99139976501465 0.120000004768372
7.09127998352051 0.129999995231628
7.19116020202637 0.129999995231628
7.29103994369507 0.120000004768372
7.39091014862061 0.120000004768372
7.49078989028931 0.110000014305115
7.59067010879517 0.110000014305115
7.89029979705811 0.0800000429153442
8.09004974365234 0.100000023841858
8.6893196105957 0.100000023841858
8.7891902923584 0.110000014305115
8.88906955718994 0.110000014305115
8.9889497756958 0.100000023841858
9.18869972229004 0.100000023841858
9.2885799407959 0.0900000333786011
9.48832988739014 0.0900000333786011
9.588210105896 0.0800000429153442
9.68809032440186 0.0800000429153442
9.7879695892334 0.0700000524520874
9.98771953582764 0.0700000524520874
};
\addlegendentry{AVG, $W=50$}
\addplot [line width=2pt, mediumturquoise64179230, dotted]
table {%
0 1
0.0998772382736206 0.995000004768372
0.199753999710083 0.995000004768372
0.29963207244873 0.990000009536743
0.599262952804565 0.960000038146973
0.799018025398254 0.930000066757202
0.898895025253296 0.910000085830688
0.998772025108337 0.894999980926514
1.0986499786377 0.875
1.19852995872498 0.850000023841858
1.29840004444122 0.829999923706055
1.49816000461578 0.779999971389771
1.59803998470306 0.759999990463257
1.79779005050659 0.710000038146973
1.89767003059387 0.690000057220459
1.99753999710083 0.664999961853027
2.09741997718811 0.644999980926514
2.19729995727539 0.620000004768372
2.59681010246277 0.539999961853027
2.69668006896973 0.524999976158142
2.89644002914429 0.485000014305115
2.99632000923157 0.470000028610229
3.09618997573853 0.450000047683716
3.79533004760742 0.345000028610229
3.8952100276947 0.335000038146973
3.99509000778198 0.319999933242798
4.09497022628784 0.309999942779541
4.19483995437622 0.294999957084656
4.39459991455078 0.274999976158142
4.49447011947632 0.269999980926514
4.79410982131958 0.240000009536743
4.99385976791382 0.230000019073486
5.09373998641968 0.220000028610229
5.29348993301392 0.210000038146973
5.39337015151978 0.200000047683716
5.49324989318848 0.194999933242798
5.59312009811401 0.184999942779541
6.39213991165161 0.144999980926514
6.49202013015747 0.144999980926514
6.59189987182617 0.139999985694885
6.69177007675171 0.139999985694885
6.79164981842041 0.129999995231628
6.99139976501465 0.120000004768372
7.29103994369507 0.120000004768372
7.39091014862061 0.115000009536743
7.59067010879517 0.115000009536743
7.69053983688354 0.110000014305115
7.89029979705811 0.110000014305115
8.09004974365234 0.100000023841858
8.28981018066406 0.100000023841858
8.4895601272583 0.0900000333786011
8.58944034576416 0.0950000286102295
8.6893196105957 0.0950000286102295
8.88906955718994 0.0850000381469727
9.38846015930176 0.0850000381469727
9.48832988739014 0.0800000429153442
9.588210105896 0.0850000381469727
9.68809032440186 0.0850000381469727
};
\addlegendentry{AVG, $W=100$}
\addplot [line width=2pt, dodgerblue0154222, dotted]
table {%
0 1
0.199880957603455 0.995000004768372
0.299821019172668 0.990000009536743
0.399760961532593 0.982500076293945
0.499701023101807 0.972500085830688
0.599642038345337 0.960000038146973
0.799521923065186 0.930000066757202
0.99940299987793 0.894999980926514
1.09933996200562 0.872499942779541
1.19928002357483 0.852499961853027
1.29921996593475 0.829999923706055
1.39916002750397 0.805000066757202
1.49909996986389 0.782500028610229
1.59904003143311 0.757499933242798
1.69897997379303 0.735000014305115
1.79892003536224 0.710000038146973
2.09875011444092 0.642499923706055
2.19868993759155 0.622499942779541
2.29862999916077 0.600000023841858
2.59844994544983 0.539999961853027
2.69839000701904 0.522500038146973
2.79833006858826 0.502500057220459
3.09815001487732 0.450000047683716
3.29802989959717 0.419999957084656
3.39796996116638 0.402500033378601
3.4979100227356 0.389999985694885
3.69778990745544 0.360000014305115
4.19749021530151 0.297500014305115
4.49731016159058 0.267500042915344
4.59724998474121 0.259999990463257
4.69718980789185 0.25
5.19688987731934 0.212499976158142
5.39677000045776 0.202499985694885
5.59665012359619 0.1875
5.79653978347778 0.177500009536743
5.89648008346558 0.169999957084656
5.99641990661621 0.164999961853027
6.096360206604 0.157500028610229
6.19630002975464 0.152500033378601
6.29623985290527 0.149999976158142
6.39618015289307 0.144999980926514
6.69600009918213 0.137500047683716
6.79593992233276 0.132499933242798
7.39557981491089 0.117499947547913
7.49552011489868 0.112499952316284
7.59545993804932 0.112499952316284
7.99522018432617 0.102499961853027
8.09515953063965 0.102499961853027
8.29504013061523 0.0974999666213989
8.39498043060303 0.0974999666213989
8.4949197769165 0.0950000286102295
8.5948600769043 0.0950000286102295
8.69480037689209 0.0924999713897705
8.79473972320557 0.0924999713897705
8.89468002319336 0.0900000333786011
8.99462032318115 0.0900000333786011
9.09455966949463 0.0874999761581421
9.19449996948242 0.0874999761581421
9.29444026947021 0.0850000381469727
};
\addlegendentry{AVG, $W=200$}
\end{axis}

\end{tikzpicture}%
	\end{subfigure}
	\hfill
	\begin{subfigure}[htbp]{0.49\textwidth}
		\centering
		% This file was created with tikzplotlib v0.10.1.
\begin{tikzpicture}

\definecolor{darkgray176}{RGB}{176,176,176}
\definecolor{dodgerblue0154222}{RGB}{0,154,222}
\definecolor{lightgray204}{RGB}{204,204,204}
\definecolor{mediumturquoise64179230}{RGB}{0,116,166}
\definecolor{skyblue128205238}{RGB}{159,217,243}

\begin{axis}[
height=\figureheight,
legend cell align={left},
legend style={
  fill opacity=0.8,
  draw opacity=1,
  text opacity=1,
  at={(0.01,0.99)},
  anchor=north west,
  draw=lightgray204
},
tick align=outside,
tick pos=left,
width=\figurewidth,
x grid style={darkgray176},
xlabel={\(\displaystyle t^{*}\)},
xmajorgrids,
xmin=0, xmax=10,
xtick style={color=black},
y grid style={darkgray176},
ylabel style={rotate=-90.0},
ylabel={\(\displaystyle w^{*}\)},
ymajorgrids,
ymin=0, ymax=28,
ytick style={color=black}
]
\addplot [very thick, black, mark=*, mark size=2, mark options={solid}, only marks]
table {%
0.409999966621399 1.11000001430511
0.839999914169312 1.22000002861023
1.19000005722046 1.44000005722046
1.42999994754791 1.66999995708466
1.62999999523163 1.88999998569489
1.83000004291534 2.10999989509583
1.98000001907349 2.32999992370605
2.20000004768372 2.55999994277954
2.3199999332428 2.77999997138977
2.50999999046326 3
2.65000009536743 3.22000002861023
2.82999992370605 3.44000005722046
2.97000002861023 3.67000007629395
3.10999989509583 3.89000010490417
3.32999992370605 4.1100001335144
4.01999998092651 5
4.44000005722046 5.8899998664856
5.09000015258789 7
5.69000005722046 8
6.30000019073486 9
6.82999992370605 10
7.44000005722046 11
8.07999992370605 12
8.67000007629395 13
9.3100004196167 14
};
\addlegendentry{Experiment~\cite{martin1952PartIVExperimental}}
\addplot [line width=2pt, skyblue128205238, dotted]
table {%
0 1
0.0998772382736206 1
0.199753999710083 1.01999998092651
0.499386072158813 1.13999998569489
0.699140071868896 1.25999999046326
0.799018025398254 1.3400000333786
0.998772025108337 1.46000003814697
1.19852995872498 1.62000000476837
1.29840004444122 1.72000002861023
1.3982800245285 1.8400000333786
1.49816000461578 1.94000005722046
1.59803998470306 2.05999994277954
1.69790995121002 2.16000008583069
1.79779005050659 2.27999997138977
1.89767003059387 2.38000011444092
2.19729995727539 2.74000000953674
2.49692988395691 3.03999996185303
2.59681010246277 3.1800000667572
2.89644002914429 3.66000008583069
3.19606995582581 4.07999992370605
3.29594993591309 4.23999977111816
3.39582991600037 4.3600001335144
3.49569988250732 4.5
3.79533004760742 4.8600001335144
3.99509000778198 5.17999982833862
4.09497022628784 5.32000017166138
4.19483995437622 5.48000001907349
4.29472017288208 5.61999988555908
4.49447011947632 5.8600001335144
4.59434986114502 6
4.79410982131958 6.23999977111816
4.89398002624512 6.48000001907349
5.19361019134521 7.01999998092651
5.39337015151978 7.30000019073486
5.49324989318848 7.46000003814697
5.59312009811401 7.59999990463257
5.79288005828857 7.84000015258789
5.89275979995728 7.98000001907349
5.99263000488281 8.07999992370605
6.09251022338867 8.22000026702881
6.29226016998291 8.53999996185303
6.39213991165161 8.81999969482422
6.49202013015747 8.97999954223633
6.59189987182617 9.15999984741211
6.79164981842041 9.4399995803833
6.89153003692627 9.61999988555908
6.99139976501465 9.77999973297119
7.09127998352051 9.96000003814697
7.19116020202637 10.0799999237061
7.39091014862061 10.3599996566772
7.49078989028931 10.539999961853
7.59067010879517 10.6999998092651
7.89029979705811 11.1199998855591
7.99018001556396 11.2399997711182
8.38969039916992 11.8000001907349
8.7891902923584 12.2799997329712
8.9889497756958 12.5600004196167
9.08882999420166 12.6800003051758
9.18869972229004 12.7799997329712
9.2885799407959 12.9200000762939
9.38846015930176 13
9.588210105896 13.3199996948242
9.7879695892334 13.6000003814697
9.98771953582764 13.9200000762939
};
\addlegendentry{AVG, $W=50$}
\addplot [line width=2pt, mediumturquoise64179230, dotted]
table {%
0 1
0.0998772382736206 1
0.199753999710083 1.01999998092651
0.29963207244873 1.04999995231628
0.599262952804565 1.20000004768372
0.799018025398254 1.3400000333786
0.998772025108337 1.5
1.0986499786377 1.5900000333786
1.19852995872498 1.69000005722046
1.29840004444122 1.77999997138977
1.3982800245285 1.88999998569489
1.49816000461578 1.99000000953674
1.59803998470306 2.09999990463257
1.69790995121002 2.22000002861023
1.79779005050659 2.32999992370605
1.89767003059387 2.46000003814697
2.09741997718811 2.70000004768372
2.29717993736267 2.96000003814697
2.49692988395691 3.24000000953674
2.59681010246277 3.39000010490417
2.69668006896973 3.52999997138977
2.79656004905701 3.69000005722046
2.99632000923157 3.99000000953674
3.69546008110046 5.1100001335144
3.79533004760742 5.28000020980835
4.09497022628784 5.76000022888184
4.19483995437622 5.8899998664856
4.29472017288208 6.03000020980835
4.39459991455078 6.17999982833862
4.49447011947632 6.3600001335144
4.59434986114502 6.51999998092651
4.69423007965088 6.67000007629395
4.89398002624512 6.98999977111816
4.99385976791382 7.09999990463257
5.09373998641968 7.23000001907349
5.19361019134521 7.40999984741211
5.39337015151978 7.78999996185303
5.49324989318848 7.96999979019165
5.59312009811401 8.07999992370605
5.69299983978271 8.22999954223633
5.89275979995728 8.56999969482422
6.09251022338867 8.94999980926514
6.19238996505737 9.10999965667725
6.39213991165161 9.39000034332275
6.49202013015747 9.55000019073486
6.59189987182617 9.69999980926514
6.69177007675171 9.84000015258789
6.89153003692627 10.1800003051758
6.99139976501465 10.3199996948242
7.09127998352051 10.4399995803833
7.19116020202637 10.6000003814697
7.29103994369507 10.789999961853
7.39091014862061 10.9700002670288
7.49078989028931 11.1899995803833
7.59067010879517 11.3500003814697
7.69053983688354 11.4799995422363
7.7904200553894 11.6000003814697
7.89029979705811 11.7299995422363
7.99018001556396 11.8400001525879
8.1899299621582 12.2200002670288
8.28981018066406 12.3800001144409
8.38969039916992 12.5100002288818
8.4895601272583 12.6599998474121
8.58944034576416 12.8199996948242
8.6893196105957 12.960000038147
8.7891902923584 13.0699996948242
9.08882999420166 13.6400003433228
9.18869972229004 13.8100004196167
9.2885799407959 13.9700002670288
9.38846015930176 14.1499996185303
9.48832988739014 14.289999961853
9.588210105896 14.4099998474121
9.68809032440186 14.5600004196167
};
\addlegendentry{AVG, $W=100$}
\addplot [line width=2pt, dodgerblue0154222, dotted]
table {%
0 1
0.099940299987793 1.00499999523163
0.199880957603455 1.02499997615814
0.299821019172668 1.05999994277954
0.399760961532593 1.10000002384186
0.499701023101807 1.14999997615814
0.599642038345337 1.21000003814697
0.699581980705261 1.27499997615814
0.899461984634399 1.42499995231628
0.99940299987793 1.50999999046326
1.19928002357483 1.69000005722046
1.29921996593475 1.78999996185303
1.39916002750397 1.89499998092651
1.49909996986389 2.00500011444092
1.59904003143311 2.11999988555908
1.69897997379303 2.24000000953674
1.79892003536224 2.35500001907349
1.9988100528717 2.60500001907349
2.19868993759155 2.875
2.49850988388062 3.26500010490417
2.69839000701904 3.55999994277954
2.89826989173889 3.84999990463257
3.09815001487732 4.16499996185303
3.39796996116638 4.67000007629395
3.59785008430481 5.00500011444092
3.69778990745544 5.15500020980835
3.89767003059387 5.47499990463257
3.99761009216309 5.61999988555908
4.29743003845215 6.13000011444092
4.39736986160278 6.32000017166138
4.49731016159058 6.4850001335144
4.59724998474121 6.65999984741211
4.69718980789185 6.80999994277954
5.0969500541687 7.48999977111816
5.19688987731934 7.67500019073486
5.39677000045776 8.01000022888184
5.4967098236084 8.18000030517578
5.59665012359619 8.375
5.69658994674683 8.5
5.79653978347778 8.65999984741211
5.89648008346558 8.82999992370605
5.99641990661621 8.98999977111816
6.29623985290527 9.52999973297119
6.69600009918213 10.1899995803833
6.79593992233276 10.375
6.89588022232056 10.5749998092651
6.99582004547119 10.7600002288818
7.19570016860962 11.0649995803833
7.49552011489868 11.5749998092651
7.59545993804932 11.7200002670288
7.89527988433838 12.25
7.99522018432617 12.4449996948242
8.09515953063965 12.6300001144409
8.39498043060303 13.1300001144409
8.4949197769165 13.3199996948242
8.5948600769043 13.4799995422363
8.69480037689209 13.6300001144409
8.89468002319336 13.9549999237061
8.99462032318115 14.1350002288818
9.09455966949463 14.3050003051758
9.29444026947021 14.625
};
\addlegendentry{AVG, $W=200$}
\end{axis}

\end{tikzpicture}%
	\end{subfigure}
	\addtocounter{figure}{-1}
	\caption{(\textit{Continued})}
\end{figure}

\clearpage

\section{Cylindrical dam break}

\begin{figure}[htbp]
	\centering
	\setlength{\figureheight}{0.3\textwidth}
	\setlength{\figurewidth}{0.95\textwidth}
	\begin{subfigure}[htbp]{0.99\textwidth}
		\centering
		% This file was created with tikzplotlib v0.10.1.
\begin{tikzpicture}

\definecolor{darkgray176}{RGB}{176,176,176}
\definecolor{lightgray204}{RGB}{204,204,204}
\definecolor{mediumaquamarine128230181}{RGB}{159,236,200}
\definecolor{mediumaquamarine64218145}{RGB}{0,128,68}
\definecolor{springgreen0205108}{RGB}{0,205,108}

\begin{axis}[
height=\figureheight,
legend cell align={left},
legend style={
  fill opacity=0.8,
  draw opacity=1,
  text opacity=1,
  at={(0.01,0.99)},
  anchor=north west,
  draw=lightgray204
},
tick align=outside,
tick pos=left,
width=\figurewidth,
x grid style={darkgray176},
xlabel={\(\displaystyle t^{*}\)},
xmajorgrids,
xmin=0, xmax=5,
xtick style={color=black},
y grid style={darkgray176},
ylabel style={rotate=-90.0},
ylabel={\(\displaystyle r^{*}\)},
ymajorgrids,
ymin=0.776347, ymax=5.696713,
ytick style={color=black}
]
\addplot [semithick, black, mark=*, mark size=2, mark options={solid}, only marks]
table {%
1.19000005722046 1.22000002861023
1.5 1.44000005722046
1.73000001907349 1.66999995708466
1.91999995708466 1.88999998569489
2.1800000667572 2.10999989509583
2.35999989509583 2.32999992370605
2.58999991416931 2.55999994277954
2.78999996185303 2.77999997138977
2.97000002861023 3
3.20000004768372 3.22000002861023
3.38000011444092 3.44000005722046
3.57999992370605 3.67000007629395
3.75999999046326 3.89000010490417
3.97000002861023 4.1100001335144
};
\addlegendentry{Experiment~\cite{martin1952PartIVExperimental}}
\addplot [line width=2pt, mediumaquamarine128230181]
table {%
0 1
0.0968506336212158 1.02286005020142
0.193701028823853 1.03618001937866
0.290552020072937 1.06209003925323
0.387402057647705 1.07738995552063
0.4842529296875 1.11576998233795
0.581104040145874 1.15838003158569
0.677953958511353 1.21512997150421
0.871655941009521 1.32088994979858
0.96850597858429 1.38689994812012
1.06535995006561 1.45061004161835
1.25906002521515 1.59710001945496
1.4527599811554 1.7537100315094
1.54961001873016 1.82871997356415
1.64646005630493 1.9172500371933
1.84016001224518 2.08051991462708
1.93701004981995 2.17027997970581
2.22756004333496 2.43247008323669
2.32440996170044 2.52584004402161
2.42126989364624 2.62300992012024
2.5181200504303 2.72603988647461
2.61496996879578 2.81713008880615
2.71181988716125 2.91303992271423
2.80867004394531 3.01963996887207
2.90551996231079 3.11072993278503
3.00237011909485 3.21027994155884
3.09922003746033 3.31354999542236
3.19606995582581 3.4242000579834
3.29292011260986 3.52501010894775
3.38977003097534 3.62908005714417
3.58347010612488 3.84839010238647
3.77716994285583 4.05600023269653
3.87402009963989 4.1699800491333
3.97088003158569 4.28163003921509
4.06772994995117 4.38526010513306
4.35828018188477 4.70484018325806
4.45513010025024 4.82678985595703
4.55198001861572 4.93756008148193
4.6488299369812 5.04385995864868
4.74567985534668 5.16454982757568
4.84252977371216 5.27234983444214
};
\addlegendentry{EQ, $D=50$}
\addplot [line width=2pt, mediumaquamarine64218145]
table {%
0 1
0.199753999710083 1.02860999107361
0.399508953094482 1.08675003051758
0.499386072158813 1.12607002258301
0.599262952804565 1.17005002498627
0.699140071868896 1.22022998332977
0.799018025398254 1.27520000934601
0.898895025253296 1.33589005470276
0.998772025108337 1.40039002895355
1.0986499786377 1.4675999879837
1.19852995872498 1.54121994972229
1.29840004444122 1.61765003204346
1.3982800245285 1.6981600522995
1.49816000461578 1.7820600271225
1.59803998470306 1.87033998966217
1.69790995121002 1.96103000640869
1.79779005050659 2.05397009849548
1.89767003059387 2.1508800983429
1.99753999710083 2.25089001655579
2.09741997718811 2.35339999198914
2.19729995727539 2.45902991294861
2.29717993736267 2.56807994842529
2.39704990386963 2.67926001548767
2.49692988395691 2.79317998886108
2.69668006896973 3.02970004081726
2.89644002914429 3.27740001678467
3.09618997573853 3.53986001014709
3.19606995582581 3.67710995674133
3.29594993591309 3.82133007049561
3.39582991600037 3.96157002449036
3.59558010101318 4.26762008666992
3.69546008110046 4.42120981216431
3.79533004760742 4.58034992218018
};
\addlegendentry{EQ, $D=100$}
\addplot [line width=2pt, springgreen0205108]
table {%
0 1
0.0998772382736206 1.00802004337311
0.199753999710083 1.02504003047943
0.29963207244873 1.05154001712799
0.399508953094482 1.08580994606018
0.499386072158813 1.1258499622345
0.599262952804565 1.17148995399475
0.699140071868896 1.22225999832153
0.799018025398254 1.27874004840851
0.898895025253296 1.33844995498657
0.998772025108337 1.40128004550934
1.0986499786377 1.46823000907898
1.19852995872498 1.5382000207901
1.29840004444122 1.61097002029419
1.3982800245285 1.68839001655579
1.49816000461578 1.77014994621277
1.59803998470306 1.85602998733521
2.19729995727539 2.38798999786377
2.29717993736267 2.48127007484436
2.39704990386963 2.5792601108551
2.49692988395691 2.68318009376526
3.09618997573853 3.33053994178772
3.19606995582581 3.44525003433228
3.8952100276947 4.27276992797852
4.59434986114502 5.11206007003784
4.89398002624512 5.473060131073
};
\addlegendentry{EQ, $D=200$}
\end{axis}

\end{tikzpicture}%
	\end{subfigure}%
	\vspace{0.5cm}
	\begin{subfigure}[htbp]{0.99\textwidth}
		\centering
		% This file was created with tikzplotlib v0.10.1.
\begin{tikzpicture}

\definecolor{burlywood248194145}{RGB}{250,209,172}
\definecolor{darkgray176}{RGB}{176,176,176}
\definecolor{darkorange24213334}{RGB}{242,133,34}
\definecolor{lightgray204}{RGB}{204,204,204}
\definecolor{sandybrown24516489}{RGB}{181,100,26}

\begin{axis}[
height=\figureheight,
legend cell align={left},
legend style={
  fill opacity=0.8,
  draw opacity=1,
  text opacity=1,
  at={(0.01,0.99)},
  anchor=north west,
  draw=lightgray204
},
tick align=outside,
tick pos=left,
width=\figurewidth,
x grid style={darkgray176},
xlabel={\(\displaystyle t^{*}\)},
xmajorgrids,
xmin=0, xmax=5,
xtick style={color=black},
y grid style={darkgray176},
ylabel style={rotate=-90.0},
ylabel={\(\displaystyle r^{*}\)},
ymajorgrids,
ymin=0.7856855, ymax=5.5006045,
ytick style={color=black}
]
\addplot [semithick, black, mark=*, mark size=2, mark options={solid}, only marks]
table {%
1.19000005722046 1.22000002861023
1.5 1.44000005722046
1.73000001907349 1.66999995708466
1.91999995708466 1.88999998569489
2.1800000667572 2.10999989509583
2.35999989509583 2.32999992370605
2.58999991416931 2.55999994277954
2.78999996185303 2.77999997138977
2.97000002861023 3
3.20000004768372 3.22000002861023
3.38000011444092 3.44000005722046
3.57999992370605 3.67000007629395
3.75999999046326 3.89000010490417
3.97000002861023 4.1100001335144
};
\addlegendentry{Experiment~\cite{martin1952PartIVExperimental}}
\addplot [line width=2pt, burlywood248194145, dash pattern=on 1pt off 3pt on 3pt off 3pt]
table {%
0 1
0.0968506336212158 1.02286005020142
0.193701028823853 1.03618001937866
0.290552020072937 1.06209003925323
0.387402057647705 1.07918000221252
0.4842529296875 1.11406004428864
0.581104040145874 1.1583399772644
0.677953958511353 1.21988999843597
0.774805068969727 1.27057003974915
0.871655941009521 1.32781994342804
0.96850597858429 1.39542996883392
1.06535995006561 1.45601999759674
1.16220998764038 1.53331005573273
1.25906002521515 1.60519003868103
1.35590994358063 1.68075001239777
1.54961001873016 1.83721005916595
1.64646005630493 1.9120899438858
1.74330997467041 1.9935599565506
1.84016001224518 2.06684994697571
1.93701004981995 2.15118002891541
2.03385996818542 2.22920989990234
2.1307098865509 2.3183000087738
2.42126989364624 2.56959009170532
2.5181200504303 2.65446996688843
2.71181988716125 2.83363008499146
2.80867004394531 2.91740989685059
3.09922003746033 3.1965799331665
3.29292011260986 3.39218997955322
3.38977003097534 3.48320007324219
3.68032002449036 3.78045988082886
3.77716994285583 3.88034009933472
3.87402009963989 3.97697997093201
3.97088003158569 4.07696008682251
4.06772994995117 4.17097997665405
4.16457986831665 4.26901006698608
4.26142978668213 4.38173007965088
4.35828018188477 4.47024011611938
4.45513010025024 4.57707977294922
4.6488299369812 4.77517986297607
4.74567985534668 4.87985992431641
4.84252977371216 4.9785099029541
4.93938016891479 5.08333015441895
5.03623008728027 5.1829400062561
};
\addlegendentry{EQ+NEQ, $D=50$}
\addplot [line width=2pt, sandybrown24516489, dash pattern=on 1pt off 3pt on 3pt off 3pt]
table {%
0 1
0.199753999710083 1.02909994125366
0.399508953094482 1.08736002445221
0.499386072158813 1.13205003738403
0.599262952804565 1.18444001674652
0.699140071868896 1.25417995452881
0.799018025398254 1.37356996536255
0.898895025253296 1.46627998352051
5.09373998641968 1.46627998352051
};
\addlegendentry{EQ+NEQ, $D=100$}
\addplot [line width=2pt, darkorange24213334, dash pattern=on 1pt off 3pt on 3pt off 3pt]
table {%
0 1
0.0998772382736206 1.00788998603821
0.199753999710083 1.02504003047943
0.29963207244873 1.05154001712799
0.399508953094482 1.08521997928619
0.499386072158813 1.1253399848938
0.599262952804565 1.17142999172211
0.699140071868896 1.22283005714417
0.898895025253296 1.33932995796204
0.998772025108337 1.40050995349884
1.0986499786377 1.46370005607605
1.19852995872498 1.5306099653244
1.29840004444122 1.600909948349
1.3982800245285 1.67360997200012
1.49816000461578 1.74828004837036
1.69790995121002 1.90714001655579
1.79779005050659 1.98941004276276
1.99753999710083 2.15936994552612
2.19729995727539 2.33471989631653
2.29717993736267 2.42314004898071
2.39704990386963 2.51345992088318
2.59681010246277 2.70412993431091
2.69668006896973 2.80232000350952
3.09618997573853 3.20538997650146
3.19606995582581 3.30735993385315
3.49569988250732 3.6279399394989
3.59558010101318 3.73668003082275
3.69546008110046 3.84795999526978
4.09497022628784 4.30402994155884
};
\addlegendentry{EQ+NEQ, $D=200$}
\end{axis}

\end{tikzpicture}%
	\end{subfigure}%
	\vspace{0.5cm}
	\begin{subfigure}[htbp]{0.99\textwidth}
		\centering
		% This file was created with tikzplotlib v0.10.1.
\begin{tikzpicture}

\definecolor{crimson2553191}{RGB}{255,31,91}
\definecolor{darkgray176}{RGB}{176,176,176}
\definecolor{hotpink255143173}{RGB}{255,171,193}
\definecolor{lightcoral25587132}{RGB}{159,19,57}
\definecolor{lightgray204}{RGB}{204,204,204}

\begin{axis}[
height=\figureheight,
legend cell align={left},
legend style={
  fill opacity=0.8,
  draw opacity=1,
  text opacity=1,
  at={(0.01,0.99)},
  anchor=north west,
  draw=lightgray204
},
tick align=outside,
tick pos=left,
width=\figurewidth,
x grid style={darkgray176},
xlabel={\(\displaystyle t^{*}\)},
xmajorgrids,
xmin=0, xmax=5,
xtick style={color=black},
y grid style={darkgray176},
ylabel style={rotate=-90.0},
ylabel={\(\displaystyle r^{*}\)},
ymajorgrids,
ymin=0.7757785, ymax=5.7086515,
ytick style={color=black}
]
\addplot [semithick, black, mark=*, mark size=2, mark options={solid}, only marks]
table {%
1.19000005722046 1.22000002861023
1.5 1.44000005722046
1.73000001907349 1.66999995708466
1.91999995708466 1.88999998569489
2.1800000667572 2.10999989509583
2.35999989509583 2.32999992370605
2.58999991416931 2.55999994277954
2.78999996185303 2.77999997138977
2.97000002861023 3
3.20000004768372 3.22000002861023
3.38000011444092 3.44000005722046
3.57999992370605 3.67000007629395
3.75999999046326 3.89000010490417
3.97000002861023 4.1100001335144
};
\addlegendentry{Experiment~\cite{martin1952PartIVExperimental}}
\addplot [line width=2pt, hotpink255143173, dash pattern=on 1pt off 3pt on 3pt off 3pt]
table {%
0 1
0.0968506336212158 1.02286005020142
0.193701028823853 1.03618001937866
0.290552020072937 1.06209003925323
0.387402057647705 1.07738995552063
0.4842529296875 1.11576998233795
0.581104040145874 1.16007995605469
0.677953958511353 1.21670997142792
0.871655941009521 1.32155001163483
0.96850597858429 1.38898003101349
1.06535995006561 1.45068001747131
1.25906002521515 1.59687995910645
1.4527599811554 1.75293004512787
1.54961001873016 1.82904994487762
1.64646005630493 1.91475999355316
1.84016001224518 2.07748007774353
1.93701004981995 2.16874003410339
2.03385996818542 2.25029993057251
2.22756004333496 2.43010997772217
2.42126989364624 2.60615992546082
2.5181200504303 2.69722008705139
2.61496996879578 2.78618001937866
2.71181988716125 2.88040995597839
2.80867004394531 2.98284006118774
2.90551996231079 3.07389998435974
3.00237011909485 3.17289996147156
3.19606995582581 3.37696003913879
3.29292011260986 3.48223996162415
3.38977003097534 3.59388995170593
3.48661994934082 3.6972599029541
3.77716994285583 4.02408981323242
3.87402009963989 4.12253999710083
3.97088003158569 4.2398099899292
4.06772994995117 4.35094976425171
4.16457986831665 4.4521598815918
4.26142978668213 4.56606006622314
4.35828018188477 4.677490234375
4.45513010025024 4.79201984405518
4.55198001861572 4.90282011032104
4.74567985534668 5.13135004043579
4.84252977371216 5.24245023727417
4.93938016891479 5.35621976852417
};
\addlegendentry{GEQ, $D=50$}
\addplot [line width=2pt, lightcoral25587132, dash pattern=on 1pt off 3pt on 3pt off 3pt]
table {%
0 1
0.199753999710083 1.02860999107361
0.29963207244873 1.05660998821259
0.399508953094482 1.08668994903564
0.499386072158813 1.12485003471375
0.599262952804565 1.16944003105164
0.699140071868896 1.21837997436523
0.799018025398254 1.27296996116638
0.898895025253296 1.33211994171143
0.998772025108337 1.39565002918243
1.0986499786377 1.46185994148254
1.19852995872498 1.53404998779297
1.29840004444122 1.6097400188446
1.3982800245285 1.68986999988556
1.49816000461578 1.77448999881744
1.59803998470306 1.862380027771
1.69790995121002 1.95278000831604
1.79779005050659 2.04664993286133
1.99753999710083 2.24027991294861
2.19729995727539 2.44106006622314
2.39704990386963 2.64749002456665
2.59681010246277 2.86091995239258
2.79656004905701 3.07940006256104
2.99632000923157 3.30328011512756
3.19606995582581 3.53225994110107
3.29594993591309 3.64823007583618
3.49569988250732 3.88633990287781
3.69546008110046 4.12908983230591
3.8952100276947 4.37735986709595
3.99509000778198 4.50324010848999
4.19483995437622 4.7609601020813
4.29472017288208 4.89234018325806
};
\addlegendentry{GEQ, $D=100$}
\addplot [line width=2pt, crimson2553191, dash pattern=on 1pt off 3pt on 3pt off 3pt]
table {%
0 1
0.0998772382736206 1.00788998603821
0.199753999710083 1.02491998672485
0.29963207244873 1.05141997337341
0.399508953094482 1.08591997623444
0.499386072158813 1.12597000598907
0.599262952804565 1.17182004451752
0.699140071868896 1.22277998924255
0.799018025398254 1.27892994880676
0.898895025253296 1.33872997760773
0.998772025108337 1.40173995494843
1.0986499786377 1.46867001056671
1.19852995872498 1.53860998153687
1.29840004444122 1.61105000972748
1.3982800245285 1.68845999240875
1.49816000461578 1.77000999450684
1.59803998470306 1.85605001449585
2.19729995727539 2.39085006713867
2.29717993736267 2.48412990570068
2.39704990386963 2.58265995979309
2.49692988395691 2.68703007698059
2.99632000923157 3.22206997871399
3.09618997573853 3.33349990844727
3.29594993591309 3.56449007987976
3.69546008110046 4.03301000595093
3.8952100276947 4.27243995666504
4.59434986114502 5.11942005157471
4.89398002624512 5.48442983627319
};
\addlegendentry{GEQ, $D=200$}
\end{axis}

\end{tikzpicture}%
	\end{subfigure}
	\caption{
		\label{fig:dam-break-cylindrical-convergence}
		Simulated cylindrical dam break with non-dimensionalized liquid column radius $r^{*}(t^{*})$ and time $t^{*}$.
		The simulations were performed with computational domain resolutions, that is, initial column diameters of $D \in \{50, 100, 200\}$ lattice cells.
		The markers represent the mean value of $r^{*}(t^{*})$.
		The EXT scheme was numerically unstable for all tested resolutions.
		All other simulations have converged moderately well.
		The EQ+NEQ and AVG scheme were unstable for $D=100$, and the EQ scheme was inaccurate at $D=100$ when compared with the experimental data~\cite{martin1952PartIVExperimental}.
	}
\end{figure}

\begin{figure}[htbp]
	\centering
	\setlength{\figureheight}{0.3\textwidth}
	\setlength{\figurewidth}{0.95\textwidth}
	\vspace{0.5cm}
	\begin{subfigure}[htbp]{0.99\textwidth}
		\centering
		% This file was created with tikzplotlib v0.10.1.
\begin{tikzpicture}

\definecolor{darkgray176}{RGB}{176,176,176}
\definecolor{lightgray204}{RGB}{204,204,204}
\definecolor{mediumorchid17588186}{RGB}{175,88,186}
\definecolor{orchid195130203}{RGB}{109,55,116}
\definecolor{plum215172220}{RGB}{225,192,229}

\begin{axis}[
height=\figureheight,
legend cell align={left},
legend style={
  fill opacity=0.8,
  draw opacity=1,
  text opacity=1,
  at={(0.01,0.99)},
  anchor=north west,
  draw=lightgray204
},
tick align=outside,
tick pos=left,
width=\figurewidth,
x grid style={darkgray176},
xlabel={\(\displaystyle t^{*}\)},
xmajorgrids,
xmin=0, xmax=5,
xtick style={color=black},
y grid style={darkgray176},
ylabel style={rotate=-90.0},
ylabel={\(\displaystyle r^{*}\)},
ymajorgrids,
ymin=0.8445, ymax=4.2655,
ytick style={color=black}
]
\addplot [semithick, black, mark=*, mark size=2, mark options={solid}, only marks]
table {%
1.19000005722046 1.22000002861023
1.5 1.44000005722046
1.73000001907349 1.66999995708466
1.91999995708466 1.88999998569489
2.1800000667572 2.10999989509583
2.35999989509583 2.32999992370605
2.58999991416931 2.55999994277954
2.78999996185303 2.77999997138977
2.97000002861023 3
3.20000004768372 3.22000002861023
3.38000011444092 3.44000005722046
3.57999992370605 3.67000007629395
3.75999999046326 3.89000010490417
3.97000002861023 4.1100001335144
};
\addlegendentry{Experiment~\cite{martin1952PartIVExperimental}}
\addplot [line width=2pt, plum215172220, dashed]
table {%
0 1
0.0968506336212158 1.02479994297028
0.193701028823853 1.03890001773834
0.387402057647705 1.09197998046875
0.4842529296875 1.12753999233246
0.581104040145874 1.17192995548248
0.677953958511353 1.21389997005463
0.774805068969727 1.21174001693726
5.03623008728027 1.21174001693726
};
\addlegendentry{EXT, $D=50$}
\addplot [line width=2pt, orchid195130203, dashed]
table {%
0 1
0.0998772382736206 1.01350998878479
0.199753999710083 1.02974998950958
0.29963207244873 1.05901002883911
0.399508953094482 1.09669995307922
0.499386072158813 1.14033997058868
3.99509000778198 1.14030003547668
5.09373998641968 1.14030003547668
};
\addlegendentry{EXT, $D=100$}
\addplot [line width=2pt, mediumorchid17588186, dashed]
table {%
0 1
0.0998772382736206 1.0081399679184
0.199753999710083 1.00952994823456
5.09373998641968 1.00952994823456
};
\addlegendentry{EXT, $D=200$}
\end{axis}

\end{tikzpicture}%
	\end{subfigure}%
	\vspace{0.5cm}
	\begin{subfigure}[htbp]{0.99\textwidth}
		\centering
		% This file was created with tikzplotlib v0.10.1.
\begin{tikzpicture}

\definecolor{darkgray176}{RGB}{176,176,176}
\definecolor{dodgerblue0154222}{RGB}{0,154,222}
\definecolor{lightgray204}{RGB}{204,204,204}
\definecolor{mediumturquoise64179230}{RGB}{0,116,166}
\definecolor{skyblue128205238}{RGB}{159,217,243}

\begin{axis}[
height=\figureheight,
legend cell align={left},
legend style={
  fill opacity=0.8,
  draw opacity=1,
  text opacity=1,
  at={(0.01,0.99)},
  anchor=north west,
  draw=lightgray204
},
tick align=outside,
tick pos=left,
width=\figurewidth,
x grid style={darkgray176},
xlabel={\(\displaystyle t^{*}\)},
xmajorgrids,
xmin=0, xmax=5,
xtick style={color=black},
y grid style={darkgray176},
ylabel style={rotate=-90.0},
ylabel={\(\displaystyle r^{*}\)},
ymajorgrids,
ymin=0.7769885, ymax=5.6832415,
ytick style={color=black}
]
\addplot [semithick, black, mark=*, mark size=2, mark options={solid}, only marks]
table {%
1.19000005722046 1.22000002861023
1.5 1.44000005722046
1.73000001907349 1.66999995708466
1.91999995708466 1.88999998569489
2.1800000667572 2.10999989509583
2.35999989509583 2.32999992370605
2.58999991416931 2.55999994277954
2.78999996185303 2.77999997138977
2.97000002861023 3
3.20000004768372 3.22000002861023
3.38000011444092 3.44000005722046
3.57999992370605 3.67000007629395
3.75999999046326 3.89000010490417
3.97000002861023 4.1100001335144
};
\addlegendentry{Experiment~\cite{martin1952PartIVExperimental}}
\addplot [line width=2pt, skyblue128205238, dotted]
table {%
0 1
0.0968506336212158 1.02286005020142
0.193701028823853 1.03618001937866
0.290552020072937 1.06209003925323
0.387402057647705 1.07918000221252
0.4842529296875 1.11576998233795
0.581104040145874 1.15927004814148
0.677953958511353 1.2182400226593
0.774805068969727 1.26953995227814
0.871655941009521 1.32483005523682
1.06535995006561 1.45685994625092
1.16220998764038 1.52789998054504
1.25906002521515 1.5960099697113
1.35590994358063 1.66588997840881
1.4527599811554 1.7408299446106
1.54961001873016 1.81272995471954
1.64646005630493 1.89272999763489
1.74330997467041 1.96944999694824
1.84016001224518 2.04998993873596
2.03385996818542 2.2182400226593
2.42126989364624 2.56016993522644
2.5181200504303 2.64154005050659
2.61496996879578 2.73624992370605
2.71181988716125 2.82564997673035
2.80867004394531 2.9124801158905
2.90551996231079 3.00836992263794
3.00237011909485 3.09411001205444
3.19606995582581 3.28600001335144
3.29292011260986 3.37914991378784
3.38977003097534 3.48011994361877
3.48661994934082 3.57447004318237
3.58347010612488 3.67586994171143
3.68032002449036 3.76947999000549
3.77716994285583 3.87250995635986
3.97088003158569 4.06587982177734
4.16457986831665 4.26697015762329
4.45513010025024 4.56350994110107
4.55198001861572 4.65996980667114
4.6488299369812 4.76221990585327
4.74567985534668 4.85246992111206
4.93938016891479 5.04469013214111
5.03623008728027 5.13581991195679
};
\addlegendentry{AVG, $D=50$}
\addplot [line width=2pt, mediumturquoise64179230, dotted]
table {%
0 1
0.199753999710083 1.02860999107361
0.399508953094482 1.08686995506287
0.499386072158813 1.12845003604889
0.599262952804565 1.17637002468109
0.699140071868896 1.23135995864868
0.799018025398254 1.29347002506256
0.898895025253296 1.36319994926453
0.998772025108337 1.45181000232697
1.0986499786377 1.58846998214722
1.19852995872498 1.67941999435425
1.3982800245285 1.68924999237061
1.49816000461578 1.68909001350403
1.69790995121002 1.68638002872467
1.79779005050659 1.68985998630524
1.89767003059387 1.69020998477936
1.99753999710083 1.69289004802704
2.09741997718811 1.69237995147705
2.19729995727539 1.6959400177002
2.29717993736267 1.69510996341705
2.39704990386963 1.69988000392914
2.49692988395691 1.69845998287201
2.59681010246277 1.70507001876831
2.69668006896973 1.70861995220184
2.79656004905701 1.7021199464798
2.89644002914429 1.71143996715546
2.99632000923157 1.71467995643616
3.09618997573853 1.70731997489929
3.19606995582581 1.70784997940063
};
\addlegendentry{AVG, $D=100$}
\addplot [line width=2pt, dodgerblue0154222, dotted]
table {%
0 1
0.0998772382736206 1.00802004337311
0.199753999710083 1.02491998672485
0.29963207244873 1.05154001712799
0.399508953094482 1.08590996265411
0.499386072158813 1.12553000450134
0.599262952804565 1.17138004302979
0.699140071868896 1.22197997570038
0.799018025398254 1.27809000015259
0.898895025253296 1.33750998973846
0.998772025108337 1.39986002445221
1.0986499786377 1.46676003932953
1.19852995872498 1.53620994091034
1.29840004444122 1.60875999927521
1.3982800245285 1.68456995487213
1.49816000461578 1.76425004005432
1.69790995121002 1.9329400062561
1.79779005050659 2.01922988891602
1.99753999710083 2.19442009925842
2.09741997718811 2.28050994873047
2.19729995727539 2.36951994895935
2.29717993736267 2.46360993385315
2.49692988395691 2.65796995162964
2.99632000923157 3.15473008155823
3.19606995582581 3.35975003242493
3.39582991600037 3.57277989387512
3.79533004760742 4.00234985351562
4.09497022628784 4.33318996429443
4.19483995437622 4.44245004653931
4.59434986114502 4.89779996871948
4.69423007965088 5.0156397819519
4.89398002624512 5.23965978622437
5.09373998641968 5.46022987365723
};
\addlegendentry{AVG, $D=200$}
\end{axis}

\end{tikzpicture}%
	\end{subfigure}
	\addtocounter{figure}{-1}
	\caption{(\textit{Continued})}
\end{figure}

\clearpage

\section{Taylor bubble}

\begin{figure}[htbp]
	\centering
	\setlength{\figureheight}{0.35\textwidth}
	\setlength{\figurewidth}{0.48\textwidth}
	\begin{subfigure}[htbp]{0.49\textwidth}
		\centering
		% This file was created with tikzplotlib v0.10.1.
\begin{tikzpicture}

\definecolor{darkgray176}{RGB}{176,176,176}
\definecolor{lightgray204}{RGB}{204,204,204}
\definecolor{mediumaquamarine128230181}{RGB}{159,236,200}
\definecolor{mediumaquamarine64218145}{RGB}{0,128,68}
\definecolor{springgreen0205108}{RGB}{0,205,108}

\begin{axis}[
height=\figureheight,
legend cell align={left},
legend style={
  fill opacity=0.8,
  draw opacity=1,
  text opacity=1,
  at={(0.01,0.01)},
  anchor=south west,
  draw=lightgray204
},
tick align=outside,
tick pos=left,
width=\figurewidth,
x grid style={darkgray176},
xlabel={\(\displaystyle r^{*}\)},
xmajorgrids,
xmin=-0.0373953402042391, xmax=0.78530214428902,
xtick style={color=black},
y grid style={darkgray176},
ylabel style={rotate=-90.0},
ylabel={\(\displaystyle z^{*}\)},
ymajorgrids,
ymin=-0.628900408744815, ymax=0.0299476385116579,
ytick style={color=black}
]
\addplot [very thick, black, mark=*, mark size=1, mark options={solid}, only marks]
table {%
0 0
0.0293359756469727 0
0.0573979616165161 -0.000633955001831055
0.0854599475860596 -0.00253605842590332
0.119897961616516 -0.00570595264434814
0.160714030265808 -0.0120450258255005
0.192602038383484 -0.0183850526809692
0.219388008117676 -0.0247249603271484
0.244897961616516 -0.0316979885101318
0.271683931350708 -0.0412089824676514
0.293367981910706 -0.0494489669799805
0.320153951644897 -0.0595920085906982
0.340561985969543 -0.0684679746627808
0.362244009971619 -0.0773429870605469
0.380102038383484 -0.0862189531326294
0.39795994758606 -0.0950939655303955
0.415816068649292 -0.10523796081543
0.429846048355103 -0.112844944000244
0.442602038383484 -0.121086955070496
0.456632018089294 -0.129328012466431
0.470664024353027 -0.138838052749634
0.48852002620697 -0.149615049362183
0.503826022148132 -0.162294030189514
0.519132018089294 -0.173071980476379
0.533164024353027 -0.183848977088928
0.543367981910706 -0.193992018699646
0.554846048355103 -0.204769968986511
0.565052032470703 -0.214913964271545
0.57397997379303 -0.225057005882263
0.581632018089294 -0.235200047492981
0.593111991882324 -0.247879028320312
0.603316068649292 -0.259925007820129
0.614795923233032 -0.273237943649292
0.623723983764648 -0.286550998687744
0.633928060531616 -0.299864053726196
0.641582012176514 -0.312543034553528
0.64795994758606 -0.325222969055176
0.656888008117676 -0.339169979095459
0.66198992729187 -0.351848959922791
0.667092084884644 -0.365162014961243
0.672194004058838 -0.376574039459229
0.679846048355103 -0.387984991073608
0.686223983764648 -0.401932001113892
0.690052032470703 -0.414610981941223
0.692601919174194 -0.424754977226257
0.697704076766968 -0.438068032264709
0.701529979705811 -0.453283071517944
0.705358028411865 -0.464694023132324
0.707906007766724 -0.476740002632141
0.71045994758606 -0.48751699924469
0.711734056472778 -0.49956202507019
0.715561985969543 -0.511608004570007
0.716835975646973 -0.523019075393677
0.718111991882324 -0.533161997795105
0.720664024353027 -0.544573068618774
0.721935987472534 -0.560423016548157
0.72448992729187 -0.572468042373657
0.72448992729187 -0.583878993988037
0.725765943527222 -0.595291018486023
};
\addlegendentry{Experiment~\cite{bugg2002VelocityFieldTaylor}}
\addplot [line width=2pt, mediumaquamarine128230181]
table {%
0 0
0.0312505960464478 0
0.0937505960464478 -0.00412511825561523
0.138541579246521 -0.0124845504760742
0.156250596046448 -0.0156874656677246
0.218750596046448 -0.0253124237060547
0.273225545883179 -0.0405311584472656
0.281250596046448 -0.0436563491821289
0.343750596046448 -0.0611872673034668
0.387679100036621 -0.0797343254089355
0.406250596046448 -0.0876874923706055
0.437456846237183 -0.103031158447266
0.468750596046448 -0.120093822479248
0.493356823921204 -0.134281158447266
0.531250596046448 -0.161593437194824
0.537619352340698 -0.165531158447266
0.568906903266907 -0.196781158447266
0.593750596046448 -0.219124794006348
0.606175541877747 -0.228031158447266
0.6273193359375 -0.259281158447266
0.639644384384155 -0.290531158447266
0.664972305297852 -0.313703060150146
0.673694372177124 -0.321781158447266
0.685744404792786 -0.353031158447266
0.684706807136536 -0.384281158447266
0.697588086128235 -0.415531158447266
0.715663075447083 -0.446781158447266
0.718944311141968 -0.478031158447266
0.741144299507141 -0.509281158447266
0.739306807518005 -0.540531158447266
0.746813058853149 -0.571781158447266
0.747906804084778 -0.587406635284424
};
\addlegendentry{EQ, $D=32$}
\addplot [line width=2pt, mediumaquamarine64218145]
table {%
0 0
0.0156247615814209 0
0.0468747615814209 -0.000452995300292969
0.0781247615814209 -0.00135898590087891
0.109374761581421 -0.00318717956542969
0.140624761581421 -0.00659370422363281
0.171874761581421 -0.0136241912841797
0.203124761581421 -0.0178747177124023
0.234374761581421 -0.0243434906005859
0.238043546676636 -0.025609016418457
0.265624761581421 -0.0325307846069336
0.296874761581421 -0.0413436889648438
0.328124761581421 -0.0507335662841797
0.359374761581421 -0.0627651214599609
0.386962175369263 -0.0741090774536133
0.404212474822998 -0.0819215774536133
0.421874761581421 -0.0901088714599609
0.453124761581421 -0.106546401977539
0.484374761581421 -0.125015258789062
0.500346660614014 -0.134984016418457
0.523965358734131 -0.150609016418457
0.539359092712402 -0.166234016418457
0.546874761581421 -0.171359062194824
0.560437202453613 -0.181859016418457
0.574087381362915 -0.197484016418457
0.578124761581421 -0.200140476226807
0.591987371444702 -0.213109016418457
0.600852966308594 -0.228734016418457
0.609374761581421 -0.236405849456787
0.618909120559692 -0.244359016418457
0.634812355041504 -0.275609016418457
0.640624761581421 -0.28068733215332
0.65069055557251 -0.291234016418457
0.655452966690063 -0.306859016418457
0.660893678665161 -0.322484016418457
0.667709112167358 -0.338109016418457
0.671874761581421 -0.341796398162842
0.682249784469604 -0.353734016418457
0.6869215965271 -0.369359016418457
0.686674833297729 -0.384984016418457
0.690446615219116 -0.400609016418457
0.70590615272522 -0.431859016418457
0.710468530654907 -0.447484016418457
0.715768575668335 -0.463109016418457
0.714524745941162 -0.478734016418457
0.718634128570557 -0.494359016418457
0.719837427139282 -0.509984016418457
0.722021579742432 -0.525609016418457
0.720168590545654 -0.556859016418457
0.728184223175049 -0.572484016418457
0.728515386581421 -0.588109016418457
0.734374761581421 -0.598952770233154
};
\addlegendentry{EQ, $D=64$}
\addplot [line width=2pt, springgreen0205108]
table {%
0 0
0.03125 -0.000117301940917969
0.0546867847442627 -0.000624656677246094
0.0703117847442627 -0.00140666961669922
0.0859367847442627 -0.00265598297119141
0.109375 -0.00574207305908203
0.117186784744263 -0.00656223297119141
0.140625 -0.00847721099853516
0.148436784744263 -0.00921916961669922
0.171875 -0.0136327743530273
0.179686784744263 -0.0148439407348633
0.195311784744263 -0.0167970657348633
0.21875 -0.0219535827636719
0.226561784744263 -0.0233592987060547
0.243249416351318 -0.0269927978515625
0.244311809539795 -0.02734375
0.28237509727478 -0.0370311737060547
0.302778959274292 -0.04296875
0.304686784744263 -0.0437498092651367
0.320311784744263 -0.0481252670288086
0.370997667312622 -0.06640625
0.386256694793701 -0.0726957321166992
0.407027244567871 -0.08203125
0.423317909240723 -0.08984375
0.445311784744263 -0.101093292236328
0.464172124862671 -0.111445426940918
0.478635311126709 -0.119882583618164
0.480708599090576 -0.12109375
0.497566223144531 -0.132851600646973
0.505377769470215 -0.138280868530273
0.514861822128296 -0.14453125
0.525735139846802 -0.15234375
0.533960103988647 -0.16015625
0.544907093048096 -0.16796875
0.552094459533691 -0.17578125
0.554686784744263 -0.177499771118164
0.561839818954468 -0.18359375
0.5696702003479 -0.19179630279541
0.570311784744263 -0.192187309265137
0.577710151672363 -0.19921875
0.583280563354492 -0.20703125
0.585936784744263 -0.208906173706055
0.592608690261841 -0.21484375
0.596686840057373 -0.22265625
0.603640079498291 -0.228750228881836
0.605717897415161 -0.23046875
0.610264778137207 -0.23828125
0.616608619689941 -0.246562480926514
0.617186784744263 -0.247031211853027
0.623249292373657 -0.25390625
0.62621808052063 -0.26171875
0.630749225616455 -0.26953125
0.632811784744263 -0.271015644073486
0.638952255249023 -0.27734375
0.640139818191528 -0.28515625
0.648139953613281 -0.29296875
0.649077415466309 -0.30078125
0.656139850616455 -0.30859375
0.657389879226685 -0.31640625
0.659405469894409 -0.32421875
0.665991544723511 -0.330351829528809
0.66792106628418 -0.33203125
0.671366214752197 -0.343749523162842
0.67238974571228 -0.34765625
0.672999143600464 -0.35546875
0.677014827728271 -0.36328125
0.681780576705933 -0.368594169616699
0.683874130249023 -0.37109375
0.685921192169189 -0.37890625
0.687374114990234 -0.38671875
0.687843084335327 -0.39453125
0.689014911651611 -0.40234375
0.692155361175537 -0.41015625
0.699264764785767 -0.41796875
0.699671030044556 -0.42578125
0.701749324798584 -0.43359375
0.703264951705933 -0.44140625
0.703999280929565 -0.44921875
0.703686714172363 -0.45703125
0.705296039581299 -0.46484375
0.708788156509399 -0.476562023162842
0.709811687469482 -0.48046875
0.710936784744263 -0.481468677520752
0.716311693191528 -0.48828125
0.715546131134033 -0.49609375
0.717741250991821 -0.507812023162842
0.718389749526978 -0.51171875
0.71870231628418 -0.523437023162842
0.718733549118042 -0.53515625
0.720389842987061 -0.55078125
0.723702430725098 -0.55859375
0.723889827728271 -0.56640625
0.727702379226685 -0.578539371490479
0.728842973709106 -0.58203125
0.729233503341675 -0.58984375
0.732483625411987 -0.59765625
};
\addlegendentry{EQ, $D=128$}
\end{axis}

\end{tikzpicture}%
	\end{subfigure}
	\hfill
	\begin{subfigure}[htbp]{0.49\textwidth}
		\centering
		\input{figures/taylor-bubble/convergence-shape-tail-eq.tex}%
	\end{subfigure}%
	\vspace{0.5cm}
	\begin{subfigure}[htbp]{0.49\textwidth}
		\centering
		\input{figures/taylor-bubble/convergence-shape-front-eq-neq.tex}%
		\caption*{Bubble front}
	\end{subfigure}
	\hfill
	\begin{subfigure}[htbp]{0.49\textwidth}
		\centering
		\input{figures/taylor-bubble/convergence-shape-tail-eq-neq.tex}%
		\caption*{Bubble tail}
	\end{subfigure}%
	\caption{
		\label{fig:taylor-bubble-shape-convergence}
		Simulated shape of the Taylor bubble's front and tail.
		The simulations were performed with computational domain resolutions, that is, tube diameters of $D \in \{32, 64, 128\}$ lattice cells.
		The comparison with experimental data~\cite{bugg2002VelocityFieldTaylor} is drawn in terms of the non-dimensionalized axial location $z^{*}$ and radial location $r^{*}$ at time $t^{*}=15$.
		All simulations have converged well.
	}
\end{figure}

\begin{figure}[htbp]
	\centering
	\setlength{\figureheight}{0.35\textwidth}
	\setlength{\figurewidth}{0.48\textwidth}
	\begin{subfigure}[htbp]{0.49\textwidth}
		\centering
		% This file was created with tikzplotlib v0.10.1.
\begin{tikzpicture}

\definecolor{crimson2553191}{RGB}{255,31,91}
\definecolor{darkgray176}{RGB}{176,176,176}
\definecolor{hotpink255143173}{RGB}{255,171,193}
\definecolor{lightcoral25587132}{RGB}{159,19,57}
\definecolor{lightgray204}{RGB}{204,204,204}

\begin{axis}[
height=\figureheight,
legend cell align={left},
legend style={
  fill opacity=0.8,
  draw opacity=1,
  text opacity=1,
  at={(0.01,0.01)},
  anchor=south west,
  draw=lightgray204
},
tick align=outside,
tick pos=left,
width=\figurewidth,
x grid style={darkgray176},
xlabel={\(\displaystyle r^{*}\)},
xmajorgrids,
xmin=-0.037402829527855, xmax=0.785459420084955,
xtick style={color=black},
y grid style={darkgray176},
ylabel style={rotate=-90.0},
ylabel={\(\displaystyle z^{*}\)},
ymajorgrids,
ymin=-0.628031229972825, ymax=0.029906249046325,
ytick style={color=black}
]
\addplot [very thick, black, mark=*, mark size=1, mark options={solid}, only marks]
table {%
0 0
0.0293359756469727 0
0.0573979616165161 -0.000633955001831055
0.0854599475860596 -0.00253605842590332
0.119897961616516 -0.00570595264434814
0.160714030265808 -0.0120450258255005
0.192602038383484 -0.0183850526809692
0.219388008117676 -0.0247249603271484
0.244897961616516 -0.0316979885101318
0.271683931350708 -0.0412089824676514
0.293367981910706 -0.0494489669799805
0.320153951644897 -0.0595920085906982
0.340561985969543 -0.0684679746627808
0.362244009971619 -0.0773429870605469
0.380102038383484 -0.0862189531326294
0.39795994758606 -0.0950939655303955
0.415816068649292 -0.10523796081543
0.429846048355103 -0.112844944000244
0.442602038383484 -0.121086955070496
0.456632018089294 -0.129328012466431
0.470664024353027 -0.138838052749634
0.48852002620697 -0.149615049362183
0.503826022148132 -0.162294030189514
0.519132018089294 -0.173071980476379
0.533164024353027 -0.183848977088928
0.543367981910706 -0.193992018699646
0.554846048355103 -0.204769968986511
0.565052032470703 -0.214913964271545
0.57397997379303 -0.225057005882263
0.581632018089294 -0.235200047492981
0.593111991882324 -0.247879028320312
0.603316068649292 -0.259925007820129
0.614795923233032 -0.273237943649292
0.623723983764648 -0.286550998687744
0.633928060531616 -0.299864053726196
0.641582012176514 -0.312543034553528
0.64795994758606 -0.325222969055176
0.656888008117676 -0.339169979095459
0.66198992729187 -0.351848959922791
0.667092084884644 -0.365162014961243
0.672194004058838 -0.376574039459229
0.679846048355103 -0.387984991073608
0.686223983764648 -0.401932001113892
0.690052032470703 -0.414610981941223
0.692601919174194 -0.424754977226257
0.697704076766968 -0.438068032264709
0.701529979705811 -0.453283071517944
0.705358028411865 -0.464694023132324
0.707906007766724 -0.476740002632141
0.71045994758606 -0.48751699924469
0.711734056472778 -0.49956202507019
0.715561985969543 -0.511608004570007
0.716835975646973 -0.523019075393677
0.718111991882324 -0.533161997795105
0.720664024353027 -0.544573068618774
0.721935987472534 -0.560423016548157
0.72448992729187 -0.572468042373657
0.72448992729187 -0.583878993988037
0.725765943527222 -0.595291018486023
};
\addlegendentry{Experiment~\cite{bugg2002VelocityFieldTaylor}}
\addplot [line width=2pt, hotpink255143173, dash pattern=on 3pt off 5pt on 1pt off 5pt]
table {%
0 0
0.0312502384185791 0
0.0937502384185791 -0.00590610504150391
0.1875 -0.0190467834472656
0.218750238418579 -0.0235624313354492
0.259312868118286 -0.0356249809265137
0.281250238418579 -0.0429997444152832
0.343750238418579 -0.0603752136230469
0.382828712463379 -0.0764689445495605
0.406250238418579 -0.0860624313354492
0.430869102478027 -0.0981249809265137
0.478290796279907 -0.124000072479248
0.487831592559814 -0.129374980926514
0.530203342437744 -0.161234378814697
0.531250238418579 -0.161843776702881
0.566100358963013 -0.191874980926514
0.593750238418579 -0.217437267303467
0.601962804794312 -0.223124980926514
0.626075267791748 -0.254374980926514
0.63687539100647 -0.285624980926514
0.656250238418579 -0.304062366485596
0.671100378036499 -0.316874980926514
0.684562683105469 -0.348124980926514
0.684050321578979 -0.379374980926514
0.697519063949585 -0.410624980926514
0.712337732315063 -0.441874980926514
0.718575239181519 -0.473297119140625
0.718750238418579 -0.473468780517578
0.739662885665894 -0.504374980926514
0.73930025100708 -0.535624980926514
0.746825218200684 -0.566874980926514
0.748056650161743 -0.598124980926514
};
\addlegendentry{GEQ, $D=32$}
\addplot [line width=2pt, lightcoral25587132, dash pattern=on 3pt off 5pt on 1pt off 5pt]
table {%
0 0
0.0156254768371582 0
0.0468754768371582 -0.000469207763671875
0.0781254768371582 -0.00142192840576172
0.109375476837158 -0.00323486328125
0.140625476837158 -0.00654697418212891
0.171875476837158 -0.0135784149169922
0.203125476837158 -0.0179224014282227
0.234375476837158 -0.0242347717285156
0.238906741142273 -0.0257816314697266
0.265625476837158 -0.0325632095336914
0.297394275665283 -0.0414066314697266
0.328125476837158 -0.0507345199584961
0.359375476837158 -0.0627660751342773
0.387206792831421 -0.0741806030273438
0.404459357261658 -0.0819921493530273
0.421875476837158 -0.0900468826293945
0.453125476837158 -0.106500625610352
0.484375476837158 -0.124969482421875
0.500587940216064 -0.135156631469727
0.524344205856323 -0.150781631469727
0.539725422859192 -0.166406631469727
0.546875476837158 -0.171266078948975
0.560666084289551 -0.182031631469727
0.574694275856018 -0.197656631469727
0.578125476837158 -0.199859619140625
0.592231750488281 -0.213281631469727
0.601387977600098 -0.228906631469727
0.609375476837158 -0.234766006469727
0.62095046043396 -0.244531631469727
0.62533175945282 -0.260156631469727
0.640625476837158 -0.277109622955322
0.648925423622131 -0.291406631469727
0.654944181442261 -0.307031631469727
0.663356781005859 -0.322656631469727
0.665763020515442 -0.338281631469727
0.671875476837158 -0.342562675476074
0.684125423431396 -0.353906631469727
0.683341145515442 -0.369531631469727
0.68763792514801 -0.385156631469727
0.690509796142578 -0.400781631469727
0.700077176094055 -0.424109935760498
0.70322859287262 -0.432031631469727
0.712169170379639 -0.447656631469727
0.714934825897217 -0.463281631469727
0.718231678009033 -0.478906631469727
0.718234896659851 -0.494531631469727
0.719831705093384 -0.525781631469727
0.720034837722778 -0.541406631469727
0.724306702613831 -0.557031631469727
0.72823166847229 -0.588281631469727
0.729961395263672 -0.596094608306885
};
\addlegendentry{GEQ, $D=64$}
\addplot [line width=2pt, crimson2553191, dash pattern=on 3pt off 5pt on 1pt off 5pt]
table {%
0 0
0.03125 -0.000117301940917969
0.0546905994415283 -0.000624656677246094
0.0703155994415283 -0.00132846832275391
0.0859405994415283 -0.00257778167724609
0.109375 -0.00570297241210938
0.117190599441528 -0.00656223297119141
0.140625 -0.00847625732421875
0.148440599441528 -0.00921916961669922
0.171875 -0.0135936737060547
0.1875 -0.0157814025878906
0.195315599441528 -0.0167970657348633
0.21875 -0.0219526290893555
0.226565599441528 -0.0233592987060547
0.243244886398315 -0.0269918441772461
0.244300127029419 -0.02734375
0.275598526000977 -0.03515625
0.295878648757935 -0.0409364700317383
0.323977470397949 -0.0494527816772461
0.359375 -0.062108039855957
0.369009494781494 -0.0656251907348633
0.376825094223022 -0.0688295364379883
0.38618540763855 -0.0727338790893555
0.406884431838989 -0.08203125
0.423153162002563 -0.08984375
0.441927909851074 -0.0994539260864258
0.460940599441528 -0.109766006469727
0.471881628036499 -0.116055488586426
0.48044228553772 -0.12109375
0.505248785018921 -0.138360023498535
0.514600038528442 -0.14453125
0.524347305297852 -0.151796340942383
0.525254726409912 -0.15234375
0.53365159034729 -0.16015625
0.544615745544434 -0.16796875
0.551545381546021 -0.17578125
0.554690599441528 -0.177890777587891
0.561578273773193 -0.18359375
0.568137645721436 -0.19140625
0.570315599441528 -0.192890167236328
0.577284336090088 -0.19921875
0.583237648010254 -0.20703125
0.585940599441528 -0.208906173706055
0.592221975326538 -0.21484375
0.596362590789795 -0.22265625
0.606174945831299 -0.23046875
0.610534429550171 -0.23828125
0.613815546035767 -0.24609375
0.617190599441528 -0.248593807220459
0.622987508773804 -0.25390625
0.625565767288208 -0.26171875
0.638112545013428 -0.27734375
0.641643762588501 -0.28515625
0.642081260681152 -0.29296875
0.648440599441528 -0.297890663146973
0.65292501449585 -0.30078125
0.653503179550171 -0.30859375
0.658128261566162 -0.31640625
0.659924983978271 -0.32421875
0.664760589599609 -0.331210613250732
0.665456295013428 -0.33203125
0.670143842697144 -0.33984375
0.671081304550171 -0.34765625
0.67412805557251 -0.35546875
0.67622184753418 -0.36328125
0.683300018310547 -0.37109375
0.68515944480896 -0.37890625
0.686565637588501 -0.38671875
0.689214468002319 -0.398438930511475
0.690018892288208 -0.40234375
0.692300081253052 -0.41015625
0.696909427642822 -0.41796875
0.698909521102905 -0.42578125
0.702643871307373 -0.43359375
0.702268838882446 -0.44140625
0.70342493057251 -0.44921875
0.703940629959106 -0.45703125
0.706424951553345 -0.46484375
0.706659317016602 -0.47265625
0.711979389190674 -0.479339599609375
0.713018894195557 -0.48046875
0.714659452438354 -0.48828125
0.716987609863281 -0.49609375
0.716722011566162 -0.50390625
0.718440771102905 -0.51171875
0.718675136566162 -0.523438930511475
0.718878269195557 -0.531251430511475
0.719565629959106 -0.54296875
0.720753192901611 -0.55078125
0.722784519195557 -0.55859375
0.724143743515015 -0.56640625
0.723846912384033 -0.57421875
0.72826886177063 -0.58203125
0.730128049850464 -0.58984375
0.730846881866455 -0.59765625
};
\addlegendentry{GEQ, $D=128$}
\end{axis}

\end{tikzpicture}%
	\end{subfigure}
	\hfill
	\begin{subfigure}[htbp]{0.49\textwidth}
		\centering
		\input{figures/taylor-bubble/convergence-shape-tail-geq.tex}%
	\end{subfigure}%
	\vspace{0.5cm}
	\begin{subfigure}[htbp]{0.49\textwidth}
		\centering
		\input{figures/taylor-bubble/convergence-shape-front-ext.tex}%
	\end{subfigure}
	\hfill
	\begin{subfigure}[htbp]{0.49\textwidth}
		\centering
		\input{figures/taylor-bubble/convergence-shape-tail-ext.tex}%
	\end{subfigure}%
	\vspace{0.5cm}
	\begin{subfigure}[htbp]{0.49\textwidth}
		\centering
		% This file was created with tikzplotlib v0.10.1.
\begin{tikzpicture}

\definecolor{darkgray176}{RGB}{176,176,176}
\definecolor{dodgerblue0154222}{RGB}{0,154,222}
\definecolor{lightgray204}{RGB}{204,204,204}
\definecolor{mediumturquoise64179230}{RGB}{0,116,166}
\definecolor{skyblue128205238}{RGB}{159,217,243}

\begin{axis}[
height=\figureheight,
legend cell align={left},
legend style={
  fill opacity=0.8,
  draw opacity=1,
  text opacity=1,
  at={(0.01,0.01)},
  anchor=south west,
  draw=lightgray204
},
tick align=outside,
tick pos=left,
width=\figurewidth,
x grid style={darkgray176},
xlabel={\(\displaystyle r^{*}\)},
xmajorgrids,
xmin=-0.0373762011528013, xmax=0.784900224208826,
xtick style={color=black},
y grid style={darkgray176},
ylabel style={rotate=-90.0},
ylabel={\(\displaystyle z^{*}\)},
ymajorgrids,
ymin=-0.628523397445682, ymax=0.0299296855926515,
ytick style={color=black}
]
\addplot [very thick, black, mark=*, mark size=1, mark options={solid}, only marks]
table {%
0 0
0.0293359756469727 0
0.0573979616165161 -0.000633955001831055
0.0854599475860596 -0.00253605842590332
0.119897961616516 -0.00570595264434814
0.160714030265808 -0.0120450258255005
0.192602038383484 -0.0183850526809692
0.219388008117676 -0.0247249603271484
0.244897961616516 -0.0316979885101318
0.271683931350708 -0.0412089824676514
0.293367981910706 -0.0494489669799805
0.320153951644897 -0.0595920085906982
0.340561985969543 -0.0684679746627808
0.362244009971619 -0.0773429870605469
0.380102038383484 -0.0862189531326294
0.39795994758606 -0.0950939655303955
0.415816068649292 -0.10523796081543
0.429846048355103 -0.112844944000244
0.442602038383484 -0.121086955070496
0.456632018089294 -0.129328012466431
0.470664024353027 -0.138838052749634
0.48852002620697 -0.149615049362183
0.503826022148132 -0.162294030189514
0.519132018089294 -0.173071980476379
0.533164024353027 -0.183848977088928
0.543367981910706 -0.193992018699646
0.554846048355103 -0.204769968986511
0.565052032470703 -0.214913964271545
0.57397997379303 -0.225057005882263
0.581632018089294 -0.235200047492981
0.593111991882324 -0.247879028320312
0.603316068649292 -0.259925007820129
0.614795923233032 -0.273237943649292
0.623723983764648 -0.286550998687744
0.633928060531616 -0.299864053726196
0.641582012176514 -0.312543034553528
0.64795994758606 -0.325222969055176
0.656888008117676 -0.339169979095459
0.66198992729187 -0.351848959922791
0.667092084884644 -0.365162014961243
0.672194004058838 -0.376574039459229
0.679846048355103 -0.387984991073608
0.686223983764648 -0.401932001113892
0.690052032470703 -0.414610981941223
0.692601919174194 -0.424754977226257
0.697704076766968 -0.438068032264709
0.701529979705811 -0.453283071517944
0.705358028411865 -0.464694023132324
0.707906007766724 -0.476740002632141
0.71045994758606 -0.48751699924469
0.711734056472778 -0.49956202507019
0.715561985969543 -0.511608004570007
0.716835975646973 -0.523019075393677
0.718111991882324 -0.533161997795105
0.720664024353027 -0.544573068618774
0.721935987472534 -0.560423016548157
0.72448992729187 -0.572468042373657
0.72448992729187 -0.583878993988037
0.725765943527222 -0.595291018486023
};
\addlegendentry{Experiment~\cite{bugg2002VelocityFieldTaylor}}
\addplot [line width=2pt, skyblue128205238, dotted]
table {%
0 0
0.0312490463256836 0
0.0937490463256836 -0.00578117370605469
0.218749046325684 -0.0237188339233398
0.260836601257324 -0.0360937118530273
0.281249046325684 -0.043062686920166
0.343749046325684 -0.0603747367858887
0.406249046325684 -0.086249828338623
0.431442737579346 -0.0985937118530273
0.478527426719666 -0.124328136444092
0.488305330276489 -0.129843711853027
0.530461549758911 -0.16154670715332
0.531249046325684 -0.162000179290771
0.566673994064331 -0.192343711853027
0.593749046325684 -0.218218803405762
0.60139274597168 -0.223593711853027
0.623411536216736 -0.254843711853027
0.642655253410339 -0.286093711853027
0.668761491775513 -0.317343711853027
0.68055534362793 -0.348593711853027
0.691317796707153 -0.379843711853027
0.694717764854431 -0.411093711853027
0.704736590385437 -0.442343711853027
0.725149273872375 -0.466234683990479
0.731549024581909 -0.473593711853027
0.728742837905884 -0.504843711853027
0.745024085044861 -0.536093711853027
0.743942737579346 -0.567343711853027
0.74752402305603 -0.598593711853027
};
\addlegendentry{AVG, $D=32$}
\addplot [line width=2pt, mediumturquoise64179230, dotted]
table {%
0 0
0.0156247615814209 0
0.0468747615814209 -0.000484466552734375
0.0781247615814209 -0.00145339965820312
0.109374761581421 -0.00323486328125
0.140624761581421 -0.006500244140625
0.171874761581421 -0.0134849548339844
0.203124761581421 -0.0179224014282227
0.234374761581421 -0.0241413116455078
0.23958420753479 -0.0259065628051758
0.265624761581421 -0.0325469970703125
0.297974824905396 -0.0415315628051758
0.328124761581421 -0.0507345199584961
0.359374761581421 -0.0627193450927734
0.387363910675049 -0.0742340087890625
0.418718576431274 -0.0884065628051758
0.421874761581421 -0.0899534225463867
0.453124761581421 -0.10646915435791
0.484374761581421 -0.124953269958496
0.508201360702515 -0.140195846557617
0.524587392807007 -0.150906562805176
0.539974927902222 -0.166531562805176
0.546874761581421 -0.171187877655029
0.560834169387817 -0.182156562805176
0.574890375137329 -0.197781562805176
0.578124761581421 -0.199875354766846
0.592287302017212 -0.213406562805176
0.601365327835083 -0.229031562805176
0.609374761581421 -0.235265731811523
0.620090484619141 -0.244656562805176
0.628059148788452 -0.260281562805176
0.628571748733521 -0.275906562805176
0.640624761581421 -0.282281398773193
0.655509233474731 -0.291531562805176
0.647843599319458 -0.307156562805176
0.661884307861328 -0.322781562805176
0.671874761581421 -0.332109451293945
0.679409265518188 -0.338406562805176
0.678534269332886 -0.354031562805176
0.68721866607666 -0.369656562805176
0.687487363815308 -0.385281562805176
0.691202878952026 -0.400906562805176
0.699815511703491 -0.432156562805176
0.703124761581421 -0.435562610626221
0.712455987930298 -0.447781562805176
0.713671684265137 -0.463406562805176
0.717852830886841 -0.479031562805176
0.718884229660034 -0.494656562805176
0.718885660171509 -0.518094062805176
0.718968629837036 -0.525906562805176
0.719702959060669 -0.541531562805176
0.724421739578247 -0.557156562805176
0.724081039428711 -0.572781562805176
0.731227874755859 -0.588406562805176
0.730010747909546 -0.596219062805176
};
\addlegendentry{AVG, $D=64$}
\addplot [line width=2pt, dodgerblue0154222, dotted]
table {%
0 0
0.015625 -0.000117301940917969
0.0302367210388184 -0.000546455383300781
0.046875 -0.00144577026367188
0.0859367847442627 -0.00382804870605469
0.109375 -0.00515651702880859
0.117186784744263 -0.00570297241210938
0.134788274765015 -0.00835895538330078
0.148436784744263 -0.0107030868530273
0.164061784744263 -0.0122652053833008
0.179686784744263 -0.0142965316772461
0.195311784744263 -0.0180463790893555
0.210936784744263 -0.0203123092651367
0.227944612503052 -0.0239839553833008
0.25 -0.0289058685302734
0.257811784744263 -0.0305461883544922
0.261631965637207 -0.0317964553833008
0.28125 -0.0369529724121094
0.289957046508789 -0.0392580032348633
0.290852308273315 -0.0396089553833008
0.310483455657959 -0.0455465316772461
0.328125 -0.0512104034423828
0.33816123008728 -0.0543746948242188
0.345972776412964 -0.0572652816772461
0.356337308883667 -0.0611715316772461
0.382811784744263 -0.0717964172363281
0.399343013763428 -0.0786714553833008
0.416661739349365 -0.0864839553833008
0.438929796218872 -0.0976161956787109
0.447517871856689 -0.102108955383301
0.46105694770813 -0.109921455383301
0.503380060195923 -0.136366844177246
0.510699272155762 -0.141171455383301
0.519463300704956 -0.148983955383301
0.527097940444946 -0.154218673706055
0.530758619308472 -0.156796455383301
0.539937496185303 -0.1640625
0.540813207626343 -0.164608955383301
0.548680543899536 -0.172421455383301
0.559139966964722 -0.180233955383301
0.565197706222534 -0.188046455383301
0.575286626815796 -0.195858955383301
0.580546140670776 -0.203671455383301
0.588062047958374 -0.209882736206055
0.590186834335327 -0.211483955383301
0.595171213150024 -0.219296455383301
0.602538347244263 -0.226327896118164
0.603514909744263 -0.227108955383301
0.611194372177124 -0.238827705383301
0.613624334335327 -0.242733955383301
0.619491577148438 -0.248241901397705
0.621796131134033 -0.250546455383301
0.624842882156372 -0.258358955383301
0.630264759063721 -0.266171455383301
0.637327432632446 -0.273983955383301
0.641171216964722 -0.281796455383301
0.642186641693115 -0.289608955383301
0.648436784744263 -0.295624732971191
0.651061773300171 -0.297421455383301
0.654014825820923 -0.305233955383301
0.658077239990234 -0.313046455383301
0.658514738082886 -0.320858955383301
0.665554046630859 -0.327499866485596
0.667046070098877 -0.328671455383301
0.671280384063721 -0.344296455383301
0.674436807632446 -0.352108955383301
0.675936698913574 -0.359921455383301
0.682921171188354 -0.367733955383301
0.687155485153198 -0.383358955383301
0.688092947006226 -0.395077228546143
0.688546180725098 -0.398983955383301
0.692077398300171 -0.406796455383301
0.697468042373657 -0.414608955383301
0.698389768600464 -0.422421455383301
0.701999187469482 -0.430233955383301
0.702913284301758 -0.441952228546143
0.703296184539795 -0.445858955383301
0.703952312469482 -0.461483955383301
0.709249258041382 -0.469296455383301
0.711764812469482 -0.477108955383301
0.711671113967896 -0.484921455383301
0.716577291488647 -0.492733955383301
0.717811822891235 -0.508358955383301
0.719585180282593 -0.535702228546143
0.720327377319336 -0.543514728546143
0.721999168395996 -0.563046455383301
0.726936817169189 -0.570858955383301
0.727233648300171 -0.578671455383301
0.730358600616455 -0.586483955383301
0.730639934539795 -0.594296455383301
0.731178998947144 -0.598202228546143
};
\addlegendentry{AVG, $D=128$}
\end{axis}

\end{tikzpicture}%
		\caption*{Bubble front}
	\end{subfigure}
	\hfill
	\begin{subfigure}[htbp]{0.49\textwidth}
		\centering
		\input{figures/taylor-bubble/convergence-shape-tail-avg.tex}%
		\caption*{Bubble tail}
	\end{subfigure}
	\addtocounter{figure}{-1}
	\caption{(\textit{Continued})}
\end{figure}

\clearpage

\bibliography{literature}

\begin{thebibliography}{43}%
\makeatletter
\providecommand \@ifxundefined [1]{%
 \@ifx{#1\undefined}
}%
\providecommand \@ifnum [1]{%
 \ifnum #1\expandafter \@firstoftwo
 \else \expandafter \@secondoftwo
 \fi
}%
\providecommand \@ifx [1]{%
 \ifx #1\expandafter \@firstoftwo
 \else \expandafter \@secondoftwo
 \fi
}%
\providecommand \natexlab [1]{#1}%
\providecommand \enquote  [1]{``#1''}%
\providecommand \bibnamefont  [1]{#1}%
\providecommand \bibfnamefont [1]{#1}%
\providecommand \citenamefont [1]{#1}%
\providecommand \href@noop [0]{\@secondoftwo}%
\providecommand \href [0]{\begingroup \@sanitize@url \@href}%
\providecommand \@href[1]{\@@startlink{#1}\@@href}%
\providecommand \@@href[1]{\endgroup#1\@@endlink}%
\providecommand \@sanitize@url [0]{\catcode `\\12\catcode `\$12\catcode
  `\&12\catcode `\#12\catcode `\^12\catcode `\_12\catcode `\%12\relax}%
\providecommand \@@startlink[1]{}%
\providecommand \@@endlink[0]{}%
\providecommand \url  [0]{\begingroup\@sanitize@url \@url }%
\providecommand \@url [1]{\endgroup\@href {#1}{\urlprefix }}%
\providecommand \urlprefix  [0]{URL }%
\providecommand \Eprint [0]{\href }%
\providecommand \doibase [0]{http://dx.doi.org/}%
\providecommand \selectlanguage [0]{\@gobble}%
\providecommand \bibinfo  [0]{\@secondoftwo}%
\providecommand \bibfield  [0]{\@secondoftwo}%
\providecommand \translation [1]{[#1]}%
\providecommand \BibitemOpen [0]{}%
\providecommand \bibitemStop [0]{}%
\providecommand \bibitemNoStop [0]{.\EOS\space}%
\providecommand \EOS [0]{\spacefactor3000\relax}%
\providecommand \BibitemShut  [1]{\csname bibitem#1\endcsname}%
\let\auto@bib@innerbib\@empty
%</preamble>
\bibitem [{\citenamefont {K{\"o}rner}\ \emph {et~al.}(2005)\citenamefont
  {K{\"o}rner}, \citenamefont {Thies}, \citenamefont {Hofmann}, \citenamefont
  {Th{\"u}rey},\ and\ \citenamefont
  {R{\"u}de}}]{korner2005LatticeBoltzmannModel}%
  \BibitemOpen
  \bibfield  {author} {\bibinfo {author} {\bibfnamefont {C.}~\bibnamefont
  {K{\"o}rner}}, \bibinfo {author} {\bibfnamefont {M.}~\bibnamefont {Thies}},
  \bibinfo {author} {\bibfnamefont {T.}~\bibnamefont {Hofmann}}, \bibinfo
  {author} {\bibfnamefont {N.}~\bibnamefont {Th{\"u}rey}}, \ and\ \bibinfo
  {author} {\bibfnamefont {U.}~\bibnamefont {R{\"u}de}},\ }\bibfield  {title}
  {\enquote {\bibinfo {title} {Lattice {{Boltzmann Model}} for {{Free Surface
  Flow}} for {{Modeling Foaming}}},}\ }\href {\doibase
  10.1007/s10955-005-8879-8} {\bibfield  {journal} {\bibinfo  {journal}
  {Journal of Statistical Physics}\ }\textbf {\bibinfo {volume} {121}}
  (\bibinfo {year} {2005}),\ 10.1007/s10955-005-8879-8}\BibitemShut {NoStop}%
\bibitem [{\citenamefont {Hirt}\ and\ \citenamefont
  {Nichols}(1981)}]{hirt1981VolumeFluidVOF}%
  \BibitemOpen
  \bibfield  {author} {\bibinfo {author} {\bibfnamefont {C.W}\ \bibnamefont
  {Hirt}}\ and\ \bibinfo {author} {\bibfnamefont {B.D}\ \bibnamefont
  {Nichols}},\ }\bibfield  {title} {\enquote {\bibinfo {title} {Volume of fluid
  ({{VOF}}) method for the dynamics of free boundaries},}\ }\href {\doibase
  10.1016/0021-9991(81)90145-5} {\bibfield  {journal} {\bibinfo  {journal}
  {Journal of Computational Physics}\ }\textbf {\bibinfo {volume} {39}}
  (\bibinfo {year} {1981}),\ 10.1016/0021-9991(81)90145-5}\BibitemShut
  {NoStop}%
\bibitem [{\citenamefont {Donath}\ \emph {et~al.}(2011)\citenamefont {Donath},
  \citenamefont {Mecke}, \citenamefont {Rabha}, \citenamefont {Buwa},\ and\
  \citenamefont {R{\"u}de}}]{donath2011VerificationSurfaceTension}%
  \BibitemOpen
  \bibfield  {author} {\bibinfo {author} {\bibfnamefont {Stefan}\ \bibnamefont
  {Donath}}, \bibinfo {author} {\bibfnamefont {Klaus}\ \bibnamefont {Mecke}},
  \bibinfo {author} {\bibfnamefont {Swapna}\ \bibnamefont {Rabha}}, \bibinfo
  {author} {\bibfnamefont {Vivek}\ \bibnamefont {Buwa}}, \ and\ \bibinfo
  {author} {\bibfnamefont {Ulrich}\ \bibnamefont {R{\"u}de}},\ }\bibfield
  {title} {\enquote {\bibinfo {title} {Verification of surface tension in the
  parallel free surface lattice {{Boltzmann}} method in {{waLBerla}}},}\ }\href
  {\doibase 10.1016/j.compfluid.2010.12.027} {\bibfield  {journal} {\bibinfo
  {journal} {Computers \& Fluids}\ }\bibinfo {series} {22nd {{International
  Conference}} on {{Parallel Computational Fluid Dynamics}} ({{ParCFD}}
  2010)},\ \textbf {\bibinfo {volume} {45}} (\bibinfo {year} {2011}),\
  10.1016/j.compfluid.2010.12.027}\BibitemShut {NoStop}%
\bibitem [{\citenamefont {Zhao}\ \emph {et~al.}(2013)\citenamefont {Zhao},
  \citenamefont {Huang}, \citenamefont {Li},\ and\ \citenamefont
  {Li}}]{zhao2013LatticeBoltzmannMethod}%
  \BibitemOpen
  \bibfield  {author} {\bibinfo {author} {\bibfnamefont {Zhuangming}\
  \bibnamefont {Zhao}}, \bibinfo {author} {\bibfnamefont {Ping}\ \bibnamefont
  {Huang}}, \bibinfo {author} {\bibfnamefont {Yineng}\ \bibnamefont {Li}}, \
  and\ \bibinfo {author} {\bibfnamefont {Junmin}\ \bibnamefont {Li}},\
  }\bibfield  {title} {\enquote {\bibinfo {title} {A lattice {{Boltzmann}}
  method for viscous free surface waves in two dimensions},}\ }\href {\doibase
  10.1002/fld.3660} {\bibfield  {journal} {\bibinfo  {journal} {International
  Journal for Numerical Methods in Fluids}\ }\textbf {\bibinfo {volume} {71}}
  (\bibinfo {year} {2013}),\ 10.1002/fld.3660}\BibitemShut {NoStop}%
\bibitem [{\citenamefont {Jan{\ss}en}\ and\ \citenamefont
  {Krafczyk}(2011)}]{janssen2011FreeSurfaceFlow}%
  \BibitemOpen
  \bibfield  {author} {\bibinfo {author} {\bibfnamefont {Christian}\
  \bibnamefont {Jan{\ss}en}}\ and\ \bibinfo {author} {\bibfnamefont {Manfred}\
  \bibnamefont {Krafczyk}},\ }\bibfield  {title} {\enquote {\bibinfo {title}
  {Free surface flow simulations on {{GPGPUs}} using the {{LBM}}},}\ }\href
  {\doibase 10.1016/j.camwa.2011.03.016} {\bibfield  {journal} {\bibinfo
  {journal} {Computers \& Mathematics with Applications}\ }\bibinfo {series}
  {Mesoscopic {{Methods}} for {{Engineering}} and {{Science}} \textemdash{}
  {{Proceedings}} of {{ICMMES-09}}},\ \textbf {\bibinfo {volume} {61}}
  (\bibinfo {year} {2011}),\ 10.1016/j.camwa.2011.03.016}\BibitemShut {NoStop}%
\bibitem [{\citenamefont {Lehmann}\ \emph {et~al.}(2021)\citenamefont
  {Lehmann}, \citenamefont {Oehlschl{\"a}gel}, \citenamefont {H{\"a}usl},
  \citenamefont {Held},\ and\ \citenamefont
  {Gekle}}]{lehmann2021EjectionMarineMicroplastics}%
  \BibitemOpen
  \bibfield  {author} {\bibinfo {author} {\bibfnamefont {Moritz}\ \bibnamefont
  {Lehmann}}, \bibinfo {author} {\bibfnamefont {Lisa~Marie}\ \bibnamefont
  {Oehlschl{\"a}gel}}, \bibinfo {author} {\bibfnamefont {Fabian~P.}\
  \bibnamefont {H{\"a}usl}}, \bibinfo {author} {\bibfnamefont {Andreas}\
  \bibnamefont {Held}}, \ and\ \bibinfo {author} {\bibfnamefont {Stephan}\
  \bibnamefont {Gekle}},\ }\bibfield  {title} {\enquote {\bibinfo {title}
  {Ejection of marine microplastics by raindrops: A computational and
  experimental study},}\ }\href {\doibase 10.1186/s43591-021-00018-8}
  {\bibfield  {journal} {\bibinfo  {journal} {Microplastics and Nanoplastics}\
  }\textbf {\bibinfo {volume} {1}} (\bibinfo {year} {2021}),\
  10.1186/s43591-021-00018-8}\BibitemShut {NoStop}%
\bibitem [{\citenamefont {Ammer}\ \emph {et~al.}(2014)\citenamefont {Ammer},
  \citenamefont {Markl}, \citenamefont {Ljungblad}, \citenamefont
  {K{\"o}rner},\ and\ \citenamefont
  {R{\"u}de}}]{ammer2014SimulatingFastElectron}%
  \BibitemOpen
  \bibfield  {author} {\bibinfo {author} {\bibfnamefont {Regina}\ \bibnamefont
  {Ammer}}, \bibinfo {author} {\bibfnamefont {Matthias}\ \bibnamefont {Markl}},
  \bibinfo {author} {\bibfnamefont {Ulric}\ \bibnamefont {Ljungblad}}, \bibinfo
  {author} {\bibfnamefont {Carolin}\ \bibnamefont {K{\"o}rner}}, \ and\
  \bibinfo {author} {\bibfnamefont {Ulrich}\ \bibnamefont {R{\"u}de}},\
  }\bibfield  {title} {\enquote {\bibinfo {title} {Simulating fast electron
  beam melting with a parallel thermal free surface lattice {{Boltzmann}}
  method},}\ }\href {\doibase 10.1016/j.camwa.2013.10.001} {\bibfield
  {journal} {\bibinfo  {journal} {Computers \& Mathematics with Applications}\
  }\bibinfo {series} {Mesoscopic {{Methods}} for {{Engineering}} and
  {{Science}} ({{Proceedings}} of {{ICMMES-2012}}, {{Taipei}}, {{Taiwan}},
  23\textendash 27 {{July}} 2012)},\ \textbf {\bibinfo {volume} {67}} (\bibinfo
  {year} {2014}),\ 10.1016/j.camwa.2013.10.001}\BibitemShut {NoStop}%
\bibitem [{\citenamefont {Peng}\ \emph {et~al.}(2016)\citenamefont {Peng},
  \citenamefont {Teng}, \citenamefont {Hwang}, \citenamefont {Guo},\ and\
  \citenamefont {Wang}}]{peng2016ImplementationIssuesBenchmarking}%
  \BibitemOpen
  \bibfield  {author} {\bibinfo {author} {\bibfnamefont {Cheng}\ \bibnamefont
  {Peng}}, \bibinfo {author} {\bibfnamefont {Yihua}\ \bibnamefont {Teng}},
  \bibinfo {author} {\bibfnamefont {Brian}\ \bibnamefont {Hwang}}, \bibinfo
  {author} {\bibfnamefont {Zhaoli}\ \bibnamefont {Guo}}, \ and\ \bibinfo
  {author} {\bibfnamefont {Lian-Ping}\ \bibnamefont {Wang}},\ }\bibfield
  {title} {\enquote {\bibinfo {title} {Implementation issues and benchmarking
  of lattice {{Boltzmann}} method for moving rigid particle simulations in a
  viscous flow},}\ }\href {\doibase 10.1016/j.camwa.2015.08.027} {\bibfield
  {journal} {\bibinfo  {journal} {Computers \& Mathematics with Applications}\
  }\textbf {\bibinfo {volume} {72}} (\bibinfo {year} {2016}),\
  10.1016/j.camwa.2015.08.027}\BibitemShut {NoStop}%
\bibitem [{\citenamefont {Lallemand}\ and\ \citenamefont
  {Luo}(2003)}]{lallemand2003LatticeBoltzmannMethod}%
  \BibitemOpen
  \bibfield  {author} {\bibinfo {author} {\bibfnamefont {Pierre}\ \bibnamefont
  {Lallemand}}\ and\ \bibinfo {author} {\bibfnamefont {Li-Shi}\ \bibnamefont
  {Luo}},\ }\bibfield  {title} {\enquote {\bibinfo {title} {Lattice
  {{Boltzmann}} method for moving boundaries},}\ }\href {\doibase
  10.1016/S0021-9991(02)00022-0} {\bibfield  {journal} {\bibinfo  {journal}
  {Journal of Computational Physics}\ }\textbf {\bibinfo {volume} {184}}
  (\bibinfo {year} {2003}),\ 10.1016/S0021-9991(02)00022-0}\BibitemShut
  {NoStop}%
\bibitem [{\citenamefont {Chikatamarla}\ \emph {et~al.}(2006)\citenamefont
  {Chikatamarla}, \citenamefont {Ansumali},\ and\ \citenamefont
  {Karlin}}]{chikatamarla2006GradApproximationMissing}%
  \BibitemOpen
  \bibfield  {author} {\bibinfo {author} {\bibfnamefont {S.~S}\ \bibnamefont
  {Chikatamarla}}, \bibinfo {author} {\bibfnamefont {S}~\bibnamefont
  {Ansumali}}, \ and\ \bibinfo {author} {\bibfnamefont {I.~V}\ \bibnamefont
  {Karlin}},\ }\bibfield  {title} {\enquote {\bibinfo {title} {Grad's
  approximation for missing data in lattice {{Boltzmann}} simulations},}\
  }\href {\doibase 10.1209/epl/i2005-10535-x} {\bibfield  {journal} {\bibinfo
  {journal} {Europhysics Letters (EPL)}\ }\textbf {\bibinfo {volume} {74}}
  (\bibinfo {year} {2006}),\ 10.1209/epl/i2005-10535-x}\BibitemShut {NoStop}%
\bibitem [{\citenamefont {Krithivasan}\ \emph {et~al.}(2014)\citenamefont
  {Krithivasan}, \citenamefont {Wahal},\ and\ \citenamefont
  {Ansumali}}]{krithivasan2014DiffusedBouncebackCondition}%
  \BibitemOpen
  \bibfield  {author} {\bibinfo {author} {\bibfnamefont {Siddharth}\
  \bibnamefont {Krithivasan}}, \bibinfo {author} {\bibfnamefont {Siddhant}\
  \bibnamefont {Wahal}}, \ and\ \bibinfo {author} {\bibfnamefont {Santosh}\
  \bibnamefont {Ansumali}},\ }\bibfield  {title} {\enquote {\bibinfo {title}
  {Diffused bounce-back condition and refill algorithm for the lattice
  {{Boltzmann}} method},}\ }\href {\doibase 10.1103/PhysRevE.89.033313}
  {\bibfield  {journal} {\bibinfo  {journal} {Physical Review E}\ }\textbf
  {\bibinfo {volume} {89}} (\bibinfo {year} {2014}),\
  10.1103/PhysRevE.89.033313}\BibitemShut {NoStop}%
\bibitem [{\citenamefont {Dorschner}\ \emph {et~al.}(2015)\citenamefont
  {Dorschner}, \citenamefont {Chikatamarla}, \citenamefont {B{\"o}sch},\ and\
  \citenamefont {Karlin}}]{dorschner2015GradApproximationMoving}%
  \BibitemOpen
  \bibfield  {author} {\bibinfo {author} {\bibfnamefont {B.}~\bibnamefont
  {Dorschner}}, \bibinfo {author} {\bibfnamefont {S.S.}\ \bibnamefont
  {Chikatamarla}}, \bibinfo {author} {\bibfnamefont {F.}~\bibnamefont
  {B{\"o}sch}}, \ and\ \bibinfo {author} {\bibfnamefont {I.V.}\ \bibnamefont
  {Karlin}},\ }\bibfield  {title} {\enquote {\bibinfo {title} {Grad's
  approximation for moving and stationary walls in entropic lattice
  {{Boltzmann}} simulations},}\ }\href {\doibase 10.1016/j.jcp.2015.04.017}
  {\bibfield  {journal} {\bibinfo  {journal} {Journal of Computational
  Physics}\ }\textbf {\bibinfo {volume} {295}} (\bibinfo {year} {2015}),\
  10.1016/j.jcp.2015.04.017}\BibitemShut {NoStop}%
\bibitem [{\citenamefont {Tao}\ \emph {et~al.}(2016)\citenamefont {Tao},
  \citenamefont {Hu},\ and\ \citenamefont
  {Guo}}]{tao2016InvestigationMomentumExchange}%
  \BibitemOpen
  \bibfield  {author} {\bibinfo {author} {\bibfnamefont {Shi}\ \bibnamefont
  {Tao}}, \bibinfo {author} {\bibfnamefont {Junjie}\ \bibnamefont {Hu}}, \ and\
  \bibinfo {author} {\bibfnamefont {Zhaoli}\ \bibnamefont {Guo}},\ }\bibfield
  {title} {\enquote {\bibinfo {title} {An investigation on momentum exchange
  methods and refilling algorithms for lattice {{Boltzmann}} simulation of
  particulate flows},}\ }\href {\doibase 10.1016/j.compfluid.2016.04.009}
  {\bibfield  {journal} {\bibinfo  {journal} {Computers \& Fluids}\ }\textbf
  {\bibinfo {volume} {133}} (\bibinfo {year} {2016}),\
  10.1016/j.compfluid.2016.04.009}\BibitemShut {NoStop}%
\bibitem [{\citenamefont {Fang}\ \emph {et~al.}(2002)\citenamefont {Fang},
  \citenamefont {Wang}, \citenamefont {Lin},\ and\ \citenamefont
  {Liu}}]{fang2002LatticeBoltzmannMethod}%
  \BibitemOpen
  \bibfield  {author} {\bibinfo {author} {\bibfnamefont {Haiping}\ \bibnamefont
  {Fang}}, \bibinfo {author} {\bibfnamefont {Zuowei}\ \bibnamefont {Wang}},
  \bibinfo {author} {\bibfnamefont {Zhifang}\ \bibnamefont {Lin}}, \ and\
  \bibinfo {author} {\bibfnamefont {Muren}\ \bibnamefont {Liu}},\ }\bibfield
  {title} {\enquote {\bibinfo {title} {Lattice {{Boltzmann}} method for
  simulating the viscous flow in large distensible blood vessels},}\ }\href
  {\doibase 10.1103/PhysRevE.65.051925} {\bibfield  {journal} {\bibinfo
  {journal} {Physical Review E}\ }\textbf {\bibinfo {volume} {65}},\ \bibinfo
  {pages} {051925} (\bibinfo {year} {2002})}\BibitemShut {NoStop}%
\bibitem [{\citenamefont {Grad}(1949)}]{grad1949KineticTheoryRarefied}%
  \BibitemOpen
  \bibfield  {author} {\bibinfo {author} {\bibfnamefont {Harold}\ \bibnamefont
  {Grad}},\ }\bibfield  {title} {\enquote {\bibinfo {title} {On the kinetic
  theory of rarefied gases},}\ }\href {\doibase 10.1002/cpa.3160020403}
  {\bibfield  {journal} {\bibinfo  {journal} {Communications on Pure and
  Applied Mathematics}\ }\textbf {\bibinfo {volume} {2}} (\bibinfo {year}
  {1949}),\ 10.1002/cpa.3160020403}\BibitemShut {NoStop}%
\bibitem [{\citenamefont {Bauer}\ \emph {et~al.}(2021)\citenamefont {Bauer},
  \citenamefont {Eibl}, \citenamefont {Godenschwager}, \citenamefont {Kohl},
  \citenamefont {Kuron}, \citenamefont {Rettinger}, \citenamefont {Schornbaum},
  \citenamefont {Schwarzmeier}, \citenamefont {Th{\"o}nnes}, \citenamefont
  {K{\"o}stler},\ and\ \citenamefont
  {R{\"u}de}}]{bauer2021WaLBerlaBlockstructuredHighperformance}%
  \BibitemOpen
  \bibfield  {author} {\bibinfo {author} {\bibfnamefont {Martin}\ \bibnamefont
  {Bauer}}, \bibinfo {author} {\bibfnamefont {Sebastian}\ \bibnamefont {Eibl}},
  \bibinfo {author} {\bibfnamefont {Christian}\ \bibnamefont {Godenschwager}},
  \bibinfo {author} {\bibfnamefont {Nils}\ \bibnamefont {Kohl}}, \bibinfo
  {author} {\bibfnamefont {Michael}\ \bibnamefont {Kuron}}, \bibinfo {author}
  {\bibfnamefont {Christoph}\ \bibnamefont {Rettinger}}, \bibinfo {author}
  {\bibfnamefont {Florian}\ \bibnamefont {Schornbaum}}, \bibinfo {author}
  {\bibfnamefont {Christoph}\ \bibnamefont {Schwarzmeier}}, \bibinfo {author}
  {\bibfnamefont {Dominik}\ \bibnamefont {Th{\"o}nnes}}, \bibinfo {author}
  {\bibfnamefont {Harald}\ \bibnamefont {K{\"o}stler}}, \ and\ \bibinfo
  {author} {\bibfnamefont {Ulrich}\ \bibnamefont {R{\"u}de}},\ }\bibfield
  {title} {\enquote {\bibinfo {title} {{{waLBerla}}: {{A}} block-structured
  high-performance framework for multiphysics simulations},}\ }\href {\doibase
  10.1016/j.camwa.2020.01.007} {\bibfield  {journal} {\bibinfo  {journal}
  {Computers \& Mathematics with Applications}\ }\bibinfo {series} {Development
  and {{Application}} of {{Open-source Software}} for {{Problems}} with
  {{Numerical PDEs}}},\ \textbf {\bibinfo {volume} {81}} (\bibinfo {year}
  {2021}),\ 10.1016/j.camwa.2020.01.007}\BibitemShut {NoStop}%
\bibitem [{\citenamefont {Schwarzmeier}\ \emph {et~al.}(2023)\citenamefont
  {Schwarzmeier}, \citenamefont {Holzer}, \citenamefont {Mitchell},
  \citenamefont {Lehmann}, \citenamefont {H{\"a}usl},\ and\ \citenamefont
  {R{\"u}de}}]{schwarzmeier2022ComparisonFreeSurface}%
  \BibitemOpen
  \bibfield  {author} {\bibinfo {author} {\bibfnamefont {Christoph}\
  \bibnamefont {Schwarzmeier}}, \bibinfo {author} {\bibfnamefont {Markus}\
  \bibnamefont {Holzer}}, \bibinfo {author} {\bibfnamefont {Travis}\
  \bibnamefont {Mitchell}}, \bibinfo {author} {\bibfnamefont {Moritz}\
  \bibnamefont {Lehmann}}, \bibinfo {author} {\bibfnamefont {Fabian}\
  \bibnamefont {H{\"a}usl}}, \ and\ \bibinfo {author} {\bibfnamefont {Ulrich}\
  \bibnamefont {R{\"u}de}},\ }\bibfield  {title} {\enquote {\bibinfo {title}
  {Comparison of free-surface and conservative {{Allen}}\textendash{{Cahn}}
  phase-field lattice {{Boltzmann}} method},}\ }\href {\doibase
  10.1016/j.jcp.2022.111753} {\bibfield  {journal} {\bibinfo  {journal}
  {Journal of Computational Physics}\ }\textbf {\bibinfo {volume} {473}},\
  \bibinfo {pages} {111753} (\bibinfo {year} {2023})}\BibitemShut {NoStop}%
\bibitem [{\citenamefont {Schwarzmeier}\ and\ \citenamefont
  {R{\"u}de}(2022)}]{schwarzmeier2022AnalysisComparisonBoundary}%
  \BibitemOpen
  \bibfield  {author} {\bibinfo {author} {\bibfnamefont {Christoph}\
  \bibnamefont {Schwarzmeier}}\ and\ \bibinfo {author} {\bibfnamefont {Ulrich}\
  \bibnamefont {R{\"u}de}},\ }\bibfield  {title} {\enquote {\bibinfo {title}
  {Analysis and comparison of boundary condition variants in the free-surface
  lattice {{Boltzmann}} method},}\ }\href {\doibase 10.48550/arXiv.2207.13962}
  {\bibfield  {journal} {\bibinfo  {journal} {arXiv preprint}\ } (\bibinfo
  {year} {2022}),\ 10.48550/arXiv.2207.13962}\BibitemShut {NoStop}%
\bibitem [{\citenamefont {Kr{\"u}ger}\ \emph {et~al.}(2017)\citenamefont
  {Kr{\"u}ger}, \citenamefont {Kusumaatmaja}, \citenamefont {Kuzmin},
  \citenamefont {Shardt}, \citenamefont {Silva},\ and\ \citenamefont
  {Viggen}}]{kruger2017LatticeBoltzmannMethod}%
  \BibitemOpen
  \bibfield  {author} {\bibinfo {author} {\bibfnamefont {Timm}\ \bibnamefont
  {Kr{\"u}ger}}, \bibinfo {author} {\bibfnamefont {Halim}\ \bibnamefont
  {Kusumaatmaja}}, \bibinfo {author} {\bibfnamefont {Alexandr}\ \bibnamefont
  {Kuzmin}}, \bibinfo {author} {\bibfnamefont {Orest}\ \bibnamefont {Shardt}},
  \bibinfo {author} {\bibfnamefont {Goncalo}\ \bibnamefont {Silva}}, \ and\
  \bibinfo {author} {\bibfnamefont {Erlend~Magnus}\ \bibnamefont {Viggen}},\
  }\href@noop {} {\emph {\bibinfo {title} {The Lattice {{Boltzmann}} Method:
  Principles and Practice}}}\ (\bibinfo  {publisher} {{Springer}},\ \bibinfo
  {address} {{Switzerland}},\ \bibinfo {year} {2017})\BibitemShut {NoStop}%
\bibitem [{\citenamefont {Bauer}\ \emph {et~al.}(2020)\citenamefont {Bauer},
  \citenamefont {Silva},\ and\ \citenamefont
  {R{\"u}de}}]{bauer2020TruncationErrorsD3Q19a}%
  \BibitemOpen
  \bibfield  {author} {\bibinfo {author} {\bibfnamefont {Martin}\ \bibnamefont
  {Bauer}}, \bibinfo {author} {\bibfnamefont {Goncalo}\ \bibnamefont {Silva}},
  \ and\ \bibinfo {author} {\bibfnamefont {Ulrich}\ \bibnamefont {R{\"u}de}},\
  }\bibfield  {title} {\enquote {\bibinfo {title} {Truncation errors of the
  {{D3Q19}} lattice model for the lattice {{Boltzmann}} method},}\ }\href
  {\doibase 10.1016/j.jcp.2019.109111} {\bibfield  {journal} {\bibinfo
  {journal} {Journal of Computational Physics}\ }\textbf {\bibinfo {volume}
  {405}} (\bibinfo {year} {2020}),\ 10.1016/j.jcp.2019.109111}\BibitemShut
  {NoStop}%
\bibitem [{\citenamefont {Guo}\ \emph {et~al.}(2002)\citenamefont {Guo},
  \citenamefont {Zheng},\ and\ \citenamefont
  {Shi}}]{guo2002DiscreteLatticeEffects}%
  \BibitemOpen
  \bibfield  {author} {\bibinfo {author} {\bibfnamefont {Zhaoli}\ \bibnamefont
  {Guo}}, \bibinfo {author} {\bibfnamefont {Chuguang}\ \bibnamefont {Zheng}}, \
  and\ \bibinfo {author} {\bibfnamefont {Baochang}\ \bibnamefont {Shi}},\
  }\bibfield  {title} {\enquote {\bibinfo {title} {Discrete lattice effects on
  the forcing term in the lattice {{Boltzmann}} method},}\ }\href {\doibase
  10.1103/PhysRevE.65.046308} {\bibfield  {journal} {\bibinfo  {journal}
  {Physical Review E}\ }\textbf {\bibinfo {volume} {65}} (\bibinfo {year}
  {2002}),\ 10.1103/PhysRevE.65.046308}\BibitemShut {NoStop}%
\bibitem [{\citenamefont {Hou}\ \emph {et~al.}(1996)\citenamefont {Hou},
  \citenamefont {Sterling}, \citenamefont {Chen},\ and\ \citenamefont
  {Doolen}}]{hou1996LatticeBoltzmannSubgrid}%
  \BibitemOpen
  \bibfield  {author} {\bibinfo {author} {\bibfnamefont {S.}~\bibnamefont
  {Hou}}, \bibinfo {author} {\bibfnamefont {J.}~\bibnamefont {Sterling}},
  \bibinfo {author} {\bibfnamefont {S.}~\bibnamefont {Chen}}, \ and\ \bibinfo
  {author} {\bibfnamefont {G.~D.}\ \bibnamefont {Doolen}},\ }\bibfield  {title}
  {\enquote {\bibinfo {title} {A {{Lattice Boltzmann Subgrid Model}} for {{High
  Reynolds Number Flows}}},}\ }in\ \href {\doibase 10.1090/fic/006} {\emph
  {\bibinfo {booktitle} {Pattern {{Formation}} and {{Lattice Gas
  Automata}}}}},\ \bibinfo {series} {Fields {{Institute Communications}}},
  Vol.~\bibinfo {volume} {6},\ \bibinfo {editor} {edited by\ \bibinfo {editor}
  {\bibfnamefont {Anna~T.}\ \bibnamefont {Lawniczak}}\ and\ \bibinfo {editor}
  {\bibfnamefont {Raymond}\ \bibnamefont {Kapral}}}\ (\bibinfo {organization}
  {{American Mathematical Society}},\ \bibinfo {year} {1996})\BibitemShut
  {NoStop}%
\bibitem [{\citenamefont {Yu}\ \emph {et~al.}(2005)\citenamefont {Yu},
  \citenamefont {Girimaji},\ and\ \citenamefont
  {Luo}}]{yu2005DNSDecayingIsotropic}%
  \BibitemOpen
  \bibfield  {author} {\bibinfo {author} {\bibfnamefont {Huidan}\ \bibnamefont
  {Yu}}, \bibinfo {author} {\bibfnamefont {Sharath~S.}\ \bibnamefont
  {Girimaji}}, \ and\ \bibinfo {author} {\bibfnamefont {Li-Shi}\ \bibnamefont
  {Luo}},\ }\bibfield  {title} {\enquote {\bibinfo {title} {{{DNS}} and {{LES}}
  of decaying isotropic turbulence with and without frame rotation using
  lattice {{Boltzmann}} method},}\ }\href {\doibase 10.1016/j.jcp.2005.03.022}
  {\bibfield  {journal} {\bibinfo  {journal} {Journal of Computational
  Physics}\ }\textbf {\bibinfo {volume} {209}} (\bibinfo {year} {2005}),\
  10.1016/j.jcp.2005.03.022}\BibitemShut {NoStop}%
\bibitem [{\citenamefont {Pohl}(2008)}]{pohl2008HighPerformanceSimulation}%
  \BibitemOpen
  \bibfield  {author} {\bibinfo {author} {\bibfnamefont {Thomas}\ \bibnamefont
  {Pohl}},\ }\emph {\bibinfo {title} {High {{Performance Simulation}} of {{Free
  Surface Flows Using}} the {{Lattice Boltzmann Method}}}},\ \href@noop {}
  {Ph.D. thesis},\ \bibinfo  {school} {Universit\"at Erlangen-N\"urnberg},
  \bibinfo {address} {{Erlangen}} (\bibinfo {year} {2008})\BibitemShut
  {NoStop}%
\bibitem [{\citenamefont
  {Th{\"u}rey}(2007)}]{thurey2007PhysicallyBasedAnimation}%
  \BibitemOpen
  \bibfield  {author} {\bibinfo {author} {\bibfnamefont {Nils}\ \bibnamefont
  {Th{\"u}rey}},\ }\emph {\bibinfo {title} {Physically Based {{Animation}} of
  {{Free Surface Flows}} with the {{Lattice Boltzmann Method}}}},\ \href
  {https://opus4.kobv.de/opus4-fau/frontdoor/index/index/docId/450} {Ph.D.
  thesis},\ \bibinfo  {school} {Universit\"at Erlangen-N\"urnberg}, \bibinfo
  {address} {{Erlangen}} (\bibinfo {year} {2007})\BibitemShut {NoStop}%
\bibitem [{\citenamefont {Bogner}(2017)}]{bogner2017DirectNumericalSimulation}%
  \BibitemOpen
  \bibfield  {author} {\bibinfo {author} {\bibfnamefont {Simon}\ \bibnamefont
  {Bogner}},\ }\emph {\bibinfo {title} {Direct {{Numerical Simulation}} of
  {{Liquid-Gas-Solid Flows Based}} on the {{Lattice Boltzmann Method}}}},\
  \href {https://opus4.kobv.de/opus4-fau/frontdoor/index/index/docId/8719}
  {Ph.D. thesis},\ \bibinfo  {school} {Friedrich-Alexander-Universit\"at
  Erlangen-N\"urnberg}, \bibinfo {address} {{Erlangen}} (\bibinfo {year}
  {2017})\BibitemShut {NoStop}%
\bibitem [{\citenamefont {Donath}(2011)}]{donath2011WettingModelsParallel}%
  \BibitemOpen
  \bibfield  {author} {\bibinfo {author} {\bibfnamefont {Stefan}\ \bibnamefont
  {Donath}},\ }\emph {\bibinfo {title} {Wetting {{Models}} for a {{Parallel
  High-Performance Free Surface Lattice Boltzmann Method}}}},\ \href
  {https://www10.cs.fau.de/publications/dissertations/Diss_2011-Donath.pdf}
  {Ph.D. thesis},\ \bibinfo  {school} {Universit\"at Erlangen-N\"urnberg},
  \bibinfo {address} {{Erlangen}} (\bibinfo {year} {2011})\BibitemShut
  {NoStop}%
\bibitem [{\citenamefont {Scardovelli}\ and\ \citenamefont
  {Zaleski}(1999)}]{scardovelli1999DirectNumericalSimulation}%
  \BibitemOpen
  \bibfield  {author} {\bibinfo {author} {\bibfnamefont {Ruben}\ \bibnamefont
  {Scardovelli}}\ and\ \bibinfo {author} {\bibfnamefont {St{\'e}phane}\
  \bibnamefont {Zaleski}},\ }\bibfield  {title} {\enquote {\bibinfo {title}
  {Direct numerical simulation of free-surface and interfacial flow},}\ }\href
  {\doibase 10.1146/annurev.fluid.31.1.567} {\bibfield  {journal} {\bibinfo
  {journal} {Annual Review of Fluid Mechanics}\ }\textbf {\bibinfo {volume}
  {31}} (\bibinfo {year} {1999}),\ 10.1146/annurev.fluid.31.1.567}\BibitemShut
  {NoStop}%
\bibitem [{\citenamefont {Bogner}\ \emph {et~al.}(2015)\citenamefont {Bogner},
  \citenamefont {Ammer},\ and\ \citenamefont
  {R{\"u}de}}]{bogner2015BoundaryConditionsFree}%
  \BibitemOpen
  \bibfield  {author} {\bibinfo {author} {\bibfnamefont {Simon}\ \bibnamefont
  {Bogner}}, \bibinfo {author} {\bibfnamefont {Regina}\ \bibnamefont {Ammer}},
  \ and\ \bibinfo {author} {\bibfnamefont {Ulrich}\ \bibnamefont {R{\"u}de}},\
  }\bibfield  {title} {\enquote {\bibinfo {title} {Boundary conditions for free
  interfaces with the lattice {{Boltzmann}} method},}\ }\href {\doibase
  10.1016/j.jcp.2015.04.055} {\bibfield  {journal} {\bibinfo  {journal}
  {Journal of Computational Physics}\ }\textbf {\bibinfo {volume} {297}}
  (\bibinfo {year} {2015}),\ 10.1016/j.jcp.2015.04.055}\BibitemShut {NoStop}%
\bibitem [{\citenamefont {Thies}(2005)}]{thies2005LatticeBoltzmannModeling}%
  \BibitemOpen
  \bibfield  {author} {\bibinfo {author} {\bibfnamefont {Michael}\ \bibnamefont
  {Thies}},\ }\emph {\bibinfo {title} {Lattice {{Boltzmann Modeling}} with
  {{Free Surfaces Applied}} to {{Formation}} of {{Metal Foams}}}},\ \href
  {https://opus4.kobv.de/opus4-fau/frontdoor/index/index/docId/201} {Ph.D.
  thesis},\ \bibinfo  {school} {Universit\"at Erlangen-N\"urnberg}, \bibinfo
  {address} {{Erlangen}} (\bibinfo {year} {2005})\BibitemShut {NoStop}%
\bibitem [{\citenamefont {Bogner}\ \emph {et~al.}(2016)\citenamefont {Bogner},
  \citenamefont {R{\"u}de},\ and\ \citenamefont
  {Harting}}]{bogner2016CurvatureEstimationVolumeoffluid}%
  \BibitemOpen
  \bibfield  {author} {\bibinfo {author} {\bibfnamefont {Simon}\ \bibnamefont
  {Bogner}}, \bibinfo {author} {\bibfnamefont {Ulrich}\ \bibnamefont
  {R{\"u}de}}, \ and\ \bibinfo {author} {\bibfnamefont {Jens}\ \bibnamefont
  {Harting}},\ }\bibfield  {title} {\enquote {\bibinfo {title} {Curvature
  estimation from a volume-of-fluid indicator function for the simulation of
  surface tension and wetting with a free-surface lattice {{Boltzmann}}
  method},}\ }\href {\doibase 10.1103/PhysRevE.93.043302} {\bibfield  {journal}
  {\bibinfo  {journal} {Physical Review E}\ }\textbf {\bibinfo {volume} {93}}
  (\bibinfo {year} {2016}),\ 10.1103/PhysRevE.93.043302}\BibitemShut {NoStop}%
\bibitem [{\citenamefont {Parker}\ and\ \citenamefont
  {Youngs}(1992)}]{parker1992TwoThreeDimensional}%
  \BibitemOpen
  \bibfield  {author} {\bibinfo {author} {\bibfnamefont {B.~J.}\ \bibnamefont
  {Parker}}\ and\ \bibinfo {author} {\bibfnamefont {D.~L.}\ \bibnamefont
  {Youngs}},\ }\href@noop {} {\emph {\bibinfo {title} {Two and Three
  Dimensional {{Eulerian}} Simulation and Fluid Flow with Material
  Interfaces}}},\ \bibinfo {type} {Technical {{Report}} 01/92}\ (\bibinfo
  {institution} {{UK Atomic Weapons Establishment}},\ \bibinfo {year}
  {1992})\BibitemShut {NoStop}%
\bibitem [{\citenamefont {Anderl}\ \emph {et~al.}(2014)\citenamefont {Anderl},
  \citenamefont {Bogner}, \citenamefont {Rauh}, \citenamefont {R{\"u}de},\ and\
  \citenamefont {Delgado}}]{anderl2014FreeSurfaceLattice}%
  \BibitemOpen
  \bibfield  {author} {\bibinfo {author} {\bibfnamefont {Daniela}\ \bibnamefont
  {Anderl}}, \bibinfo {author} {\bibfnamefont {Simon}\ \bibnamefont {Bogner}},
  \bibinfo {author} {\bibfnamefont {Cornelia}\ \bibnamefont {Rauh}}, \bibinfo
  {author} {\bibfnamefont {Ulrich}\ \bibnamefont {R{\"u}de}}, \ and\ \bibinfo
  {author} {\bibfnamefont {Antonio}\ \bibnamefont {Delgado}},\ }\bibfield
  {title} {\enquote {\bibinfo {title} {Free surface lattice {{Boltzmann}} with
  enhanced bubble model},}\ }\href {\doibase 10.1016/j.camwa.2013.06.007}
  {\bibfield  {journal} {\bibinfo  {journal} {Computers \& Mathematics with
  Applications}\ }\bibinfo {series} {Mesoscopic {{Methods}} for {{Engineering}}
  and {{Science}} ({{Proceedings}} of {{ICMMES-2012}}, {{Taipei}}, {{Taiwan}},
  23\textendash 27 {{July}} 2012)},\ \textbf {\bibinfo {volume} {67}} (\bibinfo
  {year} {2014}),\ 10.1016/j.camwa.2013.06.007}\BibitemShut {NoStop}%
\bibitem [{\citenamefont
  {Dingemans}(1997)}]{dingemans1997WaterWavePropagation}%
  \BibitemOpen
  \bibfield  {author} {\bibinfo {author} {\bibfnamefont {Maarten~W}\
  \bibnamefont {Dingemans}},\ }\href {\doibase 10.1142/1241-part1} {\emph
  {\bibinfo {title} {Water {{Wave Propagation Over Uneven Bottoms}}: {{Part}}
  1}}},\ \bibinfo {series} {Advanced {{Series}} on {{Ocean Engineering}}},
  Vol.~\bibinfo {volume} {13}\ (\bibinfo  {publisher} {{World Scientific
  Publishing Company}},\ \bibinfo {year} {1997})\BibitemShut {NoStop}%
\bibitem [{\citenamefont {Lamb}(1975)}]{lamb1975Hydrodynamics}%
  \BibitemOpen
  \bibfield  {author} {\bibinfo {author} {\bibfnamefont {Horace}\ \bibnamefont
  {Lamb}},\ }\href@noop {} {\emph {\bibinfo {title} {Hydrodynamics}}},\
  \bibinfo {edition} {sixth}\ ed.\ (\bibinfo  {publisher} {{Cambridge
  University Press}},\ \bibinfo {year} {1975})\BibitemShut {NoStop}%
\bibitem [{\citenamefont {Sato}\ \emph {et~al.}(2022)\citenamefont {Sato},
  \citenamefont {Kawasaki},\ and\ \citenamefont
  {Koshimura}}]{sato2022ComparativeStudyCumulant}%
  \BibitemOpen
  \bibfield  {author} {\bibinfo {author} {\bibfnamefont {Kenta}\ \bibnamefont
  {Sato}}, \bibinfo {author} {\bibfnamefont {Koji}\ \bibnamefont {Kawasaki}}, \
  and\ \bibinfo {author} {\bibfnamefont {Shunichi}\ \bibnamefont {Koshimura}},\
  }\bibfield  {title} {\enquote {\bibinfo {title} {A comparative study of the
  cumulant lattice {{Boltzmann}} method in a single-phase free-surface model of
  violent flows},}\ }\href {\doibase 10.1016/j.compfluid.2021.105303}
  {\bibfield  {journal} {\bibinfo  {journal} {Computers \& Fluids}\ }\textbf
  {\bibinfo {volume} {236}} (\bibinfo {year} {2022}),\
  10.1016/j.compfluid.2021.105303}\BibitemShut {NoStop}%
\bibitem [{\citenamefont {Moraga}\ \emph {et~al.}(2015)\citenamefont {Moraga},
  \citenamefont {Lemus}, \citenamefont {Saavedra},\ and\ \citenamefont
  {{Lemus-Mondaca}}}]{moraga2015VOFFVMPrediction}%
  \BibitemOpen
  \bibfield  {author} {\bibinfo {author} {\bibfnamefont {Nelson~O.}\
  \bibnamefont {Moraga}}, \bibinfo {author} {\bibfnamefont {Luis~A.}\
  \bibnamefont {Lemus}}, \bibinfo {author} {\bibfnamefont {Mario~A.}\
  \bibnamefont {Saavedra}}, \ and\ \bibinfo {author} {\bibfnamefont
  {Roberto~A.}\ \bibnamefont {{Lemus-Mondaca}}},\ }\bibfield  {title} {\enquote
  {\bibinfo {title} {{{VOF}}/{{FVM}} prediction and experimental validation for
  shear-thinning fluid column collapse},}\ }\href {\doibase
  10.1016/j.camwa.2014.11.018} {\bibfield  {journal} {\bibinfo  {journal}
  {Computers \& Mathematics with Applications}\ }\textbf {\bibinfo {volume}
  {69}} (\bibinfo {year} {2015}),\ 10.1016/j.camwa.2014.11.018}\BibitemShut
  {NoStop}%
\bibitem [{\citenamefont {Martin}\ \emph {et~al.}(1952)\citenamefont {Martin},
  \citenamefont {Moyce}, \citenamefont {Penney}, \citenamefont {Price},\ and\
  \citenamefont {Thornhill}}]{martin1952PartIVExperimental}%
  \BibitemOpen
  \bibfield  {author} {\bibinfo {author} {\bibfnamefont {J.~C.}\ \bibnamefont
  {Martin}}, \bibinfo {author} {\bibfnamefont {W.~J.}\ \bibnamefont {Moyce}},
  \bibinfo {author} {\bibfnamefont {William~George}\ \bibnamefont {Penney}},
  \bibinfo {author} {\bibfnamefont {A.~T.}\ \bibnamefont {Price}}, \ and\
  \bibinfo {author} {\bibfnamefont {C.~K.}\ \bibnamefont {Thornhill}},\
  }\bibfield  {title} {\enquote {\bibinfo {title} {Part {{IV}}. {{An}}
  experimental study of the collapse of liquid columns on a rigid horizontal
  plane},}\ }\href {\doibase 10.1098/rsta.1952.0006} {\bibfield  {journal}
  {\bibinfo  {journal} {Philosophical Transactions of the Royal Society of
  London. Series A, Mathematical and Physical Sciences}\ }\textbf {\bibinfo
  {volume} {244}} (\bibinfo {year} {1952}),\
  10.1098/rsta.1952.0006}\BibitemShut {NoStop}%
\bibitem [{\citenamefont {Rumble}(2021)}]{rumble2021CRCHandbookChemistry}%
  \BibitemOpen
  \bibinfo {editor} {\bibfnamefont {John}\ \bibnamefont {Rumble}},\ ed.,\
  \href@noop {} {\emph {\bibinfo {title} {{{CRC Handbook}} of {{Chemistry}} and
  {{Physics}}}}},\ \bibinfo {edition} {one hundred second}\ ed.\ (\bibinfo
  {publisher} {{CRC Press}},\ \bibinfo {year} {2021})\BibitemShut {NoStop}%
\bibitem [{\citenamefont
  {Pavlidis}(1982)}]{pavlidis1982AlgorithmsGraphicsImage}%
  \BibitemOpen
  \bibfield  {author} {\bibinfo {author} {\bibfnamefont {Theo}\ \bibnamefont
  {Pavlidis}},\ }\href {\doibase 10.1007/978-3-642-93208-3} {\emph {\bibinfo
  {title} {Algorithms for {{Graphics}} and {{Image Processing}}}}}\ (\bibinfo
  {publisher} {{Springer Berlin Heidelberg}},\ \bibinfo {address} {{Berlin,
  Heidelberg}},\ \bibinfo {year} {1982})\BibitemShut {NoStop}%
\bibitem [{\citenamefont {Bugg}\ and\ \citenamefont
  {Saad}(2002)}]{bugg2002VelocityFieldTaylor}%
  \BibitemOpen
  \bibfield  {author} {\bibinfo {author} {\bibfnamefont {J.D.}\ \bibnamefont
  {Bugg}}\ and\ \bibinfo {author} {\bibfnamefont {G.A.}\ \bibnamefont {Saad}},\
  }\bibfield  {title} {\enquote {\bibinfo {title} {The velocity field around a
  {{Taylor}} bubble rising in a stagnant viscous fluid: Numerical and
  experimental results},}\ }\href {\doibase 10.1016/S0301-9322(02)00002-2}
  {\bibfield  {journal} {\bibinfo  {journal} {International Journal of
  Multiphase Flow}\ }\textbf {\bibinfo {volume} {28}} (\bibinfo {year}
  {2002}),\ 10.1016/S0301-9322(02)00002-2}\BibitemShut {NoStop}%
\bibitem [{\citenamefont {Wang}\ and\ \citenamefont
  {Chen}(2000)}]{wang2000SplashingImpactSingle}%
  \BibitemOpen
  \bibfield  {author} {\bibinfo {author} {\bibfnamefont {An-Bang}\ \bibnamefont
  {Wang}}\ and\ \bibinfo {author} {\bibfnamefont {Chi-Chang}\ \bibnamefont
  {Chen}},\ }\bibfield  {title} {\enquote {\bibinfo {title} {Splashing impact
  of a single drop onto very thin liquid films},}\ }\href {\doibase
  10.1063/1.1287511} {\bibfield  {journal} {\bibinfo  {journal} {Physics of
  Fluids}\ }\textbf {\bibinfo {volume} {12}} (\bibinfo {year} {2000}),\
  10.1063/1.1287511}\BibitemShut {NoStop}%
\bibitem [{\citenamefont
  {Batchelor}(2000)}]{batchelor2000IntroductionFluidDynamics}%
  \BibitemOpen
  \bibfield  {author} {\bibinfo {author} {\bibfnamefont {G.~K.}\ \bibnamefont
  {Batchelor}},\ }\href {\doibase 10.1017/CBO9780511800955} {\emph {\bibinfo
  {title} {An {{Introduction}} to {{Fluid Dynamics}}}}},\ Cambridge
  {{Mathematical Library}}\ (\bibinfo  {publisher} {{Cambridge University
  Press}},\ \bibinfo {address} {{Cambridge}},\ \bibinfo {year}
  {2000})\BibitemShut {NoStop}%
\end{thebibliography}%

\end{document}